\documentclass[11pt]{article}
\usepackage{graphics}
\usepackage{psfrag}
\usepackage{epsfig}
\usepackage{psfig}
\usepackage{epsf}
\usepackage{float}
\usepackage[sort&compress]{natbib}
\bibpunct{[}{]}{,}{n}{}{}
\usepackage{amssymb,stmaryrd,latexsym}
\newcommand{\fet}[1]{\mbox{\boldmath $#1$}}

\newcommand{\beq}{\begin{equation}}
\newcommand{\eeq}{\end{equation}}
\newcommand{\beqa}{\begin{eqnarray}}
\newcommand{\eeqa}{\end{eqnarray}}
\newcommand{\nn}{\nonumber \\ }

\newcommand{\vs}{\vspace{-0.2cm}}

\newcommand{\Mp}{M_\pi}

\newcommand{\Mpn}{M_{\pi^0}}
\newcommand{\Mpp}{M_{\pi^+}}
\newcommand{\Mpm}{M_{\pi^\pm}}

\newcommand{\Mppz}{M_{\pi^+}^2}

\newcommand{\3}{{\ss}}
\def\palka{\hspace{-7.5pt}/}

\begin{document}
\hfill {\tiny JLAB-THY-05-371}

%\vspace{1cm}
\begin{center}

{{\Large\bf Few--Nucleon Forces and Systems in \\[0.3em] 
Chiral Effective Field Theory
}}
%\footnote{To appear in Prog.~Part.~Nucl.~Phys.}

\end{center}

%\vspace{.3in}

\begin{center}

{\large 
Evgeny Epelbaum
\footnote{email: epelbaum@jlab.org}
}

\vskip 0.2 true cm

{\it Jefferson Laboratory, Theory Division, Newport News, VA 23606, USA}
\end{center}

%\vspace{.6in}

\thispagestyle{empty} 

\begin{abstract}

\vskip -0.2 true cm
\noindent 
We outline the structure of the nuclear force in the framework of 
chiral effective field theory of QCD and review recent applications 
to processes involving few nucleons.

\vskip 0.5 true cm
\noindent
\end{abstract}

%\vfill
%\pagebreak

\vskip 0.8 true cm
\addtolength{\parskip}{-0.5\baselineskip}
\tableofcontents
\addtolength{\parskip}{0.5\baselineskip}

%\pagebreak

\vskip 1.0 true cm
%%%%%%%%%%%%%%%%%%%%%%%%%%%%%%%%%%%%%%%%%%%%%%%%%%%%%%%%%%%%%%%%%%%%%%%%%%%%%%%%%
\section{Introduction}
\def\theequation{\arabic{section}.\arabic{equation}}
\setcounter{equation}{0}
\label{sec1a}

One of the basic problems in nuclear physics is determining the nature of the interactions 
between the nucleons which is crucial for understanding the properties of nuclei. 
The standard way to describe the nuclear force is based on the meson--exchange picture,
which goes back to the seminal work by Yukawa \citep{Yukawa:1935aa}. His idea as well as experimental discovery 
of $\pi$-- and heavier mesons ($\rho$, $\omega$, $\dots$) 
stimulated the development of boson--exchange models of the nuclear force,
which still provide a basis for many modern, highly sophisticated phenomenological 
nucleon--nucleon (NN) potentials.

According to our present understanding, the nuclear force is due to residual 
strong interactions between the color--charge neutral hadrons. A direct derivation
of the nuclear force from QCD, the underlying theory of strong interactions,  is not 
yet possible due to its nonperturbative nature at low--energy. In order to provide reliable input 
for few-- and many--body calculations, research on the NN interaction 
proceeded  along (semi-)phenomenological lines with the aim of achieving 
the best possible description of the low--energy NN data. In the case of two nucleons,
the potential can be decomposed in a few different spin--space structures, and 
the corresponding radial functions can be parameterized using an extensive set of data. 
Although the resulting models provide an excellent description of experimental data in many cases, 
there are certain major conceptual deficiencies that cannot be overcome. 
In particular, one important concern is related to the problem 
of the construction of \emph{consistent} many--body forces. These can 
only be meaningfully defined in a consistent scheme with a given two--nucleon (2N) 
interaction \citep{Gloeckle:1990aa}.
Notice that because of the large variety of different possible structures in the 
three--nucleon force (3NF), following the 
same phenomenological path as in the 2N system and parametrizing the most general structure 
of the 3NF seems not to be feasible without additional theoretical guidance.   
Clearly, the same problem of consistency arises in the context of reactions
with electroweak probes, whose description requires the knowledge of the corresponding 
consistent nuclear current operator. 
%*EE
%Further, one should keep in mind, that due to the lack of systematics 
%in the organization of the dominant dynamical contributions, it is not clear how to improve on the 
%achieved theoretical results. 
Further, one lacks within phenomenological treatments a means of systematically improving 
the theory of the nuclear force in terms of the dominant dynamical contributions. 
Finally, and most important, the phenomenological scheme 
provides only loose connection to QCD.

Chiral Effective Field Theory (EFT) has become a standard tool for analyzing the properties of 
hadronic systems at low energies in a systematic and model independent way. 
It is based upon the approximate and spontaneously broken chiral symmetry of QCD, 
which governs low--energy hadron structure and dynamics.
In addition, it provides a straightforward way to improve the results by going to higher 
orders in a perturbative expansion.
In the past two decades, this framework was successfully applied to a variety of low--energy 
reactions in the meson and single--baryon sectors. Fifteen years ago Weinberg \citep{Weinberg:1990rz,Weinberg:1991um}
proposed a generalization of this approach to the few--nucleon sector, where one has to deal 
with a nonperturbative problem. He demonstrated that the strong enhancement of the 
few--nucleon scattering amplitude arises from purely nucleonic intermediate states 
and suggested to apply EFT to the kernel of the corresponding scattering   
equation, which can be viewed as an effective nuclear potential. This idea has been explored 
in the last decade by many authors. In this work, we will review the current status 
of research along these lines, focusing on the description of few--nucleon systems. 
The manuscript is organized as follows. In section \ref{sec1b} we discuss 
effective field theories which are of relevance for the topics considered in this work.  
In particular, we give a brief account of chiral EFT  for the pion 
and single--nucleon sectors and then discuss how it can be generalized 
to the few--nucleon sector. In section \ref{sec4} we consider the structure of the 
nuclear force based on chiral EFT. Applications to systems with $2\ldots 7$ nucleons are presented 
in section \ref{sec5}. Section \ref{sec7} lists some related further topics which are
not covered in this work. A brief outlook is presented in section \ref{sec8}.  
Finally, appendix \ref{sec:SF} contains 
explicit expressions for certain contributions to the 2N force.

%%%%%%%%%%%%%%%%%%%%%%%%%%%%%%%%%%%%%%%%%%%%%%%%%%%%%%%%%%%%%%%%%%%%%%%%%%%%%%%%%
\section{Effective field theories in nuclear physics}
\def\theequation{\arabic{section}.\arabic{equation}}
\setcounter{equation}{0}
\label{sec1b}

\subsection{Chiral perturbation theory}
\label{sec1}

Chiral Perturbation Theory (CHPT) is the effective theory of QCD and, more generally,
of the Standard Model, which was formulated by Weinberg \citep{Weinberg:1978kz} 
and developed in to a systematic tool for analyzing low--energy Quantum Chromodynamics 
(QCD) by Gasser and Leutwyler \citep{Gasser:1983yg,Gasser:1984gg}. In this section,
we give a brief overview  of the foundations of this approach. 

Consider the QCD Lagrangian in the two--flavor case of the light up and down quarks 
\beq
\label{LQCD}
\mathcal{L}_{\rm QCD} = \bar q \, (  i \gamma_\mu D^\mu - \mathcal{M} ) q  -  \frac{1}{4}  
G^a_{\mu \nu} G^{a\, \mu \nu}\,,
\eeq
where $D_\mu = \partial_\mu - i g_s G_\mu^a T^a$ with $T^a$, (with $a = 1 \ldots 8$)
are the  SU(3)$_{\rm color}$ Gell--Mann matrices and $q$
the quark fields. Further, $G_{\mu \nu}^a$ are the 
gluon field strength tensors, and the quark mass matrix is given by $\mathcal{M} = \mbox{diag} (m_u, \, m_d)$.  
We do not show in Eq.~(\ref{LQCD}) the 
%*EE
$\theta$-- and gauge fixing terms 
which are not relevant for our consideration. The 
left-- and right--handed quark fields 
are defined by 
$q_{\rm L, R} = 1/2 (1 \pm \gamma_5 ) q$. The chiral group $G$ is a group of 
independent SU(2)$_{\rm flavor}$ transformations of the left-- and right--handed 
quark fields, $G=$ SU(2)$_{\rm L}$ $\times$ SU(2)$_{\rm R}$.  Expressing the quark part in the QCD Lagrangian 
(\ref{LQCD}) in terms of $q_{L,R}$, 
it is easy to see that the covariant derivative term is  invariant with respect 
to global chiral rotations, while the quark mass term is not. The running quark masses 
in the $\overline{\mbox{MS}}$ scheme at the renormalization scale $\mu = 1$ GeV are 
$m_u \sim 5$ MeV and $m_d \sim 9$ MeV \citep{Leutwyler:1996sa}. Given the fact that the masses 
of the up and down quarks are much smaller than the typical hadron scale of the order of 1~GeV, 
chiral SU(2)$_{\rm L}$ $\times$ SU(2)$_{\rm R}$ symmetry can be considered as a rather 
accurate symmetry of QCD. 

There is a strong evidence on 
both experimental and theoretical sides that  chiral symmetry of QCD is 
spontaneously broken down to its vector subgroup (isospin group in the two--flavor 
case). Perhaps, the most striking evidence of the spontaneous breaking of the axial generators
is provided by the nonexistence of degenerate parity doublets in the hadron spectrum 
and the presence of the triplet of unnaturally light pseudoscalar mesons (pions). The latter are natural candidates 
for the corresponding Nambu--Goldstone bosons which acquire a small nonzero mass due to
the explicit chiral symmetry breaking by the nonvanishing quark masses. Further,
on the theoretical side, recent (quenched) QCD determinations
of the vacuum expectation value of the scalar quark condensate $\Sigma = \langle 0 | \bar q q | 0 \rangle$, a natural  
order parameter of the spontaneous chiral symmetry breaking,
yields  the nonvanishing value \citep{Hasenfratz:2002rp}: 
\beq
\Sigma = - (262 \pm 12 \mbox{ MeV})^3\,.
\eeq
This value is based on the $\overline{\mbox{MS}}$ scheme at the renormalization scale $\mu = 2$ GeV. 
Also based  on rather general arguments, it has been shown  that the vector subgroup 
of the chiral group cannot be spontaneously broken \citep{Vafa:1983tf}. Further, in the three--flavor sector,
spontaneous chiral symmetry breaking is consistent with the so-called anomaly matching condition \citep{'tHooft:1980xb}.
For arguments based on the large $N_c$ limit the reader is referred to Ref.~\citep{Coleman:1980mx}.

The low--energy dynamics of QCD can be studied using the method 
of external sources \citep{Gasser:1983yg,Gasser:1984gg}. The idea is to couple quarks 
to external classical fields which 
formally allows to compute Green functions of the corresponding 
quark currents in a straightforward way. 
The extended QCD Lagrangian takes the form 
\beq
\mathcal{L}_{\rm QCD} = \mathcal{L}_{\rm QCD}^0 + \bar q \gamma_\mu \left( v^\mu + \gamma_5 a^\mu \right) q
- \bar q \left( s - i \gamma_5 p \right) q\,,
\eeq
where the external fields $v_\mu$, $a_\mu$, $s$ and $p$ 
are hermitian, color neutral, traceless $2 \times 2$ matrices in flavor space and $\mathcal{L}_{\rm QCD}^0$
refers to the Lagrangian in Eq.~(\ref{LQCD}) in the absence of the quark mass term.
Notice that one can also include an external vector singlet field which becomes particularly 
useful for studying  electromagnetic processes. 
The transformation properties of external sources follow from the requirement that the extended
QCD Lagrangian is invariant under \emph{local} chiral rotations.
The ordinary QCD Lagrangian is recovered by setting 
$v_\mu = a_\mu = p = 0$, $s = \mbox{diag} (m_u, \, m_d)$. 
The QCD Green functions built from the associated quark currents can be
derived  by taking 
functional derivatives of the generating functional $\Gamma ( v, a, s, p)$ defined as 
\beq
e^{i \Gamma [ v, a, s, p ]} = \int [D q] [D \bar q ] [dG_\mu ]  e^{\int i \, d^4 x 
\mathcal{L}_{\rm QCD}^0 (q, \bar q, G_\mu ; \, v, a, s, p  )}
\eeq
with respect to the sources. 
It is not presently possible to evaluate the Green functions in a closed form. 
At low energy, however,
 one can calculate the generating functional within effective 
field theory formulated in terms of the observed asymptotic states. As proven by 
Leutwyler \citep{Leutwyler:1993iq},
the gauge--invariant generating functional can be represented by a path 
integral constructed with a gauge--invariant effective Lagrangian for the Goldstone bosons
$\mathcal{L}_{\rm eff} (U ; \, v, a, s, p  )$:
 \beq
e^{i \Gamma [ v, a, s, p ]} = \int [d U ] e^{\int i \, d^4 x 
\mathcal{L}_{\rm eff} (U; \, v, a, s, p  )}\,.
\eeq
Here, the $2 \times 2$ unitary matrices  $U$, which satisfy $\det \, U = 1$, collect the triplet of 
pseudoscalar Goldstone bosons. 
The Green functions can be evaluated from the effective Lagrangian $\mathcal{L}_{\rm eff}$ 
in a systematic way by expanding in powers of the external momenta $q$  and the quark 
mass matrix $\mathcal{M}$ and keeping the ratio $\mathcal{M} / q^2$ fixed. 
This procedure to evaluate $S$--matrix elements is called chiral perturbation theory and can, in principle, 
be carried out to arbitrarily high orders in the low--energy expansion. 
Since, at present, one cannot derive $\mathcal{L}_{\rm eff}$ from QCD 
directly, one  writes down its \emph{most general} form including all terms 
consistent with the required symmetry principles. This can be done following the lines 
of \citep{Coleman:1969sm,Callan:1969sn}. The effective Lagrangian takes the form 
\beq
\mathcal{L}_{\rm eff} = \mathcal{L}_\pi^{(2)} + \mathcal{L}_\pi^{(4)} + \mathcal{L}_\pi^{(6)} + \ldots\,,
\eeq
where the superscripts $d = d_\partial + 2 d_{\mathcal{M}}$  refer to the number of derivatives $d_\partial$ and/or 
quark mass matrices $d_{\mathcal{M}}$. To be specific, let us parametrize the matrix $U$ as
\beq 
\label{Udef}
U(x) = \exp \left[ i \frac{ \fet \tau \cdot \fet \pi (x)}{F} \right]\,,
\eeq
where $\tau_i$ are the Pauli isospin matrices, $\pi_i$ are pion fields and $F$ is a constant. 
Notice that one also often uses the so-called sigma--representation 
$U (x) = \sqrt{1 - \fet \pi^2 (x) / F^2} + i \, \fet \tau \cdot \fet \pi (x) / F$.
The matrix $U$ is required to transform under local chiral rotations as 
\beq 
\label{Utransf}
U \stackrel{G}{\to} R U L^\dagger\,.
\eeq
Here, the $2 \times 2$ matrices $L$ and $R$ represent local SU(2)$_{\rm L}$ and SU(2)$_{\rm R}$ 
rotations: $L(x) = \exp [ - i \fet \tau \cdot \fet \theta_L (x)/2]$,
$R(x) = \exp [ - i \fet \tau \cdot \fet \theta_R (x)/2]$. One can show from Eqs.~(\ref{Utransf}), (\ref{Udef})
that pions belong to a nonlinear realization of the chiral group \citep{Coleman:1969sm} and transform 
linearly under its vector (isospin) subgroup given by a subset of rotations with $\fet \theta_L = \fet \theta_R$.
Different realizations of the chiral group turn out to be equivalent to each other by means of nonlinear 
field redefinitions \citep{Coleman:1969sm}. It is convenient to define the quantity $u$
\beq
\label{defh}
u^2 = U \,, \quad \quad u \stackrel{G}{\to} R u  h^\dagger \,,
\eeq
where the $2 \times 2$ matrix $h(x)$ depends on $L$, $R$ and $U$. More precisely, it is given by
$h = \sqrt{L U^\dagger R^\dagger} R \sqrt{U}$.
The effective Lagrangian is constructed out of the following building blocks, see e.g.~\citep{Gasser:1999fb}:
\beqa
\label{build_blocks}
u_\mu &=& i u^\dagger D_\mu U u^\dagger = - i u D_\mu U^\dagger u = u_\mu^\dagger\,,\nn
\chi_\pm &=& u^\dagger \chi u^\dagger \pm u \chi^\dagger u \,, \nn
\chi_-^\mu &=& u^\dagger D^\mu \chi u^\dagger - u D^\mu \chi^\dagger u\,, \nn
f_\pm^{\mu \nu} &=&  u^\dagger  F_R^{\mu \nu} u \pm u F_L^{\mu \nu} u^\dagger \,.
\eeqa
Here, $\chi = 2 B (s + p)$ with $B$ being a constant, $F_{I}^{\mu \nu} =  \partial^\mu F^\nu_I -   \partial^\nu F^\mu_I 
- i [F_I^\mu , \; F_I^\nu ]$ with $I = L, R$ is the field strength tensor associated with $F_R^\mu = v^\mu + a^\mu$,
$F_L^\mu = v^\mu - a^\mu$, and the covariant derivative $D_\mu X$ is defined via
\beq
D_\mu X = \partial_\mu X - i ( v_\mu + a_\mu ) X + i X (v_\mu - a _\mu)\,.
\eeq
All quantities in Eq.~(\ref{build_blocks}) transform covariantly under $G$, i.~e. $I \stackrel{G}{\to} h I h^\dagger$.
Chiral invariant terms in the Lagrangian can therefore be easily constructed via building the traces 
(denoted in the following by $\langle \dots \rangle$) of the products of these objects.
The leading and subleading Lagrangians take the form \citep{Gasser:1983yg,Gasser:1999fb}:
\beqa
\label{Leff}
\mathcal{L}_\pi^{(2)} &=& \frac{F^2}{4} \langle u_\mu u^\mu + \chi_+ \rangle \,, \nn
\mathcal{L}_\pi^{(4)} &=& \frac{l_1}{4} \langle u^\mu u_\mu \rangle^2 + 
\frac{l_2}{4} \langle u_\mu u_\nu \rangle  \langle u^\mu u^\nu \rangle +
\frac{l_3}{16} \langle \chi_+ \rangle^2 +
i \frac{l_4}{4} \langle u_\mu \chi_-^\mu \rangle -
\frac{l_5}{2} \langle f_-^{\mu \nu}  f_{- \,\mu \nu}  \rangle 
\nn
&& {} + 
i \frac{l_6}{4} \langle f_+^{\mu \nu} [ u_\mu, \, u_\nu ] \rangle -
\frac{l_7}{16} \langle \chi_- \rangle^2\,,
\eeqa
where $l_{1, \ldots , 7}$ are low--energy constants (LECs). Further, the constant $F$ can be 
identified with the pion decay constant in the chiral limit while the constant $B$ is related 
to the scalar quark condensate via $\langle 0 | \bar u u | 0 \rangle = \langle 0 | \bar d d | 0 \rangle =
- B F^2 + \mathcal{O} (\mathcal{M} )$. Notice that we are following here the standard CHPT scenario with 
$2 B (m_u + m_d ) / M_\pi^2 \sim 1$. The generalized CHPT scenario \citep{Fuchs:1991cq}, in which 
$2 B (m_u + m_d ) / M_\pi^2 \ll 1$, is ruled out by the recent determination of the S--wave, isospin--zero   
$\pi \pi$ scattering length $a_0^0$ from the kaon decay $K \to \pi \pi e \nu$ \citep{Pislak:2001bf,Pislak:2003sv}.
We stress that there are further terms in $\mathcal{L}_\pi^{(4)}$ which do not contain Goldstone boson fields 
and are therefore not directly measurable. In addition, one has also to account for the chiral anomaly 
which can be done along the lines of Ref.~\citep{Wess:1971yu}, see also \citep{Witten:1983tw}. 
The effective Lagrangian in Eqs.~(\ref{Leff}) can be used 
to describe the interaction between pions among themselves  and
between pions and  external fields in the low--energy regime. 
For a reaction involving $N_\pi$ external pions, the transition amplitude $M$ is related to the $S$--matrix
via $S = \delta^4 (p_1 + p_2 + \ldots + p_{N_\pi} ) M$. The low--momentum dimension of $M$, i.e.~the 
power of a soft scale $Q$ associated with the pion mass or external momenta, is given by \citep{Weinberg:1978kz}
\beq
\nu = 2 + 2 L + \sum_i V_i^\pi (d_i - 2 )\,,
\eeq
where $L$ ($V_i^\pi$) is the number of loops (vertices of type $i$). 
The chiral dimension $d_i$ is given by the number of derivatives and/or quark mass insertions.  
Diagrams with loops and/or vertices with more derivatives and pion mass insertions are therefore 
suppressed by powers of $Q/\Lambda_\chi$ with $\Lambda_\chi$ being a hard scale, which is sometimes referred to 
as the chiral symmetry breaking scale. This scale sets the (maximal) range of convergence of the chiral expansion. 
The appearance in the spectrum of the $\rho$, the first meson of the non--Goldstone type, suggests 
$\Lambda_\chi \sim M_\rho \sim 770$ MeV. Another estimate 
based on consistency arguments \citep{Manohar:1983md} leads to 
$\Lambda_\chi \sim 4 \pi F_\pi$ with $F_\pi = 92.4$ MeV being the pion decay constant.  
The leading term in the low--energy expansion of the scattering amplitude results by evaluating 
tree diagrams with $\mathcal{L}_\pi^{(2)}$. The first corrections arise from tree graphs 
with exactly one insertion from $\mathcal{L}_\pi^{(4)}$ and one--loop diagrams with all vertices from $\mathcal{L}_\pi^{(2)}$.
They are suppressed by two powers of momenta or one power of the quark masses compared to the leading terms.
Notice that all ultraviolet divergences in loop diagrams are cancelled by counterterms from $\mathcal{L}_\pi^{(4)}$. 
The divergent parts of the LECs $l_i$ have been worked out in \citep{Gasser:1983yg} using the heat--kernel method.
The finite parts of the $l_i$'s are not fixed by chiral symmetry and have to be determined from measured data.
This then allows one to make predictions for other observables. At next--to--next--to--leading order,
one must include tree diagrams with one insertion from $\mathcal{L}_\pi^{(6)}$ (and all remaining vertices from 
$\mathcal{L}_\pi^{(2)}$) or two insertions from $\mathcal{L}_\pi^{(4)}$, as well as one--loop graphs with a single 
insertion from $\mathcal{L}_\pi^{(4)}$ and two--loop graphs with all vertices from $\mathcal{L}_\pi^{(2)}$. The Lagrangian 
$\mathcal{L}_\pi^{(6)}$ contains 53 independent terms plus 4 terms depending only on external sources 
and 5 terms (in the absence of a singlet external vector current) of odd intrinsic parity 
\citep{Bijnens:1999sh,Bijnens:1999hw,Ebertshauser:2001nj,Bijnens:2001bb}. The renormalization  
at this order is carried out in \citep{Bijnens:1999hw}. At present, several 
two--loop calculations (i.e.~at order $p^6$) have already been performed, see 
e.g.~\citep{Ananthanarayan:2000ht,Bijnens:2003jw} for some examples. One of the most impressive theoretical predictions is given by 
the precision calculation of the isoscalar S--wave $\pi \pi$ scattering length $a_0^0$, a fundamental 
quantity that measures explicit breaking of chiral symmetry. The results 
of the two--loop analysis \citep{Knecht:1995tr,Bijnens:1995yn,Bijnens:1997vq} combined with dispersion relations in the 
form of the Roy equations 
\citep{Roy:1971tc} allowed for an accurate prediction: $a_0^0 = 0.220 \pm 0.005$ \citep{Colangelo:2000jc}. 
To compare, the leading--order calculation by Weinberg yielded $a_0^0 = 0.16$  \citep{Weinberg:1966kf} while 
the next--to--leading value obtained by Gasser and Leutwyler is $a_0^0 = 0.20$ \citep{Gasser:1983yg}.
The results of the E865 experiment at Brookhaven beautifully confirmed the prediction of the two--loop analysis of  
Ref.~\citep{Colangelo:2000jc} 
yielding the value $a_0^0 = 0.216 \pm 0.013\mbox{ (stat)} \pm0.002\mbox{ (syst)} \pm 0.002\mbox{ (theor)}$ 
\citep{Pislak:2001bf,Pislak:2003sv}.

It is clear that  CHPT can, in principle, be carried out to an arbitrarily high order in the low--energy expansion. 
The predictive power is, however, limited due to the rapid increase of the number of new LECs.\footnote{Clearly, 
not all LECs contribute to a particular process/observable, so that there is usually no need to determine all LECs at a given order.}
It is, therefore, particularly important to be able to estimate the values of the LECs. 
One possible way is to assume that the LECs are saturated by low--lying resonances such as 
the triplet of  $\rho$--mesons. The form of their coupling to Goldsone 
bosons is dictated by chiral symmetry and can be parametrized 
in terms of a few parameters. 
At low energy, the resonance fields can be integrated out which gives rise to a series of terms
in the effective Lagrangian whose strength is given in terms of resonance couplings and masses and can be used 
as an estimation for the corresponding LECs. For more details on resonance saturation the reader is referred to 
\citep{Ecker:1988te,Ecker:1989yg,Donoghue:1988ed}, see also \citep{Fuchs:2003sh,Djukanovic:2004mm,Bruns:2004tj,Cirigliano:2004ue} 
for recent works on meson resonances in the chiral EFT framework. We also emphasize that the LECs in the effective 
Lagrangian are, in principle, calculable in QCD, see 
\citep{Heitger:2000ay,Irving:2001vy,Nelson:2001yq,Fleming:2002ms,Farchioni:2003bx} for some recent attempts using 
the framework of  lattice gauge theory. 

So far we have only discussed interactions between Goldstone bosons and external fields.
We will now consider the extension of CHPT to the single--nucleon sector. 
%*EE
%For that one first needs to enlarge the effective Lagrangian 
We enlarge the effective Lagrangian 
\beq
\mathcal{L}_{\rm eff} = \mathcal{L}_{\pi} + \mathcal{L}_{\pi N}\,,
\eeq
to include terms which couple mesons to nucleons. 
It it is convenient to introduce the
nucleon field $N$ in the isospin--doublet representation which 
transforms under the chiral SU(2)$_{\rm L}$ $\times$ SU(2)$_{\rm R}$ group as
\beq
\label{Ntransf}
N \stackrel{G}{\to} h (L,R,U) \, N\,,
\eeq
with the matrix $h$ being defined according to Eq.~(\ref{defh}). The above equation together with Eq.~(\ref{Utransf})
specifies the nonlinear realization of the chiral group in terms of pions and nucleons. 
Notice that for vector--like transformations with 
$\fet \theta_V \equiv \fet \theta_L = \fet \theta_R$, the matrix $h$ does not depend on $U$ and reduces to the usual 
isospin transformation matrix $h = \exp [ - i \fet \tau \cdot \fet \theta_V /2]$.  Eq.~(\ref{Ntransf}) is, 
therefore, consistent with the transformation properties of the nucleon field under isospin group. 
We stress that there is no loss of generality in the requirement for the nucleon field to transform 
under $G$ according to Eq.~(\ref{Ntransf}). Different realizations of the chiral group 
can be reduced to the one specified in Eqs.~(\ref{Utransf}), (\ref{Ntransf}) by means of field redefinitions 
\citep{Coleman:1969sm,Callan:1969sn}. The covariant derivative of the nucleon field is given by
\beq
D_\mu N = \partial_\mu N + \Gamma_\mu N\,, \mbox{\hskip 1 true cm}
\Gamma_\mu = \frac{1}{2} \, [ u^\dagger , \; \partial_\mu u ] - \frac{i}{2} u^\dagger  (v_\mu + a_\mu ) u
- \frac{i}{2} u (v_\mu - a _\mu ) u^\dagger \,.
\eeq
The so-called chiral connection $\Gamma_\mu$ ensures that $D_\mu$ transforms covariantly under $G$, i.e.: 
$D_\mu \stackrel{G}{\to} h D_\mu h^\dagger$. To construct the effective Lagrangian $\mathcal{L}_{\pi N}$ 
one simply combines $D_\mu$ and the building blocks in Eq.~(\ref{build_blocks}), which transform covariantly
under $G$, with the appropriate nucleon bilinears. To first order in
the  derivatives, the most general pion--nucleon 
Lagrangian takes the form \citep{Gasser:1987rb}
\beq
\label{LeffN}
\mathcal{L}_{\pi N}^{(1)} = \bar N \left( i \gamma^\mu D_\mu - m  + \frac{g_A}{2} \gamma^\mu \gamma_5 u_\mu \right) N\,,
\eeq
where $m$ and $g_A$ are the bare nucleon mass and the axial--vector coupling constant. 
Further, the superscript of $\mathcal{L}_{\pi N}$ denotes the power of the soft scale $Q$ related to a generic nucleon tree--momentum,
pion four--momentum or pion mass. Notice that contrary to the pion mass, $m$ does not vanish in the chiral 
limit and introduces an additional large scale. Consequently, terms proportional to $D_0$ and $m$ in Eq.~(\ref{LeffN})
are individually large. It can, however, be shown that $(i \gamma^\mu D_\mu - m )N \sim \mathcal{O} (Q )$ \citep{Krause:1990xc}. 
The presence of the additional hard scale associated with the nucleon mass makes the power counting 
significantly more complicated since the contributions from loops are not automatically suppressed. 
To see this consider the correction to the nucleon mass $m_N$ due to the pion loop which in the chiral limit
takes the form \citep{Gasser:1987rb}
\beq
\label{mNrelativ}
( m_N  - m )_{\rm loop}^{\rm rel} \stackrel{\mathcal{M} \to 0}{=}
 - \frac{3 g_A^2 m^3}{F^2} \left( L + \frac{1}{32 \pi^2} \ln \frac{m^2}{\mu^2} \right) + \mathcal{O} (d-4 )\,,
\eeq
where $\mu$ is the mass scale introduced by dimensional regularization (DR), $d$ is the number of dimensions and 
the quantity $L$ is given by
\beq
L = \frac{\mu ^{d-4}}{16 \pi^2} \left\{ \frac{1}{d-4} - \frac{1}{2} (\ln (4 \pi ) + \Gamma ' (1) + 1 ) \right\}\,, 
\mbox{\hskip 1 true cm} 
\Gamma ' (1) = -0.577215\ldots \,.
\eeq
The result in Eq.~(\ref{mNrelativ}) shows that $m_N$ receives an (infinite) contribution 
which is formally of the order $\sim m \, (m/4 \pi F )^2$ and is not suppressed compared to $m$. 
In addition, the parameter $m$ in the lowest--order Lagrangian $\mathcal{L}_{\pi N}^{(1)}$
needs to be renormalized. These features of the relativistic $\pi N$ EFT should be contrasted 
with the purely mesonic sector where loop contributions are always suppressed by powers of the soft 
scale and the parameters $F$ and $B$ in the lowest--order Lagrangian $L_\pi^{(2)}$ remain unchanged 
by higher--order corrections (if dimensional regularization is applied).\footnote{This statement applies for dimensionally regularized  
loop integrals.} This problem with the power counting in the baryonic sector can be dealt with using 
the heavy--baryon formalism \citep{Jenkins:1990jv,Bernard:1992qa} which is closely related to the 
nonrelativistic expansion due to Foldy and Wouthuysen \citep{Foldy:1950aa}. The idea is to decompose 
the nucleon four--momentum $p^\mu$ according to
\beq
\label{HBmomentum}
p_\mu = m v_\mu + k_\mu \,,
\eeq
with $v_\mu$ the four--velocity of the nucleon satisfying $v^2 = 1$ and $k_\mu$ its small residual momentum,
$v \cdot k \ll m$. One can now decompose the nucleon field $N$ in to the velocity eigenstates 
\beq
N_v = e^{i m v \cdot x} P_v^+ N\,, \mbox{\hskip 1.5 true cm}
h_v = e^{i m v \cdot x} P_v^- N\,,
\eeq
where $P_v^\pm = (1 \pm \gamma_\mu v^\mu )/2$ denote the corresponding projection operators.
Notice that for 
the particular choice $v_\mu = (1, 0, 0, 0 )$, the quantities $N_v$ and $h_v$ coincide with the 
usual large and small components of the
free positive--energy fields (modulo the modified time dependence), 
see e.g.~\citep{Itzykson:1980rh}. One, therefore, usually refers to $N_v$ and $h_v$ as to the large and small 
components of $N$. The relativistic Lagrangian $\mathcal{L}_{\pi N}^{(1)}$ in Eq.~(\ref{LeffN}) 
can be expressed in terms of $N_v$ and $h_v$ as:
\beq
\mathcal{L}_{\pi N}^{(1)} = \bar N_v  \mathcal{A} N_v  + \bar h_v \mathcal{B} N_v + \bar N_v  \gamma_0 \mathcal{B}^\dagger 
\gamma_0 h_v - \bar h_v \mathcal{C} h_v \,,
\eeq
where 
\beq
\mathcal{A} = i (v \cdot D ) + g_A (S \cdot u )\,,\quad 
\mathcal{B} = - \gamma_5 \left[ 2 i (S \cdot D) + \frac{g_A}{2} (v \cdot u ) \right]\,, \quad
\mathcal{C} = 2 m + i (v \cdot D ) + g_A (S \cdot u )\,.
\eeq
Here $S_\mu = i \gamma_5 \sigma_{\mu \nu} v^\nu$ refers to the nucleon spin operator. 
One can now use the equations of motion for the large and small component fields to 
completely eliminate $h_v$ from the Lagrangian. Utilizing the more elegant path integral formulation \citep{Mannel:1991mc}, 
the heavy degrees of freedom can be integrated out performing the Gaussian integration over the 
(appropriately shifted) variables $h_v$, $\bar h_v$. This leads to the effective Lagrangian of the form \citep{Bernard:1992qa}
\beq
\label{Lfin}
\mathcal{L}_{\pi N}^{\rm eff} = \bar N_v  \left[ \mathcal{A} + (\gamma_0 \mathcal{B}^\dagger \gamma_0 ) 
\mathcal{C}^{-1} \mathcal{B} \right] N_v
= \bar N_v  \left[ i (v \cdot D ) + g_A (S \cdot u ) \right] N_v  + \mathcal{O} \left(\frac{1}{m} \right)\,.
\eeq
Notice that the (large) nucleon mass term disappeared from the Lagrangian, 
and the dependence on $m$ in $\mathcal{L}_{\pi N}^{\rm eff}$
resides entirely in new vertices which can be classified according to their powers of $1/m$. 
Clearly, the formalism outlined above can be extended to the relativistic pion--nucleon Lagrangian beyond 
the leading order in derivatives/quark masses. The resulting heavy--baryon Lagrangian can be expressed as
\beq
\mathcal{L}_{\pi N} =  \mathcal{L}_{\pi N}^{(1)} +   \mathcal{L}_{\pi N}^{(2)}  +  \mathcal{L}_{\pi N}^{(3)}  + \ldots\,,
\eeq
where $\mathcal{L}_{\pi N}^{(1)}$ is given by
the terms in the right--hand side of Eq.~(\ref{Lfin}) and the superscripts 
refer to the power of the soft scale $Q$. Higher--order terms in the Lagrangian 
will be discussed in section \ref{sec4}. We stress that in the single--nucleon sector, relativistic corrections 
are usually treated on the same footing as the corresponding chiral corrections, i.e.~one counts $1/m \sim 1/\Lambda_\chi$.
Notice further that some of the $1/m$--terms in the heavy--baryon Lagrangian are protected from extra counterterm 
contributions as a consequence of the so--called reparametrization invariance associated with 
the freedom in parametrizing the nucleon momentum $p_\mu$. It relies on the fact that 
the same physics should be described using $p_\mu = m v_\mu ' + k_\mu '$, $v '\,^2 = 1$, instead of Eq.~(\ref{HBmomentum})
\citep{Luke:1992cs,Chen:1993sx,Finkemeier:1997re}.

The advantage of the heavy--baryon formulation of CHPT (HBCHPT) compared to
the relativistic one can be illustrated using the example of
the leading one--loop correction to the nucleon mass 
\beq
\label{mNHB}
( m_N  - m )^{\rm HB} = -4 c_1 M_\pi^2 - \frac{3 g_A^2 M_\pi^3}{32 \pi F^2} \,,
\eeq
where the counterterm contribution $\propto c_1$ stems from $\mathcal{L}_{\pi N}^{(2)}$. 
Contrary to the relativistic CHPT result in Eq.~(\ref{mNrelativ}), the loop correction in HBCHPT 
is finite (in DR) and vanishes in the chiral limit. The parameters in the lowest--order 
Lagrangian do not get modified due to higher--order corrections which are suppressed by powers of $Q/\Lambda_\chi$. 
Notice further that the second term in Eq.~(\ref{mNHB}) represents the leading contributions nonanalytic in quark masses
and agrees with the
relativistic CHPT result \citep{Gasser:1987rb}, see \citep{Gasser:1980sb} for an 
earlier determination of this 
correction. In general, the power $\nu$ of a soft scale $Q$ for the scattering amplitude in the single--nucleon sector 
HBCHPT is given by
\beq
\nu = 1 + 2 L + \sum_i V_i^\pi (d_i - 2 ) + \sum_i V_i^{\pi N} (d_i - 1 )  \,,
\eeq
where $V_i^{\pi N}$ is the number of vertices from $\mathcal{L}_{\pi N}$ with the chiral dimension $d_i$. 
Notice that no closed fermion loops appear in the heavy--baryon approach, so that exactly one nucleon line 
connecting the initial and final states runs through all diagrams in the single--baryon sector.

While most of the calculations in the single--nucleon sector have so far been performed in HBCHPT,
it was realized a few years ago that its range of convergence is rather limited in some 
kinematical regions. The problem can be traced back to the fact that certain analytical properties 
of the relativistic amplitude are destroyed in the heavy--baryon approach, see e.g.~\citep{Becher:1999he}. 
This can be avoided using a manifestly Lorentz invariant formulation. Various methods like e.g.~the infrared 
regularized CHPT have been developed which allow one to stay covariant and, at the same time, 
to preserve a consistent power counting \citep{Tang:1996ca,Ellis:1997kc,Gegelia:1999qt,Fuchs:2003qc,Becher:1999he}. 
In the formulation of \citep{Becher:1999he}, this is achieved by keeping the infrared 
singular contributions of the loop integrals and simultaneously discarding 
the polynomial terms that are responsible for
the breakdown of the power counting and which can be absorbed by local counter terms.
More details on the foundations and the applications of CHPT in the meson and single--nucleon sectors  
can be found in the review articles \citep{Bijnens:1993xi,Bernard:1995dp,Meissner:1993ah,Pich:1995bw,Ecker:1994gg}, 
lecture notes \citep{Leutwyler:1994fi,Manohar:1996faa} and recent conference proceedings \citep{CD2000,CD2003}.
A pedagogical introduction is given in \citep{Scherer:2002tk}.

\subsection{EFT for nucleons at very low energy}
\label{sec2}

So far we have only dealt with the low--energy processes in the mesonic and 
single--baryon sectors. Perturbation theory works well in these cases due to 
the fact that Goldstone bosons do not interact at vanishingly low energies in 
the chiral limit. In the few--nucleons sector one has to deal with a nonperturbative problem. 
Indeed, given the fact that there are shallow few--nucleon bound states, perturbation theory is 
expected to fail already at low energy. To understand how this difficulty can 
be handled in the EFT framework it is instructive to look at the two--nucleon system 
in the kinematical regime where $Q \ll M_\pi$ \citep{Lutz:1996aa,vanKolck:1998bw,Kaplan:1998tg,Kaplan:1998we,Gegelia:1998xr}. 
Then, no pions need to be taken into account explicitly, 
and the only relevant degrees of freedom are the nucleons themselves. The corresponding 
EFT is usually referred to as  pionless EFT. The most general effective Lagrangian 
consistent with Galilean invariance, baryon number conservation and the isospin symmetry 
takes in the absence of external sources the following form:
\beq
\label{Lagr_nopi}
\mathcal{L} = N^\dagger \left( i \partial_0 + \frac{\vec \nabla ^2}{2 m} \right) N
- \frac{1}{2} C_S \, ( N^\dagger N ) ( N^\dagger N )  - \frac{1}{2}  C_T \, 
( N^\dagger \vec \sigma N ) ( N^\dagger \vec \sigma N ) + \ldots\,,
\eeq
where $C_{S,T}$ are LECs and ellipses denote operators with derivatives. 
Isospin--breaking and relativistic corrections to Eq.~(\ref{Lagr_nopi}) can be included perturbatively
\citep{Chen:1999tn}.  Notice further that in certain cases it turns out to be convenient  
to introduce, in addition to the nucleon field, the auxiliary ``dimeron'' fields with the quantum numbers 
of the two--nucleon system \citep{Kaplan:1996nv}.\footnote{The auxiliary dimeron fields can be integrated out 
which leads to a completely equivalent form of the EFT with only nucleonic degrees of freedom.}  
Let us consider NN scattering in the $^1S_0$ channel. The $S$--matrix can be written as
\beq
\label{Tnorm}
S = e^{2 i \delta} = 1 - i \left( \frac{k m}{2 \pi} \right) T\,,
\eeq 
where $k$ is the magnitude of the nucleon momentum in the center--of--mass system (CMS) and $\delta$ ($T$) is the phase shift ($T$--matrix).
Utilizing the effective range expansion (ERE) for $(k \cot \delta )$, the $T$--matrix can be expressed as 
\beq
T =  - \frac{4 \pi}{m} \, \frac{1}{k \cot \delta - i k} = - \frac{4 \pi}{m} \, \frac{1}{\left(- \frac{1}{a} + \frac{1}{2} r_0 k^2 
+ v_2 k^4 + v_3 k^6 + \ldots \right)  - i k}\,,
\eeq
where $a$, $r_0$ and $v_i$ are the scattering length, effective range and shape parameters, respectively. 
While the effective range is bounded from above by the range $R$ of the nuclear potential, 
the scattering length can take any value. In particular, it diverges in the presence of a bound state at threshold. 
It is then useful to distinguish between a natural case with $|a | \sim R$ and an unnatural case with 
$|a | \gg R$, where the range of the nuclear potential is of the order 
 $M_\pi^{-1}$. In the natural case,
the $T$--matrix can be expanded in powers of $k$ as:
\beq
\label{Tnatural}
T = T_0 + T_1 + T_2 + \ldots  = \frac{4 \pi a}{m} \left[ 1 - i a k + \left(\frac{a r_0}{2} - a^2 \right) k^2 + \ldots \right]  \,,
\eeq
A natural value of the scattering length implies that there are no bound states close to threshold. 
The $T$--matrix can then be evaluated perturbatively in the EFT 
provided one uses a regularization and subtraction scheme that does not introduce an additional large scale. 
A convenient choice is given by DR with the minimal or the power divergence subtraction (PDS) 
\citep{Kaplan:1998tg,Kaplan:1998we}
or momentum subtraction at $k^2 = -\mu^2$ \citep{Gegelia:1998xr}. In the PDS scheme, the power law divergences, 
which are normally discarded in DR, are explicitly accounted for by subtracting from dimensionally regulated loop 
integrals not only $1/(d-4)$--poles but also e.g.~$1/(d-3)$--poles. The typical loop integral takes then the 
form \citep{Kaplan:1998tg,Kaplan:1998we}:
\beq
\left( \frac{\mu}{2} \right)^{4-d} \int \frac{d^{d-1}  q}{( 2 \pi )^{d-1}} \;
\frac{m \, q ^{2 n} }{p  ^2 - q  ^2 + i \epsilon}  
\stackrel{d\to 4}{\longrightarrow}
- \frac{m}{4 \pi} \, p \, ^{2n} (\mu + i p )\,,
\eeq
where $p \equiv | \vec p \, |$, $q \equiv | \vec q \, |$. The choice  $\mu = 0$ leads to the result of 
the minimal subtraction scheme (MS). Taking $\mu \sim k \ll M_\pi$, the leading and subleading terms $T_0$ 
and $T_1$ are given by the tree-- and one--loop graphs constructed with the lowest--order vertices 
from Eq.~(\ref{Lagr_nopi}). $T_2$ receives a contribution from both the two--loop graph with the lowest--order 
vertices and from the tree graph with a subleading vertex \citep{Kaplan:1998tg,Kaplan:1998we}. 
Higher--order corrections can be evaluated straightforwardly.  
Matching the resulting $T$--matrix to the ERE in Eq.~(\ref{Tnatural}) order by order in the low--momentum expansion allows to 
fix the LECs $C_i$. At next--to--next--to--leading order (N$^2$LO), for example, one finds:
\beq
\label{Cnatural}
C_0 = \frac{4 \pi a}{m} \Big[ 1 + \mathcal{O} (a \mu ) \Big] \,, 
\mbox{\hskip 2 true cm}
C_2 = \frac{2 \pi a^2}{m} \,r_0\,,
\eeq
where the LECs $C_0$ and $C_2$ are defined via the tree--level $T$--matrix: 
$T_{\rm tree} = C_0 + C_2 \, k^2 + \ldots$. The LEC $C_0$ is related to $C_{S,T}$ in Eq.~(\ref{Lagr_nopi})
as $C = C_S - 3 C_T$.
We stress that the ERE in Eq.~(\ref{Tnatural}) can also be reproduced in the cut--off EFT framework  
choosing $\Lambda \sim M_\pi$ and resumming loop diagrams to all orders (i.e.~solving the Lippmann--Schwinger 
equation with the nuclear potential given by contact interactions). 

\begin{figure*}
%\vspace{0.3cm}
\centerline{
\psfrag{xxx}{\raisebox{0.2cm}{\hskip -0.3 true cm  {\large $T_{-1}$}}}
\psfrag{yyy}{\raisebox{0.2cm}{\hskip -0.2 true cm   {\large $T_0$}}}
\psfrag{zzz}{\raisebox{0.25cm}{\hskip -0.2 true cm  where:}}
\psfig{file=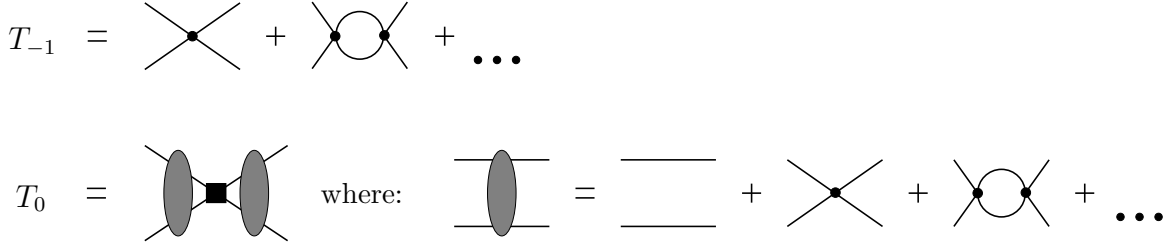,width=16cm}
}
\vspace{0.0cm}
\caption[fig1]{\label{fig1}  The leading and subleading contributions to the S--wave $T$--matrix in the case of 
unnaturally large scattering length. Solid dots (filled rectangles) refer to contact vertices without 
(with two) derivatives. Lines represent the nucleon propagators. }
\vspace{0.2cm}
\end{figure*}

For the physically interesting case of {\it np} scattering, the two S--wave scattering lengths take unnaturally large values:
\beq
a_{^1S_0} = -23.714 \mbox{ fm} \sim - 16.6 \, M_\pi^{-1}\,, 
\mbox{\hskip 2 true cm}
a_{^3S_1} = 5.42 \mbox{ fm} \sim 3.8 \, M_\pi^{-1}\,.
\eeq
Instead of using the low--momentum representation in Eq.~(\ref{Tnatural}) which is valid only for $k < 1/a$,
one can expand the $T$--matrix in powers of $k$ keeping $a k \sim 1$ \citep{Kaplan:1998tg,Kaplan:1998we}:
\beqa
\label{Tunnatural}
T &=& T_{-1} + T_0 + T_1 + \ldots  \\
&=& \frac{4 \pi}{m} \, \frac{1}{(a^{-1} + i k)} \, 
\left[ 1 +  \frac{r_0}{2 (a^{-1} + i k)} k^2 + \left(  
\frac{r_0^2}{4 (a^{-1} + i k)^2} + \frac{v_2}{(a^{-1} + i k)} \right) k^4 + \ldots \right]\,.
\nonumber
\eeqa
The EFT expansion of the $T$--matrix in the unnatural case is illustrated in Fig.~\ref{fig1}. 
The leading term $T_{-1}$ results from summing an infinite chain of bubble diagrams with the 
lowest--order vertices. The corrections are given by perturbative insertions of higher--order interactions dressed 
to all orders by the leading vertices. Matching the resulting $T$--matrix with the one in 
Eq.~(\ref{Tunnatural}) one finds at NLO:
\beq
C_0 = \frac{4 \pi}{m} \; \frac{1}{a^{-1} - \mu } \,, 
\mbox{\hskip 2 true cm}
C_2 = \frac{4 \pi}{m} \; \frac{1}{(a^{-1} - \mu )^2} \frac{r_0}{2}\,.
\eeq
Notice that since $\mu \sim k$ and $a^{-1} \ll \mu$, the LECs $C_{0,2}$ scale as $C_0 \sim 1/k$ and $C_2 \sim  1/k^2$. 
More generally, a LEC $C_{2n}$ accompanying a vertex with $2 n$ derivatives can be shown to scale as $C_{2n} \sim 1/k^{n+1}$
\citep{Kaplan:1998tg,Kaplan:1998we}.  
This has to be contrasted with the scaling $C_{2n} \sim k^0$ in the case of a natural scattering length, cf.~Eq.~(\ref{Cnatural}).
Notice further that the LECs $C_{0,2}$ take very large values in the MS scheme (i.e.~for $\mu = 0$) which 
destroys the manifest power counting \citep{Kaplan:1998tg,Kaplan:1998we}. 

\begin{figure*}
%\vspace{0.3cm}
\centerline{
\psfig{file=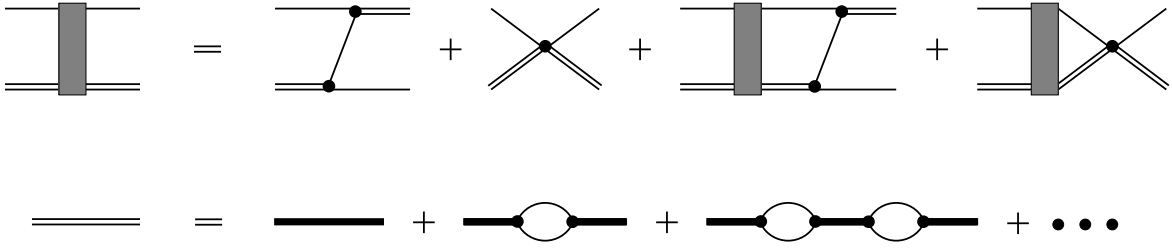,width=16cm}
}
\vspace{-0.2cm}
\caption[fig2]{\label{fig2}  The first line is the graphical representation of the integral equation describing nucleon--dimeron scattering. 
The second line shows the dressed dimeron propagator. Double (bold) lines correspond to the dressed (bare) propagator of the dimeron. 
Shaded rectangles refer to the 3N $T$--matrix. For remaining notation see Fig.~\ref{fig1}.}
\vspace{0.2cm}
\end{figure*}

The three--nucleon problem within pionless EFT has attracted a lot of scientific interest 
during the past few years, see e.g.~\citep{Bedaque:1998km,Bedaque:1998kg,Bedaque:1999ve,
Gabbiani:1999yv,Bedaque:2002yg,Blankleider:2001qv,Afnan:2003bs,Griesshammer:2004pe}. 
The ultimate question is to what extent the low--energy behavior of the 2N system constrains the 
properties of the three--nucleon (3N) system. Here, it is particularly interesting that one can identify universal properties 
of systems, where the scattering length in the two--body system is large. This situation is not only realized 
in the NN system, but also for $^4$He atoms and atomic systems close to a Feshbach resonance, see  \citep{Braaten:2004rn}
for more details. The integral equation for the $T$--matrix 
describing nucleon--dimeron scattering and including the leading three--nucleon force is schematically depicted in Fig.~\ref{fig2}. 
For simplicity, we here discuss the case of three interacting bosons,
which gathers the main aspects of the problem. For the state with total orbital angular momentum $L=0$, the integral equation 
takes the following form in the three--body  CMS:
\beq
\label{STM}
T(k, p; \, E) = \frac{16}{3 a} M (k, p; \, E) + \frac{4}{\pi} \int_0^\Lambda dq\, q^2\, 
T(k, q; \, E) \, \frac{1}{- a^{-1} + \sqrt{3 q^2/4 - m E - i \epsilon}} \, M (q, p; \, E) \,, 
\eeq 
where the inhomogeneous term reads
\beq
M (k, p; \, E) = \frac{1}{2 k p} \ln \left( \frac{k^2 + k p + p^2 - m E}{k^2 - k p + p^2 - m E} \right)
+ \frac{H (\Lambda )}{\Lambda^2}\,.
\eeq
Here $a$ ($H$) is the two--body scattering length (the strength of the three--body force),  
$k \equiv | \vec k {}|$ ($p  \equiv | \vec p \,|$) is the magnitude of the dimeron incoming (outgoing) momenta
and $E = \frac{3 k^2}{4 m} - \gamma$ 
is the total energy in the incoming state with  $\gamma \simeq (m a^2)^{-1}$ 
being the two--body binding energy. The incoming and outgoing bosons are taken on the 
energy shell. The on--shell point corresponds to $k = p$ and the phase shift can be obtained via
\beq
\frac{1}{k \cot \delta  - i k } = T(k, k; \, E)\,.
\eeq
For $H=0$, $\Lambda \to \infty$,  Eq.~(\ref{STM}) has been first derived by Skorniakov and Ter-Martirosian \citep{Skorniakov:1957aa}.
It is well known that Eq.~(\ref{STM}) has no unique solution in this limit \citep{Danilov:1961aa}.\footnote{Whether 
Eq.~(\ref{STM}) with $H=0$ and $\Lambda \to \infty$ possesses a unique solution or not depends on the value of the factor which 
multiplies the last term in Eq.~(\ref{STM}).}  
The regularized equation has a unique solution for any given (finite) value of the ultraviolet cut--off $\Lambda$ but 
the amplitude in the absence of the three--body force shows an
 oscillatory behavior on $\Lambda$. 
Cut--off independence of the amplitude is restored by an appropriate ``running'' of $H (\Lambda )$ which turns out to be of  
a limit cycle type \citep{Bedaque:1998kg,Bedaque:1998km}. Adjusting $H (\Lambda )$ to a single three--body observable for large enough $\Lambda$ 
(or even in the limit $\Lambda \to \infty$) allows to determine all other low--energy properties of the three--body system. 
Alternatively, this can also be achieved by choosing $H = 0$, tuning $\Lambda$ to reproduce a three--body data 
point and using the same cut--off to calculate other observables \citep{Kharchenko:1973aa}. 
%XXX (???)
It has also been conjectured that the behavior of the \emph{physical} amplitude  
at asymptotically large momenta has to satisfy certain constraints which might be used to extract a unique solution 
of Eq.~(\ref{STM}) in the case $H=0$ and $\Lambda \to \infty$ \citep{Blankleider:2000vi,Blankleider:2001qv}.
%XXX

The 3N problem can be considered as generalization of the bosonic case. 
For S--wave $Nd$ scattering in the spin--$3/2$ channel, the corresponding equation for
the $T$--matrix has a unique solution 
for $H=0$ and $\Lambda \to \infty$ so that there is no need to include 
a 3N force. In the spin--$1/2$ channel two nucleons 
can form both spin--$1$ and spin--$0$ dimeron fields which leads to a pair of coupled integral equations for 
the $Nd$  $T$--matrix.
Including the leading contact 3N force allows one to solve these equations in a cut--off independent way \citep{Bedaque:1999ve}. 
Thus, one needs a new parameter which is not determined in the 2N system in order to fix the (leading) low--energy 
behavior of the 3N system in this channel. Higher--order corrections to the amplitude including the ones due to 
2N effective range terms can be included perturbatively\footnote{Resumming 
the effective range corrections to 
all orders in the dimeron propagator leads to an unphysical pole which might cause problems in the solution of the 3N scattering equation
\citep{Braaten:2004rn}.}
\citep{Efimov:1991aa,Bedaque:2002yg,Griesshammer:2004pe}.
Extension to 3N channels with different quantum numbers is straightforward \citep{Gabbiani:1999yv}. For the current status of these 
applications see \citep{Beane:2000fx,Bedaque:2002mn,vanKolck:2004te} and references therein. Universal low--energy properties of few--body 
systems with short--range interactions and large two--body scattering length are reviewed in 
\citep{Braaten:2004rn}, see also \citep{Efimov:1981aa} for an early work on this subject. 
First results in the four--body sector within pionless EFT are presented in \citep{Platter:2004qn,Platter:2004zs}. 
Recently, this approach has also been applied to halo nuclei, see \citep{vanKolck:2004te} for an overview.
%***************************************************************
For more details on these and further topics including applications to a variety of electroweak processes 
in the 2N sector see the recent review articles \citep{Beane:2000fx,Bedaque:2002mn} and references therein.
%
%Pionless EFT has also been applied to  a variety of electroweak processes in the two--nucleon sector.
%In particular, deuteron form--factors and electric polarizability have been computed in \citep{Chen:1999tn} and 
%the process $n + p \to d + \gamma$ has been studied at N$^3$LO in \citep{Chen:1999vd,Chen:1999bg}
%and N$^4$LO in \citep{Rupak:1999rk}. Coulomb effects in pionless EFT have been considered in 
%\citep{Kong:1998sx,Kong:1999sf,Holstein:1999nq,Gegelia:2003ta} and the application to the proton fusion reaction
%is presented in \citep{Kong:1999tw,Kong:1999mp,Kong:2000px,Butler:2001jj}. Low--energy neutrino--deuteron reactions 
%have been analyzed in \citep{Butler:1999sv,Butler:2000zp}. For more details on these and further topics
%see the recent review articles \citep{Beane:2000fx,Bedaque:2002mn}. 

\subsection{Chiral EFT for few--nucleon systems}
\label{sec3}

So far we have considered few--nucleon processes at very low momenta $k \ll M_\pi$ which can be well treated within
pionless EFT. We now wish to go to higher momenta $k\sim M_\pi$ where the inclusion of explicit pions is mandatory. 
The interaction between pions and nucleons is governed by the spontaneously broken approximate chiral symmetry of QCD
as explained in section \ref{sec1}. One would, therefore, like to have an approach which utilizes both 
resummation of certain classes of Feynman diagrams in order to describe the nonperturbative features of few--nucleon systems 
as well as chiral expansion familiar from the single--nucleon sector. 
While the leading NN contact interaction has to be resummed to all orders at least in the case of an unnaturally large scattering length,
it is not clear a priori whether the interaction resulting from the exchange of pions between the nucleons is weak enough to be treated 
perturbatively. We will now outline two basic EFT approaches  with explicit pions to few--nucleon systems:
the one due to Kaplan, Savage and Wise (KSW) \citep{Kaplan:1998tg,Kaplan:1998we} which treats pion exchange  
in perturbation theory, and the other one due to Weinberg \citep{Weinberg:1990rz,Weinberg:1991um} based on its 
nonperturbative treatment. For yet another scheme see \cite{Lutz:1999yr}.  

\begin{figure*}
%\vspace{0.3cm}
\centerline{
\psfrag{xxx}{\raisebox{0.05cm}{\hskip -0.11 true cm  {\large $2$}}}
\psfrag{yyy}{\raisebox{0.22cm}{\hskip -0.2 true cm   {\large $T_0$}}}
\psfig{file=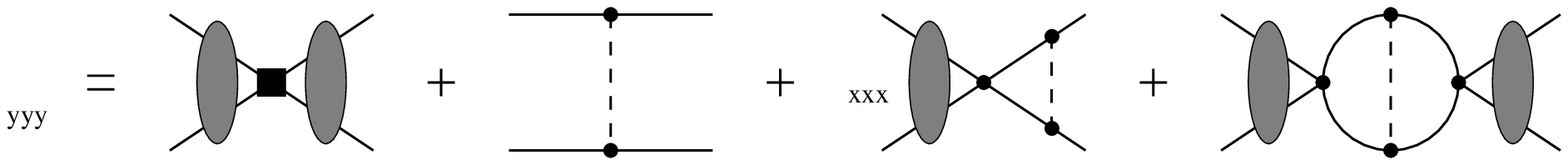,width=16cm}
}
\vspace{-0.1cm}
\caption[fig3]{\label{fig3}  The first correction to the NN scattering amplitude in the KSW approach. 
Dashed lines refer to pions, solid rectangles denote insertions of vertices with two derivatives or 
proportional to $M_\pi^2$.  For remaining notation see Fig.~\ref{fig1}.}
\vspace{0.2cm}
\end{figure*}

\begin{figure*}
\vspace{0.3cm}
  \centerline{\epsfxsize=8cm \epsfbox{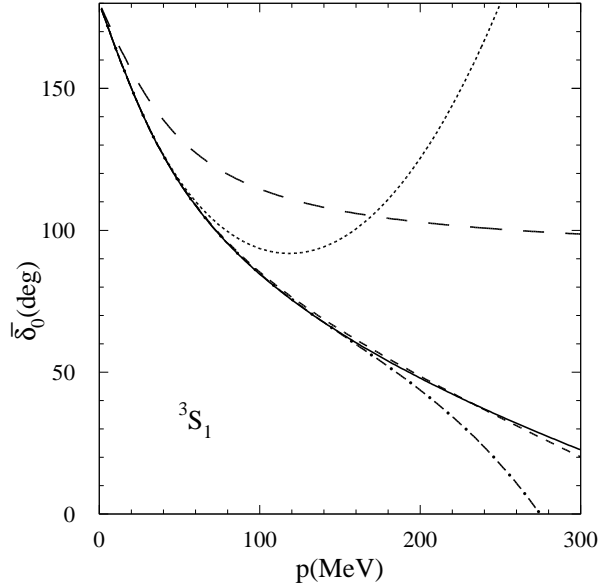} }
\vspace{0.0cm}
\caption[fig4]{\label{fig4}  
The NN $^3S_1$ phase shift $\bar \delta_0$ versus CMS momentum $p$. The solid
line is the Nijmegen multi--energy fit \citep{Stoks:1993tb,NNonline}, the long dashed line is the LO
EFT result, the short dashed line is the NLO result, and the
dotted line is the N$^2$LO result. The dash--dotted line shows the result of
including a higher--order contact interaction. Figure courtesy of Ian W.~Stewart. }
\vspace{0.2cm}
\end{figure*}

The KSW formalism represents a straightforward generalization of the pionless EFT approach for the case of large scattering length 
discussed in section \ref{sec2} to perturbatively include diagrams with exchange of one or more pions. The scaling of the contact interactions
is assumed to be the same as
in pionless EFT (provided one uses  DR with PDS or an equivalent scheme to regularize divergent 
loop integrals). For $k \sim M_\pi \sim a^{-1}$, the leading--order S--wave amplitude $T_{-1}$ is still given 
by the diagrams shown in the first line of Fig.~\ref{fig1}. The first correction $T_0$ arises from perturbative insertions of 
subleading contact interactions (i.e.~the ones with two derivatives and $\propto M_\pi^2$) and one--pion exchange (1PE) 
dressed to all orders by the leading contact interactions, see Fig.~\ref{fig3}. The undressed static 1PE contribution 
corresponding to the second graph in Fig.~\ref{fig3} and based on the Lagrangian in Eq.~(\ref{Lfin}) has the form
\beq
\label{1PE}
T_0^{\rm 1\pi, \, undressed} = - \frac{1}{4 \pi} \left( \frac{g_A}{2 F_\pi} \right)^2 \, \fet \tau_1 \cdot \fet \tau_2 \, 
\frac{(\vec \sigma_1 \cdot \vec q \, ) (\vec \sigma_2 \cdot \vec q \, )}{\vec q \,^2 + M_\pi^2}\,,
\eeq
where $\vec q = \vec p \, ' - \vec p$ is the nucleon momentum transfer and $\vec \sigma_i$ ($\fet \tau_i$) are spin (isospin) matrices 
of the nucleon $i$. The overall normalization of the $T$--matrix is consistent with Eq.~(\ref{Tnorm}). It is clear from Eq.~(\ref{1PE}) 
that this 1PE contribution as well as the contributions from the last two graphs in Fig.~\ref{fig3} are of the order $\mathcal{O} (k^0 )$. 
Notice that the coefficients of contact interactions with $2m$ derivatives and $\propto M_\pi^{2n-2m}$ 
are assumed to scale as $1/k^{n+1}$. Two--pion exchange (2PE) is suppressed 
compared to 1PE and starts to contribute at N$^2$LO. At each order in the perturbative expansion, the amplitude is made 
independent on the renormalization scale by an appropriate running of the LECs $C_i$, $D_i$. As a  nice feature, the KSW approach 
allows to derive analytic expressions for the scattering amplitude. In order to conclude on the usefulness of the 
KSW expansion with perturbative pions, it is crucial to understand the scale at which it fails. While in the single--nucleon
sector, this scale is associated with the chiral symmetry breaking scale, $\Lambda_\chi \sim M_\rho \sim 4 \pi F_\pi \sim 1$ GeV, 
the chiral expansion in the NN sector was found to entail the new scale $\Lambda_{NN}$ associated with the iterated 1PE contributions.
Estimations based on dimensional analysis yield $\Lambda_{NN} = 16 \pi F_\pi^2 /(g_A^2 m) \sim 300$ MeV \citep{Kaplan:1998tg,Kaplan:1998we}.
In \citep{Bedaque:2002mn},
an  even more conservative result was obtained: $\Lambda_{NN} = 4 \pi F_\pi^2/(g_A^2 m) \sim 70$ MeV. 
The small estimated values of $\Lambda_{NN}$ already indicate that the expansion based on perturbative pions might  
converge poorly.  Clearly, dimensional analysis only provides a fairly rough estimation for the scale $\Lambda_{NN}$.
The convergence of the KSW expansion can ultimately be
only tested in concrete calculations. The 2N system has been 
analyzed at N$^2$LO in \citep{Fleming:1999ee}. While the results for the $^1S_0$ and some other partial waves including spin--singlet channels 
were found to be in reasonable agreement with the Nijmegen partial--wave analysis (PWA), large corrections show up in spin--triplet 
channels already at momenta $\sim 100$ MeV and lead to strong disagreements with the data. This is exemplified in Fig.~\ref{fig4}. 
The perturbative inclusion of the pion--exchange contributions does not allow to increase the region of validity of the EFT
compared to the pionless theory. The failure of the KSW approach in the spin--triplet channels 
was associated in \citep{Fleming:1999ee} with the iteration of the tensor part of the 1PE potential. 
Further evidence of the poor convergence of the 
KSW expansion with perturbative pions was given by Cohen and Hansen \citep{Cohen:1998jr,Cohen:1999ia} who obtained 
predictions for the
effective range and shape coefficients in the effective range expansion at lowest nontrivial order. 
These coefficients are sensitive to pion dynamics and were found to be poorly described, which indicates that the chiral expansion 
is not converging. More details on the KSW approach with explicit pions and its applications in the two-- and three--nucleon sectors  
can be found in \citep{Beane:2000fx,Bedaque:2002mn} and references therein. For further discussion on the role of the 
pion--exchange contributions see \citep{Gegelia:1998ee,Steele:1998zc,Kaplan:1999qa,Rupak:1999aa}.

\begin{figure*}
%\vspace{0.3cm}
\centerline{
\psfig{file=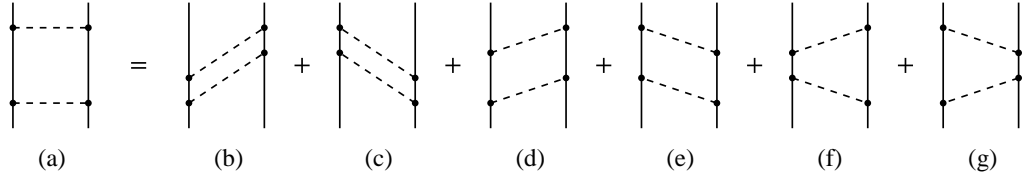,width=14cm}
}
\vspace{-0.2cm}
\caption[fig4aa]{\label{fig4aa} Two--pion exchange: Feynman diagram (a) and the corresponding time--ordered graphs (b--g). 
Solid (dashed) lines correspond to nucleons (pions). }
\vspace{0.2cm}
\end{figure*}

A suitable way of including the pion--exchange contributions nonperturbatively was proposed 
in the seminal work of Weinberg \citep{Weinberg:1990rz,Weinberg:1991um} which preceded the development of the 
KSW approach and caused a flurry of activities to apply EFT in the few--nucleon sector. Weinberg's original 
arguments are formulated in terms of ``old--fashioned'' time--ordered perturbation theory, see e.g.~\citep{Schweber:1966aa},
which is an appropriate tool since we are dealing with nonrelativistic nucleons. Consider the $S$--matrix 
for few--nucleon
 scattering 
\beq
S_{\alpha \beta} 
= \delta (\alpha - \beta) - 2 \pi i \delta ( E_\alpha - E_\beta ) T_{\alpha \beta}\,,
\eeq
where $\alpha$ and $\beta$ denote the final and initial few--nucleon states and $E_\alpha$, $E_\beta$ are the corresponding energies. 
The $T$--matrix can be evaluated in ``old--fashioned'' time--ordered perturbation theory via 
\beq
\label{old-fashioned}
T_{\alpha \beta} = (H_I)_{\alpha \beta} + \sum_a \frac{(H_I)_{\alpha a}  (H_I)_{a \beta}}{E_\beta - E_a + i \epsilon}
+ \sum_{a b} \frac{(H_I)_{\alpha a}  (H_I)_{ab}   (H_I)_{b \beta}}{(E_\beta - E_a + i \epsilon)
(E_\beta - E_{b} + i \epsilon)} + \ldots\,,
\eeq
where $H_I$ is the interaction Hamiltonian corresponding to the effective Lagrangian for pions and nucleons.\footnote{Notice
that in contrast to the purely quantum mechanical consideration in section \ref{sec2}, one has now to account 
for nucleon self--energies. This can be achieved by a proper separation between the unperturbed Hamiltonian 
and the interaction and using the formulation in terms of the corresponding ``in'' and ``out'' states, see 
e.g.~\citep{Schweber:1966aa,Gell-Mann:1954kc,Weinberg:1995aa}.} Here, we use Latin letters for intermediate states,
which, in general, may contain any number of pions, in order to distinguish them from purely nucleonic states
denoted by Greek letters. We remind the reader that no nucleon--antinucleon pairs can be created or destroyed 
due to the nonrelativistic treatment of the nucleons. Consequently, all states contain the same number of nucleons. 
It is useful to represent various contributions to the scattering amplitude in terms of time--ordered diagrams. 
For example, the Feynman box diagram for NN scattering via  $2\pi$--exchange can be expressed 
as a sum of six time--ordered graphs, see Fig.~\ref{fig4aa}, which correspond to the following term in Eq.~(\ref{old-fashioned}): 
\beq
\label{TPEold-fashioned}
T_{\alpha \beta}^{2\pi} =  \sum_{abc} \frac{(H_{\pi NN})_{\alpha a}  (H_{\pi NN})_{ab}   
(H_{\pi NN})_{bc} (H_{\pi NN})_{c \beta}}{(E_\beta - E_a + i \epsilon)
(E_\beta - E_b + i \epsilon) (E_\beta - E_c + i \epsilon)} \,,
\eeq
where $H_{\pi NN}$ denotes the $\pi NN$ vertex. It is easy to see that the contributions of diagrams (d--g) are 
enhanced due to the presence of the small (of the order $Q^2/m$) energy denominator associated with the 
purely nucleonic intermediate state $| b \rangle $ which in the CMS takes the form:\footnote{Equivalently, 
evaluation of the Feynman graph (a) in Fig.~\ref{fig4aa} using the standard heavy--nucleon propagator of the 
form $i/(p^0 + i \epsilon)$ leads to infrared divergences resulting from a pinch singularity  
associated with the poles $p^0 = \pm i \epsilon$. These infrared divergences are avoided (but still leading to the enhancement 
in the amplitude) by the inclusion of the kinetic energy term in the heavy--nucleon propagators.}  
\beq
\frac{1}{E_\beta - E_b + i \epsilon} = \frac{1}{\vec p_\beta^{\, 2} /m  - \vec p_b^{\, 2} /m  + i \epsilon}\,.
\eeq
Notice that the energy denominators corresponding to the $\pi NN$ states  
$| a \rangle$ and $| c \rangle$ are of the order $M_\pi \sim Q$. 
According to Weinberg, the failure of perturbation theory in the few--nucleon sector is caused
by the enhanced contribution of reducible diagrams, i.e.~those ones which contain purely nucleonic intermediate states.
To see how this difficulty can be dealt with, it is useful to rearrange the expansion in Eq.~(\ref{old-fashioned}) and to write 
it in the form of the Lippmann--Schwinger equation
\beq
\label{LSeq}
T_{\alpha \beta}
= (V_{\rm eff})_{\alpha \beta} + \sum_\gamma \frac{(V_{\rm eff})_{\alpha \gamma}  T_{\gamma \beta}}{E_\beta - E_\gamma + i \epsilon}\,,
\eeq
with the effective potential $(V_{\rm eff})_{\alpha \beta}$ defined as a sum of all possible irreducible diagrams (i.e. the ones 
which do not contain purely nucleonic intermediate states):
\beq
\label{ep}
(V_{\rm eff})_{\alpha \beta} = (H_I)_{\alpha \beta} + \sum_{\tilde a} \frac{(H_I)_{\alpha \tilde a}  (H_I)_{\tilde a \beta}}
{E_\beta - E_{\tilde a} + i \epsilon}
+ \sum_{\tilde a \tilde b} \frac{(H_I)_{\alpha \tilde a}  (H_I)_{\tilde a \tilde b}   (H_I)_{\tilde b \beta}}{(E_\beta - E_{\tilde a} + i \epsilon)
(E_\beta - E_{\tilde b} + i \epsilon)} + \ldots\,.
\eeq
Here, the states $| \tilde a \rangle$,  $| \tilde b \rangle$ contain at least one pion. 
The effective potential in Eq.~(\ref{ep}) does not contain small energy denominators and can be obtained within the low--momentum expansion
following the usual procedure of CHPT. The contribution of a given irreducible time--ordered diagram can be shown to be of the order 
$(Q/\Lambda )^\nu$ \citep{Weinberg:1990rz,Weinberg:1991um} with $\Lambda$ being the scale 
which enters the values of the renormalized LECs, where 
\beq
\label{powcNN}
\nu = - 2 + 2 N + 2 (L -C)  + \sum_i V_i \Delta_i \,,\quad \mbox{where} \quad 
\Delta_i = d_i + \frac{1}{2} n_i - 2\,.
\eeq  
Here $N$, $L$, $C$ and $V_i$ are the numbers of nucleons, loops, separately connected pieces and vertices of
type $i$, respectively. The quantity $\Delta_i$ gives the chiral dimension of a vertex of type $i$. 
Further, $d_i$ denotes the number of derivatives or $M_\pi$ insertions and $n_i$ is the number of 
nucleon lines at the vertex $i$. Notice that eq.~(\ref{powcNN}) is modified compared to 
the one given in Refs.~\citep{Weinberg:1990rz,Weinberg:1991um,Weinberg:1992yk} in order to account for the 
proper normalization of the $N$--nucleon states. Chiral symmetry guarantees that $\Delta_i \geq 0$. 
Consequently, the chiral order $\nu$ is bounded from below and for any given $\nu$
only a finite number of diagrams needs to be taken into account.
Notice that Eq.~(\ref{powcNN}) supports a rather natural view of nuclear dynamics, 
in which nucleons interact mainly via 2N forces while many--body 
forces provide small corrections. After the potential is obtained at a given order in the chiral expansion, few--nucleon 
observables can be calculated by solving the Lippmann--Schwinger equation (\ref{LSeq}), which leads to a nonperturbative 
resummation of the contributions resulting from reducible diagrams. It is easy to see from Eq.~(\ref{powcNN})
that the leading--order ($\nu = 0$) potential results from contact interactions 
without derivatives and the $1\pi$--exchange. This has to be contrasted with the KSW approach, where the exchange of 
pions is suppressed compared to the lowest--order contact terms. It should be understood that the 
power counting rules in Eq.~(\ref{powcNN}) apply to renormalized matrix elements.\footnote{We stress that while 
perturbative renormalization of the scattering 
amplitude in the pion and single nucleon sectors is a straightforward task, both from the conceptual and practical points of view,
nonperturbative renormalization in the few--nucleon sector still attracts the interest of many researchers, 
see e.g.~\citep{Beane:2001bc,Gegelia:2004pz,Nogga:2005hy,Birse:2005um} for some recent work. 
We will address this issue in some detail in section 
\ref{regularization}.} After removing the ultraviolet divergences 
by a redefinition of the LECs in the effective Lagrangian, the remaining integrals are effectively cut off at 
momenta of the order of the soft scale $Q$. The power counting described above is based on an assumption,
sometimes referred to as the naturalness assumption, that a renormalized coupling constant $C$ of dimension
[mass]$^{-n}$ can be written in terms of
a dimensionless coefficient $c\sim \mathcal{O} (1)$  as $C = c \Lambda^{-n}$.
\footnote{For 
LECs accompanying  NN contact interaction the expected scaling is $C = c F_\pi^{-2}  \Lambda^{-n+2}$, see e.g.~\citep{Friar:1996tj}.}
Clearly, higher--dimensional terms in the amplitude are only suppressed if the hard scale $\Lambda$ that enters the values of the 
LECs is sufficiently large, i.e.~if $Q \sim M_\pi \ll \Lambda$.  The validity of the naturalness assumption can, at present, only be
verified upon performing actual calculations.

The presence of shallow bound states in few--nucleon systems suggests that the perturbative (iterative) solution of 
Eq.~(\ref{LSeq}) does not converge. As pointed out by Weinberg \citep{Weinberg:1990rz,Weinberg:1991um}, this requires for  
the nucleon mass  to be counted as a much larger scale compared to the hard scale $\Lambda$. To see that consider the 
iteration of the leading order potential $V_{\rm eff}^{(0)}$ in the Lippmann--Schwinger equation (\ref{LSeq}) 
which, in operator form, can be written symbolically as 
\beq
\label{LSperturb}
T =  V_{\rm eff}^{(0)} + V_{\rm eff}^{(0)} \, G_0 \, V_{\rm eff}^{(0)}  + V_{\rm eff}^{(0)} \, G_0 \, V_{\rm eff}^{(0)} \, G_0 \, V_{\rm eff}^{(0)} 
+ \ldots\,,
\eeq
where $G_0$ is the free 2N resolvent operator. One can estimate the size of the leading--order potential by the 
size of the  static $1\pi$--exchange potential leading to $V_{\rm eff}^{(0)} \sim 1/F_\pi^2$. Since each momentum 
integration  in Eq.~(\ref{LSperturb}) gives an additional factor $Q^3/(4 \pi)^2$ and $G_0 \sim m/Q^2$, one finds that the $(n+1)$--th term 
in the above equation is suppressed compared to the first term by $(Q m /\Lambda^2)^n$, where 
we used the estimation $\Lambda \sim 4 \pi F_\pi$. The requirement that all terms in the right--hand side of  Eq.~(\ref{LSperturb}) are 
of the same order in order, which justifies the necessity of the nonperturbative treatment and enables to describe the physics
associated with the low--lying bound states, therefore leads to the following counting rule for the nucleon mass 
\citep{Weinberg:1990rz,Weinberg:1991um,Ordonez:1995rz}:
\beq
\label{m_counting}
\frac{Q}{m} \sim \frac{Q^2}{\Lambda^2}\,.
\eeq
This counting rule will be adopted in the present work. 
Clearly, this estimation based on the naive dimensional analysis is fairly crude. A somewhat different estimation can 
be found in \citep{Bedaque:2002mn}. Notice that it is hardly possible in such an estimation 
to keep track of various numerical factors, even of the large factors such as $4 \pi$. For example, the 
leading and subleading $2\pi$--exchange potentials, both arising from 1--loop diagrams, differ by a factor $4 \pi$, 
see section \ref{sec:piexch}. Fortunately, the particular way of counting the nucleon mass is
not crucial from the practical point of view since it only determines the relative importance 
of the relativistic corrections to the nuclear force but does not affect the lowest--order 
potential and, therefore, also
not the dominant contribution to the scattering amplitude. Finally, we stress that 
Weinberg's power counting does not explain the unnaturally large values of the NN S--wave scattering lengths
or, equivalently, the unnaturally small binding energies of the deuteron and the virtual bound state in the 
$^1S_0$ channel. This has to be achieved via an appropriate fine tuning of the lowest--order contact interactions. 

To summarize, the ``Weinberg program'' for describing the low--energy dynamics of the few--nucleon systems 
proceeds in two basic steps which will be discussed in detail in the next sections of this review.  
First, the few--nucleon potential has to be derived from the effective Lagrangian 
for pions and nucleons using the framework of chiral perturbation theory. Secondly, the corresponding dynamical 
equations with the resulting potential as an input have to be solved.

%%%%%%%%%%%%%%%%%%%%%%%%%%%%%%%%%%%%%%%%%%%%%%%%%%%%%%%%%%%%%%%%%%%%%%%%%%%%%%%%%
\section{Nuclear forces in chiral effective field theory}
\def\theequation{\arabic{section}.\arabic{equation}}
\setcounter{equation}{0}
\label{sec4}

In the previous section we have introduced the basic concept of the Weinberg approach 
to few--nucleon systems. We will now discuss the structure of the nuclear force in 
the lowest orders in the chiral expansion based on the effective Lagrangian 
\beqa
\label{lagr}
\mathcal{L}^{(0)} &=& \frac{1}{2} \partial_\mu \fet \pi \cdot \partial^\mu \fet \pi  - \frac{1}{2} M^2 \fet \pi^2  
+ N^\dagger  \bigg[  i \partial_0 +
\frac{g_A}{2 F} \fet \tau \vec \sigma \cdot \vec \nabla \fet \pi 
- \frac{1}{4 F^2} \fet \tau \cdot ( \fet \pi \times \dot{\fet \pi } ) \bigg] N  \nn
&&{} - \frac{1}{2} C_S ( N^\dagger  N  )  ( N^\dagger  N  )  - \frac{1}{2} C_T  
( N^\dagger \vec \sigma N )  ( N^\dagger \vec \sigma N )+ \ldots  \,, \nn [0.5ex]
\mathcal{L}^{(1)} &=& N^\dagger  \bigg[ 4 c_1 M^2  
- \frac{2 c_1}{F^2} M^2 \fet \pi^2 + \frac{c_2}{F^2} \dot{\fet \pi}^2  
+ \frac{c_3}{F^2} (\partial_\mu \fet \pi \cdot \partial^\mu \fet \pi )  
-  \frac{c_4}{2 F^2} \epsilon_{ijk} \, \epsilon_{abc} \, \sigma_i \tau_a (\nabla_j \, \pi_b ) (\nabla_k \, \pi_c )  
 \bigg] N  \nn
&& {} - \frac{D}{4 F}   ( N^\dagger N  )  ( N^\dagger  \vec \sigma \fet \tau  N  ) \cdot \vec \nabla \fet \pi 
- \frac{1}{2} E \,  ( N^\dagger N  )   ( N^\dagger \fet \tau N  ) \cdot  (  N^\dagger \fet \tau N  )  + \ldots \,, 
\eeqa
where the superscripts denote the vertex dimension $\Delta_i$, see Eq.~(\ref{powcNN}),  
$c_i$, $C_{S,T}$, $D$ and $E$ are LECs and ellipses refer to terms with more pion fields.
Notice that the nucleon kinetic energy contribute, according to Eq.~(\ref{m_counting}), to  $\mathcal{L}^{(2)}$.
The above terms determine the nuclear potential up to N$^2$LO (with the exception of the 
NN contact terms at NLO) in the limit of exact isospin symmetry. More complete expressions 
for the Lagrangian including higher--order terms can be found e.g.~in 
\citep{Bernard:1995dp,Fettes:2001cr,VanKolck:1993ee,Steininger:1999aa,Fettes:2000gb,Gasser:2002am,Epelbaum:2005fd}.

\subsection{Nuclear potentials from field theory}

The derivation of a potential from field theory is an 
intensively studied problem in nuclear physics. Historically, the  
important conceptual achievements in this field have been done in the fifties 
of the last century
in the context of the so called meson field theory. The problem can be formulated in 
the following way: given a field theoretical Lagrangian for interacting mesons 
and nucleons, how can one reduce the (infinite dimensional) equation of motion for 
mesons and nucleons to an effective Schr\"odinger equation for nucleonic 
degrees of freedom, which can be solved by standard methods?
It goes beyond the scope of this work to address the whole variety of
different techniques which have been developed to construct effective interactions, see Ref.~\citep{Phillips:1959aa}
for a comprehensive review. We will now briefly outline a few methods which have been used in the context of chiral EFT.

We begin with the approach developed by Tamm \citep{Tamm:1945qv} and Dancoff \citep{Dancoff:1950ud} which in the following 
will be referred to as the Tamm--Dancoff method. Consider the time--independent Schr\"odinger equation 
\beq
\label{schroed1}
(H_0 + H_I) | \Psi \rangle = E | \Psi \rangle\,,
\eeq
where $|\Psi \rangle$ denotes an eigenstate of the Hamiltonian $H$ 
with the eigenvalue $E$. 
One can divide the full Fock space in to the nucleonic subspace $|\phi \rangle$ 
and the complementary one $|\psi \rangle$  and rewrite the Schr\"odinger equation 
(\ref{schroed1}) as 
\begin{equation}
\label{schroed2}
\left( \begin{array}{cc} \eta H \eta & \eta H \lambda \\ 
\lambda H \eta & \lambda  H 
\lambda \end{array} \right) \left( \begin{array}{c} | \phi \rangle \\ 
| \psi \rangle \end{array} \right)
= E  \left( \begin{array}{c} | \phi \rangle \\ 
| \psi \rangle \end{array} \right)~,
\quad \,
\end{equation}
where we introduced the projection operators $\eta$ and 
$\lambda$ such that $|\phi \rangle = \eta | \Psi \rangle$,
$| \psi \rangle = \lambda | \Psi \rangle$.
Expressing the state $| \psi \rangle$ from the second line 
of the matrix equation (\ref{schroed2}) as 
\begin{equation}
\label{5.3}
| \psi \rangle = \frac{1}{ E - \lambda H \lambda}  H  | \phi \rangle~,
\end{equation}
and substituting this in to
the first line one obtains the Schr\"odinger--like equation for the projected 
state $| \phi \rangle$:
\begin{equation}
\label{TDschroed}
\left( H_0 + V_{{\rm eff}}^{\rm TD} ( E ) \right) | \phi \rangle  = E | \phi \rangle \,,
\end{equation}
with an effective potential $V_{\rm eff} (E)$ given by
\begin{equation}
\label{TDpot}
V_{\rm eff}^{\rm TD} (E)= \eta H_I \eta + \eta H_I \lambda 
\frac{1}{E - \lambda H \lambda} \lambda H_I \eta  \,\, .
\end{equation}
It is easy to see that the above definition of the effective potential is identical with the 
one given in Eq.~(\ref{ep}) in the context of ``old--fashioned'' time--ordered perturbation theory. 
We stress that in order to evaluate $V_{\rm eff}^{\rm TD} (E)$ one usually has to rely on perturbation theory. 
For example, for the Yukawa theory with $H_I = g H_1$, the effective potential $V_{\rm eff}^{\rm TD} (E)$ up to the fourth 
order in the coupling constant $g$ is given by
\beq
\label{TDg4}
V_{\rm eff}^{\rm TD} (E) = - \eta ' \bigg[ g^2 H_1 \frac{\lambda^1}{H_0 - E} H_1  +  g^4 
H_1 \frac{\lambda^1}{H_0 - E} H_1 \frac{\lambda^2}{H_0 - E}  H_1 \frac{\lambda^1}{H_0 - E} H_1  + \mathcal{O} (g^6) \bigg] \eta \,,
\eeq
where the superscripts of $\lambda$ refer to the number of mesons in the corresponding state. 
It is important to realize that the effective potential $V_{{\rm eff}} ( E )$ in this scheme depends explicitly on the energy,
which makes it inconvenient for practical applications. In addition, the projected nucleon states $| \phi \rangle$ have a 
normalization different from the  states $| \Psi \rangle$ we have started from, 
which are assumed to span a complete and orthonormal set in the whole Fock space:
\beq
\langle \phi_i | \phi_j \rangle =  
\langle \Psi_i | \Psi_j \rangle - \langle \psi_i | \psi_j \rangle=
\delta_{ij} - \langle \phi_i | H_I \lambda 
\left( \frac{1}{E - \lambda H \lambda} \right)^2 
\lambda H_I | \phi_j \rangle ~.
\eeq
Note that the components $\psi_i$ in this equation do, in general, not vanish.

The above mentioned deficiencies are naturally avoided in the method of unitary transformation \citep{Okubo:1954aa}, see also \citep{Fukuda:1954aa}.
In this approach, the decoupling of the $\eta$-- and $\lambda$--subspaces of the Fock space is achieved via a 
unitary transformation $U$   
\beq
\tilde H \equiv U^\dagger H U = \left( \begin{array}{cc} \eta \tilde H \eta  & 0 \\ 0 & \lambda \tilde H \lambda \end{array} \right)\,.
\eeq
Following Okubo \citep{Okubo:1954aa}, the unitary operator $U$ can be parametrized as
\begin{equation}
\label{5.9}
U = \left( \begin{array}{cc} \eta (1 +  A^\dagger  A )^{- 1/2} & - 
 A^\dagger ( 1 +  A A^\dagger )^{- 1/2} \\
 A ( 1 +  A^\dagger  A )^{- 1/2} & 
\lambda (1 +  A  A^\dagger )^{- 1/2} \end{array} \right)~,
\end{equation}
with the operator $A= \lambda  A \eta$. The operator $A$ has to satisfy 
the decoupling equation 
\begin{equation}
\label{5.10}
\lambda \left( H - \left[ A, \; H \right] - A H A \right) \eta = 0
\end{equation}
in order for the transformed Hamiltonian $\tilde H$ to be of block--diagonal form. 
The effective $\eta$--space potential $\tilde V_{\rm eff}^{\rm UT}$ can be expressed in terms of the operator $A$ as: 
\beq
\label{effpot}
\tilde{V}_{\rm eff}^{\rm UT} =  \eta (\tilde H  - H_0 )  = \eta \bigg[ (1 + A^\dagger A)^{-1/2} (H + A^\dagger H + H A + A^\dagger H A )  
(1 + A^\dagger A)^{-1/2} - H_0 \bigg] \eta~.
\eeq
For the previously considered case of the Yukawa theory, the operator $A$ and the effective potential 
$V_{\rm eff}^{\rm UT}$ can be obtained within the expansion in powers of the coupling constant $g$, which leads to:
\beqa
\label{UTg4}
V_{\rm eff}^{\rm UT} &=&  - g^2 \, \eta ' \Bigg[ \frac{1}{2} H_1 \frac{\lambda^1}{H_0 - E_\eta} H_1 + \mbox{h.~c.} \Bigg] \eta 
  - g^4 \, \eta ' \Bigg[ \frac{1}{2} H_{1} \frac{\lambda^1}{(H_0 - E_{\eta})}  H_{1} \,  \frac{\lambda^2}{(H_0 - E_{\eta})} 
\,  H_{1} \frac{\lambda^1}{(H_0 - E_{\eta} )}  H_{1}  \nn 
&& {} -
\frac{1}{2} H_{1} \frac{\lambda^1}{(H_0 - E_{\eta '})}  H_{1} \, \tilde \eta 
\,  H_{1} \frac{\lambda^1}{(H_0 - E_{\tilde \eta} )( H_0 - E_{\eta '} )}  H_{1}  \nn
&& {} 
+\frac{1}{8} H_{1} \frac{\lambda^1}{(H_0 - E_{\eta '})}  H_{1} \, \tilde \eta 
\,  H_{1} \frac{\lambda^1}{(H_0 - E_{\tilde \eta} )( H_0 - E_{\eta} )}  H_{1}  \nn
&&  {} - \frac{1}{8} H_{1} \frac{\lambda^1}{(H_0 - E_{\eta '}) ( H_0 - E_{\tilde \eta} )}  
H_{1} \, \tilde \eta 
\,  H_{1} \frac{\lambda^1}{(H_0 - E_{\tilde \eta} )}  H_{1}  + \mbox{h.~c.}
\Bigg] \eta  + \mathcal{O}(g^6)\,.
\eeqa
%\beqa
%\label{UTg4}
%V_{\rm eff}^{\rm UT} &=&  - g^2 \, \eta ' \bigg[ \frac{1}{2} H_1 \frac{\lambda^1}{H_0 - E_\eta} H_1 + \mbox{h.~c.} \bigg] \eta \\
%&&  - g^4 \, \eta ' \bigg[ \frac{1}{2} H_{1} \frac{\lambda^1}{(H_0 - E_{\eta})}  H_{1} \,  \frac{\lambda^2}{(H_0 - E_{\eta})} 
%\,  H_{1} \frac{\lambda^1}{(H_0 - E_{\eta} )}  H_{1}  \nn 
%&& \mbox{\hskip 1.2 true cm} -
%\frac{1}{2} H_{1} \frac{\lambda^1}{(H_0 - E_{\eta '})}  H_{1} \, \tilde \eta 
%\,  H_{1} \frac{\lambda^1}{(H_0 - E_{\tilde \eta} )( H_0 - E_{\eta '} )}  H_{1} \nn 
%&& \mbox{\hskip 1.2 true cm} +\frac{1}{8} H_{1} \frac{\lambda^1}{(H_0 - E_{\eta '})}  H_{1} \, \tilde \eta 
%\,  H_{1} \frac{\lambda^1}{(H_0 - E_{\tilde \eta} )( H_0 - E_{\eta} )}  H_{1}  \nn
%&&  \mbox{\hskip 1.2 true cm} - \frac{1}{8} H_{1} \frac{\lambda^1}{(H_0 - E_{\eta '}) ( H_0 - E_{\tilde \eta} )}  
%H_{1} \, \tilde \eta 
%\,  H_{1} \frac{\lambda^1}{(H_0 - E_{\tilde \eta} )}  H_{1}  + \mbox{h.~c.}
%\bigg] \eta  + \mathcal{O}(g^6)\,.
%\nonumber
%\eeqa
In contrast to $V_{\rm eff}^{\rm TD}$ given in  Eq.~(\ref{TDg4}), $V_{\rm eff}^{\rm UT}$ does not depend on 
the energy $E$. Another difference to the Tamm--Dancoff method is given by the presence of terms 
with the projection operator $\tilde \eta$ which give rise to purely nucleonic intermediate states.
These terms are needed to ensure the proper normalization of the few--nucleon states. It should be 
understood that, in spite of the presence of the purely nucleonic intermediate states, such terms are not 
generated through the iteration of the dynamical equation and are thus not reducible in the language of 
section \ref{sec3}. Since all energy denominators in Eq.~(\ref{UTg4}) correspond to intermediate states 
with at least one pion, there is no enhancement by large factors of $m/Q$, which is typical for reducible 
contributions. 

The two methods of deriving effective nuclear potentials are quite general 
and can, in principle, be applied to any field theoretical meson--nucleon Lagrangian. 
In the weak coupling case, the potential can be obtained straightforwardly via the expansion 
in powers of the corresponding coupling constant(s). Generalization to the effective chiral Lagrangian requires
the expansion in powers of the coupling constants to be replaced by the chiral expansion in powers of $Q/\Lambda$. 
For practical applications, it is helpful to use time--ordered diagrams to visualize 
the contributions to the 
potential. In ``old--fashioned'' perturbation theory or, equivalently, the Tamm--Dancoff approach, 
only irreducible diagrams are allowed. Their importance is determined by the power counting in Eq.~(\ref{powcNN})
and the explicit contributions can be found using Eq.~(\ref{TDpot}). In the method of unitary transformation
one can draw both irreducible and reducible graphs, whose importance is still given  by Eq.~(\ref{powcNN}). 
Notice that these graphs have a different meaning from the time--ordered ones arising in the context 
of ``old--fashioned'' perturbation theory and will only be used to visualize the topology associated with 
a given sequence of vertices. 
The structure of the operators contributing to the potential can, in general,  
not be guessed by looking at a given diagram and has to be determined by 
solving the decoupling equation (\ref{5.10}) for the operator $A$ and using Eq.~(\ref{effpot}). 
This is discussed in detail in Refs.~\citep{Epelbaum:1998ka,Epelbaum:2000kv}, 
where it is also demonstrated how to derive the effective potential 
from the effective chiral Lagrangian at any given order $\nu$ in the low--momentum expansion using the method
of unitary transformation, see also  \citep{Krebs:2004st} for a different but closely related scheme.
The explicit expressions for the operators contributing to the potential in
the few lowest orders can be found in these references. For issues related to renormalization within the 
method of unitary transformation see Ref.~\citep{Epelbaum:2002gb}. Another Hamiltonian approach, which 
is similar to the method of unitary transformation and is  usually referred to as the dressed particle approach,
is extensively discussed in Refs.~\citep{Greenberg:1958aa,Faddeev:1963aa,Fateev:1973aa,Sheveko:2001aa}. 

To illustrate how the above ideas work in practice, let us consider the 
contribution to the leading $2\pi$--exchange potential at order $\nu = 2$  
arising from diagram (a) in Fig.~\ref{fig4aa}. The Hamilton operator $H^{(0)}$ describing the $\pi NN$ 
vertex of the lowest possible dimension, $\Delta_i =0$, corresponds to the last term in Eq.~(\ref{LeffN}). 
In ``old--fashioned'' perturbation theory, the potential arises from diagrams (b) and (c) in Fig.~\ref{fig4aa}
and can be obtained evaluating the appropriate matrix elements of the operator
\beq
\label{pr11}
 V^{\rm TD}_{2 \pi}  = - \eta  H^{(0)} \frac{\lambda^1}{\omega} H^{(0)} 
\frac{\lambda^2}{\omega_1 + \omega_2}  H^{(0)} \frac{\lambda^1}{\omega} H^{(0)} \eta\,,
\eeq
where $\omega_i = \sqrt{\vec k \, ^2 + M_\pi^2}$ denotes the energy of
a pion with the momentum $\vec k$. 
Notice that at the order considered, it is sufficient to treat nucleons as static sources.  
Explicit evaluation of Eq.~(\ref{pr11}) yields the following result in the CMS \citep{Ordonez:1995rz}:
\beqa
\label{2PEboxTD}
V^{\rm TD}_{2 \pi} &=& - \frac{g_A^4}{4 (2 F_\pi)^4} \, \int \, \frac{d^3 l}{ (2 \pi )^3} \,
\frac{1}{\omega_+^3 \omega_- } \Bigg\{ \left( \frac{3}{\omega_-} 
+ \frac{2 \fet \tau_1 \cdot \fet \tau_2}
{\omega_+ + \omega_-} \right) \left( {\vec l \,}^2 
- {\vec q \,}^2 \right)^2 \\
&& \mbox{\hskip 2.5 true cm} + 4 \left( \frac{3}{\omega_+ + \omega_-} 
+ \frac{2 \fet \tau_1 \cdot \fet \tau_2}{\omega_-} \right)
( \vec \sigma_2 \cdot [ \vec q \times \vec l \, ] ) 
( \vec \sigma_1 \cdot [ \vec q \times \vec l \, ] ) \Bigg\} \; ,
\nonumber
\end{eqnarray}
with $\omega_\pm \equiv \sqrt{ ( \vec q \pm \vec l \, )^2  + 4 M_\pi^2}$ and 
$\vec q$ being the nucleon momentum transfer. In the method of unitary transformation, the potential 
is due to diagrams (b--g) in Fig.~\ref{fig4aa} and is given by:
\beq
\label{pr12}
V^{\rm UT}_{2 \pi} =   V^{\rm TD}_{2 \pi} + \frac{1}{2} \eta H^{(0)} \frac{\lambda^1}{\omega ^2} H^{(0)} 
\eta  H^{(0)} \frac{\lambda^1}{\omega} H^{(0)} \eta + \frac{1}{2} \eta H^{(0)} \frac{\lambda^1}{\omega } H^{(0)} 
\eta  H^{(0)} \frac{\lambda^1}{\omega ^2} H^{(0)} \eta   \,.
\eeq
The first term on the right--hand side of the above equation, $V^{\rm TD}_{2 \pi}$, gives the contribution of
the irreducible graphs (b) and (c) which,
for static nucleons, is the same as in the previously considered case. The contribution of reducible diagrams (d--g) is given by the last two 
terms in the above equation. The resulting potential has the form \citep{Epelbaum:1999dj}:
\beq
V^{\rm UT}_{2 \pi} = - \frac{g_A^4}{2 (2 F_\pi)^4} \, \int \, \frac{d^3 l}{ (2 \pi )^3} \,
\frac{\omega_+^2 + \omega_+ \omega_- + \omega_-^2}
{\omega_+^3 \omega_-^3 (\omega_+ + \omega_- )} \Bigg\{  
\fet \tau_1 \cdot \fet \tau_2
\left( {\vec l \,}^2 
- {\vec q \,}^2 \right)^2 + 6 ( \vec \sigma_2 \cdot [ \vec q \times \vec l \, ] ) 
( \vec \sigma_1 \cdot [ \vec q \times \vec l \, ] ) \Bigg\} \; . \nonumber
\eeq
Notice that the isoscalar central and isovector tensor components in Eq.~(\ref{2PEboxTD})
cancel in the method of unitary transformation against the contributions from reducible diagrams. 
We will discuss the chiral $2\pi$--exchange potential in more detail in section \ref{sec:piexch}. 
 
Before closing this section, let us mention two other methods to derive 
energy--independent potentials  used in the context of chiral EFT. 
Historically, energy--independent expressions for the chiral $2\pi$--exchange potential at order $\nu=2$ 
were first obtained by Friar and Coon \citep{Friar:1994zz} using the method described in \citep{Friar:1977xh}.
Yet another approach was applied e.g.~in Refs.~\citep{Kaiser:1997mw,Kaiser:1999ff,Kaiser:1999jg,Kaiser:2001pc}
to study chiral $2\pi$-- and $3\pi$--exchange forces, in which the potential is determined through matching 
to the $S$--matrix. 

Last but not least, one should always keep in mind that, in contrast to the on--shell scattering 
amplitude, nuclear potentials themselves are not
experimentally observable  and can always be modified via 
a unitary transformation, see e.g.~\citep{Friar:1999sj} for some explicit examples.  
This non--uniqueness of the nuclear forces should, of course, not be considered as 
a conceptual problem. A similar sort of non--uniqueness at the level of 
the Lagrangian is well 
known in quantum field theory, where one has the freedom to perform nonlinear field redefinitions. 
Notice that unitary transformations will, in general, affect not only few--nucleon 
forces but also the corresponding nuclear current operators. It is, therefore, important to have 
a \emph{consistent} (in the above mentioned sense) formulation for 2N, 3N, $\ldots$, forces and current 
operators. Such a consistent formulation is provided by the chiral EFT framework.

\subsection{Two--nucleon force}
\label{2NF}

The chiral NN force has the general form 
\beq
V_{\rm 2N} = V_{\pi} + V_{\rm cont}\,,
\eeq
where $V_{\rm cont}$ denotes the short--range terms represented by $NN$ contact interactions
and $V_{\pi}$ corresponds to the long--range part associated with the pion--exchange contributions
Both $V_{\pi}$ and $V_{\rm cont}$ are determined within the low--momentum expansion as will be discussed in 
the next sections.

\subsubsection{Regularization of the pion--exchange contributions}
\label{sec:sfr}

Let us now take a closer look at the pion--exchange contributions. The explicit form of the corresponding 
non--polynomial functions of momenta\footnote{Polynomial contributions to the potential are represented by a 
series of contact interactions $V_{\rm cont}$.} depends, to some extent,
 on the way one regularizes the corresponding loop integrals. 
Consider, for example, the isoscalar central part of the 2PE potential at order $\nu = 3$ which results from the 
triangle diagrams and is given by
\beq
\label{pot1}
V_{\rm C} (q) = \frac{3 g_A^2}{16 F_\pi^4} \int \, \frac{d^3 l}{(2 \pi)^3} 
\frac{l^2 - q^2}{\omega_-^2 \omega_+^2} 
\left( 8 c_1 M_\pi^2 + c_3 (l^2 - q^2) \right)\,,
\eeq
where  $q \equiv | \vec q \,|$, 
$l \equiv | \vec l \, |$ and $c_i$ are the corresponding LECs. 
The integral is cubically divergent and needs to be regularized. Applying 
dimensional regularization one finds:
%******************************************************************************
\beq
\label{pot2}
V_{\rm C}  (q) = - \frac{3 g_A^2}{16 \pi F_\pi^4}
\left( 2 M_\pi^2 ( 2 c_1 - c_3 ) - c_3 q^2 \right) (2 M_\pi^2 + q^2) \, A (q)\;,
\quad \quad \quad
A (q) = \frac{1}{2 q} \arctan \frac{q}{2 M_\pi} \,.
\eeq
Here, we do not show polynomial terms of the kind $\alpha + \beta q^2$
which contribute to $V_{\rm cont}$. 
%where the ellipses refer to polynomial (in $q^2$) terms of the kind $\alpha + \beta q^2$
%which contribute to $V_{\rm cont}$. 
%The corresponding loop function $A (q)$ has the form 
%\beq
%A (q) = \frac{1}{2 q} \arctan \frac{q}{2 M_\pi} \,.
%\eeq
It is instructive to express the potential using the spectral function representation:
\beq
\label{spectrfun}
V_{\rm C}  (q) = \frac{2 q^4}{\pi} \int_{2 M_\pi}^\infty d \mu \, \frac{1}{\mu^3}
\, \frac{\rho (\mu)}{\mu^2 + q^2},
\eeq
where the spectral function $\rho (\mu )$ can be obtained from 
$V_{\rm C} (q )$ in Eq.~(\ref{pot2}) via
\beq
\label{rho}
\rho (\mu ) = \Im \left[ V_{\rm C} (0^+ - i \mu ) \right]
= - \frac{3 g_A^2}{64 F_\pi^4} \left( 2 M_\pi^2 ( 2 c_1 - c_3) + c_3 \mu^2 \right)
(2 M_\pi^2- \mu^2) \frac{1}{\mu} \theta (\mu - 2 M_\pi )\,.
\eeq
In Eq.~(\ref{spectrfun}), the twice subtracted dispersion integral
is given which is needed in order to account for the large--$\mu$
behavior of $\rho (\mu)$. Eq.~(\ref{spectrfun}) shows that 
the 2PE potential resulting from Eq.~(\ref{pot1}) upon applying DR 
contains explicitly the short--range contributions associated with the integration over large values of $\mu$.  
These short--range contributions to the potential are an artifact of the chosen regularization 
procedure (i.e.~DR). They are model dependent  and cannot be predicted in the chiral EFT framework 
since the chiral expansion for the spectral function $\rho (\mu)$ is invalid for large values of $\mu$.  Instead of 
keeping the spurious short--range physics in the 2PE potential, one can perform the spectral function 
integral only over the low--$\mu$ region, where chiral EFT is applicable. This can be achieved using 
the regularized spectral function 
\beq
\label{regspectr}
\rho (\mu ) \rightarrow \rho^{\tilde \Lambda} (\mu ) = \rho (\mu ) \, \theta (\tilde \Lambda - \mu )\,,
\eeq
with the reasonably chosen finite ultraviolet cut--off $\tilde \Lambda$ 
which prevents that the 
regularized 2PE potential $V_{\rm C}^{\tilde \Lambda}$ given by 
\beq
\label{pot_sfr}
V_{\rm C}^{\tilde \Lambda} = \frac{2 q^4}{\pi} \int_{2 M_\pi}^\infty d \mu \, \frac{1}{\mu^3}
\, \frac{\rho^{\tilde \Lambda} (\mu)}{\mu^2 + q^2}
 = - \frac{3 g_A^2}{16 \pi F_\pi^4}
\left( 2 M_\pi^2 ( 2 c_1 - c_3 ) - c_3 q^2 \right) (2 M_\pi^2 + q^2) \, A^{\tilde \Lambda} (q) + \ldots \,,
\eeq
%\beqa
%\label{pot_sfr}
%V_{\rm C}^{\tilde \Lambda} &=& \frac{2 q^4}{\pi} \int_{2 M_\pi}^\infty d \mu \, \frac{1}{\mu^3}
%\, \frac{\rho^{\tilde \Lambda} (\mu)}{\mu^2 + q^2}\nn
%& =& - \frac{3 g_A^2}{16 \pi F_\pi^4}
%\left( 2 M_\pi^2 ( 2 c_1 - c_3 ) - c_3 q^2 \right) (2 M_\pi^2 + q^2) \, A^{\tilde \Lambda} (q) + \ldots 
%\eeqa
with the ellipses referring to polynomial (in $q^2$) terms 
has components with the range $r < \tilde \Lambda^{-1}$.  In the above equation, the regularized loop 
function $A^{\tilde \Lambda} (q)$ turns out to be 
\beq
\label{ASFR}
A^{\tilde \Lambda} (q) = \theta ( \tilde \Lambda - 2 M_\pi ) \frac{1}{2 q} 
\arctan \frac{q (\tilde \Lambda - 2 M_\pi )}{q^2 + 2 \tilde \Lambda M_\pi}\,.
\eeq
In what follows, we will refer to the above described regularization scheme, which has been introduced
in \citep{Epelbaum:2003gr}, as to spectral function regularization (SFR). 

What is the relation between the DR and SFR potentials in Eqs.~(\ref{pot2}) and (\ref{pot_sfr})?
To see that one can rewrite the spectral function integral in Eq.~(\ref{spectrfun}) as follows:
\beqa
\frac{2 q^4}{\pi} \int_{2 M_\pi}^\infty d \mu \, \frac{1}{\mu^3}
\, \frac{\rho (\mu)}{\mu^2 + q^2} &=& 
\frac{2 q^4}{\pi} \int_{2 M_\pi}^{\tilde \Lambda} d \mu \, \frac{1}{\mu^3}
\, \frac{\rho (\mu)}{\mu^2 + q^2} + \frac{2 q^4}{\pi} \int_{\tilde \Lambda}^\infty d \mu \, \frac{1}{\mu^3}
\, \frac{\rho (\mu)}{\mu^2 + q^2} \nn
& \stackrel{q < \tilde \Lambda}{\longrightarrow}& \frac{2 q^4}{\pi} \int_{2 M_\pi}^\infty d \mu \, \frac{1}{\mu^3}
\, \frac{\rho^{\tilde \Lambda} (\mu)}{\mu^2 + q^2} + \alpha q^4 + \beta q^6 + \ldots \,,
\eeqa
where 
\beq
\alpha =  \frac{2}{\pi} \int_{\tilde \Lambda}^\infty d \mu \, \frac{\rho (\mu)}{\mu^5} \,,
\quad \quad
\beta =  - \frac{2}{\pi} \int_{\tilde \Lambda}^\infty d \mu \, \frac{\rho (\mu)}{\mu^7} \,,
\quad \quad
\ldots \,.
\eeq
It is, therefore, obvious, that the DR and SFR potentials differ from each other 
by an infinite series of higher--order contact interactions. In fact, they might be viewed as 
two different conventions to define the non--polynomial part of the two-- and more--pion exchange 
potential. DR corresponds to the convention, according to which the non--polynomial part includes 
components of arbitrarily short range which are strongly model--dependent. In contrast, the SFR 
approach uses the convention, according to which only the components with the range $r > \tilde \Lambda^{-1}$ 
are explicitly kept in the non--polynomial part of the potential while all 
shorter--range 
contributions are represented by the contact interactions. In general, for quickly converging expansions, 
both the DR and SFR methods are completely equivalent provided the ultraviolet cut--off $\tilde \Lambda $ 
is chosen to be large enough. For example, we will see in 
section \ref{peripheral}, that both schemes lead to similar results for peripheral NN scattering at order $\nu = 2$. 
If, however, the convergence for some well understood physical reason is slow and (some) 
observables become sensitive to higher--order counter terms, it is safer to avoid the 
spurious short--distance contributions kept in DR. In such a case, SFR is a preferable choice. 
An example of such a situation will be considered in section \ref{peripheral}. 
Notice that a similar approach based on a finite momentum cut--off was 
used in \citep{Donoghue:1998aa,Donoghue:1998bs,Borasoy:2002jv} to deal with the slow convergence in 
the SU(3) baryon CHPT, see also \citep{Leinweber:2003dg} for a recent application to the chiral extrapolation
of the lattice QCD results and  \cite{Bernard:2003rp,Frink:2005ru} for a discussion on cut--off schemes in CHPT. 

It is also instructive to compare the DR and SFR potentials in configuration space. For $r > 0$, 
the  inverse Fourier--transform can be expressed in terms of the spectral function 
$\rho (\mu )$ via
\beq
\label{four}
V_{\rm C} (r ) = \frac{1}{2 \pi^2 r} \int_{2 M_\pi}^\infty d \mu
\, \mu \, e^{- \mu r} \rho (\mu ).
\eeq
Substituting $\rho (\mu )$ from Eq.~(\ref{rho}) in to Eq.~(\ref{four}), one obtains the following expression for 
the potential corresponding to DR:
\beq
V_C (r) =  \frac{3 g_A^2}{32 \pi^2 F_\pi^4} \, \frac{e^{- 2 x}}{r^6}
\bigg[ 2 c_1 \, x^2 (1 + x)^2  + c_3 (6 + 12x + 10 x^2 + 4 x^3 + x^4) \bigg]\,,
\eeq
where we have introduced  $x = M_\pi r$. Using the regularized expression for the spectral function in Eq.~(\ref{regspectr}), 
one obtains for the SFR potential:
\beqa
V_C^{\tilde \Lambda} (r) &=&  V_C (r) -  \frac{3 g_A^2}{128 \pi^2 F_\pi^4} \, \frac{e^{- y}}{r^6} \bigg[
4 c_1 x^2 \Big(2 + y ( 2 + y) - 2 x^2   \Big) \nn
&& {} + c_3 \Big(  24 +  y (24 + 12 y  + 4 y^2 + y^3   ) - 4 x^2 (2 + 2 y  + y^2 ) + 4  x^4 \Big) \bigg]\,,
\eeqa
where $y = \tilde \Lambda r$. 

\begin{figure*}
%\vspace{0.3cm}
\centerline{
\psfig{file=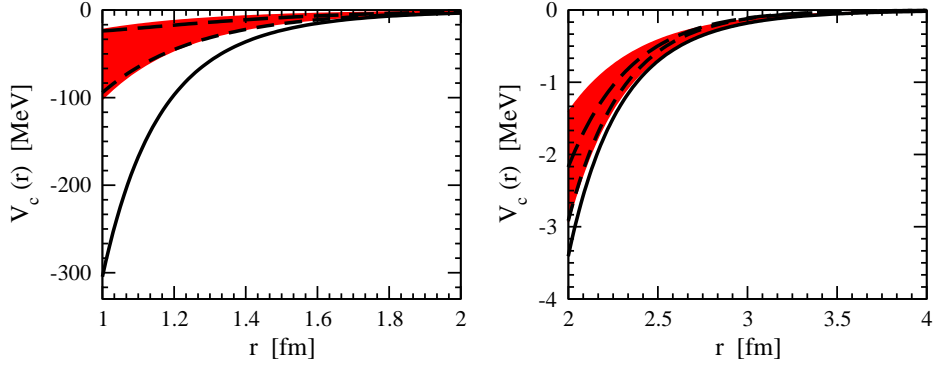,width=12.6cm}
}
\vspace{-0.1cm}
\caption[fig4a]{\label{fig4a} The potential $V_C$  in $r$--space. 
The solid line (band) shows the DR (SFR, $\tilde \Lambda = 500 \ldots 800$ MeV) result. The short--dashed (long--dashed) line 
refers to the phenomenological $\sigma$ ($\sigma + \omega + \rho$) contributions based on the
isospin triplet configuration space version (OBEPR) of the Bonn potential \citep{Machleidt:1987hj}. }
\vspace{0.2cm}
\end{figure*}

In Fig.~\ref{fig4a} we compare the isoscalar central part of 2PE obtained using DR and SFR
for the central values of the LECs $c_{1,3}$, $c_1 = -0.81$ GeV$^{-1}$ and $c_3 = -4.70$ GeV$^{-1}$, 
from Ref.~\citep{Buettiker:1999ap}. 
Clearly, the large--distance asymptotics of the potential, 
which is constrained in a nontrivial way by chiral symmetry of QCD, is unaffected by the cut--off 
procedure (provided $\tilde \Lambda \gg M_\pi$). The strongest effects of the cut--off are observed at intermediate 
and shorter distances, where 2PE becomes unphysically attractive if DR is used. In contrast,
removing the large components in the mass spectrum of the 2PE with the reasonably chosen cut--off 
$\tilde \Lambda =500 \ldots 800$ MeV greatly reduces this attraction and yields the 
potential of the same order in magnitude as the one obtained in phenomenological 
boson--exchange models. We will see in section \ref{peripheral} how a choice of regularization 
affects the results for peripheral NN scattering at N$^2$LO. 

We further stress that other regularization schemes may be applied as well. For example, 
one can regularize divergent loop integrals using an ordinary momentum--space cut--off. 
The prominent feature of the SFR scheme is given by the fact that it only affects 
the two--nucleon contact interactions. One can, therefore, directly adopt the values for 
various LECs resulting from the single--nucleon sector analyses, where dimensional 
regularization has been used. This, in general, is not the case for a  finite momentum 
cut--off regularization.

\subsubsection{Pion--exchange contributions}
\label{sec:piexch}

Consider now pion--exchange contributions to the potential 
\beq
V_{\pi} = V_{1\pi} + V_{2\pi} + V_{3\pi} + \ldots \,,
\eeq
where one--, two--  and three--pion exchange (3PE) contributions
$V_{1\pi}$,  $V_{2\pi}$ and  $V_{3\pi}$ can be written in the low--momentum expansion as  
\beqa
\label{Vschem}
V_{1\pi} &=&   V_{1\pi}^{(0)} +  V_{1\pi}^{(2)} +  V_{1\pi}^{(3)} + V_{1\pi}^{(4)} + \ldots \,, \nn
V_{2\pi} &=&   V_{2\pi}^{(2)} +  V_{2\pi}^{(3)} + V_{2\pi}^{(4)} + \ldots \,, \nn
V_{3\pi} &=&   V_{3\pi}^{(4)}  + \ldots \,. 
\eeqa
Here, the superscripts denote the corresponding chiral order and the ellipses refer to 
$(Q/\Lambda)^5$-- and higher order terms. 
Contributions due to the exchange of four-- and more pions are further suppressed:
$n$--pion  exchange diagrams start to contribute at the order $(Q/\Lambda)^{2n-2}$. 
Notice further that in this section we restrict ourselves to isopin--invariant contributions. 
Isospin--breaking corrections will be discussed in section \ref{sec:isosp}.  
The corresponding relativistic corrections will be considered in section \ref{sec:rel}.

The static 1PE potential at N$^3$LO has the form  
\beq
\label{opep}
 V_{1\pi}^{(0)} +  V_{1\pi}^{(2)} +  V_{1\pi}^{(3)} + V_{1\pi}^{(4)}  
= -\biggl(\frac{g_A}{2F_\pi}\biggr)^2 \, ( 1 + \delta )^2 \,
\fet \tau_1 \cdot \fet \tau_2 \,
\frac{\vec{\sigma}_1 \cdot\vec{q}\,\vec{\sigma}_2\cdot\vec{q}}
{\vec q \, ^2 + M_\pi^2}\,.
\eeq
Here $\delta$ denotes an isospin--conserving Goldberger--Treiman discrepancy
\beq
\label{GTD}
\delta = - \frac{2 d_{18}}{g_A} M_\pi^2 + \kappa M_\pi^4 \,,
\eeq
where $d_{18}$ is a  LEC from the dimension three $\pi N$ Lagrangian and the constant $\kappa$ 
determines the size of the first correction to the Goldberger--Treiman discrepancy. 
Here and in what follows, the expressions for the nuclear force should be understood as  
operators with respect to spin and isospin quantum numbers and matrix elements with 
respect to momentum variables. We further stress that all 
one-- and two--loop $1\pi$--exchange diagrams at this order lead to renormalization of various LECs  
without introducing any form--factor--like behavior. 
The derivation of the 1PE potential to one loop in 
the method of unitary transformation is discussed in
detail in Ref.~\citep{Epelbaum:2002gb}.

\begin{figure*}
%\vspace{0.3cm}
\hspace*{0.8cm}
\psfrag{xxx}{\raisebox{-0.15cm}{\hskip -0.8 true cm  {$\bigg( \frac{Q}{\Lambda} \bigg)^2$:}}}
\psfrag{yyy}{\raisebox{-0.15cm}{\hskip -0.8 true cm   {$\bigg( \frac{Q}{\Lambda} \bigg)^3$:}}}
\psfrag{zzz}{\raisebox{-0.15cm}{\hskip -0.8 true cm   {$\bigg( \frac{Q}{\Lambda} \bigg)^4$:}}}
\psfrag{aaa}{\raisebox{-0.15cm}{\hskip -1.1 true cm   {where:}}}
\psfig{file=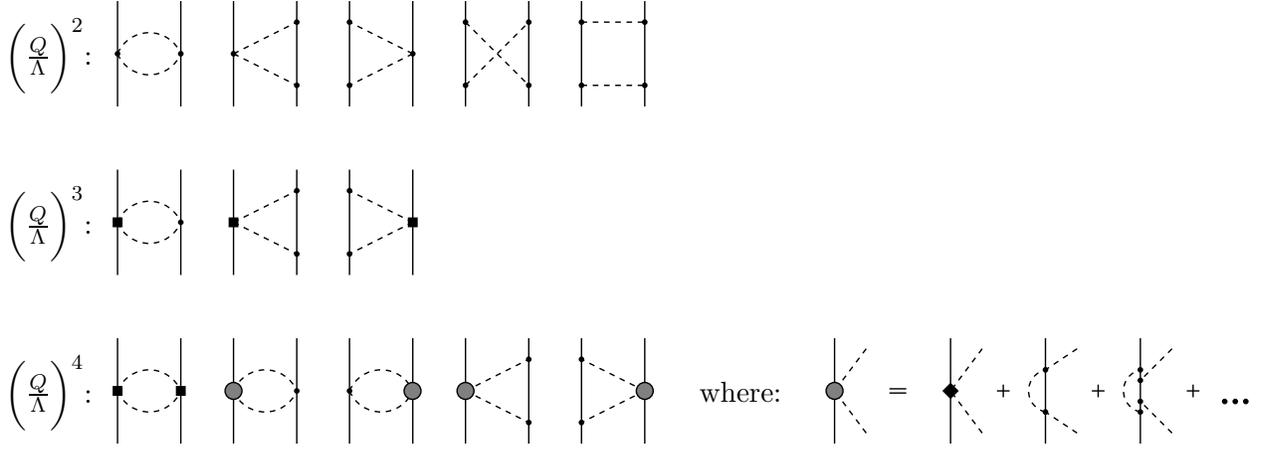,width=16.2cm}
\vspace{0.0cm}
\caption[fig5]{\label{fig5}  Leading, subleading and sub-subleading contributions to the chiral 
$2\pi$--exchange potential. Solid (dashed) lines correspond to nucleons (pions).
Solid dots, filled rectangles and filled diamonds represent vertices with $\Delta_i =0$, $1$ and $2$, respectively. 
Shaded blob denotes the next--to--next--to--leading order contribution to the pion--nucleon scattering amplitude.}
\vspace{0.2cm}
\end{figure*}

We now turn to the 2PE contributions. It is convenient to 
express $V_{2 \pi}$ in the CMS in the form:
\beqa
\label{2PEdec}
V_{2 \pi} &=& V_C + \fet \tau_1 \cdot \fet \tau_2 \, W_C + \left[   
V_S + \fet \tau_1 \cdot \fet \tau_2 \, W_S \right] \, \vec \sigma_1 \cdot \vec \sigma_2 
+ \left[ V_T + \fet \tau_1 \cdot \fet \tau_2 \, W_T \right] 
\, \vec \sigma_1 \cdot \vec q \, \vec \sigma_2 \cdot \vec q \\
&+&    \left[   
V_{LS} + \fet \tau_1 \cdot \fet \tau_2 \, W_{LS} \right] \, i ( \vec \sigma_1 + \vec \sigma_2 )
\cdot ( \vec q \times \vec k  ) 
+   \left[   
V_{\sigma L} + \fet \tau_1 \cdot \fet \tau_2 \, W_{\sigma L} \right] \,   \vec \sigma_1 
\cdot (\vec  q \times \vec k  ) \vec \sigma_2 \cdot (\vec  q \times \vec k  ) \,,
\nonumber
\eeqa
where the superscripts $C$, $S$, $T$, $LS$ and $\sigma L$ of the scalar functions 
$V_C$, $\ldots$, $W_{\sigma L}$ refer to central, spin--spin, tensor, spin--orbit and 
quadratic spin--orbit components, respectively. The chiral 2PE potential $V_{2\pi}^{(2)} +  V_{2\pi}^{(3)}$
is discussed in \citep{Friar:1994zz,Kaiser:1997mw,Epelbaum:1998ka,Epelbaum:1999dj} and in \citep{Ordonez:1995rz} 
using an energy--dependent formalism. For a related work see also \cite{Robilotta:1996ji}. 
The NLO 2PE potential is given by the contributions of the box, crossed--box, 
triangle and football diagrams shown in the first line of Fig.~\ref{fig5} which in the energy--independent formulation read
\beqa
\label{2PE_nlo}
W_C^{(2)} (q) &=& - \frac{1}{384 \pi^2 F_\pi^4}\,
L^{\tilde \Lambda} (q) \, \biggl\{4M_\pi^2 (5g_A^4 - 4g_A^2 -1)  + q^2(23g_A^4 - 10g_A^2 -1)
+ \frac{48 g_A^4 M_\pi^4}{4 M_\pi^2 + q^2} \biggr\} \,, \nn
V_T^{(2)} (q) &=& -\frac{1}{q^2} V_S^{(2)} (q)  = - \frac{3 g_A^4}{64 \pi^2 F_\pi^4} \,L^{\tilde \Lambda} (q)\,.
\eeqa
Here, the loop function $L^{\tilde \Lambda} (q)$ is given by 
\beq
\label{def_LA}
L^{\tilde \Lambda} (q) = \theta (\tilde \Lambda - 2 M_\pi ) \, \frac{\omega}{2 q} \, 
\ln \frac{\tilde \Lambda^2 \omega^2 + q^2 s^2 + 2 \tilde \Lambda q 
\omega s}{4 M_\pi^2 ( \tilde \Lambda^2 + q^2)}~, \quad \quad
\omega = \sqrt{ q^2 + 4 M_\pi^2}~,  \quad \quad
s = \sqrt{\tilde \Lambda^2 - 4 M_\pi^2}\,.
\eeq
These expressions are based on the SFR approach. The corresponding DR expressions can be obtained taking the limit  
$\tilde \Lambda \to \infty$. We further emphasize that a significant part of the NLO 2PE contributions was 
considered much earlier in the context of meson theory of nuclear forces, see 
e.g.~\citep{Taketani:1952aa,Brueckener:1953aa,Sugawara:1960aa,Sugawara:1968ty}.
At N$^2$LO one has to take into account the contributions of the triangle and 
 football  graphs in the second line of Fig.~\ref{fig5} with a single insertion of the subleading $\pi \pi NN$ vertex. One finds:
\beqa
\label{2PE_nnlo}
V_C^{(3)} (q) &=& -\frac{3g_A^2}{16\pi F_\pi^4}  \biggl\{2M_\pi^2(2c_1 -c_3) -c_3 q^2 \biggr\} 
(2M_\pi^2+q^2) A^{\tilde \Lambda} (q)\,, \nn
W_T^{(3)} (q) &=& -\frac{1}{q^2} W_S^{(3)} (q) = - \frac{g_A^2}{32\pi F_\pi^4} \,  c_4 (4M_\pi^2 + q^2) 
A^{\tilde \Lambda}(q)\,,
\eeqa
where the N$^2$LO loop function $A^{\tilde \Lambda} (q)$ has been defined in Eq.~(\ref{ASFR}). 

The N$^3$LO corrections to the 2PE potential $V_{2\pi}^{(4)}$ have been 
recently calculated by Kaiser \citep{Kaiser:2001pc} and are schematically depicted in the third line of Fig.~\ref{fig5}.  
They arise from two groups of diagrams, the one--loop  football graphs with both dimension two $\pi \pi NN$ vertices of the 
$c_{1,\ldots , 4}$--type and the diagrams which contain the third order  pion--nucleon amplitude and lead to 
one--loop and two--loop graphs. We begin with the first group of corrections, for which one finds:
\beqa
\label{2PE_nnnlo}
V_C^{(4)} (q) &=& \frac{3}{16 \pi^2  F_\pi^4} \, L^{\tilde \Lambda} (q) \, 
\left\{ \left[ \frac{c_2}{6} \omega^2 + c_3 (2 M_\pi^2 + q^2 ) - 4 c_1 M_\pi^2 
\right]^2 + \frac{c_2^2}{45} \omega^4 \right\} \,,
\nonumber \\
W_T^{(4)} (q) &=& -\frac{1}{q^2} W_S^{(4)} (q) = \frac{c_4^2}{96 \pi^2 F_\pi^4} \omega^2  
\, L^{\tilde \Lambda} (q)\,.
\eeqa
The expressions for the second group of corrections were obtained by Kaiser \citep{Kaiser:2001pc} and are 
given in appendix \ref{sec:SF} in terms of the corresponding spectral functions. 

\begin{figure*}[t]
%\vspace{0.3cm}
\centerline{
\psfig{file=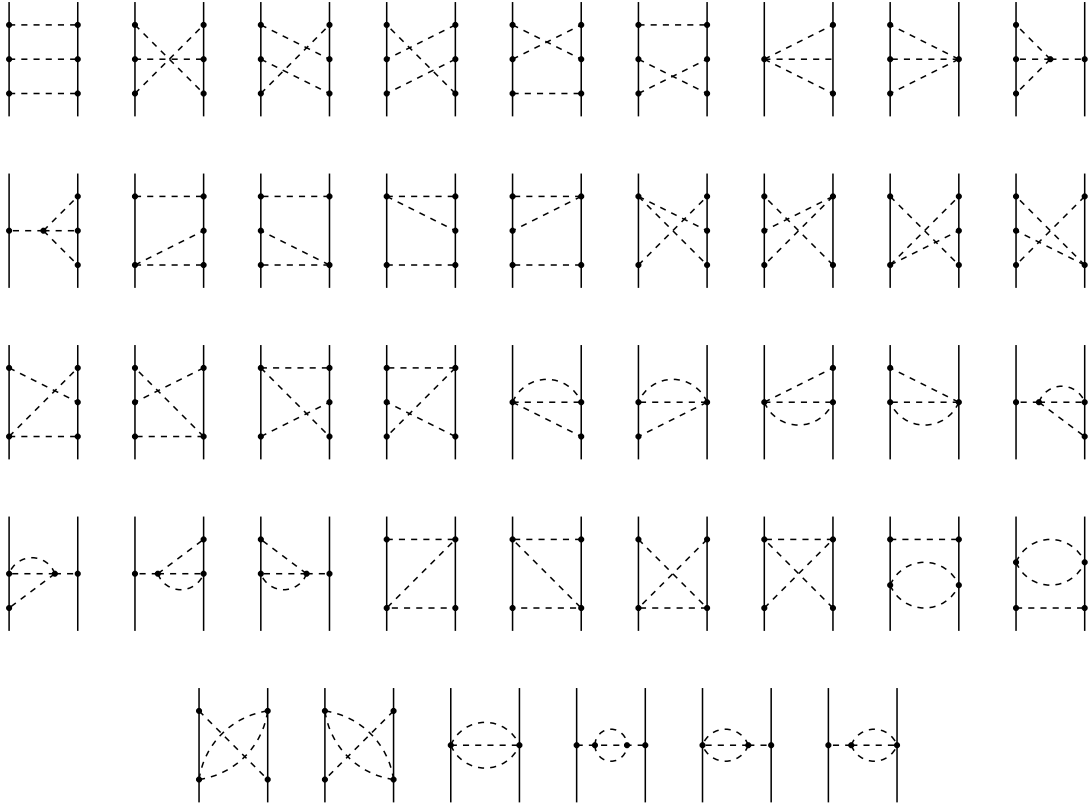,width=15.0cm}
}
\vspace{-0.1cm}
\caption[fig6]{\label{fig6}  Leading contributions to the $3\pi$--exchange potential.
For notation see Fig.~\ref{fig5}.}
\vspace{0.2cm}
\end{figure*}

\begin{figure*}[tb]
%\vspace{0.3cm}
\centerline{
\psfig{file=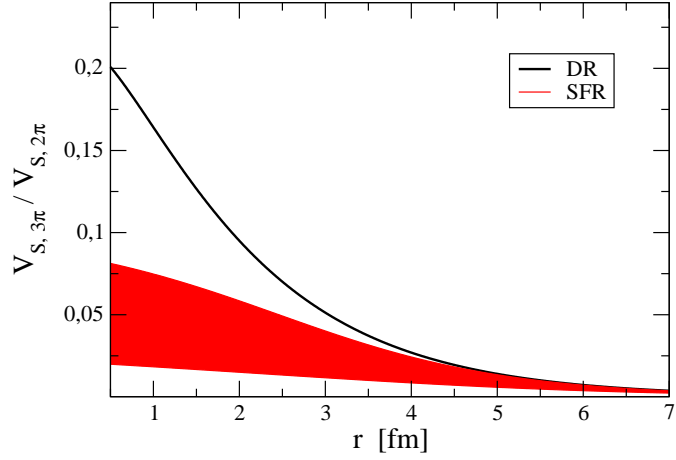,width=9cm}}
\vspace{-0.15cm}
\caption[fig7]{\label{fig7} The ratio of the isoscalar spin--spin 3PE and 
2PE N$^3$LO contributions using dimensional and spectral function 
regularization. The cut--off in the spectral function varies in the 
range $\tilde \Lambda = 500 \ldots 700$ MeV. }
\vspace{0.2cm}
\end{figure*}

Three--pion exchange starts to contribute at N$^3$LO and is given by diagrams shown in Fig.~\ref{fig6}. 
The corresponding expressions 
for the spectral functions and the potential (obtained using dimensional regularization)
have been given by Kaiser in \citep{Kaiser:1999ff,Kaiser:1999jg}, see also \citep{Pupin:1999ba} for a related work. 
It has been pointed out in these 
references that the resulting 3PE potential is much weaker than the N$^3$LO 2PE contributions
at physically interesting distances $r > 1$ fm. Having the explicit expressions for the 
3PE spectral functions, it is easy to calculate the potential in the SFR scheme.
It is obvious, even without performing the explicit calculations, that the finite--range 
part of the 3PE potential in the SFR scheme is strongly suppressed at intermediate 
and short distances compared to the result obtained using DR. This is because the 
short range components which dominate the 3PE spectrum are explicitly excluded 
in this approach. 
This is exemplified in Fig.~\ref{fig7}  for the case of the N$^3$LO isoscalar spin--spin contribution 
proportional to $g_A^4$, which has been found in \citep{Kaiser:1999ff,Kaiser:1999jg} to yield
the strongest 3PE potential for $0.6$ fm $< r <$ $1.4$ fm.
%*****************************************************************************
%In Fig.~\ref{fig7} we show the ratio of the N$^3$LO isoscalar spin--spin contributions of 3PE and 2PE
%using both regularization schemes for a wide range of $r$.
%It turns out that the 3PE contribution reaches for $r > 0.5$ fm at most $2\%-8\%$
%*****************************************************************************
For $r > 0.5$ fm, it reaches at most $2\%-8\%$ (depending on the 
choice of the spectral function cut--off)
%*****************************************************************************
of the corresponding N$^3$LO 2PE contribution \citep{Epelbaum:2004fk}. 
%depending on the choice of the spectral function cut--off. 
Similar results have been found for other leading 3PE contributions 
\citep{Kaiser:1999ff,Kaiser:1999jg}. For that reason the 3PE contributions have been neglected in the 
present N$^3$LO analyses \citep{Entem:2003ft,Epelbaum:2004fk}. Notice, however, that the potential resulting from 
subleading (i.e.~N$^4$LO) 3PE diagrams proportional to LECs $c_i$ and obtained using DR was found to be sizable  
at intermediate distances \citep{Kaiser:2001dm}.  The strength of the 2PE and 3PE contributions considered 
above and the corresponding expressions in coordinate space are discussed in detail 
in Refs.~\citep{Kaiser:1997mw,Kaiser:1999jg,Kaiser:1999ff,Kaiser:2001at,Kaiser:2001pc}.

\subsubsection{Contact terms}
\label{sec:cont}

The short--range part of the potential is represented by a series of contact interactions  
\beq
V_{\rm cont} = V^{(0)}_{\rm cont} + V^{(2)}_{\rm cont} + V^{(4)}_{\rm cont} + \ldots ~,
\eeq
where the superscripts denote the corresponding chiral order as defined in Eq.~(\ref{powcNN}). 
These terms feed into the matrix--elements of the two S--, P-- and D--waves
and the two lowest transition potentials in the following way:
\beqa\label{VC}
\langle S | V_{\rm cont}| S \rangle&=& \tilde C_{S} + C_{S} ( p^2 + p '^2) +
D_{S}^1 \, p^2 \, {p'}^2 + D_{S}^2 \, ({p}^4+{p}'^4)~,\nn
\langle P | V_{\rm cont}| P \rangle&=& C_{P} \, p \, p' +
D_{P} \,p \, p' \, ({p}^2 +  p' \, ^2)~,\nn
\langle D | V_{\rm cont}| D \rangle&=& D_{D} \, {p}^2\,{p}'^2~,\nn
\langle ^3S_1 | V_{\rm cont}| ^3D_1 \rangle&=& C_{3D1 - 3S1} \, p^2 
+ D_{3D1 - 3S1}^1 \, {p}^2\,{p}'^2 + D_{3D1 - 3S1}^2 \, p^4~,\nn
\langle ^3D_1 | V_{\rm cont}| ^3S_1 \rangle&=& C_{3D1 - 3S1} \, {p '}^2 
+ D_{3D1 - 3S1}^1 \, {p}^2\,{p}'^2 + D_{3D1 - 3S1}^2 \, {p '}^4~,\nn
\langle ^3P_2 | V_{\rm cont}| ^3F_2 \rangle&=& D_{3F2 - 3P2}\, {p}^3\,{p}'~, \nn
\langle ^3F_2 | V_{\rm cont}| ^3P_2 \rangle&=& D_{3F2 - 3P2}\, {p}\,{p '}^3~, 
\label{V4ct}
\eeqa
where $p = |\vec{p}\,|$, ${p}' = |\vec{p}\,'|$ and the subscripts $S = \{ 1S0, \, 3S1 \}$, $P = \{ 1P1, \, 3P0 , \, 3P1, \, 3P2 \}$, 
$D = \{ 1D2, \, 3D1 , \, 3D2, \, 3D3 \}$ refer to the corresponding channel. 
%The relations between the spectroscopic LECs and the ones in Eq.~(\ref{Vcon}) can be found in \citep{Epelbaum:2004fk}.
%*********************************************************************************************
The relations between the spectroscopic LECs in the above equations and the ones 
that occur in the Lagrangian can be found in \citep{Epelbaum:2004fk}.
Isospin--breaking short--range corrections will be specified in section \ref{sec:isosp}.
%*********************************************************************************************

\subsubsection{Relativistic corrections}
\label{sec:rel}

The first relativistic corrections to the nuclear force appear at order $\nu = 4$ provided that
the nucleon mass is counted according to Eq.~(\ref{m_counting}). They result from both 
$1/m^2$--corrections to the static 1PE potential and  $1/m$--corrections to the 
order   $\nu =2$ static 2PE potential. In addition, at this order one also needs to correct the 
nonrelativistic expression for the nucleon kinetic energy
\beq
E_{\rm kin} = \sqrt{\vec p \, ^2 + m^2} - m \sim \frac{\vec p \, ^2}{2 m} - \frac{\vec p \, ^4}{8 m^3}\,,
\eeq
which enters the corresponding dynamical equation. 
Equivalently, one can use the full, not expanded expression for the nucleon kinetic energy which leads to the 
following Schr\"odinger equation for two nucleons in the CMS:
\beq
\label{schroed_rel}
\left[ \left( 2 \sqrt{\vec p \, ^2 + m^2}  - 2 m \right) + V \right] \Psi
= E \Psi\,,
\eeq
where $m = m_p$ for the $pp$,  $m = m_n$ for the $nn$  and 
$m = 2 m_p m_n/(m_p + m_n)$ for the $np$ case, see \citep{Epelbaum:2004fk} for more details
on the kinematics.
%The relation between the CMS momentum $\vec p$ and $E_{\rm lab}$ reads:
%\begin{itemize}
%\vspace{-0.3cm}
%\item
%Proton--proton case: $\vec p \,^2 = \frac{1}{2} m_p E_{\rm lab}$, $m = m_p$,
%\item
%Neutron--neutron case: $\vec p \, ^2 = \frac{1}{2} m_n E_{\rm lab}$, $m = m_n$,
%\item
%Neutron--proton case: $\vec p \, ^2 = \frac{m_p^2 E_{\rm lab} (E_{\rm lab} + 2 m_n)}{(m_n+m_p)^2 + 2 E_{\rm lab} m_p}$, 
%$m = \frac{ 2 m_p m_n}{m_p + m_n}$.
%\end{itemize}
%\noindent
%The approximation made in the neutron--proton case, where $E_{\rm lab}$ refers to the neutron energy, 
%is valid modulo terms that go beyond the considered order. 
Notice that Eq.~(\ref{schroed_rel}) can be cast into equivalent nonrelativistic forms, i.e.~into the Schr\"odinger 
equations with the nucleon kinetic energy  $E_{\rm kin} = \vec p \, ^2/(2 m)$ \citep{Friar:1999sj,Kamada:1999wy}.\footnote{The 
two forms of the resulting nonrelativistic Schr\"odinger equation discussed in \citep{Epelbaum:2004fk} differ from each other 
by the definition of the nonrelativistic CMS momentum.} The advantage of Eq.~(\ref{schroed_rel}) versus the corresponding
nonrelativistic forms is that it can easily be generalized to the case of three and more nucleons.  
Changing the form of the Schr\"odinger equation causes changes in the relativistic corrections to the nuclear potential. 
Relativistic corrections to the interaction can, therefore, only be defined within a particular framework. For 
the Schr\"odinger equation (\ref{schroed_rel}), the corrections to the leading 1PE potential $V_{1 \pi}^{(0)}$ take
in the NN CMS the form:
\beq
\label{1pep_rel}
V_{1 \pi}^{(4)} = - \left( \frac{\vec p \, ^2 + \vec p \,' {}^2}{2 m^2} \right)    V_{1 \pi}^{(0)}\,,
\eeq 
where $\vec p $ and $\vec p \,'$ are the NN initial and final CMS momenta. 
This choice of corrections is sometimes called the ``minimal nonlocality'' choice, see \citep{Friar:1999sj}
and references therein. 
The corresponding $1/m$--corrections to the 2PE potential read:
\beqa
\label{tpe1m}
V_C^{(4)} (q) &=& \frac{3 g_A^4}{512  \pi m F_\pi^4} \bigg\{ \frac{2 M_\pi^5}{\omega^2}-
 3  ( 4 M_\pi^4 - q^4 )  A^{\tilde \Lambda} (q) \bigg\}\,,\nonumber \\
W_C^{(4)} (q) &=& \frac{g_A^2}{128 \pi m F_\pi^4} \Bigg\{ \frac{3 g_A^2 M_\pi^5}{\omega^2} 
- \bigg[ 4 M_\pi^2 + 2 q^2 - g_A^2 \left( 7 M_\pi^2 + \frac{9}{2} q^2 \right) \bigg] (2 M_\pi^2 + q^2 )
  A^{\tilde \Lambda} (q) \Bigg\}\,, \nn
V_T^{(4)} (q) &=& -\frac{1}{q^2} V_S^{(4)} (q) =  \frac{9 g_A^2}{512 \pi m F_\pi^4}  \left(
4 M_\pi^2 + \frac{3}{2} q^2 \right) A^{\tilde \Lambda} (q) \,, \nn
W_T^{(4)} (q) &=& -\frac{1}{q^2} W_S^{(4)} (q) = - \frac{g_A^2}{256 \pi m F_\pi^4}  \left[
8 M_\pi^2 + 2 q^2 - g_A^2 \left( 8 M_\pi^2 + \frac{5}{2} q^2 \right) \right]  A^{\tilde \Lambda} (q) \,, \nn
V_{LS}^{(4)} (q) &=& - \frac{3 g_A^4}{64 \pi m F_\pi^4} (2 M_\pi^2 + q^2 ) A^{\tilde \Lambda} (q) \,, \nn
W_{LS}^{(4)} (q) &=& - \frac{g_A^2 ( 1 - g_A^2)}{64 \pi m F_\pi^4} 
(4 M_\pi^2 + q^2 ) A^{\tilde \Lambda} (q) \,.
\eeqa
Notice that the above expressions differ from the ones given in \citep{Kaiser:1997mw} due to the different form 
of the dynamical equation and relativistic corrections to the 1PE potential employed in the present work. 
For an extensive discussion of this issue the reader is referred to 
Ref.~\citep{Friar:1999sj} where the dependence of relativistic corrections on certain kinds of unitary 
transformations is studied and the general expressions for $1/m^2$--corrections to the 1PE potential and 
$1/m$--corrections to the leading 2PE potential are obtained. We further stress that if the nucleon mass is 
counted, contrary to Eq.~(\ref{m_counting}),  as $m \sim \Lambda$, $1/m^4$--corrections to the 1PE potential,  
$1/m^2$--corrections to the order   $\nu=2$ 2PE potential and $1/m$--corrections to the order   $\nu=3$ 2PE potential 
have to be taken into account at N$^3$LO in addition to terms shown in Eqs.~(\ref{1pep_rel}), (\ref{tpe1m}). 
The corresponding expressions can be found in \citep{Kaiser:2001at}.\footnote{Notice that the expressions 
given in \citep{Kaiser:2001at} probably need to be adjusted in order to be made consistent with Eqs.~(\ref{schroed_rel}), 
(\ref{1pep_rel}) and (\ref{tpe1m}).}

\subsubsection{Isospin--breaking corrections}
\label{sec:isosp}

Within the Standard Model, isospin violation has its origin in 
the different masses of the up and down quarks and the electromagnetic interactions.
Consider first isospin--breaking in the strong interactions. 
The QCD quark mass term can be expressed in the two--flavor case as: 
\begin{equation}
\label{epsdef}
{\cal L}_{\rm mass}^{\rm QCD} = -\frac{1}{2}\bar{q} \, 
(m_{\rm u}+m_{\rm d})(1+\epsilon\tau_{3})\,q~, \quad \quad \quad \mbox{where} \quad \quad 
\epsilon \equiv {m_u-m_d \over m_u+m_d} \sim - {1 \over 3}\,.
\end{equation}
The above numerical estimation corresponds to the light quark mass values
based on a modified  $\overline{\rm MS}$ subtraction scheme
at a renormalization scale of 1~GeV \citep{Leutwyler:1996sa}. 
The isovector term ($\propto \tau_3$) in Eq.~(\ref{epsdef}) breaks isospin symmetry and generates 
a series of isospin--breaking effective interactions $\propto (\epsilon  M^2_\pi)^n$
with $n \geq 1$. It is, therefore, natural to count strong isospin violation in terms of 
$\epsilon  M^2_\pi$ \citep{VanKolck:1993ee}. 
Electromagnetic terms in the effective Lagrangian can be generated using the method of external sources, 
see e.g. \citep{Urech:1995aa,Meissner:1997fa,Meissner:1997ii,Muller:1999ww} for more details. 
All such terms are proportional to the nucleon charge matrix 
$Q_{\rm ch}= e \, (1 + \tau_3 )/2$, where $e$ denotes the electric charge.
More precisely, the vertices which contain (do not contain) the photon fields are proportional to $Q^{n}_{\rm ch}$
($Q^{2n}_{\rm ch}$), where $n=1,2,\ldots$. For processes in the absence of 
external fields, in which no photon can leave a Feynman diagram, it is convenient to introduce the 
small parameter $e^2 \sim 1/10$ for isospin--violating effects caused by the electromagnetic interactions. 
Isospin--violating terms in the effective Lagrangian at lowest orders can be found in 
\citep{VanKolck:1993ee,Steininger:1999aa,Fettes:2000gb,Gasser:2002am}, see also Ref.~\citep{Epelbaum:2005fd}.   

Isospin--breaking nuclear forces can, in principle, be derived in the EFT framework performing independent expansions in 
$Q/\Lambda$, $\epsilon$ and $e$. It is, however, convenient to relate these small parameters with each 
other in order to have a single expansion parameter. In Ref.~\citep{Epelbaum:2005fd}, the following counting 
rules were adopted:
\beq\label{CountRules}
\epsilon \sim e \sim \frac{Q}{\Lambda}, \quad \quad
\frac{e^2}{(4 \pi )^2}  \sim \frac{Q^4}{\Lambda^4}\,.
\eeq
The power counting expression in Eq.~(\ref{powcNN}) can be easily extended to include the contributions due to 
$n_{\gamma}$ virtual photons:
\beq
\label{powcNNphotons}
\nu = - 2 + 2 n_\gamma + 2 N + 2 (L -C)  + \sum_i V_i \Delta_i \,.
\eeq  
Here, the $Q$--power $\Delta_i$ of the vertex $i$ defined in Eq.~(\ref{powcNN}) has to be adjusted 
according to the rules given in Eq.~(\ref{CountRules}). 
It should be understood that the counting rules on Eq.~(\ref{CountRules}) are by no means unique and represent 
an attempt to relate the sizes of the isospin--breaking and isospin--conserving nuclear forces with
each other in a realistic way. Different rules are usually adopted in the meson and single--nucleon sectors,
see e.g.~\citep{Steininger:1999aa,Fettes:2000gb,Gasser:2002am}. Counting rules very similar to the ones in Eq.~(\ref{CountRules}) 
(but not exactly the same) have been used in applications in the 2N sector in 
Refs.~\citep{VanKolck:1993ee,vanKolck:1995cb,vanKolck:1996rm,Friar:1999zr,Friar:2003yv,Friar:2004ca}. Clearly, changing 
the counting rules shifts various nuclear force contributions to different orders but does not affect 
their explicit form. The most realistic set of counting rules, i.e.~the one that finally leads 
to the most natural values of the LECs,  can only be figured out in practical calculations.

The 2N forces fall in to four classes with respect to their isospin structure \citep{Henley:1979aa}:
\beq
\label{2NF_classes}
\begin{array}{lcl}
\mbox{Class I:} & \mbox{\hskip 1 true cm} & V_{\rm I} = \alpha_{\rm I}  + \beta_{\rm I} \,\fet \tau_1 \cdot \fet \tau_2 \,,\\ [0.6ex]
\mbox{Class II:} &                        & V_{\rm II} = \alpha_{\rm II}  \, \tau_1^3 \, \tau_2^3  \,,\\[0.6ex]
\mbox{Class III:} &                        & V_{\rm III} = \alpha_{\rm III}  \,( \tau_1^3 + \tau_2^3)  \,,\\[0.6ex]
\mbox{Class IV:} &                        & V_{\rm IV} = \alpha_{\rm IV}  \, (\tau_1^3 - \tau_2^3) + \beta_{\rm IV} \, [\fet \tau_1 \times \fet \tau_2]^3  \,, 
\end{array}
\eeq
where $\alpha_{\rm i}$, $\beta_{\rm i}$ are space and spin operators.  The operator $\beta_{\rm IV}$ has to be odd under a time reversal transformation.
While class (I) forces are isospin--invariant, all other classes (II), (III) and (IV) forces are isospin--breaking. 
Class (II) forces, $V_{\rm II}$, maintain charge symmetry but break charge independence. They are usually referred 
to as charge independence breaking (CIB) forces.
Class (III) forces break charge symmetry but do  not lead to isospin mixing in the 2N system. 
Finally, class (IV) forces break charge symmetry and cause isospin mixing in the 2N system. 

We will now discuss various contributions to the isospin--violating 2N force which have been extensively studied in 
the chiral EFT framework and worked out up to order $\nu = 5$. It can be expressed as:
\beq
V_{\rm 2N}^I = V_{\rm EM}  + V_{\pi \gamma}  + V_{1\pi}^I   + V_{2\pi}^I   + V_{\rm cont}^I\,, 
\eeq
where the terms in the right--hand side  refer to the long--range electromagnetic force, pion--photon exchange,
isospin--breaking one-- and two--pion exchange potentials and contact terms, respectively. 
The superscript $I$ is used in order to distinguish the isospin--breaking from the corresponding isospin--invariant 
contributions considered in sections \ref{sec:piexch}, \ref{sec:cont}.
Let us first comment on the long--range electromagnetic force whose dominant contribution is 
given by the static Coulomb interaction at order $\nu=2$.   The first long--range corrections are suppressed
by $m^{-2}$ (relativistic corrections to the static one--photon exchange).
At this order, the long--range electromagnetic NN interaction is given by
\beqa
\label{vc1vc2}
V_{\rm EM} (pp) &=& V_{\rm C1} +  V_{\rm C2} + V_{\rm VP} + V_{\rm MM} (pp)\,, \nn
V_{\rm EM} (np) &=& V_{\rm MM} (np)\,, \nn
V_{\rm EM} (nn) &=& V_{\rm MM} (nn)\,,
\eeqa
where $V_{\rm C1}$ and $V_{\rm C2}$ are usually referred to as ``improved Coulomb potential''.
They include the  relativistic $1/m^2$--corrections to the static 
Coulomb potential worked out in 
Ref.~\citep{Austin:1983aa}. The expressions for the vacuum polarization potential $V_{\rm VP}$ and 
magnetic moment interaction $V_{\rm MM}$ can be found in Refs.~\citep{Ueling:1935aa,Durand:1957aa} and \citep{Stoks:1990bb},
respectively. Notice that $V_{\rm EM}$ contains classes (II), (III) and (IV) forces. The class (IV) force is given 
exclusively by the magnetic moment interaction. We also stress that the effects of $V_{\rm EM}$ are enhanced 
at low energy due to the long--range nature of this force. Even the effects due to $V_{\rm MM}$, which is 
suppressed by factor $\sim (Q/m )^2$ compared to the static Coulomb interaction and thus contributes at order $\nu = 6$, 
might be large for certain scattering observables under specific kinematical conditions, see e.g.~\citep{Stoks:1990bb}.

\begin{figure*}[tb]
%\vspace{0.3cm}
\centerline{
\psfig{file=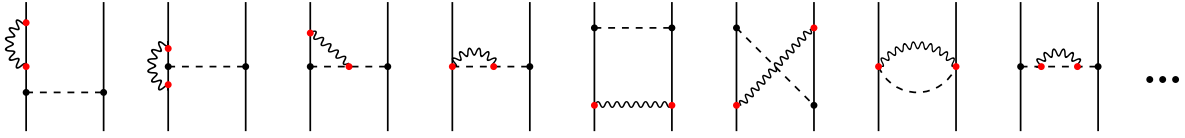,width=16cm}
}
\vspace{-0.2cm}
\caption[fig8]{\label{fig8} One--loop $\pi \gamma$ diagrams contributing to the isospin--breaking NN force
of the 1PE range. Wavy lines represent photons. Solid dots refer to the leading strong and electromagnetic vertices. 
Only one representative topology is depicted for each graph. 
For the remaining notation see Fig.~\ref{fig5}.}
\vspace{0.2cm}
\end{figure*}

The one--loop diagrams contributing to the isospin--breaking NN force of the 1PE range 
are shown in Fig.~\ref{fig8} and were considered by van Kolck et al.~\citep{vanKolck:1997fu}.
Some of the graphs depicted in this figure lead to renormalization of the isospin--breaking 1PE potential
(i.e.~change its strength). Notice further that due to isospin, only charged pion
exchange can contribute to the $\pi\gamma$ potential $V_{\pi\gamma}$
and thus it only affects the $np$ system. The resulting $\pi \gamma$ potential is CIB and has the form
\beqa\label{Vpiga}
V_{\pi\gamma} (\vec{q} \, ) &=& -{g_A^2 \over 4F_\pi^2 \Mppz} \,
(\fet{\tau}_1 \cdot \fet{\tau}_2 -\tau_1^3 \, \tau_2^3 ) \,
\vec{\sigma}_1 \cdot \vec{q} \, \vec{\sigma}_2 \cdot \vec{q} \,\, V_{\pi\gamma}
(\beta)~, \nonumber \\
V_{\pi\gamma}(\beta ) &=& {\alpha \over \pi} \biggl[ -{ (1-\beta^2)^2
  \over 2\beta^4 (1+\beta^2) } \ln(1+\beta^2) + {1\over 2\beta^2} 
- {2\bar{\gamma} \over 1+\beta^2} \biggr]~.
\eeqa 
Here, $\beta = |\vec{q} \,|/\Mpp$ and $\bar{\gamma}$ is a
regularization scheme dependent constant. 
The analytical form of
$V_{\pi\gamma}$ is similar to the one of the 1PE potential but differs in
strength by the factor $\alpha/\pi \simeq 1/400$. 

Isospin--breaking corrections to the 1PE potential were extensively studied within the EFT framework
\citep{VanKolck:1993ee,vanKolck:1996rm,Friar:2004ca,Epelbaum:2005fd}.  
It is convenient to express the static 1PE potential in Eq.~(\ref{opep}) in a more general form,
which already incorporates some (but not all) of the isospin--breaking corrections: 
\beqa
\label{OPEnijm}
V_{1 \pi} + V_{1 \pi}^I (pp)\,  &=& (1 + \delta_p )^2  \, V ( M_{\pi^0} )\,, \nn
V_{1 \pi} + V_{1 \pi}^I (nn) &=& (1 + \delta_n )^2 \, V ( M_{\pi^0} )\,, \nn
V_{1 \pi} + V_{1 \pi}^I (np) \, &=& - (1 + \delta_p ) (1 + \delta_n)  \, V ( M_{\pi^0} ) + (-1) ^{T+1}\, 2 (1 + \delta_c )^2  \, V ( M_{\pi^\pm} )\,, 
\eeqa
where we utilize the notation of Ref.~\citep{Epelbaum:2005fd}. In Eq.~(\ref{OPEnijm}), 
$T=0, 1$ denotes the total isospin of the two--nucleon system and $V (M_{_i} )$ is defined as:
\beq
V (M_{i} ) = - \biggl(\frac{g_A}{2F_\pi}\biggr)^2   \, \frac {(\vec \sigma_1 \cdot \vec q \,) 
(\vec \sigma_2 \cdot \vec q \,)}{\vec q\, ^2 + M_i^2}\,. 
\eeq
The constants $\delta_p$,  $\delta_n$ and $\delta_c$ in Eq.~(\ref{OPEnijm})
specify the isospin dependence of the pion--nucleon coupling constant. For the isospin--symmetric 1PE potential, the quantity $\delta_p=\delta_n=\delta_c$
gives an isospin--conserving Goldberger--Treiman discrepancy.\footnote{Notice that in addition to terms $\propto (m_u + m_d)$ in Eq.~(\ref{GTD}), 
one has now to include corrections  $\propto | m_u -  m_d |$ and $\propto \alpha$.} 
Charge--symmetry conservation implies $\delta_p = \delta_n \neq \delta_c$. The leading isospin--breaking contribution to the pion--nucleon 
coupling constants is of the strong origin and leads to the class (III) force at order $\nu = 3$. Up to  order $\nu = 5$, isospin--violating 
contributions to $\delta_i$'s  result from various tree-- and one--loop diagrams and lead to both classes (II) and (III) forces, 
see \citep{vanKolck:1996rm,Epelbaum:2005fd} for more details. Apart from the corrections due to $\delta_p \neq \delta_n \neq \delta_c$, 
the expressions for the 1PE potential in Eq.~(\ref{OPEnijm}) also incorporate the order   $\nu =2$ class (II) contribution due to the mass 
difference of the exchanged pions, $\delta M_\pi^2 = M_{\pi^\pm}^2 - M_{\pi^0}^2 = (36 \mbox{ MeV})^2$. This is, in fact, the dominant contribution
to the isospin--breaking nuclear force. Clearly, the corrections $\propto (\delta M_\pi^2 )^2$ at order $\nu = 4$ are also included in Eq.~(\ref{OPEnijm}). 
Further contributions to the isospin--breaking 1PE potential not included in Eq.~(\ref{OPEnijm}) arise at order $\nu = 4$ due to 
the proton--to--neutron mass difference, $\delta m = m_p - m_n = -1.29$ MeV, and are given by \citep{Friar:2004ca,Epelbaum:2005fd}
\beqa
\label{OPEP1}
V_{1 \pi}^{I}{}^{(4)} &=& - i \frac{\delta  m}{2 m} \left( \frac{g_A}{2 F_\pi} \right)^2 
 [ \fet \tau_1 \times \fet \tau_2 ]^3 \, 
\frac{1}{(\vec{q_1} ^2 + M_\pi^2 )} \Bigg[
\frac{(\vec \sigma_1 \cdot \vec q_1\, )(\vec \sigma_2 \cdot \vec q_1 \,)}{(\vec{q_1} ^2 + M_\pi^2 )} 
(\vec{p_1}^2 - \vec{p_2}^2 - {\vec{p_1} '}{}^2 + {\vec{p_2} '}{}^2 ) \nn
&& {} - \Big( (\vec \sigma_1 \cdot \vec q_1) (\vec \sigma_2 \cdot (\vec p_2 + \vec{p_2} ' )) 
+  (\vec \sigma_1 \cdot (\vec p_1 + \vec{p_1} ' ))  (\vec \sigma_2 \cdot \vec q_1) \Big)  \Bigg]\,,
\eeqa
where $\vec p_i$ ($\vec p_i \, '$) denotes the incoming (outgoing) momentum of the nucleon $i$ and 
$\vec q_1 = \vec{p_1}  ' - \vec p_1 = - (\vec{p_2} ' - \vec{p_2})$. Notice that the first term in the 
square bracket vanishes in the CMS. The potential in Eq.~(\ref{OPEP1}) represents the dominant 
class (IV) force. The dependence of $V_{1 \pi}^{I}{}^{(4)}$ on the total momentum 
of the 2N system is explained in \citep{Friar:2004ca}. In addition to the corrections linear in $ \delta m/m$
in Eq.~(\ref{OPEP1}), one has to take into account the CIB contribution $\propto (  \delta m )^2$ which 
reads \citep{Stoks:1990bb,VanKolck:1993ee,Friar:2004ca,Epelbaum:2005fd}:
\beq
\label{OPEP3}
V_{\rm 1 \pi}^I{}^{(4)} = - (\delta  m)^2 \left( \frac{g_A}{2 F_\pi} \right)^2 
\left( \fet \tau_1 \cdot \fet \tau_2  - \tau_1^3 \tau_2^3 \right) \, 
\frac{(\vec \sigma_1 \cdot \vec q\, )(\vec \sigma_2 \cdot \vec q \,)}{(\vec{q} \,^2 + M_\pi^2 )^2} \,.
\eeq
Notice that the isospin--violating piece has the same structure as the correction due to the pion mass difference at order $\nu = 2$ but is 
$\delta  M_\pi^2 / (\delta  m)^2 \sim 660$ times weaker. No new structures in the 1PE potential appear at order $\nu = 5$.

\begin{figure*}[tb]
%\vspace{0.3cm}
\centerline{
\psfig{file=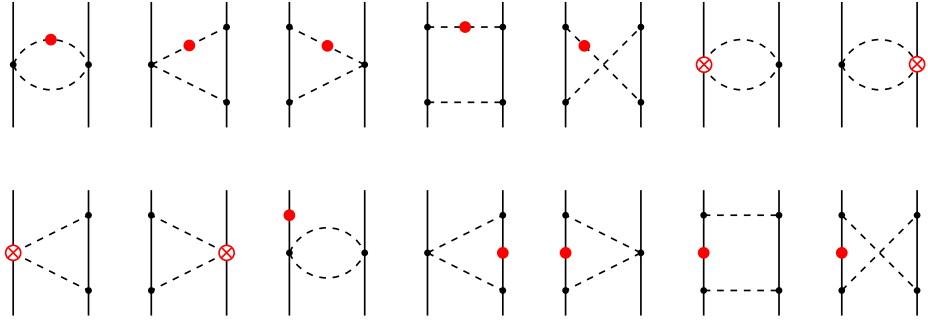,width=12.5cm}
}
\vspace{-0.1cm}
\caption[fig9]{\label{fig9}  Leading isospin--breaking corrections to the 2PE potential. 
A light--shaded circle inserted at a pion (nucleon) line refers to a single insertion of the pion (nucleon) 
mass difference. Crossed circles denote the leading isospin--breaking $\pi \pi NN$ vertices of dimension $\Delta_i =2$. For graphs with 
pion/nucleon mass difference insertions, only one representative topology is depicted. 
For remaining notation see Fig.~\ref{fig5}.}
\vspace{0.2cm}
\end{figure*}

Let us now switch to the 2PE potential. The dominant contributions arise at order $\nu = 4$ 
from 1--loop diagrams constructed from the leading $\pi NN$ and $\pi \pi NN$ vertices and a single insertion of 
%*********************************************************************************
the pion mass difference, the proton--to--neutron mass shift or the order   $\Delta_i = 2$ isospin--breaking $\pi \pi NN$ vertex.
%*********************************************************************************
%\begin{itemize}
%\vspace{-0.3cm}
%\item
%the pion mass difference, \vs
%\item
%the proton--to--neutron mass shift, \vs
%\item
%the order   $\Delta_i = 2$ isospin--breaking $\pi \pi NN$ vertex. \vs
%\end{itemize}
The corresponding diagrams are shown in Fig.~\ref{fig9}.  
Let us first discuss the class (II) contributions arising from diagrams with a single insertion of $\delta M_\pi^2$.  
As shown in Ref.~\citep{Friar:1999zr}, the potential can be expressed in terms of the corresponding 
isospin--invariant contributions given in Eq.~(\ref{2PE_nlo}) without performing any additional calculations. 
To that aim, one can first decompose the isospin--invariant 2PE potential in to the isoscalar and isovector pieces
\beq
\label{TPEdecomp}
V_{2\pi} = V_{2\pi}^0 + V_{2\pi}^1 \, \fet {\tau}_1 \cdot
\fet {\tau}_2~.
\eeq
The leading isospin--breaking effects due to $M_{\pi^\pm} \neq M_{\pi^0}$  are  incorporated properly if 
one uses $\tilde M_\pi$, defined as 
\beq\label{avmass}
\tilde \Mp = \frac{2}{3} \Mpm +  \frac{1}{3} \Mpn~,
\eeq
in the scalar part $V_{2\pi}^0$ and adopts for the isovector part:
\beq
\label{Vcib}
 \begin{array}{ll}  V_{2\pi}^1  (M_{\pi^\pm} )     & {\rm for}~~pp~~{\rm and}~~nn~, \\[1ex]
                                        V_{2\pi}^1  (M_{\pi^0} )       & {\rm for}~~np,~~T = 1~, \\[1ex]
                                        V_{2\pi}^1  (\tilde M_{\pi} )  & {\rm for}~~np,~~T = 0~. \end{array}
\eeq
These results are valid  modulo $( \delta  M_\pi^2 /M_\pi^2 )^2$--corrections which first 
contribute to the 2PE potential at order $\nu = 6$. Clearly, one can also express the corresponding potential 
in a more traditional form in terms of isospin matrices and without referring to particular isospin channels, see 
Ref.~\citep{Epelbaum:2005fd}. 

All remaining 2PE contributions at order $\nu = 4$  lead to class (III) forces
and were derived independently and using different methods by several groups, see 
\citep{Niskanen:2001aj,Friar:2003yv,Epelbaum:2005fd} and \citep{Coon:1995qh} for a related earlier work. 
One finds
\beqa
\label{2PEisosp4}
\label{WC4}
V_{2 \pi}^I{}^{(4)} &=& - \frac{g_A^2}{64 \pi F_\pi^4} \, ( \tau_1^3 + \tau_2^3 ) \, \left[ \frac{ 2 g_A^2 \, \delta  m  \, M_\pi^3}{4 M_\pi^2 + q^2} 
- \Big( 4 g_A^2 \, \delta  m - ( \delta  m )^{\rm str} \Big) (2 M_\pi^2 + q^2 ) \, A^\Lambda (q) \right]  \nn
&& {} + \frac{g_A^4 \, \delta  m}{32 \pi F_\pi^4 } \, ( \tau_1^3 + \tau_2^3 ) \, 
\left[ (\vec \sigma_1 \cdot \vec q \, )  (\vec \sigma_2 \cdot \vec q \, )  - (\vec \sigma_1 \cdot \vec \sigma_2 ) \, \vec q^{\, 2} \right] \,  A^\Lambda (q ) \,.
\eeqa
Here the LEC accompanying the order   $\Delta_i = 2$ isospin--breaking $\pi \pi NN$ vertex is expressed in terms of  
to the strong contribution $(\delta m )^{\rm str}$ to the nucleon mass shift $\delta m = (\delta m )^{\rm str} + (\delta m )^{\rm em}$, where 
\beqa
(\delta m )^{\rm str} &=& ( m_p - m_n )^{\rm str}  = -2.05 \pm 0.3 \mbox{ MeV ,}  \nn
(\delta m )^{\rm em} &=& ( m_p - m_n  )^{\rm em} = 0.76 \pm 0.3  \mbox{ MeV .}
\eeqa
These values are taken from Ref.~\citep{Gasser:1982ap} and based on an evaluation of the Cottingham sum rule. 
Notice that the vertices corresponding to  $(\delta m )^{\rm str}$ and  $(\delta m )^{\rm em}$ have, 
according to the counting rules in Eq.~(\ref{CountRules}), dimensions $\Delta_i =2$ and $3$,
respectively.   
Notice further that some of the diagrams shown in Fig.~\ref{fig9}, in particular the planar box and the  football diagrams 
with a single insertion of $\delta m$, lead to vanishing contributions.  

\begin{figure*}[t]
%\vspace{0.3cm}
\centerline{
\psfig{file=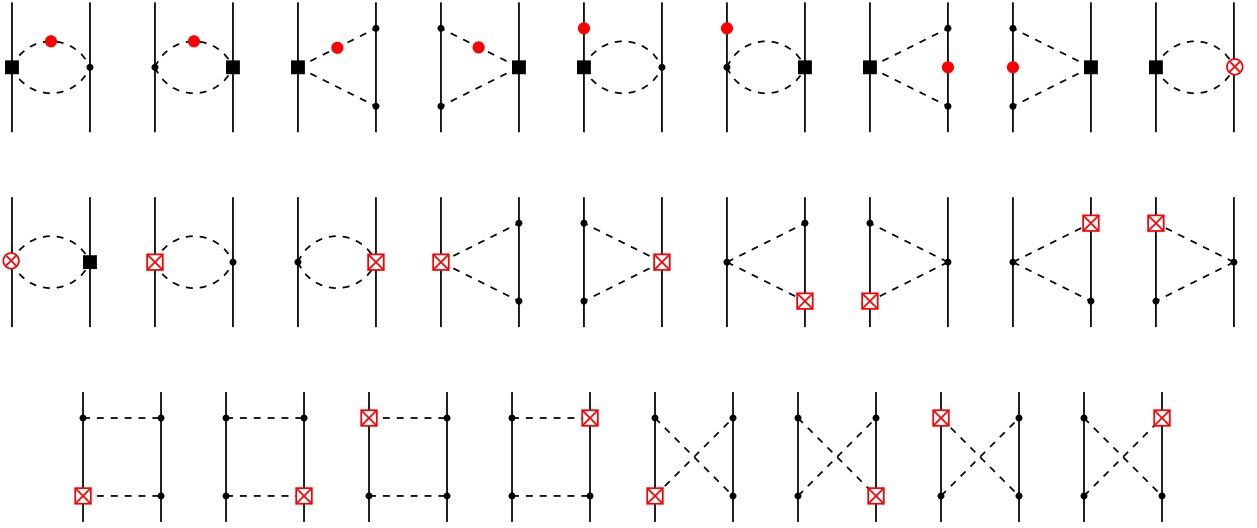,width=17.0cm}
}
\vspace{-0.2cm}
\caption[fig10]{\label{fig10}  Subleading isospin--breaking corrections to the 2PE potential. 
Crossed rectangles denote isospin--breaking vertices of dimension $\Delta_i = 3$. Graphs resulting from the 
interchange of the vertex ordering are not shown. For remaining notation see Figs.~\ref{fig5}, \ref{fig9}.}
\vspace{0.2cm}
\end{figure*}

Consider now the subleading isospin--breaking 2PE potential which is due to 1--loop diagrams 
(some of which  lead to vanishing results)
%***************************************************************************
shown in Fig.~\ref{fig10}.
%***************************************************************************
%constructed 
%from the leading $\pi NN$ and $\pi \pi NN$ vertices and a single insertion of 
%\begin{itemize}
%\vspace{-0.3cm}
%\item
%the pion mass shift and the order   $\Delta_i = 1$ isospin--invariant $\pi \pi NN$ vertex, \vs
%\item
%the proton--to--neutron mass shift and the order   $\Delta_i = 1$ isospin--invariant $\pi \pi NN$ vertex, \vs
%\item
%the order   $\Delta_i = 2$ isospin--breaking and the order   $\Delta_i = 1$ isospin--invariant 
%$\pi \pi NN$ vertices, \vs
%\item
%the order   $\Delta_i = 3$ isospin--breaking $\pi \pi NN$ or $\pi NN$ vertices. \vs
%\end{itemize}
%The corresponding diagrams, some of which  lead to vanishing results, are shown in Fig.~\ref{fig10}. 
The class (II) contributions $\propto \delta M_\pi^2$  can be obtained using 
Eq.~(\ref{Vcib}) from the corresponding isospin--invariant 2PE potential in Eq.~(\ref{2PE_nnlo}). 
The remaining terms were calculated recently using the method of unitary transformation 
\citep{Epelbaum:2005fd}. The resulting class (III) potential reads:
\beqa
\label{WC5}
V_{2 \pi}^I{}^{(5)} &=& - \frac{1}{96 \pi^2 F_\pi^4} \, (\tau_1^3 + \tau_2^3) \, L^\Lambda (q) \, \bigg\{ - g_A^2 \, \delta  m 
\frac{48 M_\pi^4 (2 c_1 + c_3)}{4 M_\pi^2 + q^2} \nn
&& \mbox{\hskip 2 true cm}  + 4 M_\pi^2 \Big[ g_A^2 \, \delta  m \, ( 18 c_1 + 2 c_2 - 3 c_3) 
+ \Big(2 \delta  m  - (\delta   m )^{\rm str} \Big) \, ( 6 c_1 - c_2 - 3 c_3 ) \Big] \nn
&& \mbox{\hskip 2 true cm}  + q^2 \Big[ g_A^2 \, \delta  m \, ( 5 c_2 - 18 c_3 ) -  
\Big( 2 \delta  m - (\delta   m )^{\rm str} \Big)\, ( c_2 + 6 c_3 ) \Big] \bigg\} \nn
&&{}  - \frac{g_A^2}{16 \pi^2 F_\pi^4} \,(\tau_1^3 + \tau_2^3) \, 
\left[ (\vec \sigma_1 \cdot \vec q \, )  (\vec \sigma_2 \cdot \vec q \, )  - (\vec \sigma_1 \cdot \vec \sigma_2 ) \, \vec q^{\, 2} \right] \,  L^\Lambda (q) \,
\Big( \delta  m \, c_4 + g_A \, \beta \Big) \,, 
\eeqa
where $\beta$ is a combination of the LECs accompanying the leading isospin--breaking $\pi NN$ vertex. 
At  order $\nu = 5$, it can be expressed in terms of the constants 
$\delta_i$ defined in Eq.~(\ref{OPEnijm}) as $\beta = (1/4) (\delta_p - \delta_n )g_A $. 
The complete expressions for the isospin--breaking 2PE potential in $r$--space can be found in 
\citep{Epelbaum:2005fd}.  

Finally, isospin--breaking contact terms up to order $\nu = 5$ feed in to the matrix--elements of the S-- and P--waves  
in the following way:
\beqa
\label{VC_iso}
\langle ^1S_0, \; pp \, | V_{\rm cont}^I| ^1S_0, \; pp \, \rangle&=& \tilde \beta_{1S0}^{pp} + \beta_{1S0}\, ( p^2 + p '^2)~,\nn
\langle ^1S_0, \, nn \, | V_{\rm cont}^I| ^1S_0, \, nn \, \rangle&=& \tilde \beta_{1S0}^{nn} - \beta_{1S0}\, ( p^2 + p '^2)~,\nn
\langle ^3P_0, \; pp \, | V_{\rm cont}^I| ^3P_0, \; pp \, \rangle&=& 
- \langle ^3P_0, \; nn \, | V_{\rm cont}^I| ^3P_0, \; nn \, \rangle = \beta_{3P0}\,  \,p \, p' ~,\nn
\langle ^3P_1, \; pp \, | V_{\rm cont}^I| ^3P_1, \; pp \, \rangle&=& 
- \langle ^3P_1, \; nn \, | V_{\rm cont}^I| ^3P_1, \; nn \, \rangle = \beta_{3P1}\,  \,p \, p' ~,\nn
\langle ^3P_2, \; pp \, | V_{\rm cont}^I| ^3P_2, \; pp \, \rangle&=& 
- \langle ^3P_2, \; nn \, | V_{\rm cont}^I| ^3P_2, \; nn \, \rangle = \beta_{3P2}\,  \,p \, p' ~,\nn
\langle ^1P_1, \, np \, | V_{\rm cont}^I| ^3P_1, \, np \, \rangle&=& \beta_{1P1-3P1}\,  \,p \, p'  ~,
\eeqa
where $\beta_i$, $\tilde \beta_i$ are the corresponding LECs. Here, we use 
the convention according to which the {\it np} matrix elements (with exception
of the last term in Eq.~(\ref{VC_iso})) do not change by switching off isospin--violating contact terms. 
Notice further that all terms quadratic in momenta are of the order  $\nu = 5$.  

Last but not least, we would like to emphasize that chiral EFT supports the 
following hierarchy of the nuclear forces \citep{VanKolck:1993ee}: class (I) $>$  class (II) $>$ class (III) 
$>$ class (IV).

\subsubsection{Towards a relativistic NN potential}

In this section, we will discuss the problem with the formal inconsistency of the heavy--baryon (HB) expansion 
mentioned in section \ref{sec1} and its impact on the nuclear potential and NN scattering.
For that, consider the contribution to the NN scattering amplitude arising from 
the triangle diagrams in the second line of Fig.~\ref{fig5}. We assume exact isospin symmetry throughout this section. 
Further, we follow closely 
the procedure of Ref.~\citep{Robilotta:2000py} and restrict ourselves to the central part of this contribution,
which is proportional to the LECs $c_{1,3}$  and can be expressed in terms of the $\pi N$ scattering amplitude and 
the nuclear scalar form factor. For a discussion of the nuclear scalar form factor in relativistic CHPT,
the reader is referred to Ref.~\citep{Gasser:1987rb}, see also \citep{Becher:1999he} for the results in the infrared 
regularized version of CHPT. Using the relativistic expression for the nucleon propagator
and utilizing the notation of Ref.~\citep{Robilotta:2000py}, 
this contribution to the scattering amplitude can be written as:
\beq 
\label{VC_rel}
T_C^{rel} (t) =\frac{3 m g_A^2}{16 \pi^2 F_\pi^4}\, \left( 2 M_\pi^2 (2 c_1 - c_3) + c_3 t \right) 
\left[ J_{c,c} ( t ) - J_{c, sN}^{(1)} (t) \right] \,.
\eeq
Here, $t = q^2$, $q = p ' - p$ and the integrals $J_{c,c} ( t )$ and $J_{c, sN}^{(1)} (t)$ read:
\beqa
J_{c,c} ( t ) &=& \frac{( 4 \pi )^2}{i} \int \frac{d^4 l }{(2 \pi )^4} 
\frac{1}{[(l - q/2 )^2 - M_\pi^2 + i \epsilon] [(l + q/2 )^2 - M_\pi^2 + i \epsilon]} \\
J_{c, sN}^{(1)} (t) &=& \frac{( 4 \pi )^2}{i\,  V^2}\int \frac{d^4 l }{(2 \pi )^4} 
\frac{2 m V \cdot l}{[(l - q/2 )^2 - M_\pi^2 + i \epsilon] [(l + q/2 )^2 - M_\pi^2 + i \epsilon]
[l^2 + 2 m V \cdot l - q^2/4+ i \epsilon]}\,,
\nonumber
\eeqa
where $V = ( p' + p )/2 m$. In the above equation, we have replaced the bare nucleon and pion masses and 
the pion decay constant by their physical values.  
The corresponding NN potential $V_C^{rel} (\vec q \, )$ is defined as \citep{Robilotta:2000py}:
\beq 
V_C^{rel} (\vec q \, ) = T_C^{rel} (t) \bigg|_{t = -\vec q \, ^2} \,.
\eeq
Neglecting terms proportional to $\vec q\, ^2 / m^2$, the spectral function $\rho_C^{rel} (\mu )$ corresponding 
to Eq.~(\ref{VC_rel}) takes the form \citep{Robilotta:2000py}:
\beq
\label{rho_c_rel}
\rho_C^{rel} (\mu ) = - \frac{3 g_A^2}{32 \pi F_\pi^4} \,  \left( 2 M_\pi^2 (2 c_1 - c_3) + c_3 \mu^2 \right)\, 
\frac{2 M_\pi^2 - \mu^2}{\mu} \,  \arctan x \,,
\eeq
where we have introduced the abbreviation
\beq 
x  =  \frac{2 m \sqrt{\mu^2 - 4 M_\pi^2}}{\mu^2 - 2 M_\pi^2} \,.
\eeq
For the sake of simplicity, we restrict ourselves to the SFR cutoff $\tilde \Lambda = \infty$. 
The choice of the cut--off is of no relevance for the present discussion. 
Since $x \sim m/Q$, the standard HB approach corresponds to the expansion of the  $\arctan x$ 
in the above equation  for large $x$ \citep{Becher:1999he}: 
\beq
\label{SFCrelEX}
\arctan x =  \frac{\pi}{2} - \frac{1}{x} + \frac{1}{3x^3} + \ldots \,.
\eeq
Keeping only the first term in this series, one reproduces the HB expression for the spectral function 
given in Eq.~(\ref{rho}). The series in Eq.~(\ref{SFCrelEX}), however, only converges for $| x | > 1$. 
This condition is not met for $\mu$ in the threshold region
$2 M_\pi \leq \mu  < 2 M_\pi (1 + \Delta )$, where $\Delta = M_\pi^2/(8 m^2) + \mathcal{O} 
(M_\pi^4/m^4 )$.\footnote{This condition is also violated for large $\mu$ of the order $\mu \sim 2 m$,
which is, however, outside of the validity region of the HB approach and therefore of no relevance.
For the role of the large $\mu$--components, see also the discussion in section \ref{sec:sfr}.}
The HB expression for the spectral function is, therefore, not correct in this region, as pointed out  
in Ref.~\citep{Bernard:1995dp}.

How does this failure of the HB expansion affect the NN potential? 
In order to answer this question, it is useful to switch to coordinate space. For $r > 0$, the inverse Fourier--transform 
can be expressed in terms of the spectral function:
\beqa
V_C^{rel} (r) &=& \frac{1}{2 \pi^2 r} \int_{2 M_\pi}^\infty d \mu \, \mu \, e^{- \mu r} \, \rho_C^{rel} (\mu )  \\
&=& \frac{1}{2 \pi^2 r}  \int_{2 M_\pi}^{2 M_\pi (1 + \Delta)} d \mu \, \mu \, e^{- \mu r} \, \rho_C^{rel} (\mu )  
+ \frac{1}{2 \pi^2 r} \int^{\infty}_{2 M_\pi (1 + \Delta)} d \mu \, \mu \, e^{- \mu r} \, \rho_C^{rel} (\mu )  \,.
\nonumber
\eeqa

The first integral in the second line of the above equation goes over the range of $\mu$, 
where the spectral function is incorrectly described in the HB approach. 
The integration in the second term goes over the interval of $\mu$, 
where the HB expansion is valid (for not too large values of $\mu$). In general, the contribution of the 
first integral is expected to be suppressed relative to the one from the second integral
by a factor $M_\pi^2/m^2$ and thus can be safely neglected at the considered order. 
This suppression, however, does not occur at asymptotically  
large distances, i.e.~for $r \gtrsim m^2 / M_\pi^3$, where the potential is determined by the threshold behavior of the 
spectral function. On the other hand, the 2PE potential falls off exponentially and becomes very weak at large 
distances. Consequently, the problem with the formal inconsistency of the HB approach is expected to be of little relevance 
for practical applications. Recent work \citep{Higa:2004cr} on peripheral NN scattering based on the relativistic approach confirms this 
expectation.    

\begin{figure*}[t]
%\vspace{0.3cm}
\centerline{
\psfig{file=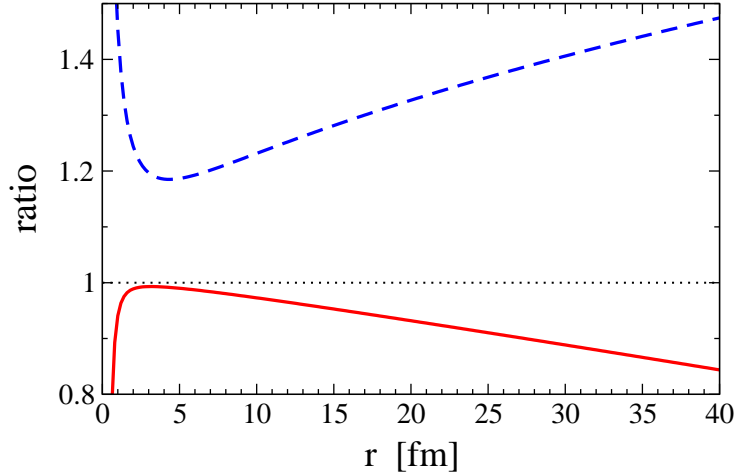,width=10.0cm}
}
\vspace{-0.2cm}
\caption[fig10a]{\label{fig10a}  Ratios of the leading (dashed line) and next--to--leading order (solid line) 
HB potentials to the relativistic result versus the distance $r$ as explained in the text. }
\vspace{0.2cm}
\end{figure*}

To see how the above qualitative arguments and estimations work in practice, we again follow
Ref.~\citep{Robilotta:2000py} and
calculate the leading and next--to--leading order HB approximations for the potential in coordinate space which correspond  
to the first and the first $+$ second terms in the series in Eq.~(\ref{SFCrelEX}). 
In Fig.~\ref{fig10a} we show the ratios of these leading and next--to--leading order HB potentials 
to the relativistic result corresponding to the spectral function in Eq.~(\ref{rho_c_rel}). Here, we adopt the 
same values for the LECs $c_{1,3}$ as in section \ref{sec:sfr}. As expected,
the deviations from the correct result increase for large values of $r$. Notice that large deviations at 
short distances $r < 1$ fm are due to the fact that the HB expansion diverges at large $\mu$. 
The leading HB approximation deviates by $\sim 20$\% from the relativistic result in the region 
of physical interest. For further discussion of these and related issues, the reader is referred to \citep{Robilotta:2000py}.

To conclude, the formal inconsistency of the HB approach strongly affects the behavior of the potential at asymptotically large 
distances. In this region, however, the potential is very weak and its contribution to observables is negligible. 
At physically interesting distances, the effects due to the formal inconsistency of the HB expansion are small and 
expected to be irrelevant at the considered order. 
%We stress, however, that the issue of the inconsistency of the HB approach 
%might become important when calculations are performed at sufficiently high orders in the chiral expansion. 
%Recently, 
The 2PE potential has been derived at order $\nu = 4$ in the relativistic version of chiral EFT \citep{Higa:2003jk,Higa:2003sz}
using the formalism based on the S--matrix, see also Ref.~\citep{Higa:2004cr} for the application to peripheral NN scattering.

\subsection{Few--nucleon forces}

Two--nucleon forces discussed above provide the most important contribution to the Hamilton operator.
Three-- and more--nucleon forces are suppressed compared to the two--nucleon ones (2NFs) by powers of $Q/\Lambda$ and thus appear
as corrections. It is important to take these corrections into account in order to understand 
the properties of few--nucleon systems at the quantitative level. The strength of the chiral EFT approach 
is that it provides a framework to derive few--nucleon forces in a systematic way and fully consistent with the 
2NF.

\subsubsection{Three-- and four--nucleon forces in the isospin limit}
\label{sec:3NFinvar}

\begin{figure*}[t]
%\vspace{0.3cm}
\centerline{
\psfig{file=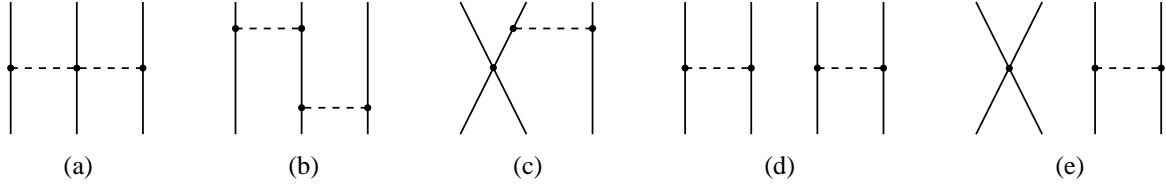,width=16.0cm}
}
\vspace{-0.3cm}
\caption[fig11]{\label{fig11}   Leading contributions to 3NF (graphs (a)--(c))  and 4NF  (graphs (d) and (e)) 
at order $\nu = 2$ that vanish. For notation see Fig.~\ref{fig5}.}
\vspace{0.2cm}
\end{figure*}

According to the power counting in Eq.~(\ref{powcNN}), three--  and four--nucleon forces  (4NFs) start to contribute at order $\nu =2$, i.e.~they 
are suppressed by a factor $(Q/\Lambda)^2$ compared to the leading 2NF. The leading 3NF contribution arises from tree diagrams constructed 
with the order   $\Delta_i = 0$ vertices, see graphs (a), (b) and (c) in Fig.~\ref{fig11}. 
The leading 4NF is represented  
by the disconnected tree diagrams (d) and (e). As pointed out by Weinberg \citep{Weinberg:1992yk}, graph (a) does 
not contribute at order $\nu = 2$ due to an additional suppression factor $Q/m$. The origin of this suppression 
is easily understood in terms of ``old--fashioned'' time--ordered perturbation theory.
Since graph (a) does not give rise 
to reducible topologies, its contribution to the 3NF is given by the sum of all possible time--ordered graphs, which  
build up the corresponding Feynman diagram. Since energy is conserved at each vertex of a Feynman graph, the time derivative,
which enters the Weinberg--Tomozawa $\pi \pi NN$ vertex,  yields a difference of nucleon kinetic energies which 
scales as $Q^2/m$ instead of $Q$. 

\begin{figure*}[t]
%\vspace{0.3cm}
\centerline{
\psfig{file=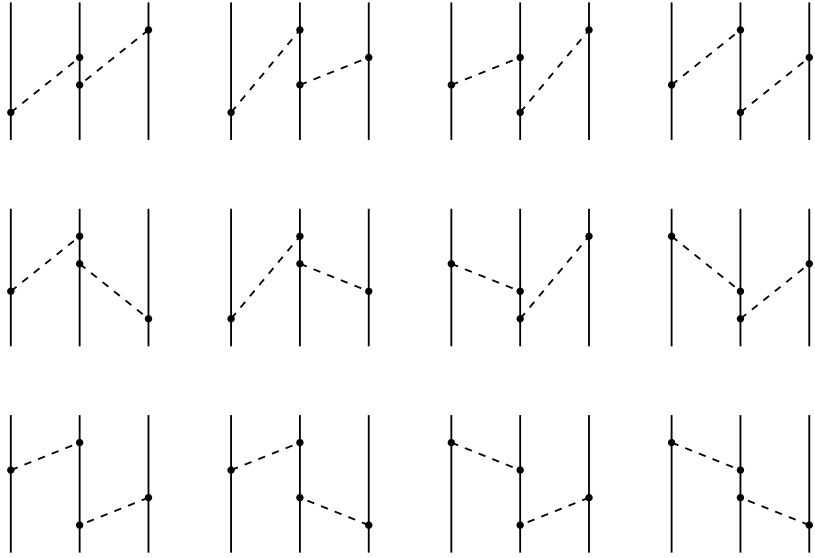,width=11.0cm}
}
\vspace{-0.1cm}
\caption[fig12]{\label{fig12}   Time--ordered 2PE diagrams at order $\nu = 2$. The first and second rows show all possible 
irreducible topologies, while the third row subsumes reducible topologies. Graphs resulting from the interchange 
of the vertex ordering at the middle nucleon line are not shown. For notation see Fig.~\ref{fig5}.}
\vspace{0.2cm}
\end{figure*}

The remaining graphs (b)--(e) in Fig.~\ref{fig11} lead to vanishing contributions to the nuclear force 
when the latter is defined within an energy--independent formulation such as the method of unitary transformation, 
see e.g.~\citep{Coon:1986kq,Eden:1996ey,Epelbaum:2000kv}. 
Consider, for example, the 2PE 3NF resulting from diagram (b) which can be obtained evaluating the corresponding matrix
elements of the operators in Eq.~(\ref{pr12}). The first term in this expression gives rise to irreducible 
time--ordered graphs (i.e.~the ones without purely nucleonic intermediate states) shown in Fig.~\ref{fig12} 
and defines the 3NF potential in the Tamm--Dancoff method, which can be expressed schematically as
\beq
\label{3n1}
- \left[ \frac{4}{\omega_1 (\omega_1 + \omega_2) \omega_2} +   
\frac{2}{\omega_1^2 ( \omega_1 + \omega_2 )} + \frac{2}{( \omega_1 
+ \omega_2) \omega_2^2}
\right] M = -2  \frac{\omega_1 + \omega_2}{\omega_1^2 \omega_2^2} M~,
\eeq
where $\omega_i = k_0 = \sqrt{\vec k \, ^2 + M_\pi^2}$ denotes the pion  energy and we pulled  out the common 
factor $M$ representing the spin, isospin and momentum structure, which is obviously
the same for all graphs in Fig.~\ref{fig12}. The remaining two terms in Eq.~(\ref{pr12}) are specific for the method 
of unitary transformation and correspond to reducible topologies shown in Fig.~\ref{fig12}. We remind the 
reader that although these diagrams contain purely nucleonic intermediate states, their contributions are 
not enhanced in the limit $m \to \infty$ and cannot be identified with the iteration of the potential in the dynamical equation. 
These diagrams are thus not truly reducible in the sense of ``old--fashioned'' time--ordered  perturbation theory.
The resulting contribution to the potential reads: 
\beq
\label{3n2}
\left[ \frac{2}{\omega_1^2 \omega_2} + \frac{2}{\omega_1 \omega_2^2} \right] M =
2 \frac{\omega_1 + \omega_2}{\omega_1^2 \omega_2^2}  M~.
\eeq
The cancellation between irreducible and reducible diagrams in the method of unitary transformation is now 
evident.\footnote{Notice that although the above cancellation at the level of the nuclear potential does not take 
place in the Tamm--Dancoff method, the resulting non--vanishing 3NF was shown to cancel against the recoil 
corrections to the 2N potential upon the iteration in the dynamical equation \citep{Yang:1986pk,vanKolck:1994yi}.} 
The same sort of cancellation occures for the remaining 
graphs (c)--(e) in Fig.~\ref{fig11}.
In other words, the entire contribution to the scattering amplitude represented by the Feynman diagrams (b)--(e) in 
Fig.~\ref{fig11} is reproduced by iteration of the leading 2NF in the corresponding dynamical equation 
with no need to introduce additional few--nucleon forces.\footnote{This statement does not apply to nuclear 
forces defined in the Tamm--Dancoff method. In that case, the 1PE 2N potential receives $1/m$--corrections, which 
are absent in the method of unitary transformation and require additional few--nucleon forces in order to cancel  
the corresponding additional contributions to the scattering amplitude.}
We therefore conclude that the 3NF and 4NF  at order $\nu = 2$ vanish completely.

\begin{figure*}[t]
%\vspace{0.3cm}
\centerline{
\psfig{file=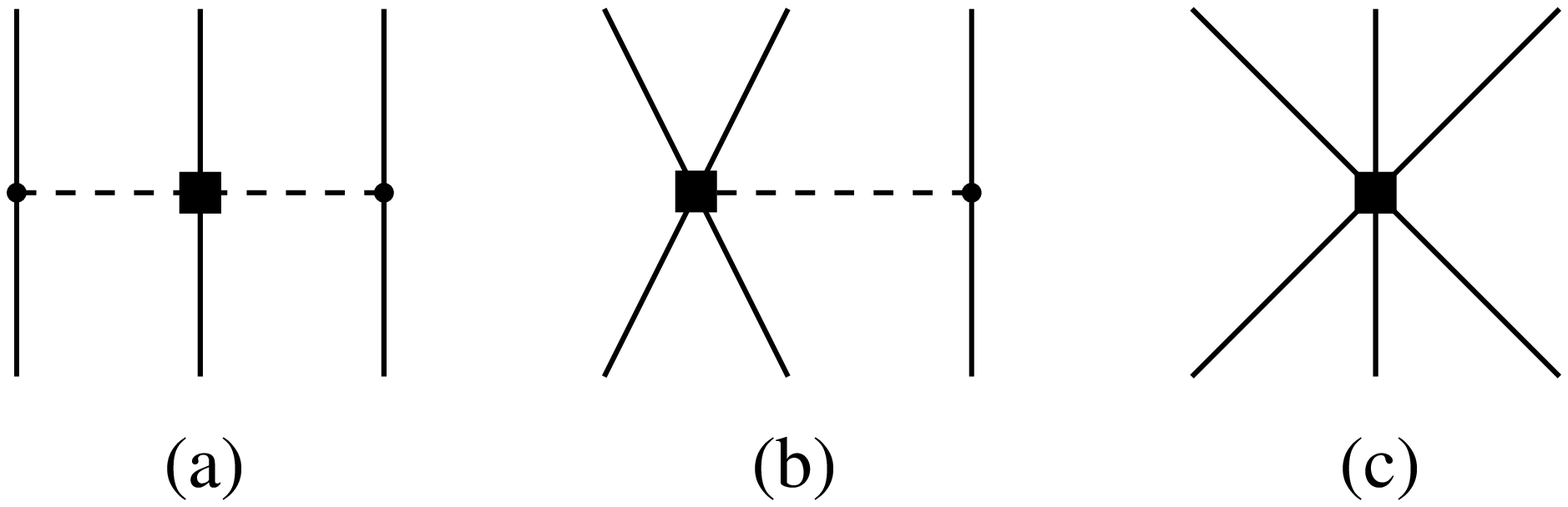,width=8.0cm}
}
\vspace{-0.1cm}
\caption[fig14]{\label{fig14}   Three--nucleon force at order $\nu = 3$.
For notation see Fig.~\ref{fig5}.}
\vspace{0.2cm}
\end{figure*}

The first non--vanishing 3NF contribution appears at order $\nu = 3$, i.e.~at N$^2$LO, and arises from 
diagrams shown in Fig.~\ref{fig14}. The contribution from graph (a) in this figure is given (in
the 3N CMS) by \citep{vanKolck:1994yi}:
\beq
\label{3nftpe}
V^{\rm (3)}_{\rm 2 \pi}=\sum_{i \not= j \not= k} \frac{1}{2}\left(
  \frac{g_A}{2 F_\pi} \right)^2 \frac{( \vec \sigma_i \cdot \vec q_{i}
  ) 
(\vec \sigma_j \cdot \vec q_j  )}{(\vec q_i\, ^2 + M_\pi^2 ) ( \vec
q_j\, ^2 + M_\pi^2)}  F^{\alpha \beta}_{ijk} \tau_i^\alpha 
\tau_j^\beta \,,
\eeq
where  $\vec q_i \equiv \vec p_i \, ' - \vec p_i$; $\vec p_i$
($\vec p_i \, '$) is the initial (final) momentum of the nucleon $i$ and 
\begin{displaymath}
F^{\alpha \beta}_{ijk} = \delta^{\alpha \beta} \left[ - \frac{4 c_1
    M_\pi^2}{F_\pi^2}  + \frac{2 c_3}{F_\pi^2}  
\vec q_i \cdot \vec q_j \right] + \sum_{\gamma} \frac{c_4}{F_\pi^2} \epsilon^{\alpha
\beta \gamma} \tau_k^\gamma  
\vec \sigma_k \cdot [ \vec q_i \times \vec q_j  ]\,.
\end{displaymath}
The subscripts of the Pauli spin and isospin matrices refer to nucleon labels. 
The form (\ref{3nftpe}) was shown to match with the
low--momentum expansion of various existing phenomenological 3NFs provided they respect chiral
symmetry. This issue is extensively discussed in \citep{Friar:1998zt}. The contributions from the 
remaining graphs (b) and (c) in Fig.~\ref{fig14} take the form  \citep{vanKolck:1994yi,Epelbaum:2002vt}
\beq
\label{3nfrest}
V^{\rm (3)}_{1 \pi, \; \rm cont} = - \sum_{i \not= j \not= k} \frac{g_A}{8
  F_\pi^2} \, D \, \frac{\vec \sigma_j \cdot \vec q_j }{\vec q_j\, ^2
  + M_\pi^2}  
\, \left( \fet \tau_i \cdot \fet \tau_j \right) 
(\vec \sigma_i \cdot \vec q_j ) \,,
\quad \quad \quad 
V^{\rm (3)}_{\rm cont} = \frac{1}{2} \sum_{j \not= k}  E \, ( \fet \tau_j \cdot \fet \tau_k ) \,,
\eeq
%\beqa
%\label{3nfrest}
%V^{\rm (3)}_{1 \pi, \; \rm cont} &=& - \sum_{i \not= j \not= k} \frac{g_A}{8
%  F_\pi^2} \, D \, \frac{\vec \sigma_j \cdot \vec q_j }{\vec q_j\, ^2
%  + M_\pi^2}  
%\, \left( \fet \tau_i \cdot \fet \tau_j \right) 
%(\vec \sigma_i \cdot \vec q_j ) \,, \\
%V^{\rm (3)}_{\rm cont} &=& \frac{1}{2} \sum_{j \not= k}  E \, ( \fet \tau_j \cdot \fet \tau_k ) \,,
%\nonumber
%\eeqa
where $D$ and $E$ are the corresponding LECs from the Lagrangian of order $\Delta=1$. 
We stress that the proper incorporation of the Pauli principle allows to substantially reduce the number of independent 
operators yielding only two terms in Eq.~(\ref{3nfrest}) \citep{Epelbaum:2002vt}. Notice further that 
no 4NFs appear at order $\nu = 3$. 

\begin{figure*}[t]
%\vspace{0.3cm}
\centerline{
\psfig{file=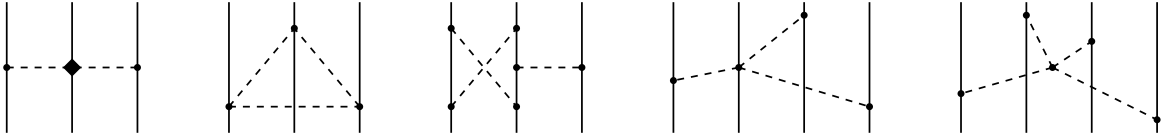,width=16.0cm}
}
\vspace{-0.4cm}
\caption[fig15]{\label{fig15}   Examples of the three-- and four--nucleon force contributions at order $\nu = 4$.
For notation see Figs.~\ref{fig5}.}
\vspace{0.2cm}
\end{figure*}

The first corrections to the 3NF as well as the first contribution to the 4NF arise at order $\nu = 4$. 
Some examples of diagrams which contribute at this order are shown in Fig.~\ref{fig15}. 
For the 3NF, one has to take into account all possible one--loop graphs constructed with the lowest--order 
vertices and tree diagrams with one insertion of the order   $\Delta_i = 2$ vertices. In addition, 
one has to include the leading $1/m$--corrections to diagrams (a)--(c) in Fig.~\ref{fig11}. 
The leading contributions to the 4NF arise at order $\nu = 4$ from connected tree diagrams with 
the lowest--order vertices. One should also consider various disconnected diagrams which might, in principle, 
also contribute to the 4NF at this order. Work along these lines is underway. The leading 
$1/m$--corrections to the static 2PE 3NF were already evaluated and 
can be found in Ref.~\citep{Friar:1994zz}.  The relative importance of various isospin--invariant contributions 
to few--nucleon forces discussed in this section and in sections \ref{sec:piexch}, \ref{sec:cont} and \ref{sec:rel}
are summarized in Table \ref{tab:isosp_symm} based on the chiral power counting in Eq.~(\ref{powcNN}). 
We stress that these results rely on the energy--independent formulation.

\begin{table*}[t] 
\vspace{0.3cm}
\begin{center}
\begin{tabular}{||c||l|l|l||}
\hline \hline 
{} & {} &  {} &  {}\\[-1.5ex]
Chiral order   &  2N force   &  3N force & 4N force \\[1ex]
\hline \hline 
   &  &  &  \\[-1.5ex]
$\nu = 0$ &  $V_{1 \pi}+V_{\rm cont}$ &  $-$ & $-$ \\[0.3ex]
$\nu = 1$ &  $-$                         &  $-$ & $-$ \\[0.3ex]
$\nu = 2$ &  $V_{1 \pi} + V_{2 \pi} + V_{\rm cont}$ &  $-$ & $-$ \\[0.3ex]
$\nu = 3$ &  $V_{1 \pi} + V_{2 \pi}$    &  $V_{2 \pi} + V_{1 \pi, \; \rm cont} + V_{\rm cont}$  & $-$ \\[0.3ex]
$\nu = 4$ &  $V_{1 \pi} + V_{2 \pi} + V_{3 \pi} + V_{\rm cont}$  &  work in progress  & work in progress \\[1ex]
\hline \hline
  \end{tabular}
%\vskip 0.1 true cm
\caption{Isospin--symmetric nuclear forces up to order $\nu =4$. \label{tab:isosp_symm}}
\end{center}
\end{table*}

\subsubsection{Isospin--breaking corrections}

Let us now discuss isospin--breaking corrections to the 3NF which were recently  
worked out up to the same order $\nu = 5$ as the corresponding 2NFs. No isospin--breaking four-- and more nucleon forces 
appear up to this order. Similar to the 2NF, it is useful to classify the 3NF with respect to its 
isospin structure. Following the lines of Ref.~\citep{Epelbaum:2004xf}, one distinguishes between the following 
three classes:  
\beq
\label{3NF_classes}
\begin{array}{lcl}
\mbox{Class I:} & \mbox{\hskip 0.3 true cm} & V_{\rm I} = \sum_{i \neq j \neq k} \left( \alpha_{I}^{ijk}  +  \beta^{ijk}_{I} \, 
\fet \tau_i \cdot \fet \tau_j +  
\gamma^{ijk}_{I} \left[ \fet \tau_i \times \fet 
\tau_j \right] \cdot \fet \tau_k \right) \,,\\ [0.6ex]
\mbox{Class II:}  &                      & V_{II} = \sum_{i \neq j \neq k} \left( \alpha^{ijk}_{II} \, \tau^3_i \tau^3_j  + \beta^{ijk}_{II} 
\, [ \fet \tau_i \times \fet \tau_j  ]^3 \tau^3_k \right)   \,,\\[0.6ex]
\mbox{Class III:}  &                     & V_{III} = \sum_{i \neq j \neq k} \left( \alpha_{III}^{ijk} \, \tau^3_i  + \beta^{ijk}_{III} \, 
[ \fet \tau_i \times \fet \tau_j ]^3  
+ \gamma^{ijk}_{III} \, \tau^3_i \, \fet \tau_j \cdot \fet \tau_k  +  \kappa_{III}^{ijk} \, \tau^3_i  \, \tau^3_j \, \tau^3_k \right) \,,
\end{array}
\eeq
where $\alpha$, $\beta$, $\gamma$ and $\kappa$ are space and spin operators and the indices $i$, $j$, $k$ refer 
to the nucleon labels. The class (I) forces are isospin invariant while the class (II) 3NFs 
break isospin but respect charge symmetry.
Finally, the class (III) forces are charge--symmetry--breaking. 
Contrary to the commonly used classification scheme in the 2N sector, see  Eq.~(\ref{2NF_classes}), conservation of the 
operator $\fet{T}^2$ with $\fet T$ being the total isospin operator, which ensures that  there is no isospin mixing,
is not used in Eq.~(\ref{3NF_classes}). This is because this property  depends on the number of particles in the 
system under consideration. For example, all isospin--breaking two--nucleon forces, which do not 
cause isospin mixing in the two--nucleon system, lead to isospin mixing in the 
three--nucleon system.  

\begin{figure*}[t]
%\vspace{0.3cm}
\centerline{
\psfig{file=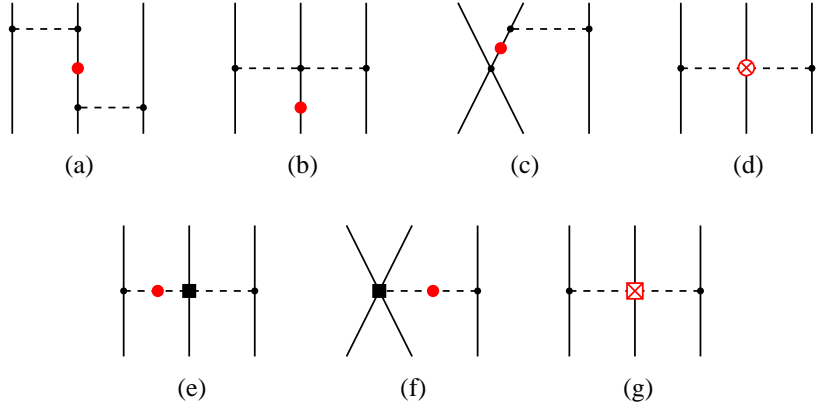,width=11.0cm}
}
\vspace{-0.2cm}
\caption[fig16]{\label{fig16}   Isospin--breaking 3NF at orders $\nu = 4$ (graphs (a)--(d)) and $\nu = 5$ 
(graphs (e)--(g)).   Graphs resulting by the 
interchange of the vertex ordering are not shown. For remaining notation see Figs.~\ref{fig5}, \ref{fig9} and \ref{fig10}.}
\vspace{0.2cm}
\end{figure*}

The leading and subleading isospin--violating 3NFs arise at orders $\nu = 4$ and $\nu = 5$
from diagrams shown in Fig.~\ref{fig16} \citep{Epelbaum:2004xf,Epelbaum:2004qe,Friar:2004rg}. Graphs (a)--(c) 
result from a single insertion of the proton--to--neutron mass difference into the order   $\nu =2$
3NF diagrams shown in Fig.~\ref{fig11}. 
Contrary to the corresponding isospin--invariant contributions discussed in section 
\ref{sec:3NFinvar}, diagrams (a)--(c) in  Fig.~\ref{fig16} yield nonvanishing 3NFs. 
The origin of this difference is explained in detail in Ref.~\citep{Epelbaum:2005fd}. 
One finds the following class (III) 3NFs resulting from diagrams (a) and (c):
\beqa
\label{3NFisosp1}
V^{\rm (4)}_{2\pi} &=& \sum_{i \not= j \not= k} \,2 \delta m  \,  \left(
  \frac{g_A}{2 F_\pi} \right)^4 \frac{( \vec \sigma_i \cdot \vec q_{i}  ) 
(\vec \sigma_j \cdot \vec q_j  )}{(\vec q_i{} ^2 + M_{\pi}^2 )^2 ( \vec
q_j{} ^2 + M_{\pi}^2)} \bigg\{ [\vec q_i \times \vec q_j ] \cdot \vec \sigma_k  \, [ \fet \tau_i \times \fet \tau_j ]^3
 \nn
&& {} + \vec q_i \cdot \vec q_j \left[ (\fet \tau_i \cdot \fet \tau_k ) \tau_j^3  -
(\fet \tau_i \cdot \fet \tau_j ) \tau_k^3 \right] \bigg\}\,, \nn
V^{\rm (4)}_{1\pi , \; \rm cont} &=& \sum_{i \not= j \not= k} 2 \, \delta m \, C_T   \left( \frac{g_A}{2 F_\pi} \right)^2
\frac{\vec \sigma_i \cdot \vec q_i}{(\vec q_i {} ^2 + M_{\pi}^2)^2} \, [ \fet \tau_k \times \fet \tau_i ]^3 \; 
[ \vec \sigma_j \times \vec \sigma_k ] \cdot \vec q_i \,.
\eeqa
The contributions from the $2\pi$--exchange diagram (b) can  naturally  be combined with the contributions from graphs (d) and (g), 
which have the same structure. The resulting 3NF reads:
\beq
\label{3NFprom}
V_{2 \pi}^{\rm (4,5)} = \sum_{i \not= j \not= k}  \bigg(
  \frac{g_A}{2 F_\pi} \bigg)^2 \, \frac{( \vec \sigma_i \cdot \vec q_{i} ) 
(\vec \sigma_j   \cdot \vec q_j  )}{(\vec q_i{}^2 + M_{\pi}^2 ) ( \vec
q_j{}^2 + M_{\pi}^2)} \,
\Bigg[ \frac{( \delta m )^{\rm str}}{4 F_\pi^2} \Big( 2 (\fet \tau_i \cdot \fet \tau_k ) \tau_j^3 - 
(\fet \tau_i \cdot \fet \tau_j ) \tau_k^3  \Big) + f_1 e^2 \tau_i^3 \tau_j^3 \Bigg]\,.
\eeq
%\beqa
%\label{3NFprom}
%V_{2 \pi}^{\rm (4,5)} &=& \sum_{i \not= j \not= k}  \bigg(
%  \frac{g_A}{2 F_\pi} \bigg)^2 \, \frac{( \vec \sigma_i \cdot \vec q_{i} ) 
%(\vec \sigma_j   \cdot \vec q_j  )}{(\vec q_i{}^2 + M_{\pi}^2 ) ( \vec
%q_j{}^2 + M_{\pi}^2)} \nn
%&& \mbox{\hskip 2 true cm} \times
%\bigg[ \frac{( \delta m )^{\rm str}}{4 F_\pi^2} \Big( 2 (\fet \tau_i \cdot \fet \tau_k ) \tau_j^3 - 
%(\fet \tau_i \cdot \fet \tau_j ) \tau_k^3  \Big) + f_1 e^2 \tau_i^3 \tau_j^3 \bigg]\,.
%\eeqa
While the term $\propto (\delta m)^{\rm str}$ breaks charge--symmetry, the contribution 
proportional to the LEC $f_1$ and arising from diagram (g) is charge--symmetry conserving, i.e.~class (II). 
Isospin--violating $\pi \pi NN$ vertex $\propto f_1$ plays an important
role in the analysis of isospin violation in pion--nucleon scattering  \citep{Fettes:2001cr} and the evaluation of the 
ground state characteristics of pionic hydrogen \citep{Gasser:2002am}. In the
2N sector, it only leads to an isospin--invariant contribution to the 2PE potential, which has the same 
form as the $c_1$--term in Eq.~(\ref{2PE_nnlo}). The charge--symmetry--breaking part of $V_{2 \pi}^{\rm (4,5)}$ in 
Eq.~(\ref{3NFprom}) was also obtained by Friar et al.~\citep{Friar:2004rg} using a completely different approach.
In their method, the neutron--to--proton mass difference, which is inconvenient to handle in practical calculations, 
is replaced by a series of new isospin--breaking vertices in the Lagrangian via an appropriate field redefinition, see 
Ref.~\citep{Friar:2004ca} for more details. 
Finally, diagrams (e) and (f) due to the pion mass difference lead to the following class (II) 3NF at order $\nu = 5$:
 \beqa
\label{3NFisosp4}
V^{\rm (5)}_{2\pi}&=&\sum_{i \not= j \not= k} \, \delta M_\pi^2 \, \left(
  \frac{g_A}{2 F_\pi} \right)^2 \frac{( \vec \sigma_i \cdot \vec q_{i})
(\vec \sigma_j   \cdot \vec q_j  )}{(\vec q_i{}^2 + M_{\pi}^2 )^2 ( \vec
q_j{}^2 + M_{\pi}^2)}  \Bigg\{ \tau_i^3 \tau_j^3 \left[  - 
\frac{4 c_1 M_\pi^2}{F_\pi^2} +    \frac{2 c_3}{F_\pi^2} (\vec q_i \cdot \vec q_j )   \right] \nn
&& \mbox{\hskip 1.7 true cm} 
+ \frac{c_4}{F_\pi^2} \, \tau_i^3 \, 
[\fet \tau_j \times \fet \tau_k ]^3 \, [\vec q_i \times \vec q_j ] \cdot \vec \sigma_k \Bigg\} \,, \nn
V^{\rm (5)}_{1\pi, \; \rm cont}&=&- \sum_{i \not= j \not= k} \, \delta M_\pi^2 \, \frac{g_A}{8 F_\pi^2} \, D \, 
\frac{\vec \sigma_i \cdot \vec q_i }{(\vec q_i{}^2  + M_\pi^2 )^2}  
\,  \tau_i^3 \tau_j^3 (\vec \sigma_j \cdot \vec q_i )\,.
\eeqa
These results are consistent with the isospin--invariant 3NFs $V^{\rm (3)}_{\rm 2 \pi}$ and $V^{\rm (3)}_{\rm 1\pi , \; cont}$ 
in Eqs.~(\ref{3nftpe}), (\ref{3nfrest}) being expressed in terms of the charged pion mass. The corresponding 
expressions in coordinate space can be found in Ref.~\citep{Epelbaum:2004xf}.  

\begin{figure*}[t]
%\vspace{0.3cm}
\centerline{
\psfig{file=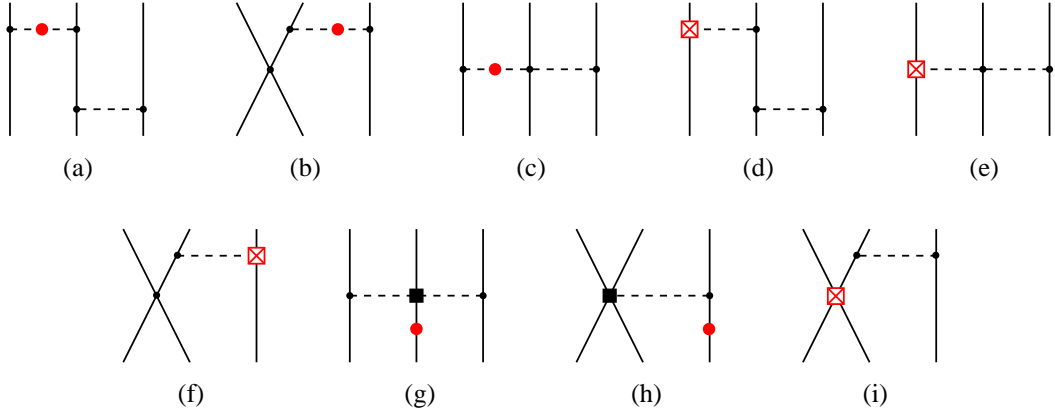,width=14.5cm}
}
\vspace{-0.2cm}
\caption[fig17]{\label{fig17} Leading (a--c) and subleading (d--i) isospin--violating contribution to the 3NF which vanish,
as discussed in the text.   Graphs resulting from the interchange of the vertex ordering are not shown. 
For notation see Figs.~\ref{fig5}, \ref{fig9} and \ref{fig10}.}
\vspace{0.2cm}
\end{figure*}

In addition to graphs shown in Fig.~\ref{fig16}, 
diagrams (a)--(c) and (d)--(i) in Fig.~\ref{fig17} formally contribute to the leading ($\nu = 4$) and subleading 
($\nu = 5$) isospin--breaking 3NF, respectively. Their pertinent contributions, however, vanish.  
In particular, for graphs (a), (b), (d), (f) and (i) one observes the same sort of cancellation between various 
time orderings as in the case of the corresponding isospin--invariant 3NFs, see section \ref{sec:3NFinvar} for more details.
The contributions of diagrams (c) and (e) are suppressed by a factor of $Q/m$ due to the time derivative 
entering the Weinberg--Tomozawa vertex. Explicit evaluation of the remaining contributions of graphs (g) and (h) can be performed 
along the lines of ref.~\citep{Epelbaum:2004xf} and yields vanishing result.

Finally, we point out that there are many $1/m$--corrections to the obtained results which  
start to contribute at order $\nu = 6$. Notice, however, that if one would adopt the counting rule 
$m \sim \Lambda$, various $1/m$--corrections (including the ones due to virtual photons) would contribute already at  
order $(Q/\Lambda)^5$. Some 3NF diagrams due to virtual photon exchange were considered 
by Yang and found to provide relatively small contributions of the order of $\sim 7$ keV 
to the $^3$He--$^3$H binding--energy difference \citep{Yang:1979aa,Yang:1983pd}. Furthermore, we remind the reader that the 
long--range electromagnetic 3NFs might, potentially, cause large effects in scattering at low energy. 
The relative sizes of various isospin--breaking contributions to the two-- and three--nucleon forces discussed in this 
section and in section \ref{sec:isosp} are summarized in Table \ref{tab:isosp_break}. 
We stress that these results rely on the energy--independent formulation and on the counting rules specified 
in Eqs.~(\ref{m_counting}), (\ref{CountRules}) and (\ref{powcNNphotons}). 
\begin{table*}[t] 
\vspace{0.3cm}
\begin{center}
\begin{tabular}{||c||clc|clc||}
\hline \hline 
{} & {} & {} &  &  &  &   {}\\[-1.5ex]
Chiral order   & &  2N force  &   &  &  3N force &  \\[1ex]
\hline \hline 
   &  & &  &  & &    \\[-1.5ex]
$\nu = 0$ & &  $-$ & & &    $-$ &   \\[0.3ex]
$\nu = 1$ & &  $-$ &  &  &  $-$ &   \\[0.3ex]
$\nu = 2$ & &  $V_{1 \gamma} + V_{1 \pi}$    &  &  &  $-$ &  \\[0.3ex]
$\nu = 3$ & &  $V_{1 \pi} + V_{\rm cont}$    &  &  &  $-$ &  \\[0.3ex]
$\nu = 4$ & &  $V_{\pi \gamma} + V_{1 \pi} + V_{2 \pi} + V_{\rm cont}$  &  &  & $V_{2 \pi}+ V_{1\pi, \; \rm cont}$ &   \\[0.3ex]
$\nu = 5$ & &  $V_{1 \pi} + V_{2 \pi} + V_{\rm cont}$                   &  &  & $V_{2 \pi}+ V_{1\pi, \; \rm cont}$ &   \\[1ex]
\hline \hline
  \end{tabular}
%\vskip 0.1 true cm
\caption{Isospin--breaking two-- and three--nucleon forces up to order $\nu =5$. \label{tab:isosp_break}}
\end{center}
\end{table*}

\subsection{Role of the $\Delta$--excitation}
\label{sec:delta}

The $\Delta$--isobar is well known to play an important role in hadronic and nuclear physics.
This is due to its low excitation energy, $\Delta m \equiv m_\Delta - m = 293$ MeV, and strong 
coupling to the $\pi N$ system.  Because of the small value of $\Delta m$, 
it is not clear a priori whether in EFT it should be included explicitly, 
treating $\Delta m$ as a small quantity \cite{Jenkins:1991es}, or integrated out.
On the one hand, inclusion of $\Delta$ yields a scheme, which 
%*EE
%is not strictly rooted in QCD 
differs from the chiral expansion
since $\Delta m$ does not vanish in the chiral limit. On the other hand, it might provide a useful phenomenological 
extension and a systematic power counting has already been worked out assuming 
$\Delta m \sim M_\pi$ \citep{Hemmert:1997ye}, see also \citep{Pascalutsa:2002pi} 
for an alternative scheme. In the chiral EFT discussed so far, the effects of the $\Delta$'s 
are only taken into account implicitly, 
i.e.~through  the values of the corresponding LECs \citep{Bernard:1996gq}. 
We will now discuss the implications of treating the $\Delta$ as a dynamical degree of freedom in the EFT. 

The leading contributions to the nuclear force due to the $\Delta$--excitation can be obtained 
in the heavy--baryon approach from the following terms in the Lagrangian \citep{Ordonez:1995rz,Hemmert:1997ye}  
\beq
\mathcal{L} = \Delta^\dagger \left( i \partial_0 - \Delta m \right) \Delta + \frac{h_A}{2 F_\pi} 
\left( N^\dagger \vec S \fet T \Delta +  \mbox{h.~c.} \right) \cdot \vec \nabla \fet \pi - D_T N^\dagger \vec \sigma \fet \tau N
\cdot \left( N^\dagger \vec S \fet T \Delta +  \mbox{h.~c.} \right) \,,
\eeq
where $h_A$ and $D_T$ are LECs and $\Delta$ is a four--component spinor in both spin and isospin spaces that represents the delta. 
Further, $S_i$ ($T_a$) are the $2 \times 4$ spin (isospin) transition matrices which satisfy the relations
$S_i S_j^\dagger = (2 \delta_{ij} - i \epsilon_{ijk} \sigma_k)/3$ ($T_a T_b^\dagger = (2 \delta_{ab} - i \epsilon_{abc} \tau_c)/3$)
\cite{Ericson:1988aa}.
Throughout this section, we will assume exact isospin symmetry. 
Notice that the large mass scale $m_\Delta$ disappears from the Lagrangian after performing the heavy--baryon expansion. 
Only the small scale $\Delta m$ enters the resulting Lagrangian.

\begin{figure*}[t]
%\vspace{0.3cm}
\centerline{
\psfig{file=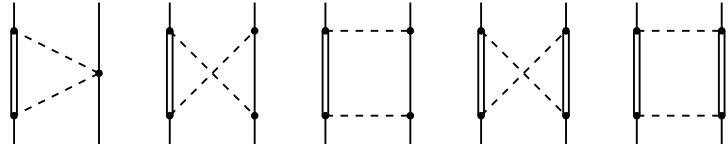,width=10.0cm}
}
\vspace{-0.1cm}
\caption[fig18]{\label{fig18}  Leading contributions to the $2\pi$--exchange potential with single and 
double $\Delta$--excitations. Double lines represent the $\Delta$--isobar.
For remaining notation see Fig.~\ref{fig5}.}
\vspace{0.2cm}
\end{figure*}

The leading contributions to the 2NF due to intermediate $\Delta$ excitations
arise at NLO, $\nu = 2$, and are shown in Fig.~\ref{fig18}. 
In the context of chiral EFT, 
they were first discussed by Ord\'o\~nez et al.~\citep{Ordonez:1995rz} based on ``old--fashioned'' time--ordered 
perturbation theory.\footnote{Certain contributions to the potential were considered much earlier, see e.g.~\citep{Sugawara:1968ty}.}
These contributions were reconsidered by Kaiser et al.~\citep{Kaiser:1998wa} using the 
Feynman graph technique. For completeness, we collect here the expressions for the corresponding non--polynomial pieces 
of the 2PE potential from \citep{Kaiser:1998wa} (generalized to SFR with arbitrary $\tilde \Lambda $) for the three distinct groups of terms: 
\begin{itemize}
\item
$\Delta$--excitation in the triangle graphs:
\beq
\label{del_tri}
V^{\rm 2\pi}_{\Delta, \, {\rm triangle}} = -\frac{h_A^2}{864 \pi^2 F_\pi^4}\,
(\fet \tau_1 \cdot \fet \tau_2) \, \biggl\{ (6E-\omega^2) L^{\tilde \Lambda}(q) +
12 (\Delta m) ^2 E D^{\tilde \Lambda} (q) \biggr\}~,
\eeq
with
\beq
D^{\tilde \Lambda}(q) = \frac{1}{\Delta m} \, \int_{2M_\pi}^{\tilde \Lambda}
\frac{d\mu}{\mu^2+q^2} \, \arctan \frac{\sqrt{\mu^2-4M_\pi^2}}{2\Delta m}~,
\quad \quad
E = 2M_\pi^2 + q^2 -2(\Delta m)^2~.
\eeq 
\item
Single $\Delta$--excitation in the box graphs:
\beqa
\label{del_sing}
V^{\rm 2 \pi}_{\Delta, \, {\rm box}-1} &=& -\frac{g_A^2 \, h_A^2}{48 \pi
  F_\pi^4 \Delta m}\, (2M_\pi^2 +q^2)^2 \,  A^{\tilde \Lambda} (q)\nn
&-& \frac{g_A^2 \, h_A^2}{864 \pi^2 F_\pi^4}\, 
(\fet \tau_1 \cdot \fet \tau_2) \,
\biggl\{(12(\Delta m)^2-20 M_\pi^2-11q^2) L^{\tilde \Lambda}(q) + 6E^2D^{\tilde \Lambda}(q) \biggr\}
\nn
&-& \frac{g_A^2 \, h_A^2}{192 \pi^2 F_\pi^4}\,  
\biggl( (\vec \sigma_1 \cdot \vec q\,)(\vec \sigma_2 \cdot \vec q\,) 
-q^2 (\vec \sigma_1 \cdot\vec \sigma_2 )\biggr) \,
\biggl\{ -2L^{\tilde \Lambda}(q) + (\omega^2-4(\Delta m)^2) D^{\tilde \Lambda}(q) \biggr\} \nn
&-& \frac{g_A^2 \, h_A^2}{576 \pi F_\pi^4 \Delta m}\, (\fet \tau_1 \cdot \fet
\tau_2) \, \biggl( (\vec \sigma_1 \cdot \vec q\,)(\vec \sigma_2 \cdot \vec q\,) 
-q^2 (\vec \sigma_1 \cdot\vec \sigma_2 )\biggr) \, \omega^2 \, A^{\tilde \Lambda}(q)~.
\eeqa
\item
Double $\Delta$--excitation in the box graphs:
\beqa
\label{del_doub}
V^{\rm 2\pi}_{\Delta, \, {\rm box}-2} &=& -\frac{h_A^4}{432 \pi^2
  F_\pi^4}\, \biggl\{-4(\Delta m)^2 L^{\tilde \Lambda}(q) + E[ H^{\tilde \Lambda}(q) + (E+8(\Delta m)^2)D^{\tilde \Lambda}(q) ] \biggr\}
\\
&-& \frac{h_A^4}{7776 \pi^2 F_\pi^4}\, 
(\fet \tau_1 \cdot \fet \tau_2) \,
\biggl\{(12E- \omega^2) L^{\tilde \Lambda}(q) + 3E [H^{\tilde \Lambda}(q)+ (8(\Delta m)^2 -E)D^{\tilde \Lambda}(q)] \biggr\}
\nn
&-& \frac{h_A^4}{3456 \pi^2 F_\pi^4}\,  
\biggl( (\vec \sigma_1 \cdot \vec q\,)(\vec \sigma_2 \cdot \vec q\,) 
-q^2 (\vec \sigma_1 \cdot\vec \sigma_2 )\biggr) \,
\biggl\{ 6L^{\tilde \Lambda}(q) + (12(\Delta m)^2- \omega^2) D^{\tilde \Lambda}(q) \biggr\} \nn
&-& \frac{h_A^4 \,(\fet \tau_1 \cdot \fet \tau_2) }{20736 \pi^2 F_\pi^4} \, \biggl( (\vec \sigma_1 \cdot \vec q\,)(\vec \sigma_2 \cdot \vec q\,) 
-q^2 (\vec \sigma_1 \cdot\vec \sigma_2 )\biggr) \, \biggl\{
2L^{\tilde \Lambda}(q) + (4(\Delta m)^2 + \omega^2 ) D^{\tilde \Lambda}(q) \biggr\}~,
\nonumber
\eeqa
with
\beq
H^{\tilde \Lambda}(q) = \frac{2E}{\omega^2-4(\Delta m)^2} \biggl[ L^{\tilde \Lambda}(q) - L^{\tilde \Lambda} (2\sqrt{(\Delta m)^2 -
  M_\pi^2}) \biggr] \quad .
\eeq
\end{itemize}
Notice that some of the above contributions, in particular, those ones corresponding to the first and the last lines 
in Eq.~(\ref{del_sing}), have precisely the same structure as the corresponding N$^2$LO terms in EFT without $\Delta$'s, see Eq.~(\ref{2PE_nnlo})
provided one chooses 
\beq
\label{delta_satur}
-c_3=2c_4=\frac{h_A^2}{9 \Delta m}\,.
\eeq
Using the large--$N_c$ value $h_A = 3 g_A/\sqrt{2}$ and 
comparing the LECs in Eq.~(\ref{delta_satur}) with the ones in EFT without explicit $\Delta$'s 
\citep{Buettiker:1999ap}, one concludes that the $\Delta$ provides 
the dominant (significant) contribution to $c_3$ ($c_4$).  
Resonance saturation for these and other $\pi N$ LECs is discussed in detail in \citep{Bernard:1996gq}. 
The remaining contributions to the NN potential due to intermediate $\Delta$--excitation in Eqs.~(\ref{del_tri})--(\ref{del_doub}) show a 
non--trivial dependence on the $N \Delta$ mass splitting. We further emphasize that some of the 
subleading corrections to the 2PE potential with intermediate 
$\Delta$--excitations are discussed in \citep{Ordonez:1995rz} while the leading relativistic corrections of order $\nu = 4$
can be found in \citep{Kaiser:1998wa}. 

\begin{figure*}[t]
%\vspace{0.3cm}
\centerline{
\psfig{file=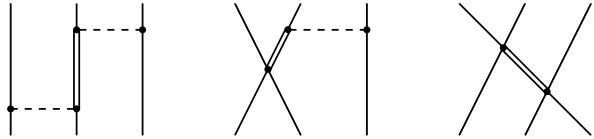,width=8.0cm}
}
\vspace{-0.1cm}
\caption[fig19]{\label{fig19}  Leading contributions to the 3NF due to explicit $\Delta$'s. 
For notation see Figs.~\ref{fig5} and \ref{fig18}.}
\vspace{0.2cm}
\end{figure*}

Let us now switch to the leading 3NF contributions due to intermediate $\Delta$ excitations, which 
also arise at NLO and are shown in Fig.~\ref{fig19}. 
In the context of chiral EFT, they
were first discussed in Ref.~\citep{vanKolck:1994yi}. We remind the reader that in the EFT 
without explicit $\Delta$'s, the first nonvanishing contribution to the 3NF in the energy independent 
formulation only arises at N$^2$LO, $\nu = 3$, see section \ref{sec:3NFinvar}. The  2PE contribution from the 
first graph in Fig.~\ref{fig19} has the same form as the corresponding terms in the theory without delta,
see Eq.~(\ref{3nftpe}), provided one chooses $c_1=0$ and  $c_{3,4}$ according to Eq.~(\ref{delta_satur}). 
Interestingly, the contributions of the two other diagrams  in Fig.~\ref{fig19} vanish 
as a consequence of the Pauli principle, see \citep{Epelbaum:2002vt} for a related discussion.\footnote{Antisymmetric 
(with respect to an interchange of nucleons $i$ and $j$) nature of few--nucleon states is not fully 
incorporated in \citep{vanKolck:1994yi}, which results in the presence of redundant terms.}   

To summarize, the effects due to intermediate $\Delta$ excitations in the EFT 
with explicit $\Delta$'s first show up at NLO (NLO-$\Delta$).  At this order, one finds important 
contributions to the 2PE 2N and 3N potentials.  In the EFT without explicit $\Delta$'s,
the major part of these terms is shifted to N$^2$LO (N$^2$LO-$\Delta \palka$) and represented by the 
contributions proportional to LECs $c_{3,4}$ which are saturated by the $\Delta$. 
In particular, the leading isoscalar central and isovector spin--spin and tensor components  
in Eq.~(\ref{del_sing}) are exactly reproduced at N$^2$LO-$\Delta \palka$. 
For the 3NF, the functional form 
of the NLO-$\Delta$  contribution 
is completely recovered at N$^2$LO-$\Delta \palka \, $. 
We stress, however, that the strength of the 2PE 3NF at N$^2$LO-$\Delta \palka$
is overestimated by a factor of 4/3 compared to the one at NLO-$\Delta$ if 
the values of the LECs $c_{3,4}$ are fixed from $\pi N$ scattering at threshold \citep{Pandharipande:2005sx}. 
Clearly, higher--order counter terms remove this discrepancy. 

Despite the fact that many of the NLO-$\Delta$ contributions to the nuclear force are 
reproduced at N$^2$LO-$\Delta \palka$, it might be advantageous to treat the delta isobar as an explicit 
degree of freedom. This might help to increase the range of applicability of the EFT. 
Further, one expects that such an approach would lead to the contributions to the 
nuclear force of a more natural size.
For example, the large portion of the 2PE potential $\propto c_{i}$ at N$^2$LO-$\Delta \palka$, which is 
known to be very strong, is shifted to NLO in the EFT with explicit $\Delta$'s. 
Similar effects are also observed for isospin--breaking terms, see \citep{Epelbaum:2004xf,Epelbaum:2005fd}. 
We emphasize, however, that such an approach with explicit $\Delta$'s is much more complicated
since one has to deal with more structures and also needs e.g. to reanalyze
pion-nucleon scattering (see Ref.~\citep{Fettes:2000bb} for an attempt). For 
some recent related work in the single--nucleon sector the reader is referred to 
\citep{Pascalutsa:1998pw,Bernard:2003xf,Bernard:2005fy}.

%%%%%%%%%%%%%%%%%%%%%%%%%%%%%%%%%%%%%%%%%%%%%%%%%%%%%%%%%%%%%%%%%%%%%%%%%%%%%%%%%
\section{Applications to few--nucleon systems}
\def\theequation{\arabic{section}.\arabic{equation}}
\setcounter{equation}{0}
\label{sec5}

\subsection{The two--nucleon system}

\subsubsection{Peripheral nucleon--nucleon scattering}
\label{peripheral}

Nucleon--nucleon scattering in peripheral partial waves within the chiral EFT framework  has attracted a 
special interest in the past few years \citep{Kaiser:1997mw,Entem:2002sf,Epelbaum:2003gr,Birse:2003nz,Higa:2004cr},
see also Ref.~\citep{Ballot:1997ht} for a related work. This has several reasons. First, the contribution 
of the short--range contact interactions is suppressed due to the centrifugal barrier in partial waves with 
large values of the orbital angular momentum. For example, no contact terms contribute to D-- (F--) and higher 
partial waves up to N$^3$LO (N$^5$LO). Consequently, peripheral phase shifts are entirely
determined by the long--range part of the nuclear force and thus are expected to provide a sensitive test 
of the chiral 2PE potential. Secondly, the smallness of the corresponding phase shifts suggests that they should be describable in 
perturbation theory. This allows one to avoid the additional complication related to the nonperturbative treatment of the Schr\"odinger equation.  
The validity of perturbation theory is confirmed by the conventional scenario of nuclear forces based on the existing 
boson--exchange models and various phenomenological potentials, see \citep{Epelbaum:2002ji}. It should be understood that the weakness 
of the chiral NN force in peripheral channels, which justifies the use of perturbation theory, may only be verified 
assuming a particular regularization scheme for the chiral potential, which shows a singular behavior at the origin.
Using, for example, dimensional regularization, the iterated 1PE potential was found to be numerically small in 
most peripheral channels \citep{Kaiser:1997mw}. Notice, however, that the weakness of the NN interaction for high values 
of $l$ is not related to the chiral expansion and does not follow from the chiral power counting. We further stress that 
the role of peripheral NN scattering and the resulting constraints on the values of the LECs should not 
be overestimated since the observables at low and moderate energies are, in general, only weakly affected  
by peripheral phase shifts. 

We will now consider D-- and higher partial waves up to N$^2$LO in chiral EFT following the lines of Ref.~\citep{Epelbaum:2003gr}. 
Using Born approximation for the scattering amplitude, 
the phase shifts and mixing angles in the convention of Stapp et al.~\citep{Stapp:1956mz} are determined by the 2N potential as:
\beq
\delta^{sj}_l = - \frac{m q}{(4 \pi)^2}  \langle lsj | V_{\rm 2N} | lsj \rangle \,, \quad \quad
\epsilon_j = - \frac{m q}{(4 \pi)^2}  \langle j-1,sj | V_{\rm 2N} | j+1,sj \rangle \,,
\eeq
where $s$ and $j$ refer to the total spin and angular momentum, respectively, and $q$ is the nucleon CMS momentum. 
Clearly, such an approximation violates unitarity. 
This violation is small provided that the corresponding phase shifts are small. Alternatively,
one can use the $K$--matrix approach in order to restore unitarity, see e.g.~\citep{Entem:2002sf}. 
Here and in what follows, we adopt the same notation for the matrix elements in the $|lsj \rangle$ basis as 
in Ref.~\citep{Epelbaum:2004fk}. The formulae for the partial wave decomposition can be found in appendix B of this reference. 
Up to N$^2$LO, the potential $V_{\rm 2N}$ consists of the $1\pi$--exchange and the leading and subleading $2\pi$--exchange 
terms given in Eqs.~(\ref{opep}), (\ref{2PE_nlo}) and (\ref{2PE_nnlo}), respectively. The contact interactions 
do not contribute to D--waves at this order. For the sake of simplicity, we  
neglect isopin--violating effects in this section. 
In what follows, we adopt the same values of the LECs as in \citep{Epelbaum:2004fk}: $g_A = 1.26$, $F_\pi =92.4$ MeV 
and $M_\pi = 138.03$ MeV. For the Goldberger--Treiman discrepancy in Eq.~(\ref{opep}),
we use $\delta = 0.03$, which leads to $g_{\pi N} \simeq 13.2$. In addition, we have to specify the 
values of the LEC $c_{1,3,4}$ which determine the strength of the subleading 2PE potential. 
Table~\ref{tab:ci} summarizes some of the modern determinations of these LECs, see also Ref.~\citep{Mojzis:1997tu}
where a different notation was used. We also include the LEC $c_2$ in Table~\ref{tab:ci},
which first contributes to the potential at order $\nu = 4$.  
For a precise meaning of the indicated errors in the various determinations, the reader is referred to the 
original publications. Presumably, the most reliable determination of the $c_i$'s from the $\pi N$ system 
in the third order HBCHPT is performed in Ref.~\citep{Buettiker:1999ap}. In this work, the $\pi N$ scattering 
amplitude is reconstructed in the unphysical region using dispersion relations, where the chiral expansion is expected to 
converge rapidly. Unfortunately, this method does not allow for a reliable determination of $c_2$. 
\begin{table*}[t] 
%\vspace{0.6cm}
\begin{center}
\begin{tabular*}{1.00\textwidth}{@{\extracolsep{\fill}}||lc|c|c|c|c||}
\hline \hline 
   &  &  &  &  &   \\[-1.5ex]
   & Ref. & $c_1$  &   $c_2$  &   $c_3$  &   $c_4$   \\[1ex]
\hline \hline 
   &  &  &  &  &   \\[-1.5ex]
$\pi N$, $Q^2$ & \citep{Bernard:1995gx} & $-$0.64(14) & 1.78(20) & $-$3.90(9) & 2.25(9)  \\[0.3ex]
$\pi N$, $Q^3$ & \citep{Bernard:1996gq} & $-$0.93(10) & 3.34(20) & $-$5.29(25) & 3.63(10)  \\[0.3ex]
$\pi N$, $Q^3$ (fit 1) & \citep{Fettes:1998ud} & $-$1.23(16) & 3.28(23) & $-$5.94(9) & 3.47(5)   \\[0.3ex]
$\pi N$, $Q^3$ (fit 1) & \citep{Buettiker:1999ap} & $-$0.81(15) & 9.35(66.7) & $-$4.69(1.34) & 3.40(4)  \\[0.3ex]
$NN$    & \citep{Entem:2002sf} & $-$0.81$^\star$ & 3.28$^\star$ & $-$3.40 & 3.40$^\star$   \\[0.3ex]
$pp$    & \citep{Rentmeester:1999vw} & $-$0.76$^\star$ & -- & $-$5.08(28) & 4.70(70)   \\[0.3ex]
$NN$    & \citep{Rentmeester:2003mf} & $-$0.76$^\star$ & -- & $-$4.78(10) & 3.96(22)   \\[1ex]
\hline \hline
  \end{tabular*}
%\vskip 0.1 true cm
\caption{Recent determinations of the LECs $c_i$. All values are in GeV$^{-1}$. 
The values of the LECs used as input are marked by the star. 
\label{tab:ci}}
\end{center}
\end{table*}
Recently, the Nijmegen group incorporated the chiral 2PE potential up to order $\nu = 3$ in an energy--dependent 
PWA of the $pp$ data \citep{Rentmeester:1999vw} and $pp\, + \, np$ data \citep{Rentmeester:2003mf} and was able to extract 
the values of the LECs $c_{3,4}$, see Table~\ref{tab:ci}. The resulting LECs $c_{3,4}$ are consistent 
with the $Q^3$--determinations in the $\pi N$ system (with the value of $c_4$ being on the upper side). 
We stress that the errors indicated in these references are statistical. 
It is not fully clear how the complete theoretical uncertainty can be estimated in this approach, which is not entirely 
based on EFT, see also \citep{Entem:2003cs} for a related discussion. A somewhat smaller value 
of the LEC $c_3$, $c_3 = -3.4$ GeV$^{-1}$, was found in Ref.~\citep{Entem:2002sf} to be consistent with empirical 
NN phase shifts as well as the results from dispersion and conventional meson theories. 
In that work, F-- and higher partial waves were studied in chiral EFT at order $\nu = 4$.
Notice that the phase shifts depend most sensitively on $c_3$ and are less sensitive on variations of $c_{1,2,4}$. 
Similar smaller values for $c_3$ were also found by Higa \citep{Higa:2004cr}, who looked at peripheral NN waves in the relativistic 
version of chiral EFT at order  $\nu = 4$. Interestingly, similar 
values for the LEC $c_3$ were also extracted recently from matching the chiral expansion of the nucleon mass to lattice 
gauge theory results at pion masses between 500 and 800 MeV \citep{Bernard:2003rp}. Given the present uncertainty in the value of $c_3$, 
we will show the results corresponding to the following two choices: the central value from 
\citep{Buettiker:1999ap}, $c_3 = -4.7$ GeV$^{-1}$, and the value from Ref.~\citep{Entem:2002sf}, $c_3 = -3.4$ GeV$^{-1}$.
For the LECs $c_{1,4}$ we adopt the central values from the $Q^3$--analysis of the $\pi N$ system  \citep{Buettiker:1999ap}:
$c_1=-0.81$ GeV$^{-1}$, $c_4=3.40$ GeV$^{-1}$.

\begin{figure*}[t]
%\vspace{0.3cm}
\centerline{
\psfig{file=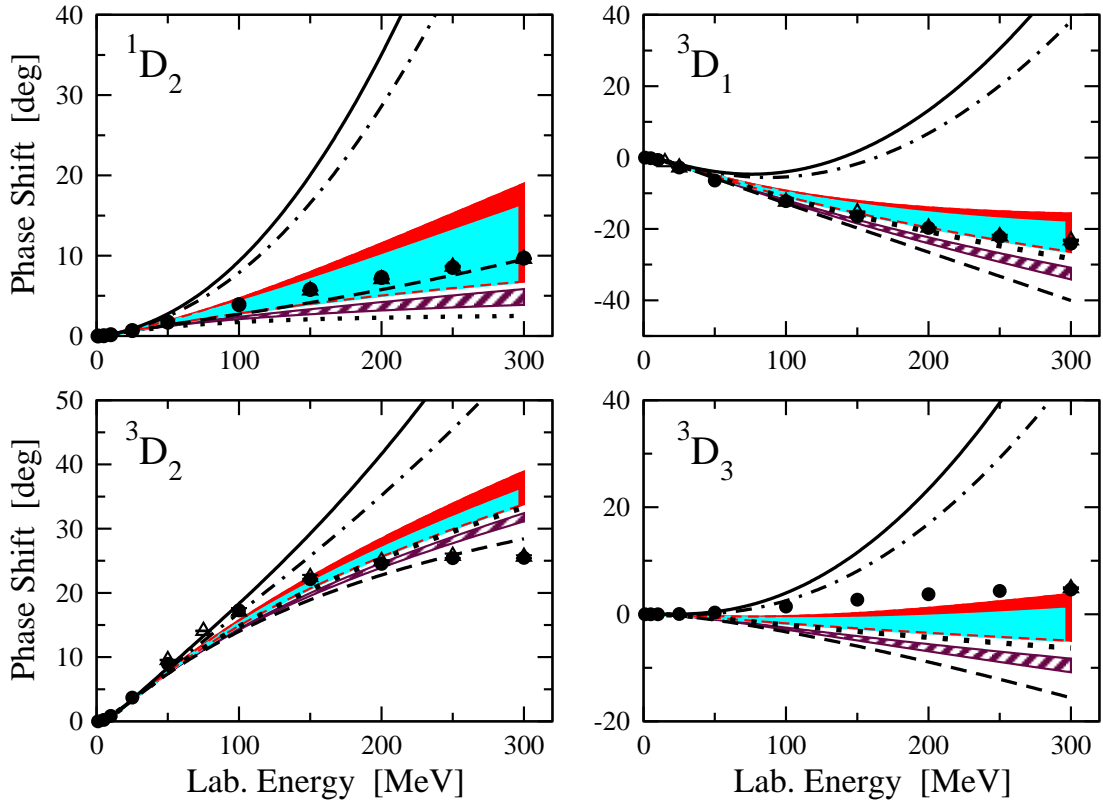,width=15.0cm}
}
\vspace{0.0cm}
\caption[fig20]{\label{fig20} D--wave NN phase shifts. 
The dotted curve is the LO prediction (i.e.~based on the pure 1PE potential).  Dashed, dash--dotted and solid 
curves are the NLO, N$^2$LO with  $c_3 = -3.4$ GeV$^{-1}$ and N$^2$LO with  $c_3 = -4.7$ GeV$^{-1}$ results based on the DR
potential. The dashed, light shaded and dark shaded bands refer to the NLO, N$^2$LO with  $c_3 = -3.4$ GeV$^{-1}$
and N$^2$LO with  $c_3 = -4.7$ GeV$^{-1}$ results  using the SFR with $\tilde \Lambda =500 \ldots 800$ MeV. 
The filled circles (open triangles) depict the results from the Nijmegen multi--energy PWA \citep{Stoks:1993tb,NNonline} 
(Virginia Tech single--energy PWA \citep{SAID}).  }
\vspace{0.2cm}
\end{figure*}

Let us begin with the D--waves which are shown in Fig.~\ref{fig20}. The LO result represented by the
pure 1PE potential already provides a good approximation to the phase shifts in the $^3D_1$ and $^3D_2$ partial
waves. It is too weak in the $^1D_2$ channel and does not describe
properly the $^3D_3$ phase shift.  The latter one appears to be quite small ($| \delta | \sim 4.6 ^\circ$ at 
$E_{\rm lab} = 300$ MeV) compared to the other D--wave phase shifts ($| \delta | \sim 9.7 ^\circ
\ldots 25.5 ^\circ$). The reason is that the partial--wave projected 1PE, taken on the energy shell, 
is strongly suppressed in this channel. Consequently, the  $^3D_3$ phase shift is sensitive
to 2PE but also to the iteration of the potential which is neglected here.
The NLO predictions show a visible improvement for the $^1D_2$ phase shift but go into the wrong 
direction in the $^3D_1$ and  $^3D_3$ channels. The N$^2$LO results based on 
DR are depicted by the dash--dotted and solid lines for the two choices of 
the LEC $c_3$ specified above. The reasonable agreement with the 
data observed at LO and NLO is destroyed in all partial waves for energies $E_{\rm lab} > 50$ 
MeV and the chiral expansion does not seem to converge. Notice that similar deviations from the data are also observed
for mixing angle $\epsilon_2$ which is not shown here \citep{Kaiser:1997mw,Epelbaum:2003gr}.
This disagreement with the data for the DR potential 
was first pointed out in \citep{Kaiser:1997mw}\footnote{There are several minor differences between the results 
presented here, which are based on Ref.~\citep{Epelbaum:2003gr}, and the ones shown in \citep{Kaiser:1997mw}. In particular, 
this latter work uses relativistic kinematics when calculating phase shifts, slightly different values of the LECs 
and also incorporates the contribution from the once iterated 1PE potential. The numerical results of \citep{Epelbaum:2003gr} 
and \citep{Kaiser:1997mw} are, however, rather similar.} and then also reported  in Ref.~\citep{Epelbaum:2003gr}. 
We stress that the results presented here are parameter--free. As pointed out by Kaiser et al.~\citep{Kaiser:1997mw}, 
the origin of the strong disagreement with
the data can be traced back to the central part of the subleading 2PE potential, see Eq.~(\ref{2PE_nnlo}), which is 
proportional to the LEC $c_3$ and shows an unphysically strong attraction at intermediate and short distances
when DR is used to regularize the divergent integrals. This is demonstrated explicitly in Fig.~\ref{fig4a}, where the 
chiral potential is compared to phenomenological $\sigma$ ($\sigma + \omega + \rho$) contributions. 
Notice that another consequence of the strong attraction of the subleading 2PE potential is 
given by the unphysical bound states in low partial waves which could not be avoided in the N$^2$LO 
analysis of Ref.~\citep{Epelbaum:1999dj}. Although such spurious bound states do not influence NN 
observables at low energies, they are at least inconvenient for the application to few--nucleon systems and 
might influence processes like e.g.~{\it Nd}
\citep{Epelbaum:2002ji} or {\it $\pi$d} \citep{Beane:2002wk} scattering.

The strong attraction of the subleading 2PE potential can partially be explained by the large values of the LECs $c_i$. 
For example, the absolute value of the dimensionless coupling corresponding to the LEC $c_3$ is of the order $\sim 10$ 
which has to be compared with its expected natural size of the order $\sim 1$ \citep{Bernard:1996gq}. The origin of this enhancement is well 
understood \citep{Bernard:1996gq}: the LECs $c_{3,4}$ are, to a large extent, saturated by the $\Delta$--excitation.
This implies that a new and smaller scale, namely $m_\Delta - m \sim 293$ MeV, enters these LECs 
in EFT without explicit $\Delta$, see also the discussion in section \ref{sec:delta}. 
As shown in  Fig.~\ref{fig4a}, the large numerical value of $c_3$ appears to be
consistent with the results based on phenomenological potential models at large distances, where 
the chiral potential is unaffected by the regularization procedure. On the other hand, the behavior of the 2PE potential at 
intermediate and short distances depends, to a large extent, on the way one regularizes the corresponding 
loop integrals, see section \ref{sec:sfr}. The unphysically strong attraction of the 2PE potential in the region 
$r \lesssim 1/M_\pi \sim 1.4$ fm arises from high--momentum components of the exchanged pions, which cannot be properly 
treated in an EFT but whose contribution is, nevertheless, included in the potential obtained using DR. 
In the SFR approach, these high--momentum components in the mass spectrum are explicitly removed. 
As depicted in Fig.~\ref{fig4a}, using SFR with the reasonably chosen cut--off 
$\tilde \Lambda =500 \ldots 800$ MeV greatly reduces this unphysical attraction and the resulting 
potential is of the same order in magnitude as the one obtained in phenomenological 
boson--exchange models. The results for D--waves are strongly improved 
using the spectral function regularization instead of dimensional one,
as documented in Fig.~\ref{fig20}. It should be understood that dimensional regularization 
is by no means ruled out by such consideration. In general, for quickly converging expansions, it can and 
should be the method of choice. If, however, the convergence for some well understood physical
reason is slow and (some) observables become sensitive to spurious short--distance physics kept in DR,
it is preferable to use SFR. Choosing DR, one picks up a portion of spurious short--distance physics which, 
under normal circumstances, can be compensated by the counter terms included in the potential at a given order. 
In the case at hand, however, unnaturally large counter terms (contact interactions) are required
due to the large values of the LECs $c_{3,4}$. The neglect of higher--order contact terms 
is, therefore, not justified and results in a large disagreement with the data. It is known that a counter 
term at order $\nu = 4$ is able to remove the bulk of the disagreement observed for D--waves in the DR 
based approach \citep{Epelbaum:1999dj,Richardson:1999hj,Entem:2003ft}. The unnaturally large size of the 
leading counter term in the $^1D_2$ partial wave was also found in Ref.~\citep{Birse:2003nz} based on the 
distorted wave method. On the other hand, applying SFR with the cut--off $\tilde \Lambda$ of the order of the separation scale allows to 
remove the spurious short--range physics from the potential, which results in a more natural size of the counter terms. 
As demonstrated in Fig.~\ref{fig20}, no N$^3$LO counter terms need then to be taken into account at N$^2$LO.  
In other words, the convergence of the chiral expansion is substantially improved using SFR with the reasonably chosen cut--off 
$\tilde \Lambda$ instead of DR. We refer the reader to Refs.~\citep{Donoghue:1998aa,Donoghue:1998bs,Borasoy:2002jv} where a similar 
method has been applied to improve the convergence of the SU(3) baryon CHPT. 

\begin{figure*}[t]
%\vspace{0.3cm}
\centerline{
\psfig{file=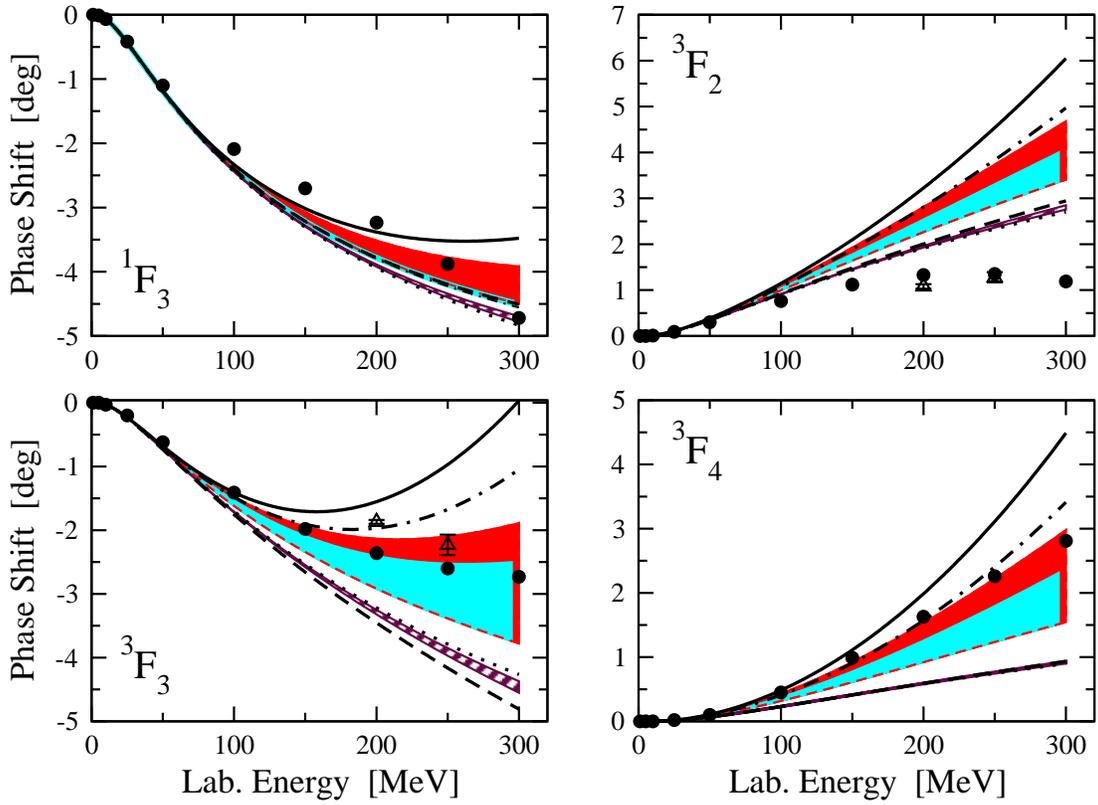,width=15.0cm}
}
\vspace{0.0cm}
\caption[fig21]{\label{fig21} F--wave NN phase shifts. 
%versus the nucleon laboratory energy. 
For notation see Fig.~\ref{fig20}.}
\vspace{0.2cm}
\end{figure*}

The results for F--waves  are presented in Fig.~\ref{fig21}. 
Although the situation with the DR subleading 2PE potential is much less dramatic compared to  D--waves, a too strong
attraction is clearly visible in the $^3F_2$, $^3F_3$ and $^3F_4$ partial waves, especially if one uses $c_3 = -4.7$ GeV$^{-1}$,
see also \citep{Kaiser:1997mw}.  
Removing the short--distance components of the 2PE potential with the cut--off regularization improves the 
results in the $^3F_3$ channel, while additional 
repulsion is still missing in the $^3F_2$ partial wave. A much smaller sensitivity of F--waves to the choice of regularization
and thus to short--range physics is, of course, the consequence of the stronger centrifugal barrier. 
The leading contact terms in F--waves are suppressed by $(Q/\Lambda )^2$ compared to D--waves and 
appear first at order $\nu = 6$. Therefore, even if the corresponding LECs are large in the approach based on DR, neglecting 
those terms leads to much smaller effects in F--waves. 

The dependence of the phase shifts on the SFR cut--off being varied in a certain 
reasonable range reflects the influence of the higher--order contact terms and thus provides a natural estimation of the 
theoretical uncertainty at a given order. The typical uncertainty of $\sim 30$\% observed in 
the N$^2$LO predictions at $E_{\rm lab} =300$ MeV is consistent with the power counting, see \citep{Epelbaum:2003gr}
for more details. In view of this theoretical uncertainty, it appears to be impossible to make any preference in favor of 
the choice $c_3 = -4.7$ GeV$^{-1}$ or  $c_3 = -3.4$ GeV$^{-1}$ based on the D-- and F--waves at N$^2$LO. 

Let us now briefly summarize the main results for D-- and F--waves at N$^2$LO: 
\begin{itemize} 
\vspace{-0.5cm}
\item
The subleading 2PE potential obtained using DR leads to strong disagreement with the data in D--waves (F--waves) 
for $E_{\rm lab} \gtrsim 50$ MeV ($E_{\rm lab} \gtrsim 150$ MeV) and for both choices $c_3 = -4.7$ GeV$^{-1}$ and  $c_3 = -3.4$ GeV$^{-1}$
(for the choice $c_3 = -4.7$ GeV$^{-1}$). \vs
\item
Using SFR instead of DR with the cut--off $\tilde \Lambda = 500 \ldots 800$ MeV strongly (sizably) improves the 
description of the phase shifts in D--waves (F--waves). \vs
\item
The theoretical uncertainty in the description of D-- and F--waves is sizable at large energy.  \vs
\end{itemize}
 
\begin{figure*}[t]
%\vspace{0.3cm}
\centerline{
\psfig{file=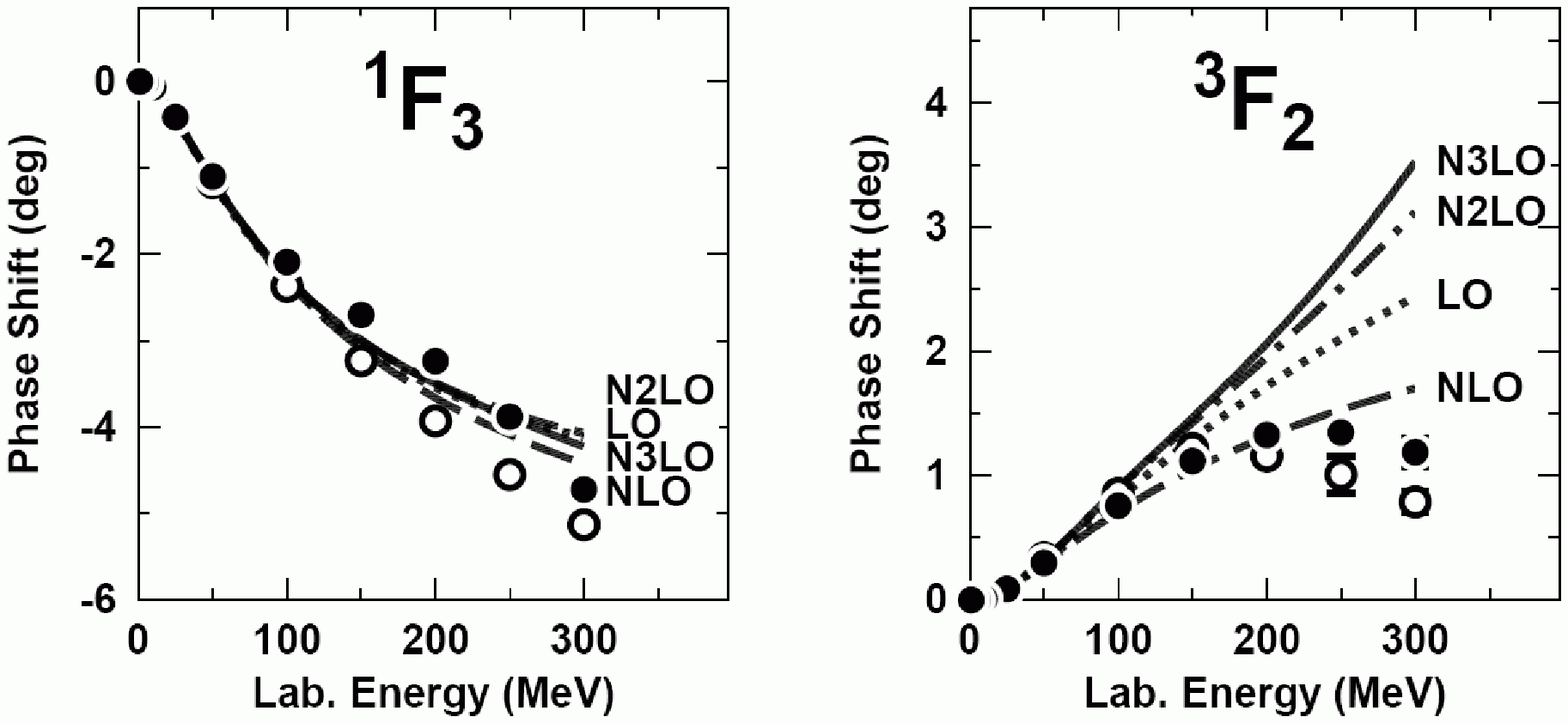,width=8.5cm}
\psfig{file=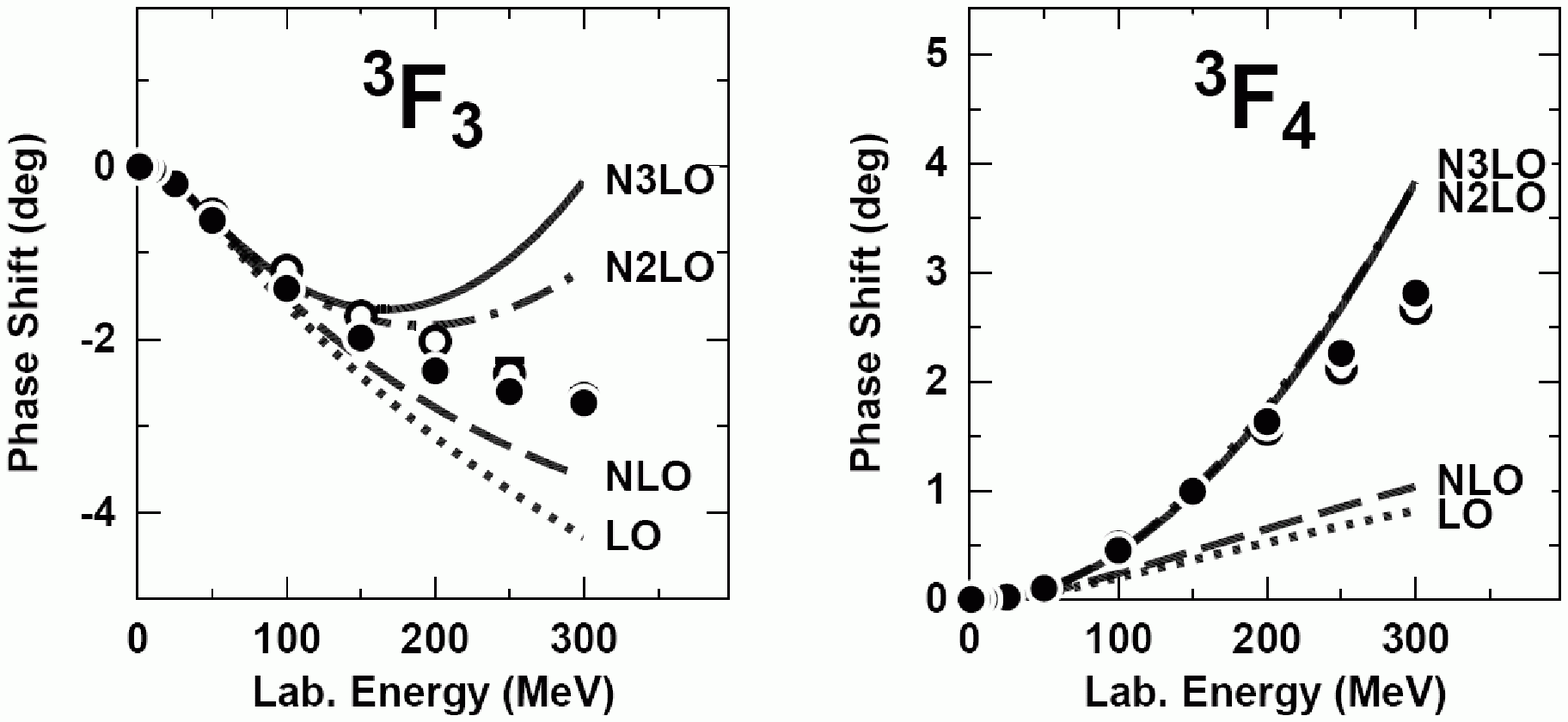,width=8.5cm}
}
\vspace{0.0cm}
\caption[fig22]{\label{fig22} F--wave neutron--proton phase shifts from Ref.~\citep{Entem:2002sf}. 
The solid dots and open circles are the results from Nijmegen PWA \citep{Stoks:1993tb,NNonline} and Virginia Tech analysis SM99 \citep{SAID}.  
Figure courtesy of Ruprecht Machleidt.}
\vspace{0.2cm}
\end{figure*}

As already mentioned before,  F-- and higher NN phase shifts were also studied at order $\nu = 4$ in both the 
standard heavy--baryon \citep{Entem:2002sf} and relativistic framework \citep{Higa:2004cr}.  
The authors of Ref.~\citep{Entem:2002sf} use the perturbative approach similar to the one described above but also take into 
account the contribution of the once iterated 1PE potential and adopt a different counting rule for the nucleon 
mass ($m \sim \Lambda$). F--wave phase shifts obtained in this work and based on the choice $c_3 = -3.4$ GeV$^{-1}$, 
are shown in Fig.~\ref{fig22}. The results at LO, NLO and N$^2$LO are similar to the ones shown in Fig.~\ref{fig21}. 
The N$^3$LO corrections are found to be small in the  $^1F_3$ and  $^3F_4$ channels and noticeable 
in the $^3F_2$ and  $^3F_3$ partial waves, where they slightly increase the disagreement with the data. We will discuss 
NN phase shifts at order $\nu = 4$ in more detail in section \ref{NN_n3lo}.

Last but not least, we would like to comment on an early attempt of Ref.~\citep{Epelbaum:2002ji} to avoid the above mentioned difficulties 
caused by the strong attraction of the subleading 2PE potential.  
In this work based on the DR version of the 2PE potential, smaller in magnitude values of the LECs $c_{3,4}$,
namely $c_3 = -1.15$ GeV$^{-1}$,  $c_4 = 1.20$ GeV$^{-1}$,
were adopted in order to reduce the unphysical attraction at N$^2$LO. 
This allowed for a fairly good description of the NN data with no spurious bound states and enabled applications 
to few--nucleon systems, see e.g.~\citep{Epelbaum:2002vt}. Clearly, such an approach is still not satisfactory 
due to its incompatibility with $\pi$N scattering. The framework based on SFR allows to avoid the 
above mentioned difficulties \emph{and}, at the same time, stay consistent with $\pi$N scattering.  
It should, therefore, be the method of choice. We will show some results based on the reduced $c_{3,4}$ from \citep{Epelbaum:2002ji}
in section \ref{2Nand4N}.

\subsubsection{Regularization and renormalization of the Schr\"odinger equation}
\label{regularization}

In the previous section, we considered peripheral NN scattering 
in perturbation theory. To properly describe low partial waves, where the interaction is strong, 
the Lippmann--Schwinger (LS) equation for the scattering amplitude has to be solved nonperturbatively. 
The nuclear potential derived in chiral EFT is only valid for small momenta and becomes 
meaningless in the large--momentum region. The results presented in section \ref{2NF} show that it 
grows with increasing momenta. Consequently, the LS equation is ultraviolet divergent 
and needs to be regularized (and renormalized). 
The problem of renormalization in the nonperturbative regime in the context of chiral EFT 
attracted a lot of interest in the past years, see e.g.~\citep{Gegelia:1998xr,Gegelia:1998iu,Frederico:1999ps,
Phillips:1999bf,Gegelia:2001ev,Beane:1997pk,Phillips:1997xu,Birse:1998dk,Cohen:1998bv,
Phillips:1998uy,Yang:2004zg,Yang:2004ss,Nogga:2005hy,Birse:2005um}. Here,   
one difficulty is due to the fact that the two--nucleon scattering amplitude 
can only be obtained numerically if pion--exchange contributions are treated nonperturbatively. 
In addition, since the nuclear potential is nonrenormalizable (in the traditional sense),
an infinite number of counter terms is needed to remove all ultraviolet divergences generated 
by the iteration of the potential in the LS equation. 
Notice that this feature is in strong contrast to the pion and single--nucleon sectors as well as to the KSW approach
discussed in section \ref{sec3}, where all divergences at a given order in the chiral expansion can be removed explicitly
by a redefinition of a finite number of parameters in the Lagrangian.  
At first sight, this seems to be in contradiction with the power counting discussed in section \ref{sec3},
which tells us that a finite number of counter terms have to be included in the potential at any finite order
in the chiral expansion. Moreover, the infinite number of counter terms needed to remove all divergences in the LS equation 
raises the question, whether the EFT in the 2N sector still has predictive power. 
Fortunately, this indeed appears to be the case. We recall that the 
power counting outlined in section \ref{sec3} is formulated for renormalized contributions to the amplitude.  
Provided that the renormalized LECs are of natural size as explained in section \ref{sec3}, 
the contribution of higher--order counter terms is suppressed by 
powers of the low momentum scale $Q$ and thus does not need to
be taken into account. 

The standard procedure to renormalize the LS equation is based on Wilson's method \citep{Wilson:1971bg} 
and implies the following two steps. First, 
one solves the LS equation regularized with the finite momentum (or coordinate space) cut--off
and with the kernel represented by the potential truncated at a given order in the chiral expansion.    
Secondly, the LECs, which accompany the short--range contact interactions entering the potential, are 
determined by matching the resulting phase shifts to experimental data. 
Based on the  ``naive dimensional analysis'', see e.g.~\citep{Beane:2000fx}, one expects the 
resulting LECs to be of natural size.\footnote{The determined LECs are, strictly speaking, the bare ones
(with respect to the ultraviolet cut--off in the LS equation) and do not necessarily need to be of natural size.} 
%It is, however, easy to see that e.g.~in HBCHPT, 
%the bare LECs take natural values provided that the renormalized LECs are of natural size and the ultraviolet 
%cut--off $\Lambda$ is chosen of the order of the scale that governs the renormalized LECs. For explicit examples, 
%the interested reader is referred to \citep{Donoghue:1998bs,Bernard:2003rp}. In the following, we will assume 
%that the above arguments also hold true in the nonperturbative case at hand so that one 
%can make use of the ``naive dimensional analysis'' \citep{Beane:2000fx} to understand the role of various terms
%in the potential. Notice further that the LECs of natural size were also found in \citep{Ordonez:1995rz}.}
The contributions of the neglected higher--order 
terms in the potential are expected to be small and irrelevant at the considered order.\footnote{This, 
however, does not mean that adding an additional higher--order term to the potential and keeping the 
values of the LECs unchanged will result in a small correction to the scattering amplitude. 
Due to mixing between operators of different chiral dimension \citep{Lepage:2000}, the LECs have to be readjusted. 
Only then the effect of adding the higher--order term may be expected to be small.}
Notice that the exact form of the regulator used in the LS equation is irrelevant (provided 
it does not affect the long--distance behavior of the potential). There is a certain freedom in choosing the value 
of the cut--off. For the sake of definiteness, let us consider the momentum space cut--off $\Lambda$. It is clear that 
taking a too small $\Lambda$ will remove the truly long--distance physics and reduce the predictive power of the EFT. 
On the other hand, as argued in \citep{Lepage:1997,Lepage:2000,Beane:2000fx}, taking too large values of $\Lambda$ leads 
to a highly nonlinear behavior and should be avoided as well. An example of such a nonlinear behavior 
is given by the appearance and accumulation of bound states for regularized $1/r^n$  attractive singular potentials.
In this case, the renormalization--group behavior of the LECs appears to be of a limit cycle type
(this feature depends on the regularization employed \citep{Braaten:2004rn}), so that the 
``naive dimensional analysis'' is not applicable anymore,\footnote{This should not be 
misunderstood in terms of impossibility to renormalize the amplitude in the presence of the $1/r^n$ potentials 
with the regulator being completely removed (i.e.~in the limit $\Lambda \to \infty$). In fact, the opposite has been demonstrated 
in Ref.~\citep{Beane:2000wh}.} 
%The relation of the power counting in that case to the one described in section \ref{sec3} 
%is, however, unclear.} 
see Ref.~\citep{Nogga:2005hy} for a recent related discussion. This example is also of relevance 
for the 1PE potential, whose tensor component behaves at short distances as $1/r^3$. Taking the 
limit  $\Lambda \to \infty$ in the LS equation will, therefore, generate bound states in partial waves with arbitrarily high values of 
the orbital angular momentum $l$ and lead to strong deviations from the data.  
As a consequence, an infinite number of higher--order counter terms will be needed in this limit in order to restore 
the agreement with the data. 
%Taking the limit $\Lambda \to \infty$ thus provides a highly inefficient way to organize 
%the chiral EFT expansion for the two--nucleon system.  
This problem is, clearly, an artifact of the unphysical behavior of the point--like static 1PE potential at large momenta (or short--distances). 
In this region, the chiral expansion is not applicable, and the 1PE potential does not properly describe the true interaction between 
the nucleons. Using a finite cut--off $\Lambda$ of the order of the characteristic hard scale of the theory allows to avoid this problem
in a natural way. 
It should also be understood that no improvement in the description of the data can, in general, be expected increasing 
$\Lambda$ beyond the pertinent hard scale unless new physics corresponding to this scale is properly incorporated in the theory.   
For a much more extensive discussion of these issues including many explicit examples, the 
reader is referred to \citep{Lepage:2000,Lepage:1997,Beane:2000fx}, see also \citep{Nogga:2005hy,Birse:2005um} 
for a recent related work.

Let us further mention that the consistency of Weinberg's approach has been questioned in Refs.~\citep{Kaplan:1998tg,Kaplan:1998we}, where 
it was pointed out, based on perturbative arguments, that the order   $\nu =2$ contact term $\propto M_\pi^2$ 
needs to be taken into account in order to absorb the ultraviolet divergence generated by the iteration of the order   $\nu =0$ potential. 
This problem with the formal inconsistency of the Weinberg scheme was also  addressed in Ref.~\citep{Beane:2001bc}
based on a nonperturbative approach and found to be present in the $^1S_0$--channel while absent in the coupled $^3S_1$--$^3D_1$--channel.
In that work, a new power counting scheme was invented based on the expansion of the potential about the chiral limit, see \citep{Beane:2001bc}  
for more details. It remains, however, unclear whether the problem with the formal inconsistency of Weinberg's approach 
raised in \citep{Kaplan:1998tg,Kaplan:1998we,Beane:2001bc} and based on the requirement that the amplitude is renormalizable with the regulator being removed, 
is relevant for calculations within the finite cut--off regularization framework as outlined above. Stated differently, it has not been shown 
that the renormalized LECs corresponding to contact terms $\propto M_\pi^2$ take unnaturally large values within this renormalization scheme. 
On the contrary, the phenomenological success of Weinberg's approach and the qualitative arguments of Ref.~\citep{Gegelia:2004pz} 
based on perturbation theory suggest that these LECs are of natural size.

Finally, we would like to emphasize that various alternative regularization schemes were also considered in the context
of NN scattering. In particular, subtractive procedures are discussed in 
\citep{Weinberg:1990rz,Weinberg:1991um,Gegelia:1998iu,Frederico:1999ps,Gegelia:2001ev}, boundary condition regularization scheme 
is applied in \citep{PavonValderrama:2003np,PavonValderrama:2004nb,PavonValderrama:2005gu,PavonValderrama:2005uj} and dimensional regularization
is considered in \citep{Phillips:1999bf}. In addition, a higher--derivative regularization scheme was applied in 
Ref.~\citep{Djukanovic:2004px}, see also \citep{Slavnov:1971} for an early application of this method in the context 
of the nonlinear sigma model.

\subsubsection{Two nucleons up to N$^3$LO}
\label{NN_n3lo}

The first quantitative application of chiral EFT in Weinberg's formulation in the 2N sector 
was performed by Ordon\'{e}z et al.~\citep{Ordonez:1995rz}. This early study was based on 
the energy--dependent NN potential and incorporated the effects of the $\Delta$--resonance. 
The calculations were performed in configuration space up to N$^2$LO. Exponential cut--offs
were used in order to regularize the divergent loop integrals entering the NN potential
and the ones arising from iterations in the LS equation. A local form of the 
employed regulator induced effects of contact interactions in all partial waves. 
Global fits to Nijmegen PWA with 26\footnote{This number includes also the LECs which 
are usually taken from the pion and single--nucleon sectors, such as $g_A$ and $F_\pi$.}
parameters, many of them being redundant since Fierz reordering was not used,
were performed for several choices of the cut--off and first numerical results for phase shifts and deuteron 
properties were obtained. The determined values for the the axial pion coupling and the pion decay constants
are in good agreement with the experimental numbers. 
For more details on this work by Ordon\'{e}z et al., which is an important milestone,
the reader is referred to the original publication \citep{Ordonez:1995rz}.

Park et al.~\citep{Park:1998cu} considered a series of interesting applications,
mostly related to the deuteron properties. They restricted themselves to 
one--pion  exchange and contact interactions for the relevant phases
$^1S_0, ^3S_1, ^3D_1$ and the mixing parameter $\epsilon_1$.  

The first complete N$^2$LO
analysis of neutron--proton scattering and the deuteron properties in the EFT without explicit 
$\Delta$'s and based on the energy--independent potential, obtained using the method of unitary 
transformation \citep{Epelbaum:1998ka}, was performed in Ref.~\citep{Epelbaum:1999dj},
see also \citep{Epelbaum:2000kv} for more details. 
In this work, the DR expressions for the chiral 2PE potential in Eqs.~(\ref{2PE_nlo}), (\ref{2PE_nnlo}) were adopted. 
For $M_\pi$, $F_\pi$,  $g_A$ and the constant $\delta$ in Eq.~(\ref{opep}) the following values were used:
$M_\pi = 138.08 \mbox{ MeV}$, $F_\pi = 93 \mbox{ MeV}$, $g_A = 1.26$, $\delta = 0$.
For $c_{1,3,4}$, the central values from Ref.~\citep{Buettiker:1999ap} were adopted. No isospin--breaking 
interactions were included. Performing anti--symmetrization of the short--range part of the potential,
the total number of independent contact terms was reduced to 9, see Eq.~(\ref{VC}). 
The potential was multiplied by an exponential regulator function, which did not introduce any angular 
dependence, so that the contact interactions only contributed to S-- and P--waves and to the mixing  
angle $\epsilon_1$. The corresponding LECs were determined fitting each low partial wave separately 
to the Nijmegen PWA. As explained in sections \ref{sec:sfr} and \ref{peripheral}, the subleading 2PE
potential calculated using dimensional regularization shows unphysically strong attraction at intermediate 
distances $r \sim 1-2$ fm. The perturbative description of D-- and F--waves based on DR potential 
leads to strong deviations from the data at rather low energies, see section \ref{peripheral} for more details.
The nonperturbative analysis in Ref.~\citep{Epelbaum:1999dj} demonstrated, that phase shifts in S--, P-- and D--waves 
could only be described simultaneously taking the momentum space cut--off $\Lambda$
in the LS equation at least of the order of 1 GeV (unless one includes higher--order counter terms), 
which has to be compared with $\Lambda = 500$ MeV used at NLO.  
With such large cut--offs, the isoscalar central 2PE potential  
is already so strongly attractive that unphysical bound 
states in D-- and in the lower partial waves are generated. 
In addition, phase shifts in D--waves, where the interaction is strong enough to produce bound states and no counter 
terms appear at N$^2$LO, are strongly cut--off dependent. 
In spite of these difficulties, the N$^2$LO potential from \citep{Epelbaum:1999dj} allowed for a good description of the NN data, which 
was also visibly improved compared to the NLO results. In addition, 
NLO analysis with explicit $\Delta$ degrees of freedom was also presented in
Ref.~\citep{Epelbaum:1999dj}. Using the large--$N_c$ value for the $\pi N \Delta$ coupling constant, 
the results were found to be similar to the ones at N$^2$LO in the theory without 
isobars (including the appearance of the spurious  bound states and strong cut--off dependence in D--waves).  
%The results of Ref.~\citep{Epelbaum:1999dj} for the $^3S_1$-- and $\epsilon_1$--channels 
%corresponding to $\Lambda = 500$ MeV
%at NLO and $\Lambda = 875$ MeV at N$^2$LO and for the  theory with explicit $\Delta$ isobars 
%are shown in Fig.~\ref{fig:delta}.

%\begin{figure}[t]
%\vspace{0.3cm}
%\begin{center}
%\psfrag{3S1}{\raisebox{-1.0cm}{\hskip 1.7 true cm $^3S_1$}}
%\psfrag{E1}{\raisebox{-1.0cm}{\hskip -1.9 true cm $\epsilon_1$}}
%\psfrag{1D2}{\raisebox{-1.0cm}{\hskip -1.9 true cm $^1D_2$}}
%\parbox{8.0cm}{\psfig{file=3S1d.ps,width=8.0cm}} 
%\hskip 1 true cm
%\parbox{8.0cm}{\psfig{file=E1d.ps,width=8.0cm}}
%\end{center}
%\vspace{-0.3cm}
%\caption{\label{fig:delta}
%Phase shifts (in degrees) for the theory with explicit $\Delta$ isobars at NLO (solid lines)
%in comparison to the NLO (dotted lines) and  N$^2$LO (dashed lines) results without explicit $\Delta$'s  versus the nucleon laboratory energy  
%(in GeV). Diamonds  depict the results from the Nijmegen multi--energy PWA \protect \citep{Stoks:1993tb,NNonline}.}
%\vspace{0.2cm}
%\end{figure}

In order to get rid of the spurious bound states and enable few--nucleon calculations,
a new version of the N$^2$LO potential was introduced in \citep{Epelbaum:2002ji} based on the numerically reduced values of the LECs $c_{3,4}$:
$c_3 = -1.15 \mbox{ GeV}^{-1}$ and $c_4 = 1.20 \mbox{ GeV}^{-1}$. This choice allowed for 
a fairly good description of the {\it np} data without unphysical bound states and using $\Lambda \sim 500$ MeV.   
The subleading 2PE potential then yields small corrections to the amplitude in most channels.
The values of LECs $c_{3,4}$ quoted above are, however, incompatible with $\pi$N scattering.

Entem and Machleidt studied the 2N system based on the chiral potential at N$^2$LO (in the energy--independent formulation) 
obtained using DR and including contact terms up to N$^3$LO \citep{Entem:2001cg}. The values of the LECs $c_i$ were 
taken consistent with $\pi N$ scattering. The inclusion of higher--order 
counter terms at N$^2$LO appears to be unavoidable in order to compensate for the unphysically strong attraction of the 
2PE potential obtained using DR. 
This, therefore, raises the question about the convergence of the chiral expansion for NN scattering in that framework. 
In their later work \citep{Entem:2003ft}, Entem and Machleidt also incorporated the 2PE potential at N$^3$LO. 
Independently, the 2N system was studied up to N$^3$LO based on the SFR scheme \citep{Epelbaum:2003gr,Epelbaum:2003xx,Epelbaum:2004fk}. 
In the following, we will show the results for various NN observables based on the most recent analyses of 
Refs.~\citep{Entem:2003ft,Epelbaum:2004fk}. 

As detailed in section \ref{2NF}, the chiral potential at N$^3$LO includes the 1PE, 2PE and 3PE contributions, 24  
isospin--invariant contact terms as well as isospin--violating corrections. The leading 3PE potential turns out to be rather weak
(especially in the SFR framework), see section \ref{sec:piexch}, and was neglected in \citep{Entem:2003ft,Epelbaum:2004fk}.
Both analyses use $F_\pi= 92.4$ MeV and $g_A (1 + \delta ) =1.29$. 
In Table \ref{tab:ci2} we summarize the values of the LECs $c_i$ and $\bar d_i$ adopted in these studies in comparison 
with the ones extracted from $\pi N$ scattering in Ref.~\citep{Buettiker:1999ap} in the case of  $c_{1,3,4}$ 
and in Ref.~\citep{Fettes:1998ud} in the case of  $c_{2}$ and $\bar d_i$. The LECs $c_{2,3,4}$ were fine 
tuned in Ref.~\citep{Entem:2003ft}, which resulted, in particular, in the large value for $c_4$
incompatible with $\pi N$ scattering. In the analysis of \citep{Epelbaum:2004fk}, the central values from the $\pi N$ system
are used for all LECs with the only exception of $c_3$, for which the value determined in \citep{Entem:2002sf}
is adopted. This choice is on the lower side but still consistent with the result from \citep{Buettiker:1999ap}.
Using this value for $c_3$ turns out to be important at N$^2$LO in order to properly describe the 
$^3P_0$ phase shift \citep{Epelbaum:2003xx}. 

\begin{table*}[t] 
%\vspace{0.6cm}
\begin{center}
\begin{tabular*}{1.0\textwidth}{@{\extracolsep{\fill}}||l|c|c|c|c|c|c|c|c||}
\hline \hline 
   &  &  &  &  &   &  &  &   \\[-1.5ex]
   &  $c_1$  &   $c_2$  &   $c_3$  &   $c_4$ & $\bar d_1 + \bar d_2$ & $\bar d_3$ & $\bar d_5$ & $\bar d_{14} - \bar d_{15}$  \\[1ex]
\hline \hline 
   &  &  &  &  &  &  &  &    \\[-1.5ex]
$NN^a$ &  $-$0.81 & 2.80 & $-$3.20 & 5.40 & 3.06 & $-$3.27 & 0.45 & $-$5.65  \\[0.3ex]
$NN^b$ &  $-$0.81 & 3.28 & $-$3.40 & 3.40 & 3.06 & $-$3.27 & 0.45 & $-$5.65  \\[0.3ex]
$\pi N$  & $-$0.81(15) & 3.28(23) & $-$4.69(1.34) & 3.40(4) & 3.06(21) & $-$3.27(73) & 0.45(42) & $-$5.65(41) \\[1ex]
\hline \hline
  \end{tabular*}
%\vskip 0.1 true cm
\caption{LECs used in the N$^3$LO analyses $NN^a$ (Ref.~\protect\citep{Entem:2003ft}) and  $NN^b$ (Ref.~\protect\citep{Epelbaum:2004fk})
compared to the values obtained from $\pi N$ scattering \protect\citep{Buettiker:1999ap,Fettes:1998ud}. 
The $c_i$ ($\bar d_i$) are in units of GeV$^{-1}$ (GeV$^{-2}$).  
\label{tab:ci2}}
\end{center}
\end{table*}

Isospin--breaking corrections are treated differently in the analyses of \citep{Entem:2003ft} and \citep{Epelbaum:2004fk}.
In \citep{Entem:2003ft}, they are incorporated following the lines of Ref.~\citep{Walzl:2000cx}.
Specifically, the authors of \citep{Entem:2003ft} include the pion mass difference in 1PE, see Eq.~(\ref{OPEnijm}),
the Coulomb potential in {\it pp} scattering, pion mass difference in the 
order   $\nu=2 $ 2PE potential as defined in Eq.~(\ref{Vcib}), the $\pi \gamma$--exchange potential in Eq.~(\ref{Vpiga})
and the two lowest--order isospin--breaking 
contact terms proportional to $\tilde \beta_{1S0}^{pp}$ and $\tilde \beta_{1S0}^{nn}$ in Eq.~(\ref{VC_iso}). 
In Ref.~\citep{Epelbaum:2004fk}, the long--range isospin--breaking corrections are treated in the same way as in the 
Nijmegen PWA \citep{Stoks:1993tb}, i.e.~are based on the pion mass difference in 1PE and the electromagnetic interactions 
in Eq.~(\ref{vc1vc2}). In addition, the two leading contact interactions proportional 
to $\tilde \beta_{1S0}^{pp}$ and $\tilde \beta_{1S0}^{nn}$ were included. Since phase shifts from the Nijmegen PWA were used in 
\citep{Epelbaum:2004fk} to fix the values of the unknown LECs, treating isospin--breaking corrections differently 
from Ref.~\citep{Stoks:1993tb} would lead to inconsistency.\footnote{The major problem is that the {\it np} isovector 
phase shifts (except in the $^1S_0$ channel) in the Nijmegen PWA \protect\citep{Stoks:1993tb} 
are not obtained independently from {\it np} data but rather extracted from the {\it pp} 
phase shifts by an appropriate change in the 1PE potential and switching off the electromagnetic interaction.} 
  
Also the treatment of the relativistic effects is rather different in \citep{Entem:2003ft} and \citep{Epelbaum:2004fk}.
The work of \citep{Epelbaum:2004fk} is based on the relativistic Schr\"odinger/Lippmann--Schwinger equation 
as explained in section \ref{sec:rel}.\footnote{As outlined in section \ref{sec:rel}, the relativistic 
Schr\"odinger equation can be cast in to equivalent nonrelativistic forms. The corresponding phase--equivalent 
potentials are also discussed in \citep{Epelbaum:2004fk}.}  On the contrary, the analysis of Ref.~\citep{Entem:2003ft} is based on the 
nonrelativistic Schr\"odinger equation and uses the static 1PE potential and the $1/m$-- and $1/m^2$--corrections to the 
2PE potential from Refs.~\citep{Kaiser:1997mw,Kaiser:2001at}, where no particular dynamical equation was specified. 
It is, therefore, not clear whether the relativistic corrections to the potential used in \citep{Entem:2003ft}
are consistent with the nonrelativistic Schr\"odinger equation. 
%We further emphasize that the authors of \citep{Entem:2003ft} also 
%incorporate the $1/m^2$--corrections to the 2PE potential obtained in \citep{Kaiser:2001at}. 

\begin{figure}[t]
%\vspace{0.3cm}
\centerline{
\psfig{file=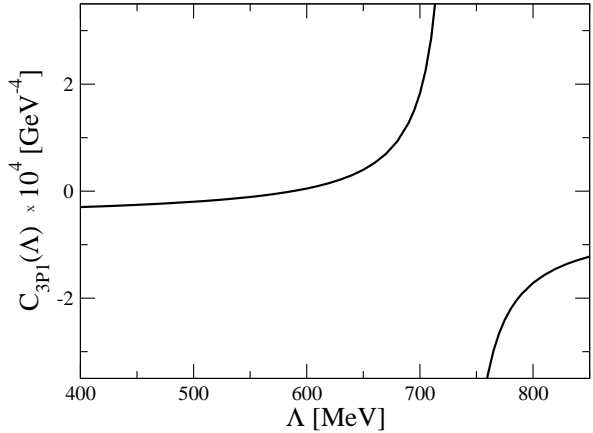,width=8.0cm}}
\vspace{-0.15cm}
\caption{\protect \small
``Running'' of the LEC $C_{3P1}$ with the cut--off $\Lambda$ at N$^2$LO.
The cut-off $\tilde \Lambda$ in the spectral function representation
is fixed at the central value $\tilde \Lambda = 600$ MeV.
}\label{fig:3p1}
\vspace{0.2cm}
\end{figure}

Further differences between the two analyses can be attributed to different regularization schemes, 
fitting procedures and error estimations. In Ref.~\citep{Epelbaum:2004fk}, SFR with the cut--off $\tilde \Lambda$ 
was employed in order to regularize divergent loop integrals entering the potential. 
Following the standard procedure, 
see e.g.~\citep{Epelbaum:1998ka,Epelbaum:1999dj,Epelbaum:2003gr,Epelbaum:2003xx,Epelbaum:2002ji},  
the resulting potential $V (\vec p, \; \vec p \, ')$ is multiplied with a regulator function 
$f^\Lambda$,
\beq
\label{pot_reg}
V (\vec p, \; \vec p \, ') \rightarrow f^\Lambda ( p ) \, 
V (\vec p, \; \vec p \, ')\, f^\Lambda (p ' )\,,
\eeq 
in order to remove the divergences in the LS equation, where the exponential regulator function 
\beq
\label{reg_fun}
f^\Lambda (p ) = \exp [- p^6/\Lambda^6 ]
\eeq 
was used.
The following cut--off combinations (all values in MeV) were adopted in \citep{Epelbaum:2004fk}:
\beq
\label{cutoffs}
\{ \Lambda, \; \tilde \Lambda \} = \{ 450, \; 500 \},  \; \{ 600, \; 600 \},  \; 
\{ 450, \; 700 \},  \; \{ 600, \; 700 \}\,.
\eeq
We remind the reader that while one, in principle, could further decrease the value of $\Lambda$ 
(at the cost of the reduced accuracy), increasing $\Lambda$ beyond $\sim 650$ MeV leads to the appearance 
of spurious bound states and one enters the regime, where the Weinberg power counting 
and ``naive dimensional analysis'' are not applicable, see 
the discussion in section \ref{regularization}. For $\tilde \Lambda = 500$ MeV, the value 
$\Lambda = 600$ MeV was already found to be close to its critical value corresponding to the appearance 
of a bound state. A typical dependence of the LECs on the cut--off $\Lambda$ is exemplified in Fig.~\ref{fig:3p1}, where 
we show the ``running'' of the LEC $C_{3P1}$ at N$^2$LO from Ref.~\citep{Epelbaum:2003xx}. 
The discontinuity in values of $C_{3P1}$ 
for $\Lambda \sim 700$ MeV corresponds to the accumulation of the first spurious  bound state in this channel. 
Further increasing the cut--off $\Lambda$ leads to discontinuities in the values of $C_{3P1}$ corresponding 
to the accumulation of the second, third, etc.~bound states. The limit cycle behavior of $C_{3P1}$
is typical for singular $1/r^n$ potentials\footnote{The N$^2$LO 2PE potential in the SFR scheme behaves at short 
distances as $1/r^5$, see \protect\citep{Epelbaum:2003gr}.} and is similar to the cut--off dependence of the strength of the 
contact 3N force observed in Refs.~\citep{Bedaque:1998kg,Bedaque:1998km}. 

The analysis of Ref.~\citep{Entem:2003ft} is based on the DR potential. The LS equation is 
regularized multiplying the potential with a regulator function
\beq
\label{regulator_EM}
f^\Lambda (p ) = \exp [ - p^{2n}/\Lambda^{2n} ]\,, 
\eeq
where the exponent $2 n$ was chosen to be sufficiently large, so that the regulator generates 
powers which are beyond the order $\nu = 4$. Notice that $n$ takes different values for various terms in the  
potential \citep{Machleidt:2003} (the actual values of $n$ are not specified in  Ref.~\citep{Entem:2003ft}). 
The choice $\Lambda = 500$ MeV was adopted in this work. 

The 24 isospin--invariant $+$ 2 isospin violating LECs corresponding to contact interactions (cf.~Eqs. (\ref{VC})
and (\ref{VC_iso})) were fixed from a fit to the phase shifts with the orbital angular momenta $l \leq 2$. 
In Ref.~\citep{Entem:2003ft}, the fit was subsequently refined by minimizing the $\chi^2$ obtained from a direct 
comparison with the data. The process of determination of the LECs is explained in detail in 
Ref.~\citep{Epelbaum:2004fk} and deserves some comments. First, we note that, in general, 
one has to expect multiple solutions for the LECs. This problem has already been discussed 
in \citep{Epelbaum:1999dj} at NLO and N$^2$LO. It is difficult to select out the true solution 
in the $^1S_0$ channel, where
five LECs $\tilde \beta_{1S0}^{pp}$, $\tilde C_{1S0}$, $C_{1S0}$, $D_{1S0}^1$, $D_{1S0}^2$ need to be fixed 
from a fit to the Nijmegen {\it pp} and {\it np} phase shifts\footnote{The LEC $\tilde \beta_{1S0}^{nn}$ was 
then obtained from the requirement to reproduce the {\it nn} S--wave scattering length.} as well as for 
the $^3S_1$--$^3D_1$--channels, where eight LECs have to be determined simultaneously. 
In the simpler case of P--waves, where only two LECs need to be determined simultaneously (except in the coupled 
$^3P_2$--$^3F_2$ channels where one has 3 LECs), the choice of the solution is usually unambiguous if 
one makes use of the naturalness assumption.  For example, the following two solutions for the LECs 
$C_{3P1}$ and $D_{3P1}$ were found for the cut--off combination $\{ \Lambda, \; \tilde \Lambda \} = \{ 450, \; 500 \}$:
$C_{3P1} = -0.6334 \times 10^4$ GeV$^{-4}$, $D_{3P1} = 4.2359 \times 10^4$ GeV$^{-6}$ and 
$C_{3P1} = 5.9620 \times 10^4$ GeV$^{-4}$, $D_{3P1} = -20.6154 \times 10^4$ GeV$^{-6}$ \citep{Epelbaum:2004fk}.
Both sets of LECs lead to a similarly accurate description of the data. The first 
solution, however, fulfills the naturalness assumption while the second does not, i.e.~the corresponding dimensionless
coefficients are of the order $\sim 10$. Notice further that 
$C_{3P1} = -0.6334 \times 10^4$ GeV$^{-4}$ is close to the NLO and N$^2$LO values 
(for the same cut--off combination) $C_{3P1} = -0.4932 \times 10^4$ GeV$^{-4}$ and 
$C_{3P1} = -0.7234\times 10^4$ GeV$^{-4}$, respectively.
The results for other P--waves are similar. No multiple solutions arise in  D--waves, where a single  LEC 
needs to be determined in each channel.  

Let us now comment on the size of the obtained LECs. 
In general, the natural size for the LECs can be (roughly) estimated as follows:
\beq
\label{natur_units}
\tilde C_i \sim \frac{4 \pi}{F_\pi^2}\,, \quad \quad 
C_i \sim \frac{4 \pi}{F_\pi^2 \Lambda_{\rm LEC}^2}\,, \quad \quad 
D_i \sim \frac{4 \pi}{F_\pi^2 \Lambda_{\rm LEC}^4}\,,
\eeq
where $\Lambda_{\rm LEC}$ is the scale entering the values of the LECs and 
the factor $4 \pi$ results from the angular integration in the partial wave decomposition.
All LECs determined in \citep{Epelbaum:2004fk} except $D_{1S0}^1$ and $D_{3S1}^1$ were found to be of natural size
for all cut--off combinations. For example, the $D_i$'s expressed in the natural units defined in Eq.~(\ref{natur_units})
take the values in the range $-1.1 \ldots 2.0$ if one adopts $\Lambda_{\rm LEC} = 500$ MeV. 
For the LECs $D_{1S0}^1$ and $D_{3S1}^1$ this range is, however, $-5.2 \ldots -11.2$. 
Still, the higher--order contact interactions are suppressed compared 
to the lower--order operators at low momenta. 
For example, for the cut--off combination $\{ 450, \; 500 \}$ and 
$p = p' = M_\pi$, the contributions of the contact operators at various orders are given by:
\beqa
\langle ^1S_0 | V_{\rm cont}^{\rm np} (p, \; p' \,)| ^1S_0 \rangle \bigg|_{p,p' = M_\pi}
&=& \bigg[\tilde C_{1S0}^{\rm np} + C_{1S0} ( p^2 + p '^2) +
\Big( D_{1S0}^1 \, p^2 \, {p'}^2 + D_{1S0}^2 \, ({p}^4+{p}'^4) \Big) \bigg]_{p, p' = M_\pi} \nn
&=& \Big[-0.091 + 0.057 +(-0.010+0.003) \Big] \times 10^4 \mbox{ GeV}^{-2}\,.
\eeqa 
For more details on the determination of various LECs and for their explicit values the reader is referred to 
Ref.~\citep{Epelbaum:2004fk}. We further stress that the LECs of natural size were also found in \citep{Ordonez:1995rz}.
 
We now turn to the discussion of phase shifts.
Before showing the results, we would like to make a simple estimate for the 
expected theoretical uncertainty. Following the rules of the ``naive dimensional analysis'', 
we expect for the uncertainty of a scattering observable at CMS momentum $k$ at N$^3$LO to be of
the order $\sim ( \max [ k, \, M_\pi ] /\Lambda_{\rm LEC} )^{5}$. To provide a fair estimate, 
we identify the hard scale with the smallest value of the ultraviolet 
cut--off, i.e.~we adopt $\Lambda_{\rm LEC} \sim 450$ MeV. This value is consistent with the 
natural size of the determined LECs and yields the following estimations for the (maximal) theoretical 
uncertainty: $\sim$ 0.5\%  for $E_{\rm lab} \simeq 50$ MeV and below, $\sim$ 7\%  for $E_{\rm lab} \simeq 150$ MeV
and $\sim$ 25\%  at $E_{\rm lab} \simeq 250$ MeV.
%
%\bigskip 
%\vskip -0.6 true cm
%\hskip 3 true cm
%\begin{minipage}{10cm}
%\begin{itemize}
%\item[]
%$\sim$ 0.5\%  at $E_{\rm lab} \sim 50$ MeV and below, \vs
%\item[]
%$\sim$ 7\%  at $E_{\rm lab} \sim 150$ MeV, \vs
%\item[]
%$\sim$ 25\%  at $E_{\rm lab} \sim 250$ MeV. \vs
%\end{itemize}
%\end{minipage}
%
%\noindent
One should keep in mind that the above estimations are fairly rough. 
For a more detailed discussion on the theoretical uncertainty
%, especially at NLO and N$^2$LO, 
the reader is referred to \citep{Epelbaum:2003xx}. 

\begin{figure*}[t]
%\vspace{0.0cm}
\centerline{
\psfig{file=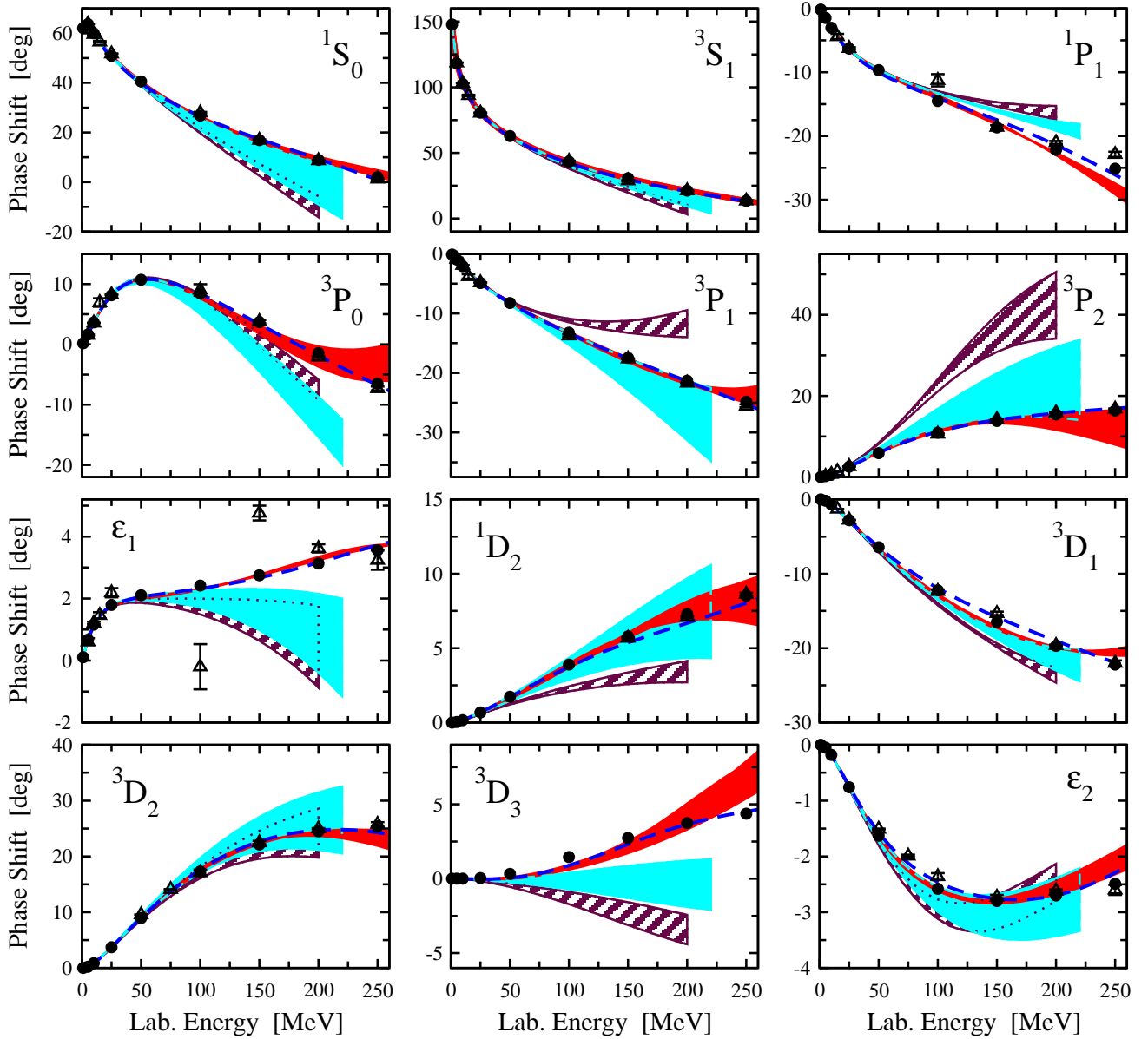,width=18.0cm}
}
\vspace{-0.15cm}
\caption[fig23]{\label{fig23} S--, P-- and D--waves {\it np} phase shifts. 
%versus the nucleon laboratory energy.
The dashed, light shaded and dark shaded bands show the NLO, N$^2$LO and N$^3$LO  \citep{Epelbaum:2004fk} results, respectively. 
The dashed line is the N$^3$LO result of Ref.~\citep{Entem:2003ft}. 
The filled circles (open triangles) depict the results from the Nijmegen multi--energy PWA \citep{Stoks:1993tb,NNonline} 
(Virginia Tech single--energy PWA \citep{SAID}).  }
\vspace{0.2cm}
\end{figure*}

The results at NLO, N$^2$LO and N$^3$LO for S--, P-- and D--waves and mixing angles $\epsilon_1$ 
and $\epsilon_2$ from Ref.~\citep{Epelbaum:2004fk} are shown in Fig.~\ref{fig23}
in comparison with the PWA results of Refs.~\citep{Stoks:1993tb,NNonline,SAID}. The bands correspond to 
variation of the cut--offs as specified in Eq.~(\ref{cutoffs}). 
Low--energy observables should not depend on the cut--off(s) provided that all terms
in the EFT expansion are included. In practice, however, calculations are performed at a 
finite order, so that some (small) residual dependence of observables on the cut--off(s) 
remains. One, in general, expects that this cut--off dependence gets weaker when   
higher order terms are included. While this is indeed the case for 
N$^3$LO, the bands at NLO and N$^2$LO are of comparable width. This does, however, not indicate 
any inconsistency as the following arguments show. 
The cut--off dependence of the scattering amplitude at both NLO and N$^2$LO has to be 
compensated by inclusion of the counter terms (contact interactions) at order $\nu = 4$ 
and higher. The contact interactions appear only at even orders 
$\nu = 2 n, \; n \geq 0$ in the low--momentum expansion while pion exchanges contribute, in general,
at both even and odd orders. Since the same contact terms enter the expression 
for the effective potential at NLO and N$^2$LO, a similar cut--off dependence
for the observables at these orders should be expected. The results shown in  Fig.~\ref{fig23} confirm 
these expectation. 

\begin{figure*}[t]
%\vspace{0.0cm}
\centerline{
\psfig{file=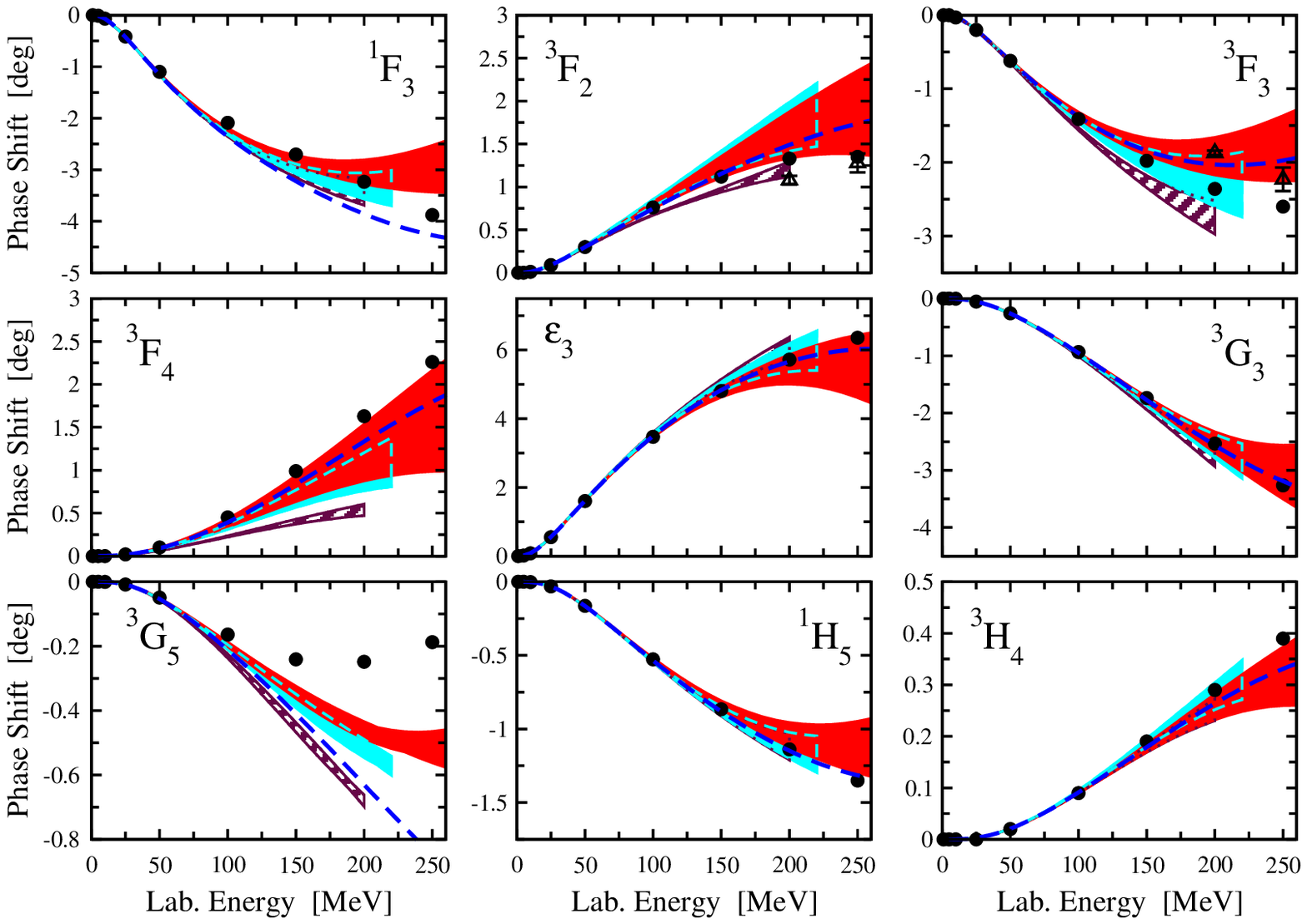,width=18.0cm}
}
\vspace{-0.15cm}
\caption[fig24]{\label{fig24} Selected peripheral {\it np} phase shifts. 
%versus the nucleon laboratory energy.
For notation see Fig.~\ref{fig23}.}
\vspace{0.2cm}
\end{figure*}

The uncertainty due to the cut--off variation in S--, P-- and D--waves 
at N$^3$LO agrees well with the estimation given above. For example, at $E_{\rm lab} = 250$ MeV, the P--wave phase shifts 
at N$^3$LO deviate from the data by an amount of up to $\sim 8^\circ$, which has to be compared with the 
typical  size of the P--wave phase shifts at this energy of the order $\sim 25^\circ$.
We further emphasize that all phase shifts at N$^3$LO are visibly improved compared to N$^2$LO (and NLO). 
The results at N$^3$LO provide an accurate description of the data up to $E_{\rm lab} \sim 200$ MeV.
It is comforting  to see that the bands at N$^2$LO and N$^3$LO overlap (except in some channels at higher energies).
As pointed out before, one cannot expect the same for the NLO bands, which underestimate the theoretical 
uncertainty at this order.  Finally, we also show the N$^3$LO results from Ref.~\citep{Entem:2003ft}
based on $\Lambda = 500$ MeV. The two N$^3$LO analyses agree with each other in all cases 
except some minor deviations in the partial waves $^1P_1$ and $^3D_1$ at intermediate 
and higher energies. 

F-- and selected higher partial waves, which are parameter--free at the considered orders, are shown 
in Fig.~\ref{fig24}. In most channels, the predictions at N$^3$LO are in agreement with the data. 
Contrary to the previously considered case of low partial waves, the bands  do not get thinner at N$^3$LO.  
This has to be expected due to the lack 
of the short--range contact terms in these channels. Such terms start to contribute to 
F--waves at N$^5$LO ($\nu = 6$) and to G--waves at N$^7$LO ($\nu =8$).  Consequently, 
a similar uncertainty due to the cut--off variation should be expected up to these high 
orders in the chiral expansion. The largest deviation from the data is observed for the $^3G_5$ 
partial wave. In this channel, the 1PE and the leading 2PE potentials are not sufficient to reproduce 
the PWA result at energies beyond $\sim 100$ MeV. The 2PE corrections at N$^2$LO and N$^3$LO improve the description 
of the data, but the effects are not big enough. This disagreement should, however, not be taken too seriously
because of the exceptionally small size of the corresponding phase shift (more than 10 times 
smaller in magnitude compared to other G--waves). The N$^3$LO predictions from Ref.~\citep{Entem:2003ft}
are also plotted in Fig.~\ref{fig24}. They lie inside the N$^3$LO bands from Ref.~\citep{Epelbaum:2004fk}
in all cases with the exception of the $^1F_3$ and $^3G_5$ partial waves. The relatively large deviations 
in these two channels might be caused by the different values of $c_i$ (especially of the LEC $c_4$)
adopted in \citep{Entem:2003ft}.  

\begin{figure*}[t]
%\vspace{0.3cm}
\parbox{8.9cm}{\psfig{file=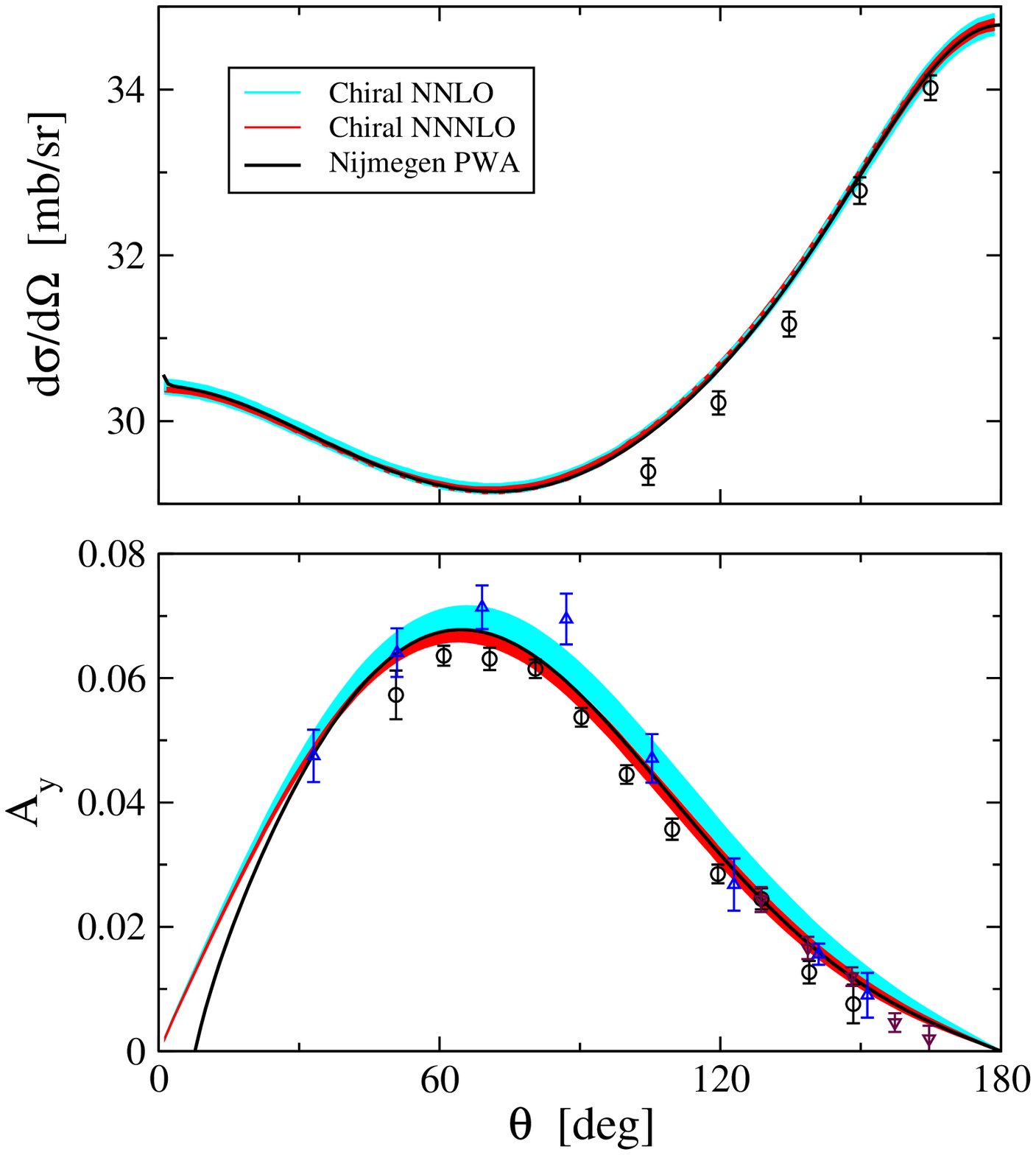,width=8.8cm}}
\parbox{8.9cm}{\psfig{file=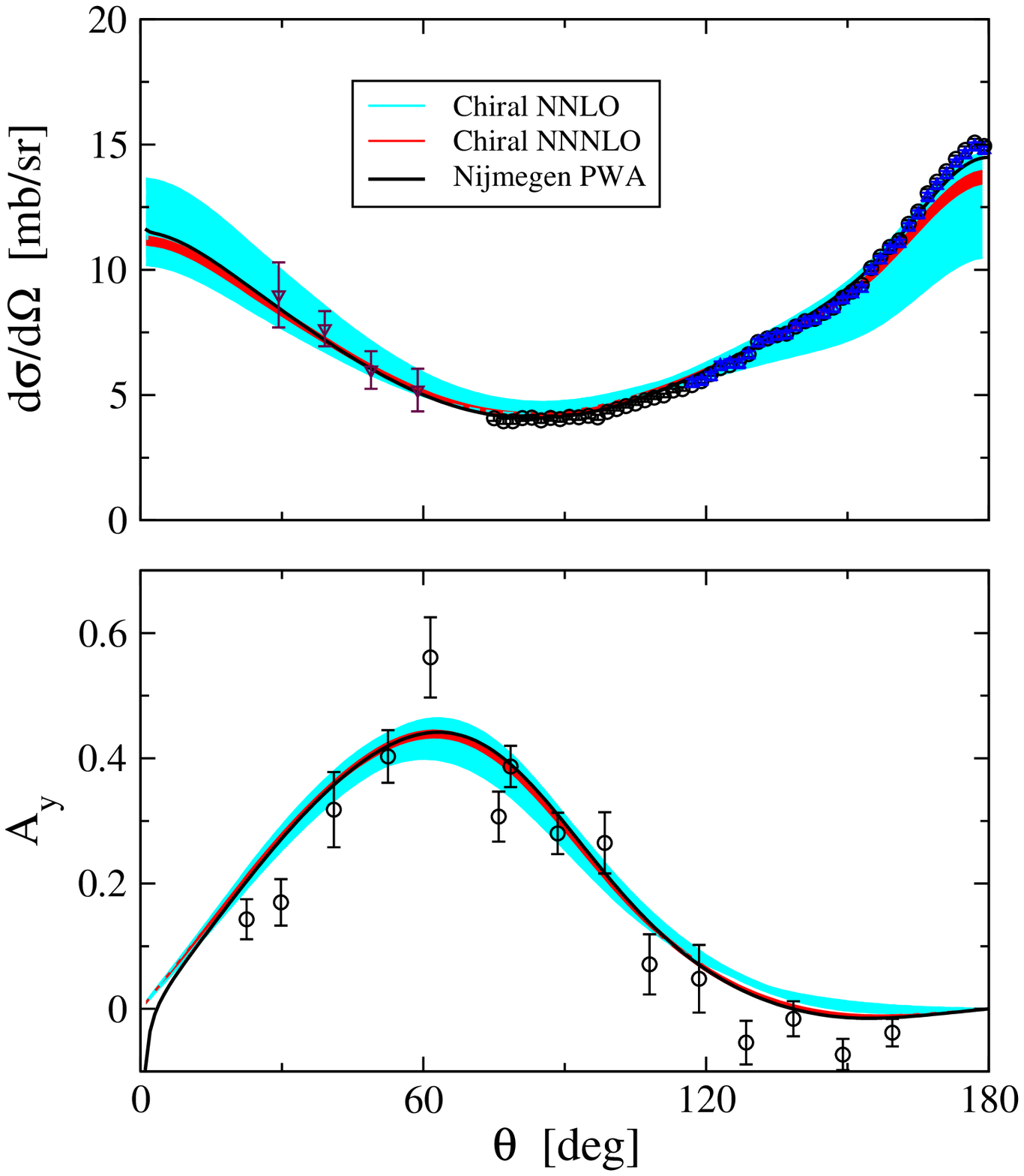,width=8.8cm}}
\vspace{-0.15cm}
\caption[fig25]{\label{fig25} {\it np} differential cross section and vector analyzing power at $E_{\rm lab} = 25$ MeV
(left panel) and $E_{\rm lab} = 96$ MeV (right panel). The Nijmegen PWA result is taken from \citep{NNonline}.
%Data for the cross section are taken from \citep{Fink:1990aa,Griffith:1958aa,Rahm:2001my,Roennqvist:1992aa} and for the 
%analyzing power from \citep{Sromicki:1986aa,Wilczynski:1984aa,Stafford:1957aa}. 
The cut--offs $\Lambda$ and $\tilde \Lambda$ are varied as specified in Eq.~(\ref{cutoffs}).
For data see \citep{Epelbaum:2004fk}.}
\vspace{0.2cm}
\end{figure*}

Once the NN phase shifts are calculated, all two--nucleon scattering observables
can be obtained in a straightforward way using e.g.~the formulae collected in \citep{Stoks:1990bb}.
In Fig.~\ref{fig25} the {\it np} 
differential cross section and vector analyzing power at $E_{\rm lab} =$25 and 96 MeV at N$^2$LO and N$^3$LO from \citep{Epelbaum:2004fk} 
are shown in comparison with the data and the Nijmegen PWA results. This calculation is based on 
{\it np} partial waves up to $j \leq 8$ and does not include the magnetic moment interaction. 
It is comforting to see that N$^2$LO and N$^3$LO results overlap and are in agreement with the Nijmegen PWA. 
The deviations in the analyzing power at forward direction are due to the neglected magnetic moment 
interaction. The small but visible deviations from the Nijmegen PWA for the differential cross section at
forward and backward angles at N$^3$LO are probably due to the neglect of partial waves with 
$j>8$.

We now regard the {\it np} S--wave scattering length and effective range parameters. 
The results at various orders in the chiral expansion are summarized in Table \ref{tab:ER}. All observables are 
improved when going from NLO to N$^2$LO and from N$^2$LO to N$^3$LO. The N$^3$LO result for the $^1S_0$ scattering length fills a 
small gap between the N$^2$LO prediction and the value of the Nijmegen PWA. For the {\it pp} $^1S_0$ scattering length
and effective range, the values $a_{pp} = -7.795 \ldots -7.812$ fm and 
$r_{pp} = 2.73 \ldots 2.76$ fm were obtained at  N$^3$LO \citep{Epelbaum:2004fk}, which agree nicely with  
the data \citep{Bergervoet:1990zy}:  $a_{pp}^{\rm exp} = -7.8149 \pm 0.0029$ fm and
$r_{pp}^{\rm exp} = 2.769 \pm 0.014$ fm.  The ``standard value'' for the {\it nn} scattering length, $a_{nn}^{\rm std} = -18.9$ fm,
was used in \citep{Epelbaum:2004fk} as 
an input parameter in order to pin down the LEC $\tilde \beta_{1S0}^{nn}$. The resulting prediction 
for the effective range, $r_{nn} = 2.76 \ldots 2.80$ fm, agrees with the experimental value, $r_{nn}^{\rm exp} = 2.75 \pm 0.11$ fm.

Finally, we would like to discuss various deuteron properties. The results at NLO, N$^2$LO from \citep{Epelbaum:2003xx} and 
N$^3$LO from \citep{Epelbaum:2004fk} are collected in Table \ref{tab:D} together with the experimental values.  
Notice that none of the deuteron properties were used in fits to determine the LECs. The predicted 
binding energy at N$^3$LO is within 0.4\% of the experimental value, which has to be  
compared with 1\%--1.5\%  ($\sim$2\%--2.5\%) deviation at N$^2$LO (NLO).
The asymptotic S--wave normalization strength $A_S$ is also visibly improved at N$^3$LO and 
deviates from the experimental (central) value by 0.3\% as compared to $\sim$1.1\% ($\sim$1.9\%)
at N$^2$LO (NLO). The results  for the asymptotic D/S--ratio at all orders are in
agreement with the data within the experimental uncertainty. The N$^3$LO result for $\eta_{\rm d}$ 
agrees well with the one of the Nijmegen PWA \citep{deSwart:1995ui}, $\eta_{\rm d} = 0.0253(2)$. 
No improvement is observed for the quadrupole momentum $Q_{\rm d}$ at N$^3$LO, which shows an even 
larger deviation from the data compared to N$^2$LO (6\%--8\% versus 4\%--5\%). One should, however, 
keep in mind that the effects  of the two--nucleon currents and relativistic corrections 
to this observable were not accounted for in the calculation of Ref.~\citep{Epelbaum:2004fk}. 
The situation with  the deuteron rms-radius is similar to the one with the quadrupole moment:
one observes a larger deviation from the data at N$^3$LO as compared to the NLO and N$^2$LO results.
Notice however that the  N$^3$LO result for this quantity is still within 0.5\% of the value from Ref.~\citep{Friar:1997aa}.

\begin{table*}[t] 
%\vspace{0.6cm}
\begin{center}
\begin{tabular*}{1.0\textwidth}{@{\extracolsep{\fill}}||l||c|c|c||c||}
\hline \hline
{} & {} &  {} & {} & {}\\[-1.5ex]
  & {NLO}     \protect\citep{Epelbaum:2003xx}  & {N$^2$LO}  \protect\citep{Epelbaum:2003xx}  & {N$^3$LO} \protect\citep{Epelbaum:2004fk} &{Nijmegen PWA} \\[1ex]
\hline  \hline
{} & {} &  {} & {} & {}\\[-1.5ex]
$a_{1S0}$ [fm]        & $-23.447 \ldots -23.522$  & $-23.497 \ldots -23.689$  & $-23.585 \ldots -23.736$  & $-$23.739  \\ [0.3ex]
$r_{1S0}$ [fm]        & $2.60 \ldots  2.62$       & $ 2.62  \ldots     2.67$  & $2.64 \ldots 2.68$  & 2.68       \\[0.3ex]
$a_{3S1}$ [fm]        & $5.429 \ldots 5.433$  & $5.424 \ldots 5.427$    & $5.414 \ldots 5.420$  & $5.420$  \\[0.3ex] 
$r_{3S1}$ [fm]        & $1.710 \ldots 1.722$  & $1.727   \ldots 1.735$  & $1.743 \ldots 1.746$  & $1.753$  \\[1.0ex]
\hline  \hline
  \end{tabular*}
%\vspace{0.1cm}
\caption{\label{tab:ER} {\it np} scattering length and effective range for the $^1S_0$ and $^3S_1$ partial waves 
at NLO, N$^2$LO and N$^3$LO compared to the Nijmegen PWA results \protect\citep{deSwart:1995ui,Rentmeester:1999}.}
\end{center}
\end{table*}

\begin{table*}[t] 
%\vspace{0.6cm}
\begin{center}
\begin{tabular*}{1.0\textwidth}{@{\extracolsep{\fill}}||l||c|c|c||c||}
    \hline \hline
{} & {} &  {} &  {} & {}\\[-1.5ex]
    & {NLO} \protect\citep{Epelbaum:2003xx}   & {N$^2$LO} \protect\citep{Epelbaum:2003xx}  & {N$^3$LO} \protect\citep{Epelbaum:2004fk}  &  {Exp} \\[1ex]
\hline  \hline
{} & {} &  {} &  {} & {}\\[-1.5ex]
$E_{\rm d}$ [MeV] &   $-2.171 \ldots  -2.186$   & $-2.189 \ldots -2.202$   & $-2.216 \ldots -2.223$  & $-$2.224575(9) \\[0.3ex]
%    \hline
$Q_{\rm d}$ [fm$^2$] &  $0.273 \ldots 0.275$    & $0.271  \ldots  0.275$   & $0.264 \ldots 0.268$     & 0.2859(3) \\[0.3ex]
%    \hline
$\eta_{\rm d}$ &     $0.0256 \ldots   0.0257$     & $0.0255 \ldots 0.0256$   & $0.0254 \ldots 0.0255$    & 0.0256(4)\\[0.3ex]
%    \hline
$\sqrt{\langle r^2 \rangle^{\rm d}_m}$ [fm] 
                     &   $1.973  \ldots  1.974$     & $1.970  \ldots  1.972$   &  $1.973 \ldots 1.985$ 
& 1.9753(11) \\[0.3ex]
%    \hline
$A_S$ [fm$^{-1/2}$] & $0.868\ldots 0.873$ & $0.874  \ldots  0.879$   &  $0.882 \ldots 0.883$    & 0.8846(9)\\[0.3ex]
%    \hline
$P_{\rm d}\; [\%]$ &     $3.46 \ldots  4.29$      & $3.53   \ldots  4.93 $   &  $2.73 \ldots 3.63$   &  -- \\[1ex]
    \hline \hline
  \end{tabular*}
%\vspace{0.1cm}
\caption{Deuteron properties at NLO, N$^2$LO and N$^3$LO compared to the data. Here, $E_{\rm d}$ is the
    binding energy, $Q_{\rm d}$ the quadrupole moment, $\eta_{\rm d}$ the asymptotic
    $D/S$ ratio, $\sqrt{\langle r^2 \rangle^{\rm d}_m}$  the root--mean--square matter radius, $A_S$ the 
    strength of the asymptotic S--wave normalization and $P_{\rm d}$ the D-state 
    probability. The data for $E_{\rm d}$ are from \protect\citep{Leun:1982aa}, for $Q_{\rm d}$ from \protect\citep{Bishop:1979aa,Ericson:1983aa},
    for $\eta_{\rm d}$ from \protect\citep{Rodning:1990aa} and for $A_S$ from \protect\citep{Ericson:1983aa}. For the rms--radius we 
    actually show the deuteron ``point--nucleon'' rms--radius from \protect\citep{Friar:1997aa}.  
    \label{tab:D}}
\end{center}
\end{table*}

Last but not least, we would like to mention that the  N$^3$LO analysis of Ref.~\citep{Entem:2003ft}
based on the cut--off in the LS equation $\Lambda = 500$ MeV  was recently extended to $\Lambda = 600$ MeV \citep{Machleidt:2005uz}.
In most partial waves, very close results for both choices of $\Lambda$ with no visible differences for energies up to $300$ MeV 
were reported. This seems to be in contradiction with the expectations based on the arguments and estimations presented above:
even if $\Lambda$ is chosen to be 500 MeV instead of 450 MeV as in Ref.~\citep{Epelbaum:2004fk}, one can still expect only a very slow 
convergence of the low--momentum expansion at $E_{\rm lab} = 300$ MeV, which corresponds to the CMS momentum $k \sim 375$ MeV. 
It would be interesting to see how sensitive the results of \citep{Entem:2003ft,Machleidt:2005uz} are to a particular 
form of the regulator in Eq.~(\ref{regulator_EM}) with $n$ being chosen differently for different contributions to the potential.
%The sensitivity of the results of \citep{Entem:2003ft,Machleidt:2005uz} to a particular 
%form of the regulator in Eq.~\ref{regulator_EM} with $n$ being chosen differently for different contributions to the potential 
%is unclear. 

\subsubsection{Resonance saturation for NN contact interactions}

In this section we would like to confront the LECs
determined from chiral effective field theory with the highly
successful phenomenological/meson exchange models of the nuclear force
following the lines of Refs.~\citep{Epelbaum:2001fm,Epelbaum:2003xx}
and restricting ourselves to N$^2$LO. 
Consider, for example, the Bonn--B \citep{Machleidt:1989tm} and 
Nijmegen 93 \citep{Stoks:1994wp} potentials which are 
genuine one--boson--exchange (OBE) models.
The long--range part of the interaction in these models is given
by 1PE (including a pion--nucleon form factor) whereas shorter
distance physics is expressed as a sum over heavier mesons exchange 
contributions. 
For nucleon momentum transfer below the masses of the exchanged mesons,
one can interpret such exchange diagrams as a sum of local contact operators
with an increasing number of derivatives (momentum insertions).
The LECs accompanying the resulting contact interactions are given within 
each model in terms of the meson masses, meson--nucleon coupling constants and corresponding 
form--factors. We now wish to compare these LECs to the ones entering 
the chiral NN potential in order to see, whether they can be understood in terms of 
resonance saturation \citep{Epelbaum:2001fm}.
In order to allow for a meaningful comparison with the OBE models,
one needs to properly account for the chiral 2PE potential, which is absent 
in the OBE models.  To achieve that, the 2PE potential at NLO and N$^2$LO 
is power expanded and the corresponding contributions to the LECs are identified, 
see \citep{Epelbaum:2003xx} for more details on this procedure and for explicit 
analytical expressions. The second and third columns in Table  \ref{tab:res} show the 
corresponding numerical results at NLO and N$^2$LO for 
the SFR cut--off $\tilde \Lambda = 600$ MeV. 
\begin{table*}[t] 
%\vspace{0.6cm}
\begin{center}
\begin{tabular*}{1.0\textwidth}{@{\extracolsep{\fill}}||l||c|c||c|c||c|c||}
    \hline \hline
{} & {} &  {} & {} & {} & {} & {}\\[-1.5ex]
 LEC   & 2PE (NLO)  & 2PE (N$^2$LO) & $C_i$ (NLO) & $C_i$ (N$^2$LO) & Bonn B & Nijm-93 \\[1ex]
\hline  \hline
{} & {} &  {} & {} & {} & {} & {}\\[-1.5ex]
$\tilde{C}_{1S0}$  & $-0.004 {{+0.000} \atop {-0.001}}$ & $-0.004 {{+0.000} \atop {-0.001}}$ 
& $-0.117 {{+2.271} \atop {-0.042}}$ & $-0.158 {{+0.178} \atop {-0.004}}$ &  $-0.117$ 
&$-0.061$ \\
{} & {} &  {} & {} & {} & {} & {}\\[-1.ex]
${C}_{1S0}$        & $-0.570 {{+0.036} \atop {-0.022}}$ & $-0.443 {{+0.078} \atop {-0.057}}$ 
& $1.294 {{+2.873} \atop {-0.322}}$  & $1.213 {{+0.408} \atop {-0.084}}$ &  $1.276$
&$1.426$ \\
{} & {} &  {} & {} & {} & {} & {}\\[-1.ex]
$\tilde{C}_{3S1}$  & $0.013 {{+0.001} \atop {-0.000}}$  & $-0.004 {{+0.000} \atop {-0.001}}$ 
& $-0.135 {{+0.025} \atop {-0.021}}$ & $-0.137 {{+0.017} \atop {-0.027}}$ &  $-0.101$
& $-0.014$ \\
{} & {} &  {} & {} & {} & {} & {}\\[-1.ex]
${C}_{3S1}$        & $0.638 {{+0.025} \atop {-0.044}}$  & $-0.443 {{+0.078} \atop {-0.057}}$ 
& $0.231 {{+0.112} \atop {-0.007}}$  & $0.523 {{+0.197} \atop {-0.039}}$  &  $0.660$
& $0.940$ \\
{} & {} &  {} & {} & {} & {} & {}\\[-1.ex]
${C}_{\epsilon 1}$ & $-0.190 {{+0.012} \atop {-0.006}}$ & $0.205 {{+0.024} \atop {-0.035}}$ 
& $-0.325 {{+0.000} \atop {-0.036}}$ & $-0.395 {{+0.007} \atop {-0.072}}$ &  $-0.410$
& $-0.343$ \\
{} & {} &  {} & {} & {} & {} & {}\\[-1.ex]
${C}_{1P1}$        & $-0.067 {{+0.007} \atop {-0.005}}$ & $-0.090 {{+0.013} \atop {-0.009}}$ 
& $0.146 {{+0.005} \atop {-0.010}}$  & $0.126 {{+0.023} \atop {-0.017}}$  &  $0.454$ 
&$0.119$ \\
{} & {} &  {} & {} & {} & {} & {}\\[-1.ex]
${C}_{3P0}$        & $-0.425 {{+0.025} \atop {-0.014}}$ & $0.006 {{+0.003} \atop {-0.003}}$ 
& $0.923 {{+0.142} \atop {-0.103}}$  & $0.920 {{+1.063} \atop {-0.109}}$  &  $0.921$
& $0.802$ \\
{} & {} &  {} & {} & {} & {} & {}\\[-1.ex]
${C}_{3P1}$        & $0.246 {{+0.009} \atop {-0.016}}$  & $0.247 {{+0.032} \atop {-0.044}}$ 
& $-0.260 {{+0.003} \atop {-0.005}}$ & $-0.108 {{+2.364} \atop {-0.176}}$ &  $-0.075$
& $-0.197$ \\
{} & {} &  {} & {} & {} & {} & {}\\[-1.ex]
${C}_{3P2}$        & $-0.022 {{+0.000} \atop {-0.000}}$ & $0.151 {{+0.020} \atop {-0.028}}$ 
& $-0.262 {{+0.032} \atop {-0.073}}$ & $-0.421 {{+0.074} \atop {-0.052}}$ &  $-0.396$
& $-0.467$
  \\[1.5ex]
\hline  \hline
  \end{tabular*}
%\vspace{0.3cm}
\caption{The LECs $C_i$ at NLO and N$^2$LO compared with the results from 
the Bonn B and Nijmegen 93 OBE potential models. Also shown are contributions 
from chiral 2PE as explained in text. The $\tilde C_i$ are in 10$^4$ GeV$^{-2}$ 
and the $C_i$ in 10$^4$ GeV$^{-4}$.
}\label{tab:res}
\end{center}
\end{table*}
The indicated
uncertainty refers to the cut--off variation $\tilde \Lambda = 500 \ldots
700$ MeV. The fourth and fifth columns contain the values of the LECs 
$C_i$ at NLO and N$^2$LO, where the just discussed contributions  
from 2PE have already been added. 
The numbers are presented for the cut--off combination  $\{ \Lambda , \; \tilde \Lambda \} = \{ 550, \; 600\}$ 
with the uncertainties referring to the variations: 
$\tilde \Lambda = 500 \ldots 700$ MeV and $\Lambda = 450 \ldots 600$
MeV ($\Lambda = 450 \ldots 650$ MeV) at NLO (N$^2$LO).
Notice that in certain cases, $C_i$'s show a rather strong cut--off dependence. 
This happens if the cut--off $\Lambda$
becomes too large (i.e.~close to the critical value, at which spurious bound 
states arise) and one leaves the plateau--region for the corresponding LEC $C_i (\Lambda )$.
This situation is exemplified in Fig.~\ref{fig:3p1}.
Clearly, it only makes sense to discuss resonance saturation of the $C_i$'s 
in the plateau--region of the first branch, where they only change modestly
and where the effective potential is, at least, not strongly non--phase--equivalent 
to the OBE models. 
The last two columns in Table \ref{tab:res} show the 
LECs as predicted by resonance saturation based upon the Nijmegen 93 
and Bonn B potential models. One observes a remarkable
agreement between the LEC values obtained from fit to NN phase shifts 
in the EFT approach and the ones resulting from the OBE models. 
For a related study based on a toy--model, the reader is referred to Ref.~\citep{Epelbaum:1998na}.

\subsubsection{Quark mass dependence of the nuclear force}
\label{quarkmass}

Since the absolute values of the running up and down quark masses at the scale 1 GeV, $m_u \simeq 5$ MeV, 
$m_d \simeq 9$ MeV, are rather small \citep{Gasser:1982ap}, one expects that hadronic properties at low energies 
do not change strongly in the chiral limit  (CL) of $M_\pi \rightarrow 0$. This feature is crucial for the 
chiral expansion to make sense and is certainly true for the pion and pion--nucleon systems, where the 
interaction becomes arbitrarily weak in the CL and for vanishing external momenta.
The situation in the few--nucleon sector is significantly more complicated  due to the nonperturbative 
nature of the problem and also due to the fact that the interaction between nucleons does not become weak in the chiral limit. 
The  $M_\pi $--dependence of the nuclear force can naturally be studied in the chiral EFT framework. 
It is not only of academic interest, but 
also relevant for interpolating the results from lattice gauge theory, see also \citep{Beane:2001bc}. For example, the S--wave 
scattering lengths have been calculated on the lattice using the quenched approximation \citep{Fukugita:1994ve}. 
Another interesting application is related to imposing bounds on the time--dependence
of fundamental couplings from the two--nucleon sector, as discussed in \citep{Beane:2002vs}.

The first, pioneering study of the NN system for vanishing quark masses was performed
by Bulgac et al.~\citep{Bulgac:1997ji} based upon the explicit $M_\pi$--dependence of the 1PE potential. 
More advanced studies and extensive discussion on this topic can be found in 
Refs.~\citep{Beane:2001bc,Beane:2002nu,Epelbaum:2002gb,Beane:2002xf,Epelbaum:2002gk}.
Here, we follow the lines of Ref.~\citep{Epelbaum:2002gb},
where the $M_\pi $--dependence of the nuclear force on the pion mass was analyzed at NLO based on the DR 
potential\footnote{As pointed out in section \ref{peripheral}, both DR and SFR lead to similar results at NLO.} 
in the limit of exact isospin symmetry.  We remind the reader that the potential at this 
order is given by the 1PE and 2PE contributions and contact interactions with  
up to two derivatives or one $M_\pi^2$--insertion as detailed in section \ref{sec:piexch}.
In addition, one has to include the corrections to 1PE and the leading contact terms
at the one--loop level, which lead to renormalization of the corresponding LECs and therefore induce implicit 
quark mass dependence. These corrections are discussed in detail in \citep{Epelbaum:2002gb}. For the 1PE potential 
\begin{equation}
\label{ope_gpiN}
V_{1\pi} = - \left( \frac{g_{\pi N}}{2 m_N} \right)^2
\, \fet \tau_1 \cdot \fet \tau_2 \, \frac{(\vec \sigma_1 \cdot \vec q \,) 
( \vec \sigma_2 \cdot \vec q\,)}
{\vec q\, ^2 + M_\pi^2}\,,
\end{equation}
with $g_{\pi N}$ being the pion--nucleon coupling constant, one has to account not only for explicit $M_\pi$--dependence 
in the denominator, but also for implicit dependence of the ratio $g_{\pi N}/m_N$ on the pion mass. For an arbitrary 
value $\tilde M_\pi$ of the pion mass one obtains \citep{Epelbaum:2002gb}:
\beq
\label{ope_str}
\frac{g_{\pi N}}{m_N} = \frac{g_A}{F_\pi} \Bigg( 1 - 
\frac{g_A^2 \tilde M_\pi^2}{4 \pi^2 F_\pi^2} \ln \frac{\tilde M_\pi}{M_\pi} 
- \frac{2 \tilde M_\pi^2}{g_A} \bar d_{18}
+  \left( \frac{g_A^2 }{16 \pi^2 F_\pi^2} - \frac{4 }{g_A}
\bar{d}_{16} + \frac{1}{16 \pi^2 F_\pi^2} \bar{l}_4 \right) (M_\pi^2 - \tilde M_\pi^2)
\Bigg)\,, 
\eeq
where $g_A=1.26$, $F_\pi=92.4$ MeV  and $M_\pi=138$ MeV denote the physical values of the 
nucleon axial vector coupling, pion decay constant and pion mass, respectively.
Further, $\bar l_4$, $\bar d_{18}$ and $\bar d_{16}$ are LECs related to 
pion and pion--nucleon interactions. Following Ref.~\citep{Epelbaum:2002gb}, we use
$\bar l_4=4.3$ \citep{Gasser:1983yg}, $\bar d_{16}=-1.23 {{+0.32} \atop {-0.53}}$ GeV$^{-2}$ \citep{Fettes:1999wp,Fettes:2000aa} 
and $\bar d_{18} = -0.97$ GeV$^{-2}$. The constant $\bar d_{18}$ is determined from the observed
value of the Golberger--Treiman discrepancy with $g_{\pi N} = 13.2$ \citep{Matsinos:1998wp}.
Notice further that for the LEC $\bar d_{16}$, we use an average of three values 
given in \citep{Fettes:2000aa}, which result from different fits. The shown uncertainty is defined 
in the way to cover the whole range of values from \citep{Fettes:2000aa}.

The remaining $\tilde M_\pi$--dependence of the nuclear force at NLO is given by 2PE, see Eq.~(\ref{2PE_nlo}), and
by the short--range terms of the form 
\beq
\label{sr}
V^{\tilde M_\pi}_{\rm cont} = \tilde M_\pi^2 \, \Bigg( \bar D_S + \bar D_T (\vec \sigma_1 \cdot \vec \sigma_2 )
 - \left(\beta_S
+ \beta_T (\vec \sigma_1 \cdot \vec \sigma_2 ) \right) \ln \frac{\tilde M_\pi}{M_\pi} \Bigg)\,,
\eeq
where the constants $\beta_{S,T}$ are given in terms of $g_A$, $F_\pi$ and $C_T$ \citep{Epelbaum:2002aa}.
All other contact terms do not depend on the pion mass and the corresponding LECs 
were adopted in \citep{Epelbaum:2002gb} from the analysis of \citep{Epelbaum:1999dj}, performed for the physical value 
$\tilde M_\pi = M_\pi$. 
The essential difficulty in extrapolating the nuclear forces in the pion mass
is due to the fact that the LECs  $\bar D_S$,  $\bar D_T$ cannot be fixed from the NN data.\footnote{They can be determined in the 
processes including pions such as e.g.~pion--deuteron scattering. Such an analysis is however not yet available.}
In order to proceed further, natural values for these LECs were assumed in \citep{Epelbaum:2002gb}
\beq
\bar D_{S,T} = \frac{\alpha_{S,T}}{F_\pi^2 \Lambda_{\rm LEC}^2}\, ,  
\quad \mbox{where} \quad -3.0 < \alpha_{S,T} < 3.0 \,,
\eeq
based on $\Lambda_{\rm LEC} \simeq 1$ GeV. This estimation is consistent with the size of other contact terms with  
the corresponding dimensionless coefficients $\alpha_i$ lying in the range  $-2.1 \ldots 3.2$ \citep{Epelbaum:2001fm}. 

\begin{figure}
%\vspace{0.3cm}
\centerline{
\psfig{file=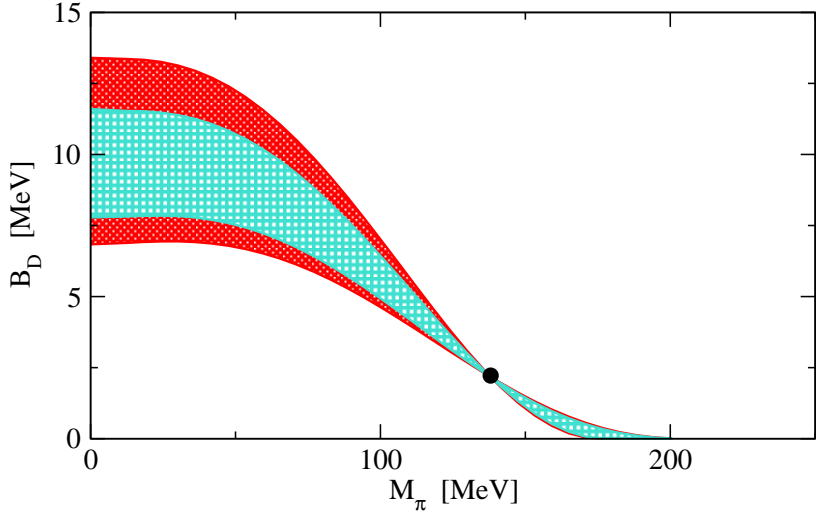,width=12cm}}
\vspace{-0.25cm}
\caption[fig25a]{\label{fig25a} Deuteron BE versus the pion mass. The shaded areas show 
allowed values. The light (dark) shaded band depicts  
the uncertainty due to the unknown LECs $\bar D_{S,T}$ ($\bar d_{16}$).
The heavy dot shows the BE for the physical case $\tilde M_\pi = M_\pi$.}
\vspace{0.2cm}
\end{figure}

The resulting deuteron binding energy (BE) as a function of the pion mass is shown in Fig.~\ref{fig25a}. 
The deuteron becomes unbound for $\tilde M_\pi \gtrsim 170$ MeV and is found to be stronger bound in the chiral limit with 
the BE  $B_{\rm D}^{\rm CL} =  9.6 \pm 1.9 {{+ 1.8} \atop  {-1.0}}$ MeV \citep{Epelbaum:2002gb}.
Here, the first indicated error refers to the uncertainty in the value of $\bar D_{^3S_1}$ and 
$\bar d_{16}$ being set to the average value  $\bar d_{16}=-1.23$ GeV$^{-2}$
while the second indicated error shows the additional uncertainty due to the uncertainty
in the determination of $\bar d_{16}$ as described above. Other deuteron properties in 
the chiral limit are discussed in \citep{Epelbaum:2002aa}.  
The resulting values for the two S--wave scattering lengths in the chiral limit are smaller in magnitude 
and more natural: $a_{\rm CL} (^1S_0) = -4.1 \pm 1.6
{{+ 0.0} \atop 
{-0.4}} \,$fm and 
$a_{\rm CL} (^3S_1) = 1.5 \pm 0.4
{{+ 0.2} \atop \\ 
{ -0.3}}\,$fm. As pointed out in \citep{Epelbaum:2002gb}, one needs lattice data for pion 
masses below 200 MeV to be able to perform a stable extrapolation to its physical value. 
Last but not least, we emphasize that the analysis of \citep{Beane:2002xf}, see also \citep{Beane:2004ks} for 
a related work, assumes a significantly 
larger uncertainty in the values of the LECs $\bar D_{S,T}$ and $\bar d_{16}$, which is the 
main reason of different results for the $^3S_1$ channel reported in that work. 
For an extended discussion on this issue, the reader is referred to \citep{Epelbaum:2002gk}. 
We also emphasize that the work of \citep{Hemmert:2003cb} seems to indicate the importance of including 
$\Delta$'s as explicit degrees of freedom in order to properly describe the quark mass dependence 
of the nucleon axial--vector coupling constant. Notice further that effects due to radiative pions might 
become important close to the CL. Finally, the possibility of an infrared renormalization group 
limit cycle in the 3N sector is discused in \citep{Braaten:2003eu}.

\subsection{The three-- and four--nucleon systems up to N$^2$LO}
\label{2Nand4N}

We now turn to systems with three and four nucleons,
which serve as an excellent testing ground for chiral forces and allow for a  
highly nontrivial check of the consistency of this approach since most of the 
unknown LECs are already determined in the  NN system. 

The first quantitative application of chiral forces to $Nd$ scattering in the 3N and 4N sectors 
was performed in \citep{Epelbaum:2000mx} at NLO. To that aim, the Faddeev--Yakubovsky equations were solved rigorously 
for the 3N and 4N systems and the corresponding binding energies as well as various 3N scattering observables 
were computed. Since no 3NF appears at this order in the chiral expansion, the study
of \citep{Epelbaum:2000mx} was entirely based on the 2NF yielding parameter--free results for the 3N and 4N systems. 
This work used the DR version of the chiral 2N potential with the LS cut--off varied in the range $\Lambda = 540 \ldots 600$ MeV and 
demonstrated a good description of the $nd$ elastic scattering data at $E_{\rm lab} = 3$ MeV
and $E_{\rm lab} = 10$ MeV as well as of some break--up observables at $E_{\rm lab} = 13$ MeV. 
The predictions for the triton and $\alpha$--particle BE were found to be in a similar range 
as the ones based on phenomenological NN potentials.

The NLO analysis of \citep{Epelbaum:2000mx} was extended to incomplete N$^2$LO in \citep{Epelbaum:2002ji} where the 2N potential 
at N$^2$LO was used without taking into account the corresponding 3NF. The first complete analysis of $nd$ scattering at this order 
including the 3NF was presented in Ref.~\citep{Epelbaum:2002vt}. Both calculations of \citep{Epelbaum:2002ji}
and  \citep{Epelbaum:2002vt} used the DR 2N potential with the numerically reduced values of $c_{3,4}$  
in order to avoid the appearance of spurious bound states, see section \ref{NN_n3lo} for more details. 
In addition, certain $nd$ scattering observables were studied in \citep{Entem:2001tj} using the 2N potential 
from \citep{Entem:2001cg} with no 3NF included. 

Recently, the 3N and 4N systems were reanalyzed at NLO and N$^2$LO within the SFR framework \citep{Epelbaum:prep}. 
We remind the reader that the SFR 2N potential at N$^2$LO is based on the LECs $c_i$ which are consistent with $\pi N$ scattering. 
This is an important advantage compared to the previous study in Ref.~\citep{Epelbaum:2002vt}. 
Notice that these LECs also determine the strength of the 2PE 3NF. 
In addition, larger cut--off variation adopted in \citep{Epelbaum:2004fk} is expected to provide a more 
realistic estimation of the theoretical uncertainty. In the following sections we will present some results for 3N 
and 4N observables based on the SFR 2NF. Further details and more results will be published elsewhere \citep{Epelbaum:prep}.
In selected cases, we will compare the results based on the SFR and DR 2N potentials from Refs.~\citep{Epelbaum:2004fk} and 
\citep{Epelbaum:2002ji}. Further applications of chiral nuclear forces to 3N scattering 
can be found in Refs.~\citep{Ermisch:2003zq,Ermisch:2005kf,Duweke:2004xv,Kistryn:2005aa,Witala:2005aa}.

\subsubsection{The formalism}
\label{formalism}

To describe the properties of the 3N and 4N systems, a corresponding Schr\"odinger equation
needs to be solved. As described in section \ref{NN_n3lo}, we use a specific form for the regularization 
of the potential, which allows us to determined the LECs of the NN force 
partial wave by partial wave. This, however, implies that the interactions 
are strongly non--local. Therefore, a formulation in momentum space is most 
natural. 

In the past, the techniques were developed for solving reliably the Schr\"odinger 
equation in momentum space using Faddeev-- or Yakubovsky equations \citep{Gloeckle:1995jg,Kamada:1992aa}.  
For the 3N bound state problem, the Faddeev equations have the form  \citep{Nogga:1997aa}
\beq
\label{eq1}
\psi = G_0 \, t \, P \, \psi + (1 + G_0\, t) \,G_0 \,V_{123}^{(1)}
\,(1 + P)\, \psi\,.
\eeq
Here $V_{123}^{(i)}$ is that part of the 3N force
which singles out the particle $i$ and which is symmetrical under the exchange of the other
two particles. The complete 3NF is decomposed as
\beq
\label{decomp} 
V_{123} = V_{123}^{(1)}+ V_{123}^{(2)}+V_{123}^{(3)}\,.
\eeq
Further, $\psi$ denotes the corresponding Faddeev component, $t$ is
the two--body $t$--operator, 
$G_0=1/(E-H_0)$ is the free propagator of three nucleons and $P$ is a
sum of a cyclical and anticyclical permutation 
of the three particles. In case of $nd$ scattering we follow the formulation 
described in Refs.~\citep{Gloeckle:1995jg,Huber:1996cg} and first calculate a quantity $T$
related to the 3N break--up process via the Faddeev--like equation:
\beq
\label{eq2}
T = t \,  P  \, \phi + (1 + t  \, G_0)  \, V_{123}^{(1)}  \, (1 +
P)  \, \phi + t  \, P  \, G_0  \, T + (1 + t  \, G_0)   
\, V_{123}^{(1)} \,  (1 + P)  \, G_0  \, T\,,
\eeq
where the initial state $\phi$ is composed of a deuteron and a
momentum eigenstate of the projectile nucleon.  
The elastic $nd$ scattering operator is then obtained as
\beq
U=P  \, G_0^{-1} + P  \, T + V_{123}^{(1)} \,  (1 + P) \,  (1 + G_0  \, T)\,,
\eeq
and the break--up operator via
\beq
U_0 = (1 + P)  \, T\,.
\eeq
These equations are accurately solved in momentum space using a
partial wave decomposition, see \citep{Gloeckle:1983aa,Witala:1988aa,Huber:1996td} for details. 
The partial wave decomposition of the chiral 3NF is given in the appendix of Ref.~\citep{Epelbaum:2002vt}.

Similarly, we use Yakubovsky equations (YE)  to solve the 4N bound state problem. 
We rewrite the Schr\"odinger equation into two YEs and thus  
decompose the wave function $\Psi$ into two independent  Yakubovsky components
(YCs): $\psi_1$ and $\psi_2$ \citep{Kamada:1992aa,Nogga:2000aa,Nogga:2002aa}.  
The wave function then reads 
\begin{equation} 
\label{wavefu} 
\Psi = (1 - (1+P) \ P_{34} ) (1 + P ) \psi_1  + (1+P) \ (1 + \tilde P ) \psi_2 
\end{equation}
and is again expressed with the help of the permutations  
$ P $  and  $\tilde P =   P_{13} P_{24}$ 
where $P_{ij}$ denotes transpositions of particles $i$ and $j$.  In the case
of a bound state it is,
in principle, possible to solve directly the Schr\"odinger equation. 
The usage of two YCs is, however, advantageous since it naturally introduces two kinds of
Jacobi coordinates which accelerates the convergence of the partial wave
decomposition \citep{Nogga:2002aa}. The YEs reduce to two coupled integral equations 
\begin{eqnarray}
\label{eq:yakueq1}
\psi_{1} & = & G_0 \ t \  P \left[ 
         (1-P_{34}) \psi_{1}+\psi_{2} \right] + (\ 1+G_0 \ t \ ) G_0  V_{123}^{(3)} \Psi \\ 
\label{eq:yakueq2}
\psi_{2} & = & G_0 \  t  \ \tilde P \ \left[ 
         (1-P_{34}) \ \psi_{1}+\psi_{2} \right] \, ,
\end{eqnarray}
which can be solved by similar techniques as the 3N problem. 
The high dimensionality (up to $10^8 \times 10^8$) of the discretized  
YEs, however, requires massively parallel computers.  

Because these techniques are formulated in momentum space, the 
application of long--range interactions, like the $pp$ Coulomb force, 
is a major difficulty. For the scattering problem, we therefore neglect 
the Coulomb interaction completely. At higher energies (above 50~MeV 
nucleon lab energy), it is expected to contribute mainly in forward direction. 
Here, the comparison of our results to the data should be less affected 
by the missing electromagnetic forces. 
For lower energies, we will compare our results to modified 
data, which have been corrected for the electromagnetic interactions 
using calculations of Alejandro Kievsky \citep{Kievsky:2002aa} 
based on the AV18 \citep{Wiringa:1994wb}
interactions with and without the electromagnetic forces. 
For the bound states, we include the $pp$ Coulomb interactions 
in the $t$--matrices.  Since the bound nucleons are confined to
a small space region, we can put  the Coulomb 
interaction to zero outside a radius of $10 - 20$~fm, effectively 
making it short--ranged. Then the Fourier transformation 
of this interaction is nonsingular. The results are cut--off
independent and numerically stable.

\subsubsection{Elastic Nd scattering}
\label{nd_elastic}

Consider now elastic $nd$ scattering. At N$^2$LO, one has to take into account the 3NF which is 
discussed in section \ref{sec:3NFinvar}. Similarly to the 2N potential, the chiral 3NF behaves unphysically at 
large momenta and leads to ultraviolet divergences in the Faddeev--Yakubovsky equations.
In \citep{Epelbaum:prep}, regularization was performed in the way analogous to the one adopted in
the analysis of the 2N system:
\beq
V^{\rm 3NF} (\vec p, \vec q;\, \vec p\, ' , \vec q\, ') \rightarrow 
f^\Lambda (\vec p, \vec q) \, V^{\rm 3NF} (\vec p, \vec q;\, \vec p\, ' , \vec q\, ') \,
f^\Lambda (\vec p\, ', \vec q\, ')\,,\quad \quad 
f^\Lambda (\vec p, \vec q \, ) = e^{ - 
\left(\frac{4 p^2 + 3 q^2}{4 \Lambda^2} \right)^3} \,,
\eeq
where $\vec p$ and $\vec q$ ($\vec p\,'$ and $\vec q\,'$) are Jacobi momenta of the 
two--body subsystem and spectator nucleon before (after) the interaction. 
The regulator function $f^\Lambda (\vec p, \vec q \, )$ is chosen 
%in the form 
%\beq
%f^\Lambda (\vec p, \vec q \, ) = e^{ - 
%\left(\frac{4 p^2 + 3 q^2}{4 \Lambda^2} \right)^3} \,,
%\eeq
%
%\beq
%f^\Lambda (\vec p, \vec q \, ) = \exp \left[ - 
%\left(\frac{4 p^2 + 3 q^2}{4 \Lambda^2} \right)^3 \right]\,,
%\eeq
so that for $\vec q =0$ it coincides with the function 
$f^\Lambda (\vec p\, )$  in Eq.~(\ref{reg_fun}).\footnote{In Refs.~\citep{Epelbaum:2002ji,Epelbaum:2002vt},
a slightly different form of the regulator functions was employed.} The results presented here and in what follows are based on the 
variation of the LS cut--off $\Lambda$ in the range $\Lambda = 400 \ldots 550$ at NLO and  $\Lambda = 450 \ldots 600$ at N$^2$LO. The 
SFR cut--off $\tilde \Lambda$ is varied in both cases in the range $\tilde \Lambda = 500 \ldots 700$ MeV.\footnote{The somewhat smaller 
values of $\Lambda$ as compared to \protect\citep{Epelbaum:2003xx} were chosen in order to avoid the appearance of unnaturally large LECs, 
cf.~Table \protect\ref{tab:res}.} 
Specifically, the following cut--off combinations $\{ \Lambda, \; \tilde \Lambda \}$ were used at NLO and N$^2$LO:
\beqa
\label{cutoffs_3N}
\mbox{NLO}: &&  \{ \Lambda, \; \tilde \Lambda \} = \{ 400, \; 500 \},  \; \{ 550, \; 500 \},  \; 
\{ 550, \; 600 \},  \;\{ 400, \; 700 \},  \; \{ 550, \; 700 \}\,, \nn
\mbox{N$^2$LO}: &&  \{ \Lambda, \; \tilde \Lambda \} = \{ 450, \; 500 \},  \; \{ 600, \; 500 \},  \; 
\{ 550, \; 600 \}, \; \{ 450, \; 700 \},  \; \{ 600, \; 700 \}\,.
\eeqa

As explained in section \ref{sec:3NFinvar}, the chiral 3NF at N$^2$LO is given by the two--pion exchange, one--pion exchange  
with the pion emitted (or absorbed) by 2N contact interactions and 3N contact interactions.
While the 2PE contribution given in Eq.~(\ref{3nftpe}) does not introduce any new parameters, 
the two other terms depend on two new LECs, $D$ and $E$, cf.~Eq.~(\ref{3nfrest}), which are not determined in the 
2N system and thus need to be fixed e.g.~from the 3N data. As demonstrated in  Ref.~\citep{Epelbaum:2002vt},
they can be determined using the $^3$H BE and the $nd$ doublet scattering length $^2a_{nd}$, which are
bona fide low--energy observables. In the present analysis, we use the coherent $nd$ scattering length $b_{nd}$
instead of the $^2a_{nd}$. This quantity is defined in terms of the doublet and quartet $nd$ scattering 
lengths  $^2a_{nd}$ and  $^4a_{nd}$ and the neutron/deuteron masses as 
\beq
b_{nd} = \frac{m_n + m_d}{m_d} \left[ \frac{1}{3} \, ^2 a_{nd} +  \frac{2}{3} \, ^4 a_{nd} \right]\,,
\eeq
where $^4a_{nd}$  is the $nd$ quartet scattering length and $m_n$ ($m_d$) refers to the neutron (deuteron) mass. 
The coherent $nd$ scattering length is much better known experimentally than   $^2a_{nd}$,  
which allows to reduce the uncertainty in the determination of the LECs $D$ and $E$. 
The resulting LECs are found to be of natural size, i.e.~the constants $c_D$ and $c_E$ defined as 
\beq
D = \frac{c_D}{F_\pi^2 \Lambda_{\rm LEC}}\, , \quad \quad E = \frac{c_E}{F_\pi^4 \Lambda_{\rm LEC}}\,,
\eeq
are of order one. The only exception is given by the cut--off combination 
$\{ 600, \; 500 \}$, for which the magnitude of $c_D$ is rather large ($c_D=-10.0$ for $\Lambda_{\rm LEC} = 700$ MeV).
Notice  further that the values for both LECs depend strongly on the cut--offs and differ significantly from 
the ones found in an earlier study \citep{Epelbaum:2002vt}. Finally, we would like to emphasize that the $^3$H BE  and the central experimental value 
of $b_{nd}$ could be reproduced simultaneously only for $\{\Lambda_i, \; \tilde \Lambda_i \}$ in Eq.~(\ref{cutoffs_3N}) with $i=1,3,4$. For two other cut--off 
combinations, it was not possible to find a solution for  $c_D$ and $c_E$, which would describe both observables at the same time. In these cases,
the scattering length $b_{nd}$ (the triton BE) was required to be reproduced exactly (as good as possible). 
We will discuss the resulting $^3$H BE in section \ref{sec:BS}. 
More details on fixing the values of $c_D$ and $c_E$ will be given in 
\citep{Epelbaum:prep}. 

\begin{table*}[t] 
%\vspace{0.6cm}
\begin{center}
\begin{tabular*}{1.0\textwidth}{@{\extracolsep{\fill}}||c||c||c|c||r||}
\hline \hline
{} & {} &  {} & {}  & {} {}\\[-1.5ex]
  &  $\pi \palka$ EFT, N$^2$LO &  NLO   &   N$^2$LO & Exp    \\[1ex]
\hline  \hline
{} & {} &  {} & {}  & {}\\[-1.5ex]
$^2a_{nd}$       & $-$            & $0.61 \ldots 1.19$   & $0.61 \ldots 0.63$     & $0.65 \pm 0.04$ \protect\citep{Dilg:1971aa} \\ [0.3ex]
$^4a_{nd}$       & $6.33 \pm 0.05$ & $6.360 \ldots 6.366$   & $6.353 \ldots 6.362$ & $6.35 \pm 0.02$ \protect\citep{Dilg:1971aa} \\ [0.3ex]
$b_{nd}$         & $-$            &  $6.656 \ldots 6.959$   & $6.669^\star$        & $6.669 \pm 0.003$ \protect\citep{Black:2003ba} \\ [1.0ex]
\hline  \hline
  \end{tabular*}
%\vspace{0.3cm}
\caption{\label{tab:scattL} $nd$ scattering lengths (in fm) in EFT in comparison with the data. For NLO and N$^2$LO in the EFT with explicit
pions, the cut-offs $\Lambda$ and $\tilde \Lambda$ are varied in the range given in Eq.~(\ref{cutoffs_3N}). The value of $b_{nd}$ used as input
is marked by the star.}
\end{center}
\end{table*}

In Table \ref{tab:scattL} we summarize the EFT results for various $nd$ scattering lengths in comparison with the 
experimental numbers. In the case of pionless EFT, the value shown in this table is taken from Ref.~\citep{Bedaque:2002mn},
where more discussion on this approach and further references to earlier determinations of $^4a_{nd}$ can be found. 

\begin{figure*}[t]
%\vspace{0.2cm}
\centerline{
\psfig{file=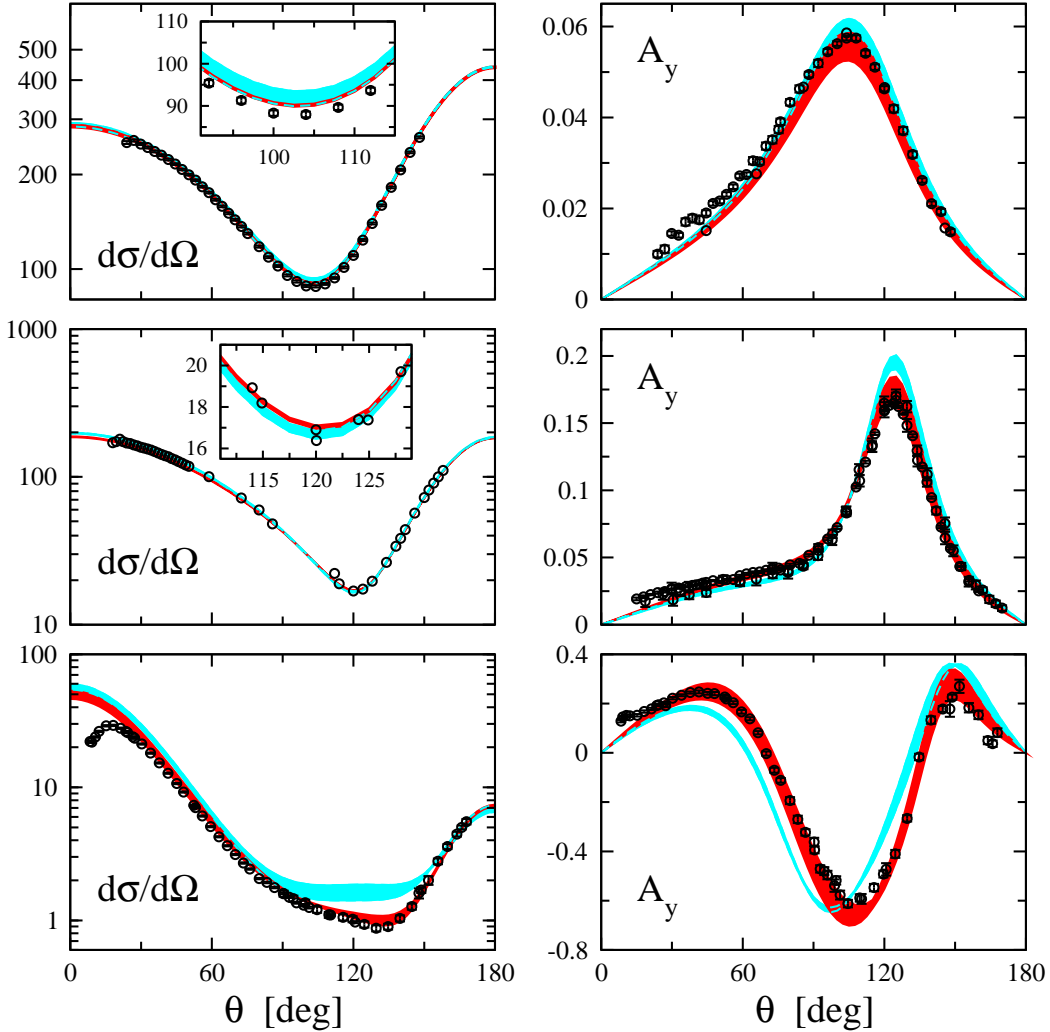,width=14.0cm}
}
\vspace{-0.3cm}
\caption[fig24]{\label{fig:DsAy_nd} Differential cross section (in mb/sr) and vector analyzing power for elastic $nd$ scattering 
at 3 MeV (upper panel) 10 MeV (middle panel) and 65 MeV (bottom panel) at NLO (light--shaded bands) and N$^2$LO (dark--shaded bands)
in the SFR framework.
%The circles are $nd$ pseudo data based on \protect\citep{Shimizu:1995aa,Sagara:1994aa,Rauprich:1988aa,Sperisen:1984aa} as well as  
%true $nd$ data \protect\citep{mcAnich:1993aa,Howell:1987aa}. 
The bands correspond to the cut--off variation as specified in Eq.~(\ref{cutoffs_3N}).
For data see \citep{Epelbaum:2002vt}.}
\vspace{0.2cm}
\end{figure*}

\begin{figure*}[t]
%\vspace{0.2cm}
\centerline{
\psfig{file=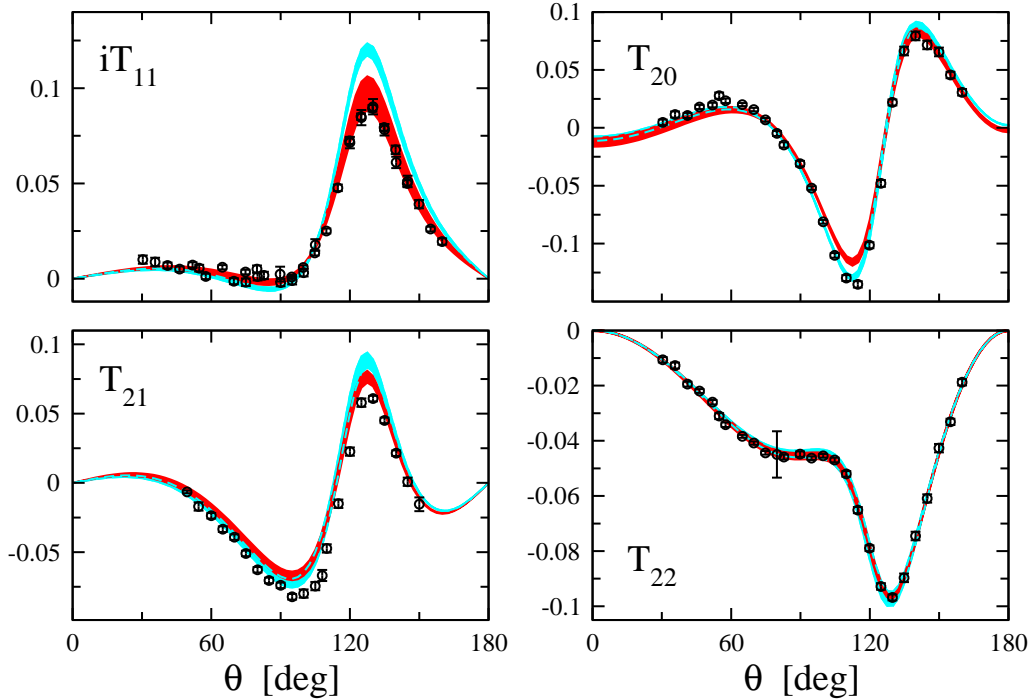,width=14.0cm}
}
\vspace{-0.3cm}
\caption[fig24]{\label{fig:10mev} Tensor analyzing powers for elastic $nd$ scattering at 10 MeV at NLO (light--shaded bands) and
N$^2$LO (dark--shaded bands). 
%The circles are $nd$ pseudo data based on  \protect\citep{Shimizu:1995aa,Sperisen:1984aa}. 
The bands correspond to the cut--off variation  as specified in Eq.~(\ref{cutoffs_3N}).
For data see \citep{Epelbaum:2002vt}.}
\vspace{0.2cm}
\end{figure*}

With the LECs $c_D$ and $c_E$ being determined as described above, we are now in the position to  
predict various $nd$ elastic scattering observables. In Fig.~\ref{fig:DsAy_nd} we show the results for 
the differential cross section (in the left column) and vector analyzing power (in the right column) 
at $3$, $10$ and $65$ MeV. 
One observes a good agreement with the data at two lowest energies at both NLO and N$^2$LO.  
This also holds true for tensor analyzing powers shown in  Fig.~\ref{fig:10mev} for 
$E_{\rm lab} = 10$ MeV. The only exceptions are given by the minima of $T_{20}$ and $T_{21}$ at N$^2$LO, where 
some deviations from the data are visible. 
We remind the reader that the data at energies 3 and 10 MeV have 
been corrected to account for the missing Coulomb force \citep{Kievsky:2002aa}. 
The uncertainty  due to the cut--off variation  in the differential cross section 
is significantly reduced at N$^2$LO compared to NLO at all energies considered.  
At 65 MeV, the results at NLO show large deviations 
from the data in the minimum of the cross section as well as for $A_y$. Similarly to the 2N system,
the bands at NLO seem to underestimate the true theoretical uncertainty at this order, see section \ref{NN_n3lo}
for more discussion. The N$^2$LO results at 65 MeV are in agreement with the data for both observables. 
Notice that the deviation from the data for $d\sigma / d\Omega$ in forward directions is due to 
the Coulomb force which is missing in the calculations. At this energy, the uncertainty at N$^2$LO 
appears to be rather large. For the tensor analyzing powers at the same energy, the situation is similar: 
while the NLO results deviate significantly from the data, the N$^2$LO predictions are in agreement with the 
data but the uncertainty due to the cut--off variation is large. A more detailed discussion 
will be given in \citep{Epelbaum:prep}.

Let us now discuss the nucleon vector analyzing power $A_y$, which  is the most problematic observable in $nd$ elastic scattering
at low energy. This particular observable is underpredicted in the maximum by an amount of $\sim 25 \ldots 30$\% by modern high--precision 
nuclear potentials, which is known in the literature as  $A_y$ puzzle \citep{Koike:1987aa,Witala:1994aa}. 
Augmenting the NN potentials with 3NF models such as
the TM99' 3NF \citep{Coon:2001pv} or the Urbana-IX 3NF \citep{Pudliner:1997ck}, which are frequently used in modern few--body calculations, 
does not substantially improve the description of this observable. The only exception is given by the phenomenological spin--orbit 3NF
introduced by Kievsky \citep{Kievsky:1999nw}, which allows to describe the data.  
Similar discrepancies (but less pronounced compared to $A_y$) are also observed for the tensor analyzing power $i T_{11}$.
As demonstrated in Fig.~\ref{fig:DsAy_nd}, The NLO result for $A_y$ is in agreement with the data at 3 MeV and  
even slightly overpredicts the data at 10 MeV. 
Very similar results based on the DR NN potential at NLO were obtained in \citep{Epelbaum:2000mx,Epelbaum:2002ji}. 
%see also Fig.~\ref{fig:10p0_DR}. 
%\begin{figure*}[t]
%\vspace{0.3cm}
%\centerline{
%\psfig{file=10p0.ps,width=16.0cm}
%}
%\vspace{-0.3cm}
%\caption[fig24]{\label{fig:10p0_DR} $nd$ elastic scattering observables at 10 MeV at NLO (light--shaded bands) and
%N$^2$LO (dark--shaded bands) from Ref.~\protect\citep{Epelbaum:2002vt} based on the DR 2NF introduced in \protect\citep{Epelbaum:2002ji}
%and the corresponding 3NF. The circles are $nd$ pseudo data based on 
%\protect\citep{Shimizu:1995aa,Sperisen:1984aa,Sagara:1994aa,Rauprich:1988aa}. The bands correspond 
%to the $\Lambda$--variation between 500 and 600 MeV. The unit of the cross section is mb/sr. }
%\vspace{0.2cm}
%\end{figure*}
While this looks encouraging, 
one cannot conclude that the $A_y$--puzzle has been solved. 
This observable is well known to be very sensitive to the spin--orbit 2NF and, therefore, to the triplet P--waves, 
see e.g.~\citep{Gloeckle:1995jg,Entem:2001tj}, which need to be reproduced accurately in order to have conclusive results. 
It is instructive to look at the spin--orbit phase shift combination $\Delta_{LS}$ defined as \citep{Entem:2001tj}:
\beq
\label{LScomb}
\Delta_{LS} = \frac{1}{12} \left( \ 2 \delta_{^3P_0} - 3 \delta_{^3P_1} + 5 \delta_{^3P_2} \right)\,.
\eeq
In Table \ref{tab:SO}, we show the results for this quantity at NLO and N$^2$LO compared to the ones from Nijmegen PWA. 
Clearly, the spin--orbit force at NLO is enhanced compared to Nijmegen PWA, which also explains the enhancement for 
$nd$ $A_y$ at this order. Notice that the overestimation of $\Delta_{LS}$ at NLO (and, to a less extend, also at N$^2$LO) is largely 
due to the failure to properly describe the $^3P_2$ partial wave, cf.~Fig.~\ref{fig23}.
We emphasize, however, that the quantity $\Delta_{LS}$ is more accurately reproduced 
at N$^2$LO, where the calculated $nd$ $A_y$ is in a reasonable agreement with the data (although the uncertainty due to 
the cut--off variation appears to be quite sizable). A more detailed discussion including the role of the 3NF will be given in \citep{Epelbaum:prep},
see also \citep{Huber:1999bi} for a related earlier work.   

%Finally, in Fig.~\ref{fig:10p0_DR} we also show various $nd$ elastic scattering observables at NLO and N$^2$LO based on the DR NN potential 
%with the numerically reduced values of the LECs $c_{3,4}$ from Ref.~\citep{Epelbaum:2002ji}. 
%***********************************************************************************
Finally, we emphasize that at low energy 
%***********************************************************************************
%At this energy, 
the results for $nd$ elastic scattering observables at NLO and N$^2$LO 
in both SFR and DR \citep{Epelbaum:2002ji} frameworks are very similar. 
The strongest differences are observed for $A_y$ and $i T_{11}$, where the N$^2$LO correction is larger in the DR approach. 
Very different values of $c_{3,4}$ adopted in these analyses, which determine the strength of the 2PE 3NF, 
have only a little impact on the considered $nd$ elastic scattering observables. At higher energies such as 65 MeV, 
the differences between the two sets of calculations, however, become quite significant. For further results in  
$nd$ elastic scattering based on the NN potential of Ref.~\citep{Entem:2001cg} the reader is referred to 
\citep{Entem:2001tj}. 

\begin{table*}[t] 
%\vspace{0.6cm}
\begin{center}
\begin{tabular*}{0.78\textwidth}{@{\extracolsep{\fill}}||c||c|c||c||}
\hline \hline
{} & {} &  {} & {} {}\\[-1.5ex]
$E_{\rm lab}$  (MeV)  &  NLO &  N$^2$LO   &   Nijmegen PWA    \\[1ex]
\hline  \hline
{} & {} &  {} & {} \\[-1.5ex]
$10$       & $0.201 \ldots 0.210$  & $0.183 \ldots 0.199$   & $0.202$  \\ [0.3ex]
$20$       & $0.694 \ldots 0.721$  & $0.622 \ldots 0.673$   & $0.641$  \\ [0.3ex]
$30$       & $1.46 \ldots 1.51$  & $1.28 \ldots 1.39$   & $1.25$  \\ [0.3ex]
$40$       & $2.43 \ldots 2.54$  & $2.10 \ldots 2.28$   & $1.97$  \\ [0.3ex]
$50$       & $3.57 \ldots 3.76$  & $2.97 \ldots 3.30$   & $2.73$  \\ [1.0ex]
\hline  \hline
  \end{tabular*}
%\vspace{0.3cm}
\caption{\label{tab:SO} Spin--orbit combination $\Delta_{LS}$ for $np$ phase shifts (in degrees) at various energies 
at NLO and N$^2$LO in comparison with the result of Nijmegen PWA \protect\citep{NNonline}. The cut-offs 
are chosen as in Eq.~(\ref{cutoffs_3N}).}
\end{center}
%\vspace{-0.8cm}
\end{table*}

\subsubsection{Nd break--up}

Let us now switch to break--up observables.
In the following, we will show some results at two energies, 13 and 65~MeV, where 
a lot of $pd$ data exist. 
Since we do not have reliable Coulomb corrections in  the case of break--up,  
we show the non--corrected $pd$ data in comparison to our $nd$ calculations.  

\begin{figure*}[t]
%\vspace{0.3cm}
\begin{center}
\psfig{file=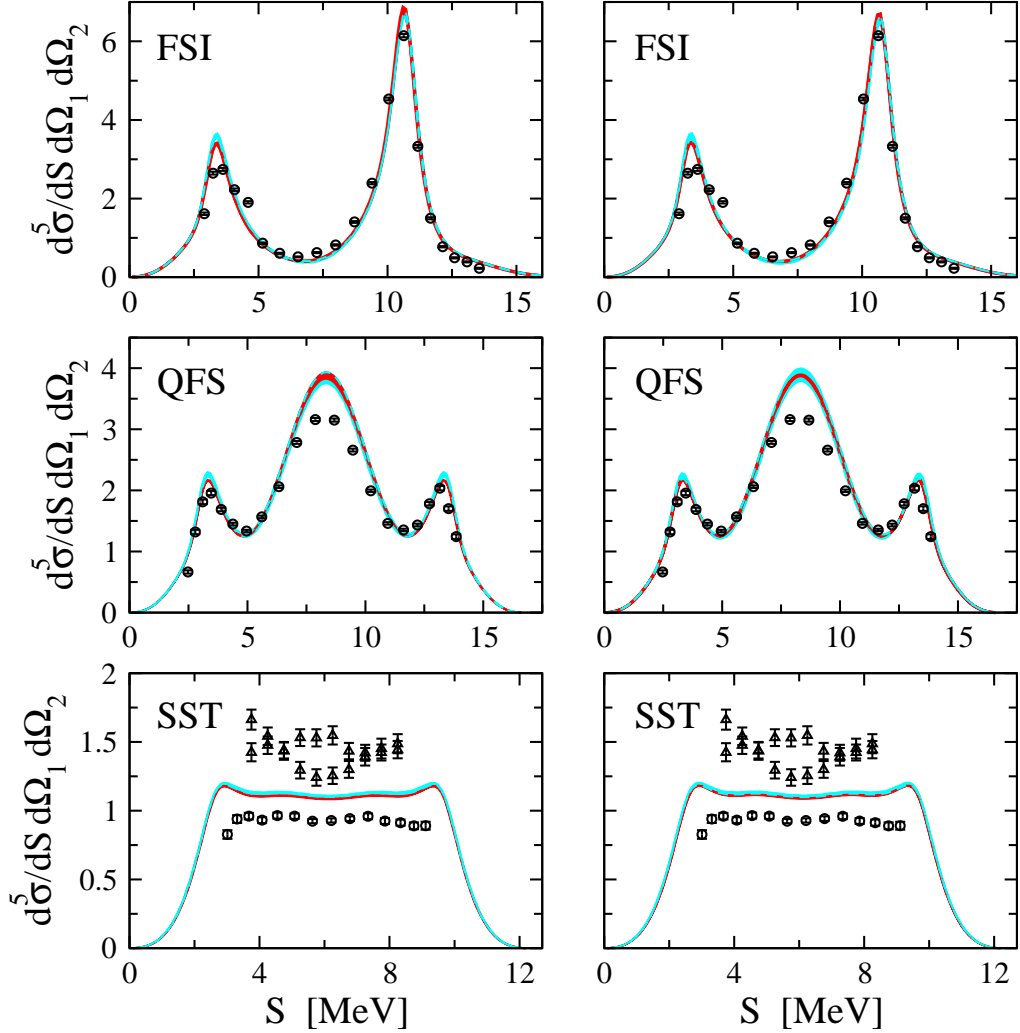,width=14.0cm}
\end{center}
\vspace{-0.7cm}
\caption[fig24]{\label{fig:br13mev} $nd$ break--up cross section in [mb MeV$^{-1}$ sr$^{-2}$] along the kinematical locus $S$ 
at 13 MeV in comparison to predictions at NLO (light shaded band) and N$^2$LO (dark shaded band) in the chiral EFT. In the left (right) panel, 
the results based on the SFR (DR) scheme are shown. In the upper row a final state interaction configuration is depicted, in the middle one 
a quasi--free scattering configuration (both in comparison to $pd$ data) and in the lower one a space star configuration 
(upper data $nd$, lower data $pd$). 
%*********************************************************************************
The precise kinematical description and references to data can be found in Ref.~\citep{Gloeckle:1995jg}. 
%$pd$ data are from \citep{Rauprich:1991aa}, $nd$ data from \citep{Setze:1996aa,Strate:1989aa}.
}
\vspace{0.2cm}
\end{figure*}

In Fig.~\ref{fig:br13mev} we present the results for a few
often investigated configurations, the space-star (SST), a final 
state interaction (FSI) peak configuration and a quasi--free scattering (QFS) configuration, respectively.  
For a general discussion on various break--up observables and configurations the 
reader is referred to \citep{Gloeckle:1995jg}. As demonstrated in Fig.~\ref{fig:br13mev}, 
the results at 13 MeV are very robust and essentially the same at NLO and N$^2$LO 
in both the SFR and DR approaches. 
The description of the configuration dominated by FSI peaks is rather accurate (for a more
elaborated procedure the angular openings of the detectors have
to be taken into account, see \citep{Gloeckle:1995jg}). 
The present theory for the break--up
configuration including a QFS geometry fails in the central maximum.
This might be due to missing Coulomb force effects. The third
configuration, the so called space--star, is one of the long standing 
puzzles of 3N scattering, see e.g.~\citep{Howell:1998aa}.
Similar to the phenomenological interactions, we even fail to describe 
the $nd$ data. The strong deviation between the  $nd$ and $pd$ data indicates 
the importance of the Coulomb effects for this particular configuration. 
It is still an open question  whether the $pd$ data can be reproduced by the theory
if Coulomb force is taken into account. 

\begin{figure*}[t]
%\vspace{0.3cm}
\centerline{
\psfig{file=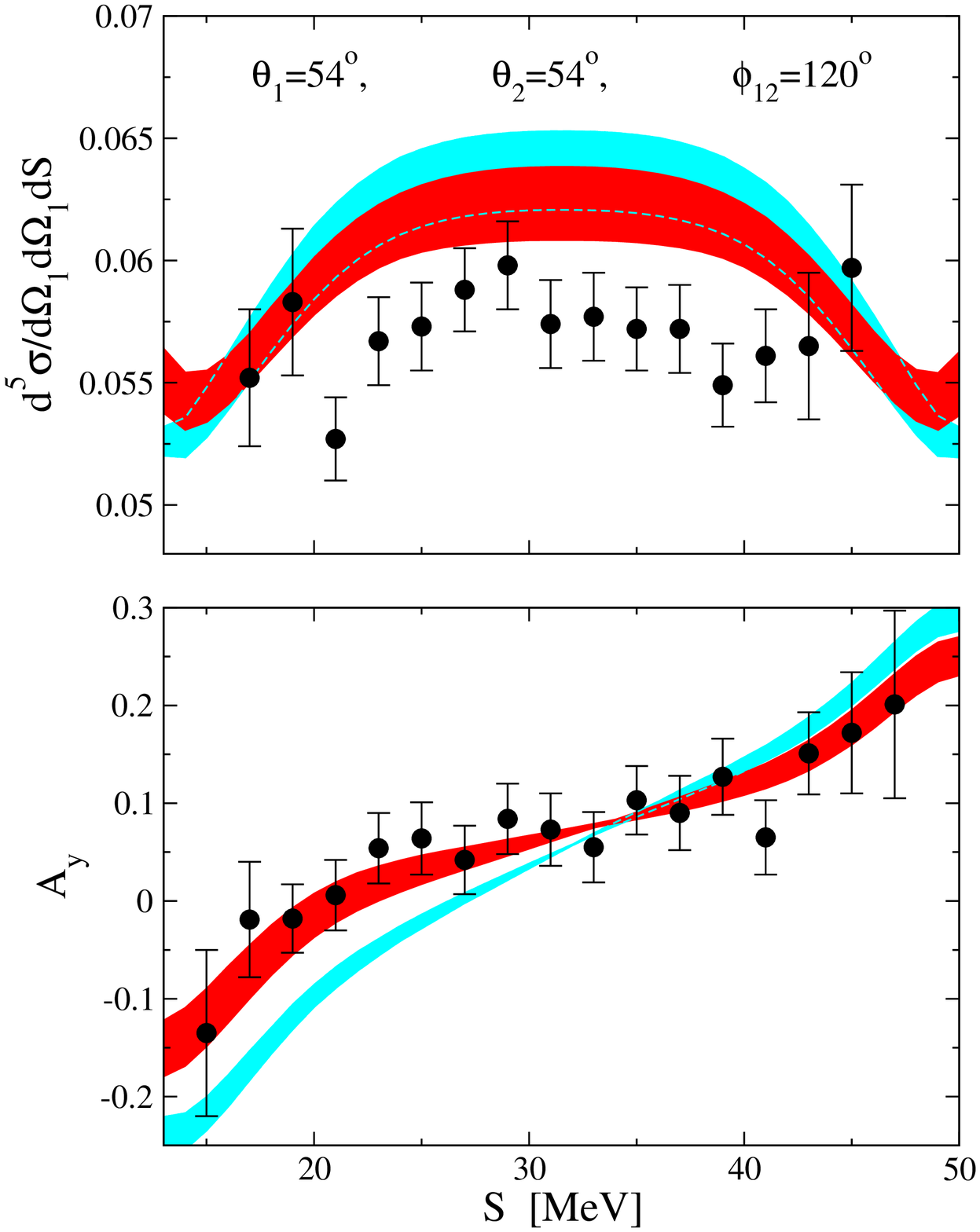,width=7.8cm}
\hskip 0.5 true cm
\psfig{file=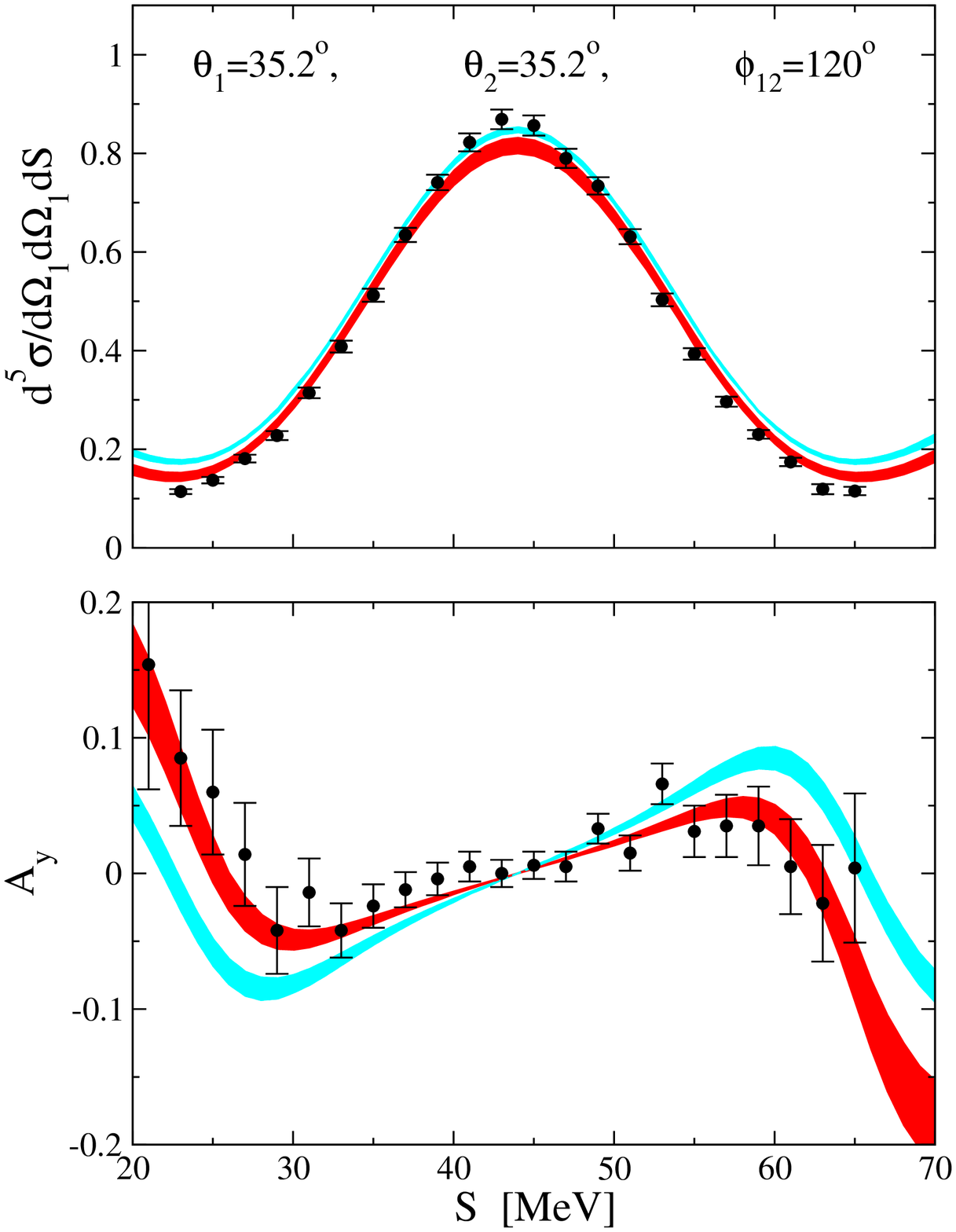,width=7.8cm}}
\vspace{-0.15cm}
\caption[fig24]{\label{fig:65p0_breakup} $nd$ break--up cross section in [mb MeV$^{-1}$ sr$^{-2}$] and nucleon analyzing power 
along the kinematical locus $S$ (in MeV) 
at 65 MeV in comparison to predictions at NLO (light shaded band) and N$^2$LO (dark shaded band) in chiral effective
field theory.  The cut-offs $\Lambda$ and $\tilde \Lambda$ are chosen as in Eq.~(\protect\ref{cutoffs_3N}). 
The left and right panels show the symmetric space star and symmetric forward star configurations, respectively. 
$pd$ data are from \protect\citep{Zejma:1997aa}.}
\vspace{0.2cm}
\end{figure*}

At 65~MeV we show in Fig.~\ref{fig:65p0_breakup} the results for the cross section 
and nucleon vector analyzing power for two selected configurations. These and other configurations
were studied in the context of phenomenological nuclear forces
\citep{Kuros-Zonierczuk:2002uv} as well in the chiral EFT based on the DR potential \citep{Epelbaum:2002vt}. 
The situation at 65 MeV seems, in general, to be promising. In most cases, a clear improvement 
in the description of the data is observed when going from NLO to N$^2$LO.
We stress, however, that the uncertainty due to the cut--off variation is rather large at this energy. 
More results for break--up observables at this energy will be given in \citep{Epelbaum:prep}, see also \citep{Kistryn:2005aa}.  

In addition, we would like to mention some further recent studies, where the results based on chiral EFT were shown. 
Differential cross section of the $^2$H$(p, \, pp)n$ reaction at the proton energy 16 MeV in three kinematical
configurations, the $np$ FSI, the co-planar star, and an intermediate-star geometry, was analyzed based on the 
conventional nuclear forces as well as on chiral EFT at NLO and (incomplete) N$^2$LO and N$^3$LO. 
Various proton--to--proton and proton--to--deuteron polarization transfer coefficients in 
$d (\vec p , \; \vec p \, )d$ and $d (\vec p , \; \vec d \, )p$ reactions at the proton energy 22.7 MeV are considered 
in \citep{Witala:2005aa}. For these observables, the restriction to the forces in NLO is shown 
to be insufficient. At N$^2$LO a satisfactory description of the data is observed, similar to the one obtained
with the (semi) phenomenological interactions. 
In addition, differential cross section of the deuteron--proton break--up reaction at the deuteron energy 130 MeV
was studied recently in Ref.~\citep{Kistryn:2005aa}. In this work, which is mainly focused on the 3NF effects,
the results based on modern NN  potentials combined with 3NF models and on chiral EFT at N$^2$LO 
are compared with the new high--precision data for 72 kinematically complete configurations. The description of 
the data at  N$^2$LO is found to be of a similar quality as the one based on realistic high--precision nuclear force models.

\subsubsection{Bound states}
\label{sec:BS}

The results for the triton and $\alpha$--particle BEs are summarized in Table \ref{tab:3H4He}. 
These observables are, in general, very sensitive 
to small changes of the interaction, as they come out as the difference 
of the large kinetic and potential energies. As a consequence, we 
found a rather large dependence of the BEs on the cut--off 
at NLO, see also \citep{Epelbaum:2000mx,Epelbaum:2002vt} for similar results obtained in the DR scheme.  
At N$^2$LO the $^3$H BE, which was used as input in order to fix the LECs in the corresponding 3NF
as explained in section \ref{nd_elastic},  is exactly reproduced for three cut--off combinations.
In the two other cases, the deviation from the experimental value is less than 1\%. 
Because of the strong correlation of the 3N and 4N BEs known as Tjon--line \citep{Tjon:1975aa}, one can expect 
a good description of the $\alpha$--particle BE. We, however, emphasize that 3NFs break this 
correlation, see e.g.~\citep{Nogga:2001cz}. Indeed, we observe a significant dependence of the $\alpha$--particle 
BE on the value of the LEC $c_D$ with $c_E$ being fixed to reproduce the triton BE. For a recent
work on the correlation between the 3N and 4N BEs in the context of pionless EFT the reader
is referred to \citep{Platter:2004zs}.
The predicted values for the $\alpha$--particle BE are within 5\% of the experimental value
for all cut--off combinations considered. 

\begin{table*}[t] 
%\vspace{0.6cm}
\begin{center}
\begin{tabular*}{1.00\textwidth}{@{\extracolsep{\fill}}||c||c|c||c|c||r||}
\hline \hline
{} & {} &  {} & {}  &  {} & {} \\[-1.5ex]
   & NLO, DR  \protect{\citep{Epelbaum:2002vt}} & N$^2$LO, DR  \protect{\citep{Epelbaum:2002vt}} & NLO, SFR   & N$^2$LO, SFR & Exp    \\[1ex]
\hline  \hline
{} & {} &  {} & {}  &  {} & {} \\[-1.5ex]
$E_{^3\rm H}$       &  $-7.53 \ldots -8.54$   & $-8.68^\star$          & $-7.71 \ldots -8.46$     & $-8.48\ldots -8.56^\star$     & $-8.482$  \\ [0.3ex]
$E_{^4\rm He}$      &  $-23.87 \ldots -29.57$ & $-29.51 \ldots -29.98$ & $-24.38 \ldots -28.77$   & $-27.77\ldots -29.61$         & $-28.30$  \\ [1.0ex]
\hline  \hline
  \end{tabular*}
%\vspace{0.3cm}
\caption{$^3$H and $^4$He BEs (in MeV) at NLO and N$^2$LO of the chiral expansion 
(for the cut--off range considered throughout) compared to experimental
BEs. The values of $E_{^3\rm H}$ used as input
is marked by the star. \label{tab:3H4He}}
\end{center}
\end{table*}

We would like to emphasize that the large numerical value of $c_D$ 
for the cut--off combination $\{600, \; 500 \}$ is reflected 
in the large expectation value of the 3NF in the triton and $\alpha$--particle.  
The situation is similar for the cut--off combination $\{600, \; 700 \}$, although the 
expectation values of the 3NF are smaller in magnitude. 
This might indicate that the cut--off $\Lambda = 600$ MeV
is already too close to its critical value, where 
spurious bound states appear and the naturalness assumption is violated. 
Further details on this topic will be given in \citep{Epelbaum:prep}.

\subsection{More nucleons}
\label{moreN}

Due to the fast increase of the computational abilities, one is now able 
to solve the Schr\"odinger equation for light $p$-shell nuclei including 3NFs \citep{Pieper:2001mp,Navratil:2003ef}.  

Using phenomenological forces, it has been established that the binding energies and 
spectra do depend on the structure of the 3NF, even if models 
describe the triton and $\alpha$--particle binding energy equally well \citep{Pieper:2001ap}.
Moreover, they also depend on the isospin 
$T=3/2$ components of the 3NF, to which the previously discussed 3N and 4N observables are not sensitive.  
This makes the application of chiral interactions to light nuclei even more interesting. 

Chiral interactions are low momentum interactions. As discussed throughout 
this review, the unknown short 
distance part of the force is absorbed in a tower of contact terms. 
It turns out that one obtains a decent description of 
the NN data using rather small cutoffs. 
The experience with traditional models indicated, however, a need for 
hard cores not only in the NN interaction, but also in the 3NF in order to 
prevent strong overbinding in systems beyond the $s$-shell. 
This seems to be in contradiction with the basic EFT philosophy.
In the following, we will discuss some results for systems with six and seven nucleons, 
which do not show any indication of unphysically increasing densities 
or binding energies.

\begin{figure}[t]
%\vspace{0.3cm}
\parbox{12.2cm}{
\psfig{file=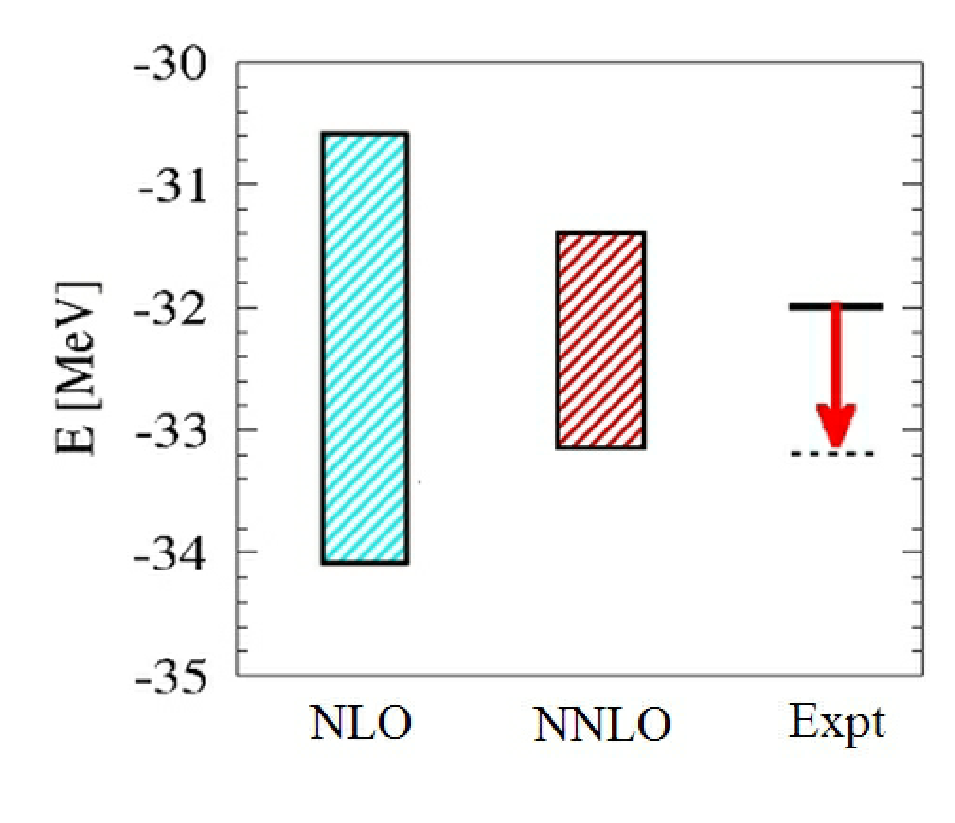,width=6.0cm}
\hskip 0.3 true cm
\psfig{file=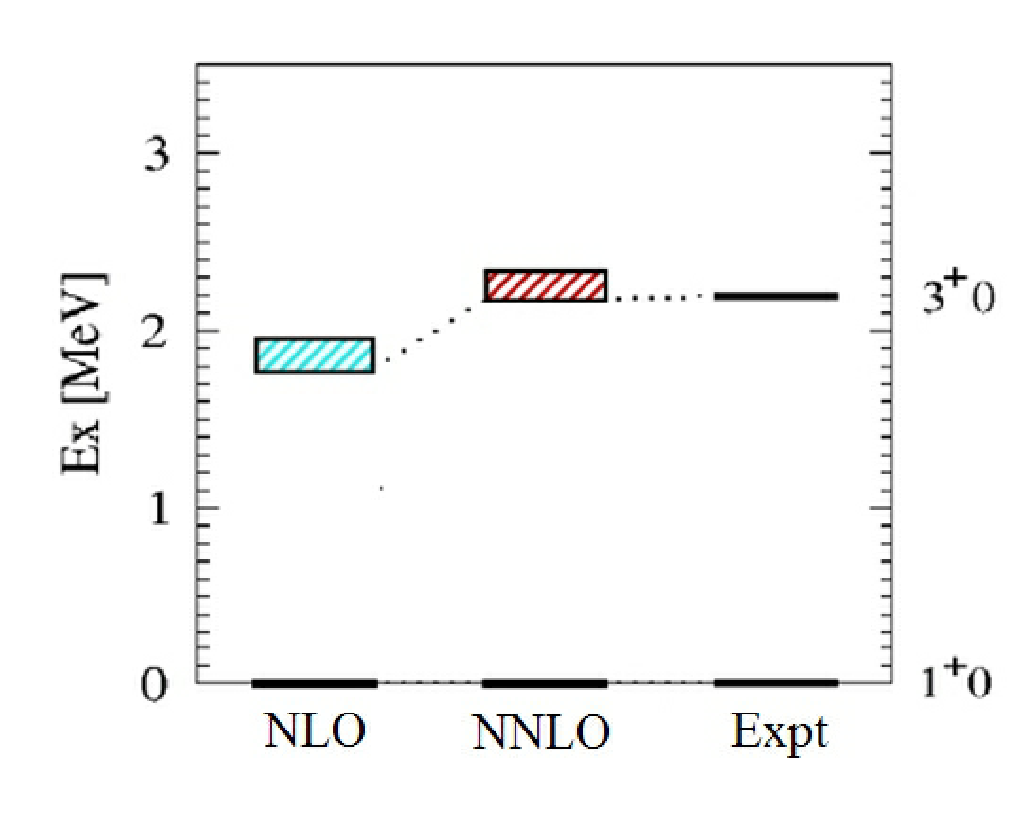,width=6.0cm}}
\parbox{5.2cm}{
\vspace{-0.25cm}
\includegraphics[width=5.2cm]{li7.idaho-n3lo.spectr.3nfdep.eps}}
\vspace{-0.15cm}
\caption{\label{fig:Li} 
$^6$Li binding (left panel) and excitation (midle panel) energies at NLO and N$^2$LO 
based upon the DR 2NF from \protect\citep{Epelbaum:2002ji} and the corresponding 3NF from 
\protect\citep{Epelbaum:2002vt} in comparison with experimental values.
The bands correspond to the $\Lambda$ variation between 500 and 600 MeV.
Right panel shows the results for $^7$Li based upon the N$^3$LO Idaho 2NF combined with 
the N$^2$LO 3NF (sets A and B) in comparison with experimental values. Figure courtesy of Andreas Nogga.}
\vspace{0.2cm}
\end{figure}

Let us begin with $^6$Li. In Fig.~\ref{fig:Li} we show the parameter--free results for the $^6$Li binding and 
excitation energies from Ref.~\citep{Nogga:2004aa} obtained within the no--core shell model framework
and based upon the DR NLO and N$^2$LO chiral forces from Ref.~\citep{Epelbaum:2002ji}.  
Going from NLO to N$^2$LO one observes a reduction of the 
cut--off dependence. At N$^2$LO, the uncertainty due to the cut--off variation is $1.8$ MeV or $5.7$\% ($170$ keV or $7.6$\%) 
of the binding (excitation) energy. Notice that for the BE, the estimated correction 
of the experimental value due to the missing CIB is also shown in Fig.~\ref{fig:Li}.
Calculation based on the N$^3$LO Idaho NN potential \citep{Entem:2003ft}
and N$^2$LO 3NF led to similar results \citep{Nogga:2004aa}. In this study, 
the LECs $c_D$ and $c_E$ were fixed from the $^3$H and $^4$He BEs 
which yielded two different sets: $c_D = -1.11$, $c_E = -0.66$ (3NF-A) 
and  $c_D = 8.14$, $c_E = -2.02$ (3NF-B).  

We now turn to $^7$Li. Here, the only calculations performed so far are
based on the N$^3$LO Idaho NN potential augmented with the N$^2$LO 3NF. 
The results for the binding energy are summarized and compared 
to other calculations and the experiment in Table~\ref{tab:li7bind} \citep{Nogga:2005aa}.
\begin{table}[b]
%\vspace{0.6cm}
\begin{tabular*}{1.00\textwidth}{@{\extracolsep{\fill}}||l||r|r|r||r|r||r||}
\hline \hline
{} & {} &  {} & {}  & {} & {}  & {} \\[-1.5ex]
                &  NN   & NN $+$ 3NF-A &  NN $+$ 3NF-B & AV18+Urbana IX &  AV18+IL2 & Exp  \\[1ex]
\hline
\hline
{} & {} &  {} & {}  & {} & {}  & {} \\[-1.5ex]
 $E_{gs}$ [MeV] &  34.6 &  38.0        & 36.7          &  37.5          & 38.9      & 39.2 \\   [0.3ex] 
 $r$ [fm]       &  2.40 &  2.19        & 2.31          &  2.33          & 2.25      & 2.27 \\ [1.0ex]
\hline \hline 
\end{tabular*}
%\vspace{0.3cm}
\caption{\label{tab:li7bind} Comparison of the ground state binding 
energy  results $E_{gs}$ and the point proton 
rms radius $r$ for $^7$Li 
for  chiral interactions and several phenomenological 
combinations to the experiment.  }
\end{table}
As one can expect, $^7$Li is underbound for the NN interaction only. Both  
sets of 3NFs  provide more binding. However, in both cases 
the final binding energy result is still short of the experiment by 1.2 and 2.5~MeV,
respectively. This slight underbinding is encouraging in 
view of the general expectation that strong repulsion at short distances is required 
to avoid a collapse of nuclei. Clearly, we do not observe 
any sign of overbinding, though neither the NN force nor the 3NF have a 
strong repulsive core. Both interactions are very soft, yet the binding energies 
are reasonable. So far, it was believed that only the addition of 
a repulsive core, like in the Urbana-IX and Illinois models 
\citep{Pudliner:1997ck,Pieper:2001ap}, can cure this overbinding problem. 
Here, it is demonstrated that the additional structures of chiral 3NFs also prevent 
overbinding.

Table~\ref{tab:li7bind} also shows the radii obtained in this way
in comparison with the experimental values and other 
calculations.  The results for the chiral  
interactions are comparable to the ones based on phenomenological models 
Urbana-IX and  Illinois. No indication is observed that soft, chiral 
interactions fail to saturate nuclear systems with a realistic binding energy 
and density. 

Finally, the predictions for the excitation energies are summarized in 
Fig.~\ref{fig:Li}. All combinations of the  
interactions, Idaho N$^3$LO NN force alone, with 3NF-A or 3NF-B, 
do predict the right ordering for these states. The splitting of the 
$3/2^-$ and $1/2^-$ states is small. The agreement 
to experiment for 3NF-A seems to be superior. Because the 
splitting itself is very small, this might be accidental. More 
significant deviations are observed for 
the $7/2^-$ and $5/2^-$ states. Both, the position of this 
multiplet and the splitting are strongly affected by the 3NFs
and the agreement with the experimental results 
is clearly best for 3NF-B, which is opposite to the case of BE. 
Clearly, further studies are needed in order to clarify the situation. 
As discussed before, the two N$^3$LO NN interactions available at present 
use different values for the LECs $c_i$ which determine the strength of the 2PE 3NF. 
The results presented here are based on the choice of Ref. \citep{Entem:2003ft} and it is conceivable 
that the detailed description of the binding energies and 
spectrum is affected by the choice of $c_i$'s.  For more details the reader is referred to \citep{Nogga:2005aa}. 

To summarize, the first results for the 6N and 7N systems based on chiral forces 
look promising. No hard repulsive core is found to be necessary to provide realistic 
binding energies and densities of $p$-shell nuclei.

%%%%%%%%%%%%%%%%%%%%%%%%%%%%%%%%%%%%%%%%%%%%%%%%%%%%%%%%%%%%%%%%%%%%%%%%%%%%%%%%%
\section{Miscellaneous omissions}
\def\theequation{\arabic{section}.\arabic{equation}}
\setcounter{equation}{0}
\label{sec7}

In this section we give a list of related topics which were not discussed
in this work. We stress that this list and the given references are not meant to be complete
and should merely provide the reader some guidance to further studies.  
\begin{itemize}
\item
\emph{Electroweak and pionic probes in the nuclear environment.}
The exchange vector and axial--vector NN currents were considered in chiral EFT 
by  Park, Min and Rho at one--loop level \citep{Park:1993jf,Park:1995pn} using the Feynman graph technique 
and dimensional regularization. Most of the practical applications including Compton scattering on 
the deuteron \citep{Beane:2004ra}, radiative $np$ capture (at threshold) \citep{Park:1995pn,Park:1999sz}, 
solar fusion \citep{Park:1998wq,Park:2001ig} and other solar neutrino--deuteron reactions \citep{Ando:2001es,Ando:2002pv}, 
{\it hep} \citep{Park:2002yp} and {\it hen} \citep{Park:2003sq}
processes, pion--deuteron scattering at threshold \citep{Weinberg:1992yk,Beane:1997yg,Beane:2002wk}, 
pion photo-- \citep{Beane:1997iv,Lensky:2005hb} and electroproduction \citep{Bernard:1999ff}, 
pion production in NN collisions, see \citep{Hanhart:2003pg}
and references therein, and others were performed in the so--called 
``hybrid'' approach.  In such a scheme, 
the interaction kernel is derived within chiral EFT while the wave functions for 
few--nucleon initial and final states are calculated using phenomenological potentials.
Certain reactions were already studied in a consistent way based on the few--nucleon wave functions 
obtained in chiral EFT, see e.~g.~Refs.~\citep{Beane:2002wk,Walzl:2001vb,Krebs:2002qr,Phillips:2005vv,Krebs:2004ir},
as well as in the pionless framework \citep{Meissner:2005bz}.

\item
\emph{Nuclear parity violation.}
A systematic study of nuclear parity--violation within the framework of effective field theory
was recently carried out by Zhu et al.~\citep{Zhu:2004vw}, who derived the parity--violating 1PE and 2PE 
potentials and discussed the ways of fixing the corresponding unknown LECs. 

\item
\emph{Nuclear forces and the large $N_c$ limit.}
The role of the large $N_c$--limit of QCD for the nucleon--nucleon interaction was originally addressed
in the seminal work by Witten \citep{Witten:1979kh} and then more recently by Kaplan et al.~\citep{Kaplan:1995yg,Kaplan:1996rk}.
The consistency of the large $N_C$--limit with the meson--exchange picture of the nuclear force 
is discussed in \citep{Banerjee:2001js,Cohen:2002qn,Belitsky:2002ni,Cohen:2002im}.

\item
\emph{Chiral effective field theory and nuclear matter.} Applications of chiral EFT to 
the nuclear many--body problem were considered by several groups, see 
\citep{Oller:2001sn,Meissner:2001gz,Girlanda:2003cq,Vretenar:2003bt,Furnstahl:2001hs} 
for some recent references. This is presently an active research field.

\item
\emph{Nucleons in (partially) quenched QCD.} 
The NN potential in quenched and partially quenched QCD is discussed in \citep{Beane:2002nu}. 
Application of a partially--quenched extension of an effective field theory to nucleon--nucleon scattering was  
performed in \citep{Beane:2002np}. These studies will help to relate lattice simulations in the 
2N sector to experimental data.

\item
\emph{Low--momentum NN interaction $V_{low \; k}$.}
An effective interaction $V_{low \; k}$ acting in the Hilbert space of low--momentum modes 
was first explicitly constructed in Refs.~\citep{Epelbaum:1998hg,Epelbaum:1998na} using 
the method of unitary transformation and based on a potential of the Malfliet--Tjon type.  
A different method was employed in Refs.~\citep{Bogner:2001gq,Bogner:2001yi} to construct  $V_{low \; k}$ for various  
high--precision NN potentials. The universality of the resulting low--momentum interactions
and the implications for the nuclear many--body problem are discussed in \citep{Bogner:2003wn}. Some 
recent applications of $V_{low \; k}$ in the few--nucleon sector are presented in 
\citep{Fujii:2004dd,Nogga:2004ab}. 
\end{itemize}

%%%%%%%%%%%%%%%%%%%%%%%%%%%%%%%%%%%%%%%%%%%%%%%%%%%%%%%%%%%%%%%%%%%%%%%%%%%%%%%%%
\section{Outlook}
\def\theequation{\arabic{section}.\arabic{equation}}
\setcounter{equation}{0}
\label{sec8}

In this work we outlined in detail the structure of the nuclear forces 
in the framework of chiral effective field theory and discussed recent applications 
in the few--nucleon sector. We also focused on some related topics including isospin 
violating effects and chiral extrapolations of the two--nucleon observables.  

In the future, these studies should be generalized in various ways. 
First, it is important to extend the N$^3$LO analyses \citep{Epelbaum:2004fk,Entem:2003ft} 
of the 2N system to few--nucleon systems which requires the derivation and numerical implementation 
of the corresponding N$^3$LO contributions to the 3N and 4N forces. It remains to be seen 
whether the presently observed difficulties in the theoretical description of certain 
low--energy 3N scattering observables like e.~g.~the nucleon vector analyzing power $A_y$ can 
be overcome at this order in the chiral expansion. 
Second, the electroweak nuclear current operators should be constructed to the same accuracy as 
the nuclear forces and applied to a rich variety of electroweak reactions in the nuclear environment
without using the ``hybrid'' approximation. Furthermore, given the recent theoretical progress in 
understanding  isospin and parity--violating corrections to the nuclear force within the chiral EFT 
framework, a systematic study of these effects in few--nucleon systems should be pursued.  
In addition to these fairly straightforward extentions,
a further effort is called for to achieve a better understanding of the nonperturbative 
renormalization in the context of the few--nucleon problem. This issue is being currently 
investigated by several groups. 
Increasing the range of applicability of the chiral EFT approach
forms another challenging direction of future research. This might require the inclusion of
$\Delta$--isobar as an explicit degree of freedom. One particular difficulty in such an approach
is related to our lack of knowledge of the values of the corresponding LECs, since 
applications of the EFT with the $\Delta$'s in the single--baryon 
sector were not yet performed to the same detail as in the $\Delta$--less 
theory. Finally, it would be highly desirable to incorporate in the EFT more constraints from 
QCD using for example the large--$N_c$ expansion or lattice gauge theory.  

%Summarizing, one may say that Weinberg's original proposal to apply the framework of chiral EFT 
%to the nuclear few--body problem captured in his seminal papers \cite{} has generated enormous 
%research activity over the past fifteen years. 

%%%%%%%%%%%%%%%%%%%%%%%%%%%%%%%%%%%%%%%%%%%%%%%%%%%%%%%%%%%%%%%%%%%%%%%%%%%%%%%%%
\section*{Acknowledgments}

It is a great pleasure to thank my collaborators, Walter Gl\"ockle, Hiroyuki Kamada, Ulf--G.~Mei{\3}ner, Andreas Nogga
and Henryk Wita{\l}a  for sharing their insights into the various 
topics discussed here. I am especially grateful to Andreas Nogga for his major contributions to 
sections \ref{formalism} and \ref{moreN}. I would like to thank Walter Gl\"ockle,
Ulf--G.~Mei{\3}ner and Andreas Nogga for carefully and critically reading the manuscript 
and their numerous suggestions for improvement. I am also grateful to  
Renato Higa, Bachir Moussallam and Mark Paris for helpful comments on the manuscript.
This work has been supported by the 
U.S.~Department of Energy Contract No.~DE-AC05-84ER40150 under which the 
Southeastern Universities Research Association (SURA) operates the Thomas Jefferson 
National Accelerator Facility and by the NATO grant No.~PST.CLG.978943. 

%%%%%%%%%%%%%%%%%%%%%%%%%%%%%%%%%%%%%%%%%%%%%%%%%%%%%%%
\appendix
\def\theequation{\Alph{section}.\arabic{equation}}
\setcounter{equation}{0}
\section{2PE potential from the  third order  $\pi N$ amplitude}
\label{sec:SF}

In this appendix, we give the expressions for the 2PE 2NF resulting from 
the diagrams which contain the third order  pion--nucleon amplitude, see Fig.~\ref{fig5},
and were obtained by Kaiser \citep{Kaiser:2001pc}. It appears to be convenient to  
use the (subtracted) spectral function representation:
\beqa
\label{SFRintegr}
V_{C,S} (q) &=& - \frac{2 q^6}{\pi} \int_{2M_\pi}^\infty \, d \mu 
\frac{\rho_{C,S} (\mu )}{\mu^5 ( \mu^2 + q^2 )}\,, \quad \quad
V_T (q) = \frac{2 q^4}{\pi} \int_{2M_\pi}^\infty \, d \mu 
\frac{\rho_{T} (\mu )}{\mu^3 ( \mu^2 + q^2 )}\,, \nn
W_{C,S} (q) &=& - \frac{2 q^6}{\pi} \int_{2M_\pi}^\infty \, d \mu 
\frac{\eta_{C,S} (\mu )}{\mu^5 ( \mu^2 + q^2 )}\,, \quad \quad
W_T (q) = \frac{2 q^4}{\pi} \int_{2M_\pi}^\infty \, d \mu 
\frac{\eta_{T} (\mu )}{\mu^3 ( \mu^2 + q^2 )}\,. 
\eeqa
For the spectral functions $\rho_i (\mu)$ ($\eta_i (\mu)$) one finds 
\citep{Kaiser:2001pc}:
\beqa
\label{TPE2loop}
\rho_C^{(4)} (\mu ) &=& - \frac{3 g_A^4 (\mu^2 - 2 M_\pi^2 )}{\pi \mu (4 F_\pi)^6}
\, \theta ( \tilde \Lambda - \mu ) \, 
\bigg\{ (M_\pi^2 - 2 \mu^2 ) \bigg[ 2 M_\pi + \frac{2 M_\pi^2 - \mu^2}{2 \mu} 
\ln \frac{\mu + 2 M_\pi}{\mu - 2 M_\pi } \bigg] \nn
&& \mbox{\hskip 5 true cm} + 4 g_A^2 M_\pi (2 M_\pi^2 - \mu^2 )
\bigg\}\,, \nn
\eta_S^{(4)} (\mu ) &=& \mu^2 \eta_T^{(4)} (\mu ) = - 
\frac{g_A^4 (\mu^2 - 4 M_\pi^2 )}{\pi (4 F_\pi)^6}
\, \theta ( \tilde \Lambda - \mu ) \, 
\left\{ \left(M_\pi^2 - \frac{\mu^2}{4} \right) \ln \frac{\mu +  2 M_\pi}{\mu - 2 M_\pi }
+ (1 + 2 g_A^2 ) \mu M_\pi \right\}\,, \nn
\rho_S^{(4)} (\mu ) &=& \mu^2 \rho_T^{(4)} (\mu ) = - 
\theta ( \tilde \Lambda - \mu ) \,  \left\{
\frac{g_A^2 r^3 \mu}{8 F_\pi^4 \pi} 
(\bar d_{14} - \bar d_{15} ) - 
\frac{2 g_A^6 \mu r^3}{(8 \pi F_\pi^2)^3} \left[ \frac{1}{9} - J_1  + J_2 \right] \right\}\,, \nn
 \eta_C^{(4)} (\mu ) &=&  \theta ( \tilde \Lambda - \mu ) \, \Bigg\{
\frac{r t^2}{24 F_\pi^4 \mu \pi} \left[ 2 (g_A^2 - 1) r^2 - 3 g_A^2 t^2 \right] (\bar d_1 + \bar d_2 ) \nn
&& {}+ \frac{r^3}{60 F_\pi^4 \mu \pi} \left[ 6 (g_A^2 - 1) r^2 - 5 g_A^2 t^2 \right] \bar d_3
- \frac{r M_\pi^2}{6 F_\pi^4 \mu \pi} \left[ 2 (g_A^2 - 1) r^2 - 3 g_A^2 t^2 \right] \bar d_5 \nn
&& {} - \frac{1}{92160 F_\pi^6 \mu^2 \pi^3} \Big[ - 320 (1 + 2 g_A^2 )^2 M_\pi^6 + 
240 (1 + 6 g_A^2 + 8 g_A^4 ) M_\pi^4 \mu^2 \nn
&& {}   \mbox{\hskip 3 true cm} - 60 g_A^2 (8 + 15 g_A^2 ) M_\pi^2  \mu^4
+ (-4 + 29 g_A^2 + 122 g_A^4 + 3 g_A^6 ) \mu^6 \Big] \ln \frac{2 r + \mu}{2 M_\pi} \nn
&& {} - \frac{r}{2700 \mu ( 8 \pi F_\pi^2 )^3} \Big[ -16 ( 171 + 2 g_A^2 ( 1 + g_A^2) 
(327 + 49 g_A^2)) M_\pi^4 + 4 (-73 + 1748 g_A^2 \nn
&& {}   \mbox{\hskip 3 true cm} + 2549 g_A^4 + 726 g_A^6 ) M_\pi^2 \mu^2 
- (- 64 + 389 g_A^2 + 1782 g_A^4 + 1093 g_A^6 ) \mu^4  \Big] \nn
&& {} + \frac{2 r}{3 \mu ( 8 \pi F_\pi^2 )^3} \Big[ 
g_A^6 t^4 J_1 - 2 g_A^4 (2 g_A^2 -1 ) r^2 t^2 J_2 \Big] \Bigg\}\,,
\eeqa
where we have introduced the abbreviations
\beq
r = \frac{1}{2} \sqrt{ \mu^2  - 4 M_\pi^2}\,, \quad \quad \quad
t= \sqrt{\mu^2 - 2 M_\pi^2}\,,
\eeq
and 
\beq
J_n = \int_0^1 \, dx \, x^{2n-2} \, \bigg\{ \frac{M_\pi^2}{r^2 x^2} - \bigg( 1 + \frac{M_\pi^2 }{r^2 x^2} \bigg)^{3/2}
\ln \frac{ r x + \sqrt{ M_\pi^2 + r^2 x^2}}{M_\pi} \bigg\}\,.
\eeq
%\beqa
%J_1 &=& \int_0^1 \, dx \,  \bigg\{ \frac{M_\pi^2}{r^2 x^2} - \bigg( 1 + \frac{M_\pi^2 }{r^2 x^2} \bigg)^{3/2}
%\ln \frac{ r x + \sqrt{ M_\pi^2 + r^2 x^2}}{M_\pi} \bigg\}\,,\nonumber \\
%J_2 &=& \int_0^1 \, dx \, x^2 \bigg\{ \frac{M_\pi^2}{r^2 x^2} - \bigg( 1 + \frac{M_\pi^2 }{r^2 x^2} \bigg)^{3/2}
%\ln \frac{ r x + \sqrt{ M_\pi^2 + r^2 x^2}}{M_\pi} \bigg\}\,.
%\eeqa
We use the scale--independent LECs $\bar d_{1}, \; \bar d_{2}, \; \bar d_{3}, \; \bar d_{5}, \; \bar d_{14}$ and 
$\bar d_{15}$ defined in \citep{Fettes:1998ud}. In the limit $\tilde \Lambda \to \infty$ 
some of the integrations in Eqs.~(\ref{SFRintegr}) 
with the spectral functions given in Eqs.~(\ref{TPE2loop}) have been carried out analytically in Ref.~\citep{Entem:2002sf}.

\setlength{\bibsep}{0.2em}
\bibliographystyle{h-elsevier3.bst}
\bibliography{/home/evgeny/refs_h-elsevier3}

\end{document}